# AI Governance InternationaL Evaluation Index

## AGILE Index 2025

***(Released in July 2025)***

- Center for Long-term Artificial Intelligence (CLAI)
- Beijing Institute of AI Safety and Governance (Beijing-AISI)
- Beijing Key Laboratory of Safe AI and Superalignment
- International Research Center for AI Ethics and Governance, Institute of Automation, Chinese Academy of Sciences

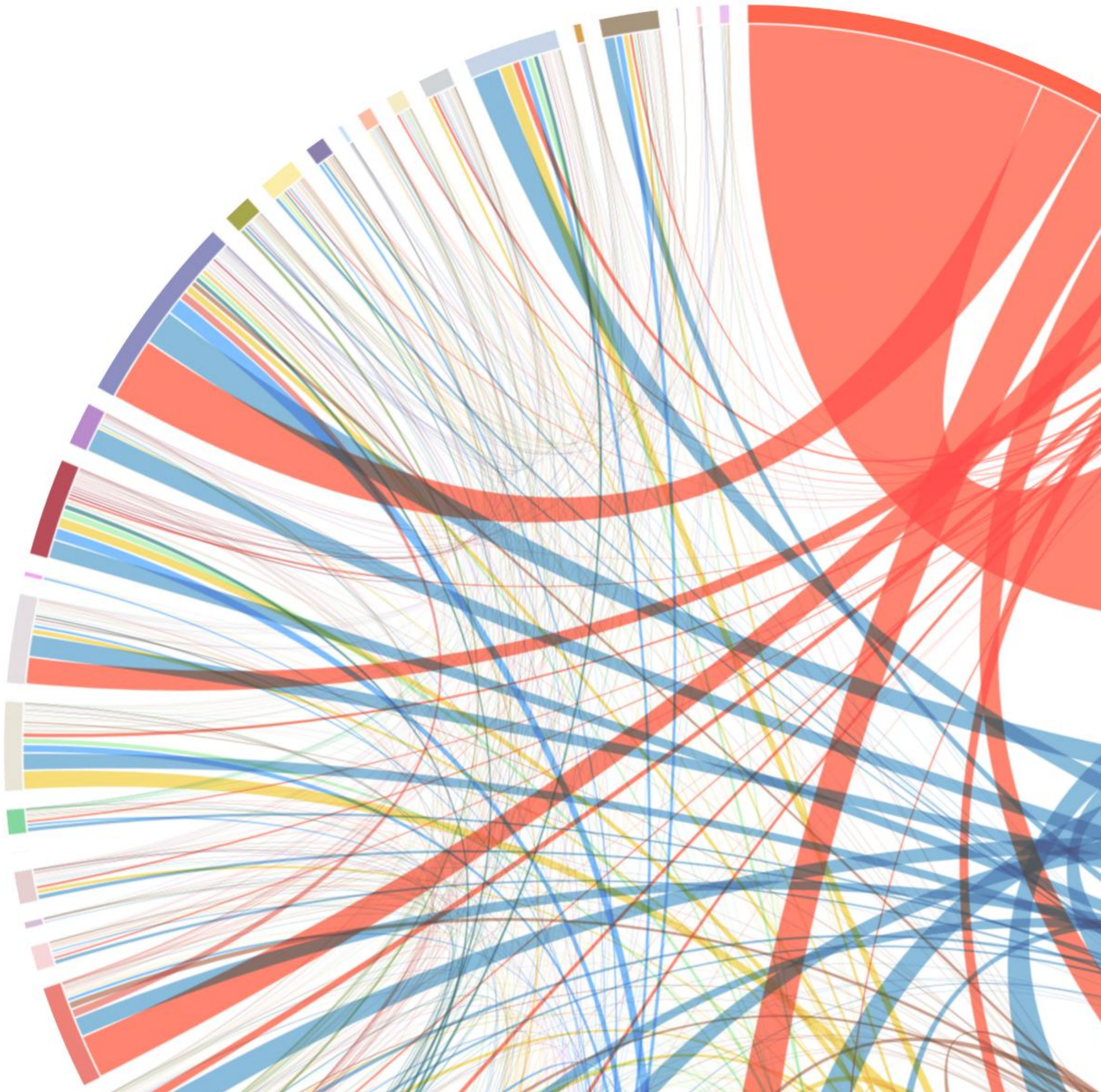

**AGILE Index Website:**

https://agile-index.ai/ (with supporting data & updates across time)

**Contributing Institutes:**

Center for Long-term Artificial Intelligence (CLAI)
https://long-term-ai.center/

Beijing Institute of AI Safety and Governance (Beijing-AISI)
https://beijing-aisi.ac.cn/

Beijing Key Laboratory of Safe AI and Superalignment
https://beijing.safe-ai-and-superalignment.cn/

International Research Center for AI Ethics and Governance,
Institute of Automation, Chinese Academy of Sciences
https://ai-ethics-and-governance.institute/

**Funding information:**

This research was financially supported by the National Science and Technology Major Project of China (Grant No. 2022ZD0116202) and by the Beijing Municipal Science & Technology Commission.

**Contact information:**

For more information or with any comments and feedback, please contact us at: contact@long-term-ai.cn .

**Report Version:**

**Copyright information:**

# Research Team

## Chief Scientist

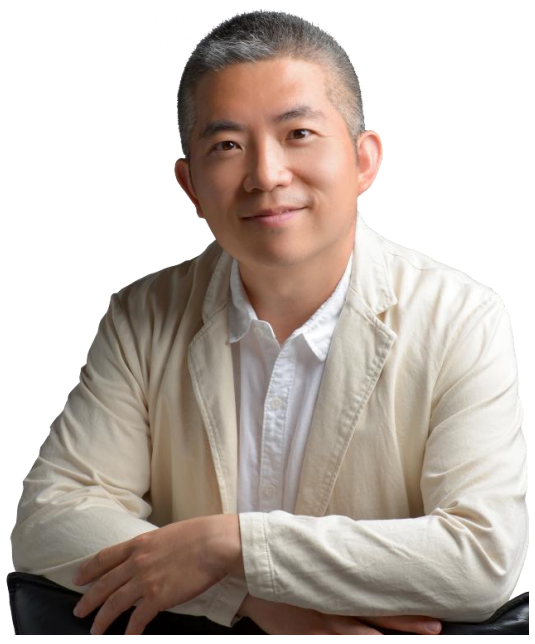

Yi Zeng

Dean and Professor

Yi Zeng is a Professor from Institute of Automation, Chinese Academy of Sciences, serving as the Director of the International Research Center for AI Ethics and Governance, the PI of the Brain-inspired Cognitive AI Lab, and the Director of Beijing Key Laboratory of Safe AI and Superalignment. He is the founding Dean of Beijing Institute of AI Safety and Governance (Beijing-AISI), the founding Director of the Center for Long-term AI, and leads the AI for SDGs Cooperation Network. He is also a member of the United Nations High-Level Advisory Body on AI, an expert in the UNESCO Ad Hoc Expert Group on the Ethics of AI, an expert in the WHO Expert Group on Ethics and Governance of AI for Health, a senior fellow at the UN Institute for Disarmament Research, a board member of the National Governance Committee of Next Generation AI and a board member of the AI Ethics Sub-Committee of the National Science and Technology Ethics Committee in China, and a Co-Chair of World Internet Conference Specialized Committee on AI. His areas of interest include Brain and Mind-inspired AI, AI Ethics, Safety and Governance, and AI for Global Sustainable Development.

## Researchers

| | |
|---|---|
| Enmeng Lu | Beijing Institute of AI Safety and Governance |
| Xiaoyang Guo | Beijing Key Laboratory of Safe AI and Superalignment |
| Cunqing Huangfu | Institute of Automation, Chinese Academy of Sciences |
| Jiawei Xie | Beijing Key Laboratory of Safe AI and Superalignment |
| Yu Chen | Beijing Key Laboratory of Safe AI and Superalignment |
| Zhengqi Wang | Beijing Institute of AI Safety and Governance |
| Dongqi Liang | Institute of Automation, Chinese Academy of Sciences |
| Gongce Cao | Beijing Institute of AI Safety and Governance |
| Jin Wang | Beijing Institute of AI Safety and Governance |
| Zizhe Ruan | Institute of Automation, Chinese Academy of Sciences |
| Xin Guan | Center for Long-term AI |
| Ammar Younas | Center for Long-term AI |

# What's New in AGILE Index 2025

## Dual Expansion of Indicators & Country Coverage:

Compared to the 2024 edition of AGILE Index, the country coverage in AGILE Index 2025 has expanded from 14 to 40, and the number of indicators has increased from 39 to 43. This broader scope provides a stronger foundation for horizontal comparisons and trend analysis.

## Data Update & Data Source Expansion:

All data from the previous edition have been thoroughly reviewed and updated where newer information was available through March, 2025. More than 20 additional data sources have also been integrated to enhance coverage and reliability.

## Inclusion of Frontier AI Technologies and Research Themes:

New data on cutting-edge AI technologies, such as generative AI, have been incorporated. Meanwhile, the literature analysis now covers a broader spectrum of AI safety and governance themes with refined keywords, providing more comprehensive overview of research advancements.

## Year-on-Year Comparisons:

For indicators with consistent data definitions, year-over-year comparisons have been enhanced to better capture trends and highlight key changes over time.

## Enhanced Missing Data Imputation Strategy:

In response to the expanded country coverage and increasing requirement of data quality and scoring, the strategy for handling missing data has been further refined. Imputation now combines historical performance trends and correlations between indicators to ensure greater rationality and consistency. Please refer to the Methodology section in the appendix for details.

# Executive Summary

Artificial Intelligence (AI) represents both an unprecedented technological revolution and a systemic governance challenge confronting the global community. The year 2024 witnessed accelerated global AI governance advancements, marked by strengthened multilateral frameworks and proliferating national regulatory initiatives. This acceleration underscores an unprecedented need to systematically track governance progress—an imperative that drove the launch of the AI Governance InternationaL Evaluation Index (AGILE Index) project since 2023.

The inaugural AGILE Index, released in February 2024 after assessing 14 countries, established an operational and comparable baseline framework. Building on pilot insights, AGILE Index 2025 incorporates systematic refinements to better balance scientific rigor with practical adaptability. The updated methodology expands data diversity while enhancing metric validity and cross-national comparability. Reflecting both research advancements and practical policy evolution, AGILE Index 2025 evaluates 40 countries across income levels, regions, and technological development stages, with 4 Pillars, 17 Dimensions and 43 Indicators. In compiling the index, the team integrates multi-source evidence including policy documents, governance practices, research outputs, and risk incidents to construct a unified comparison framework. This approach maps global disparities while enabling countries to identify governance strengths, gaps, and systemic constraints.

Through continuous refinement and iterations, we hope the AGILE Index will fundamentally advance transparency and measurability in global AI governance, delivering data-driven assessments that depict national AI governance capacity, assist governments in recognizing their maturation stages and critical governance issues, and ultimately provide actionable insights for enhancing AI governance systems nationally and globally.

# Findings of AGILE Index 2025

## Overall Observations

1. 40 evaluated countries can be categorized into three tiers based on their AGILE Index 2025 Scores, with AI Development Level (Pillar 1) and AI Governance Instruments (Pillar 3) showing relatively wide disparities across countries. (Page 14)
2. Among the 14 countries evaluated last year, the ranking changes reveal a pattern of dynamic shifts among top-tier countries and relative stability among lower-ranked ones, with notable changes such as the position swap between the US and China. (Page 16)
3. There remains a generally positive correlation between the AGILE Index total score and GDP per capita across 40 evaluated countries. Countries left behind need to raise the level of preparedness for AI Governance to be more resilient. (Page 16)
4. The performance of the 40 countries across the four pillars of AGILE Index reveals four distinct types of AI development and governance status: All-round Leaders, Governance Overachievers, Governance Shortfallers, and Foundation Seekers. (Page 17)
5. High-income countries group show a clear advantage over upper-middle and lower-middle income countries group in both AI Development Level (Pillar 1) and AI Governance Instruments (Pillar 3), while the latter tend to perform slightly better with lower AI risk exposure and higher social acceptance of AI. (Page 19)

## Pillar 1: AI Development Level

1. While the United States and China each exhibit distinct strengths in AI development, other countries collectively enrich the global AI ecosystem through specialized advantages. (Page 22)
2. Since 2010, the number of Generative AI (GenAI) patents globally has shown exponential growth, with total patents increased by about 30 times and applications by about 25 times. The growth rate accelerated significantly after 2018. (Page 24)

## Pillar 2: AI Governance Environment

1. Documented AI incidents rose sharply in 2024, up roughly 100% from 2023. The US saw an 1.8-fold increase, while other countries grew faster at 2.1-fold. (Page 25)

2. Risk incidents pertaining to AI Safe & Security, Human Rights, and Data Governance are more numerous, accounting for half of all reported AI incidents. (Page 26)
3. Although high-income countries tend to score higher in overall governance — indicating stronger innate readiness and basis for AI governance, practical experience still reveals significant room for proactive efforts to advance AI governance across diverse national contexts. (Page 27)

## Pillar 3: AI Governance Instruments

1. All 40 countries evaluated have published national-level AI strategies, with varying structural approaches to strategy formulation. (Page 29)
2. Since 2024, the legislation on AI has shown a clear accelerating trend. Some countries have enacted national general regulations on AI, while others have formulated special regulations for vertical fields of AI. (Page 30)
3. All 40 countries have participated in various forms of global AI governance mechanisms, with France, Japan, South Korea, and Singapore showing the highest levels of involvement. (Page 31)

## Pillar 4: AI Governance Effectiveness

1. The public's level of awareness, trust, and optimism regarding AI applications and services shows an overall negative correlation with GDP per capita. (Page 33)
2. A growing share of women are participating in AI research, evidenced by a continuous decline in the male-to-female ratio among AI researchers, which reached 2.2:1 in 2025. (Page 35)
3. Economic development, as reflected by GDP per capita, is generally positively correlated with the digital inclusion of social vulnerable groups to a certain extent, but higher economic development does not necessarily result in high digital inclusiveness. (Page 36)
4. China and the United States represent the highest shares in the openness of impactful AI models and datasets, together account for more than 70% of the total observed among all 40 countries. (Page 37)
5. North America and Western Europe lead the open-source AI ecosystem, with the United States far ahead at 30.43% of the 40 countries' total. Meanwhile, China and India also show strong engagement. (Page 38)

6. The AI research community is focusing more on governance-related issues, as shown by a rising share of AI governance publications in total AI publications—up to about 14% in 2024. Among the 40 countries, China and the US together account for about 54% of contributions, over half. (Page 39)
7. International co-authorship in AI governance research is most prevalent among China, the US, Canada, Germany, the UK, and Australia, accounting for about 70% of all co-authored publications within the 40 countries. (Page 41)
8. The United States and China lead in AI research for the Sustainable Development Goals (SDGs), together accounting for approximately 60% of all SDG-related AI publications among the 40 countries, with other countries also made significant contributions to various goals. (Page 42)
9. SDG3 (Good Health and Well-Being), SDG11 (Sustainable Cities and Communities), and SDG9 (Industry, Innovation and Infrastructure) have received widespread research attention. Meanwhile, High-Income Countries have allocated higher AI research attention to social-level SDGs, while Upper-Middle&Lower-Middle Income Countries have relatively higher AI research attention to economically and environmentally related SDGs. (Page 43)

# Table of Contents

# I.
# AGILE Index

## 1.1. Theoretical Framework

The AI Governance InternationaL Evaluation Index (AGILE Index) is anchored in the design principle that **"the level of governance should match the level of development."** It posits that as AI progresses through distinct innovation phases, governance mechanisms should dynamically adapt—establishing a symbiotic relationship between technological advancement and societal stewardship. This framework aims to balance the acceleration of AI-driven value creation with the mitigation of emerging risks, ensuring that governance maturity keeps pace with technological complexity.

Structurally, AGILE Index 2025 is organized into a multi-layered architecture:

- 4 Pillars, including AI Development Level, AI Governance Environment, AI Governance Instruments, and AI Governance Effectiveness.
- 17 Dimensions, with each dimension explores specific governance focus areas, bridging high-level principles with evaluation objectives.
- 43 Indicators, which provide granular measurement points and translating evaluation objectives into quantifiable evidences and actionable metrics.

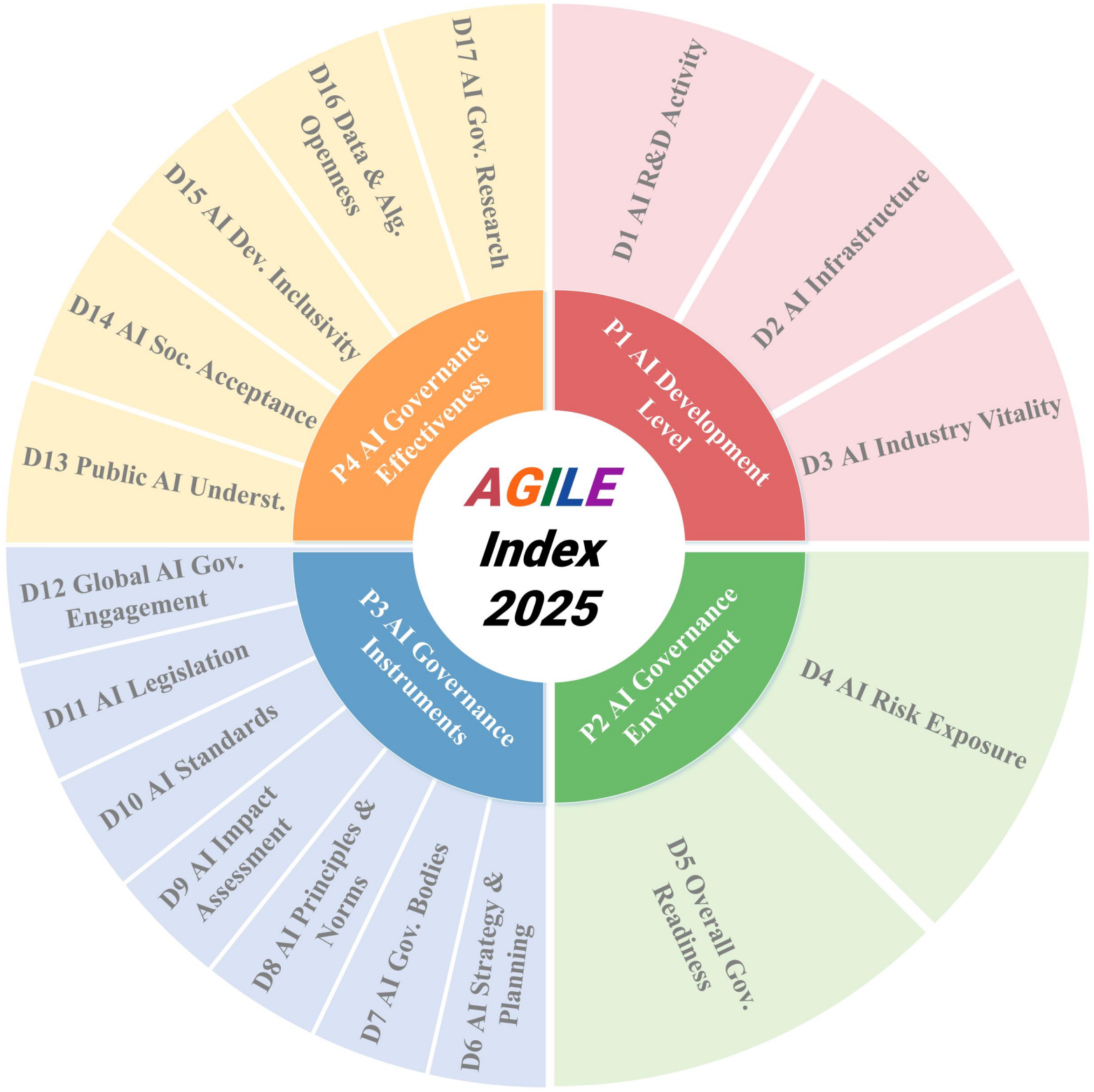

***Figure 1 AGILE Index 2025: Pillars and Dimensions***

# II.
# Overview

## 2.1. Score Composition

*Table 1 AGILE Index 2025: Total Score, Pillar Score, and Dimension Score*

| Country | AGILE INDEX | Rank | D1 | D2 | D3 | P1 | D4 | D5 | P2 | D6 | D7 | D8 | D9 | D10 | D11 | D12 | P3 | D13 | D14 | D15 | D16 | D17 | P4 |
|---|---|---|---|---|---|---|---|---|---|---|---|---|---|---|---|---|---|---|---|---|---|---|---|
| China | 70.1 | 1 | 75.5 | 50.6 | 55.4 | 60.5 | 48.5 | 68.7 | 58.6 | 100.0 | 100.0 | 100.0 | 0.0 | 100.0 | 83.3 | 87.5 | 81.5 | 100.0 | 99.5 | 45.4 | 91.0 | 63.8 | 79.9 |
| US | 69.9 | 2 | 82.6 | 100.0 | 100.0 | 94.2 | 0.0 | 80.0 | 40.0 | 100.0 | 100.0 | 100.0 | 100.0 | 100.0 | 50.0 | 87.5 | 91.1 | 48.2 | 21.7 | 27.0 | 100.0 | 74.7 | 54.3 |
| Germany | 68.8 | 3 | 52.5 | 78.8 | 43.6 | 58.3 | 73.0 | 86.7 | 79.8 | 100.0 | 100.0 | 100.0 | 0.0 | 100.0 | 100.0 | 93.5 | 84.8 | 27.8 | 42.9 | 56.0 | 77.3 | 57.3 | 52.2 |
| S.Korea | 67.8 | 4 | 59.8 | 46.0 | 43.3 | 49.7 | 58.5 | 86.0 | 72.3 | 100.0 | 100.0 | 100.0 | 100.0 | 100.0 | 50.0 | 99.5 | 92.8 | 96.4 | 51.0 | 50.6 | 45.8 | 38.8 | 56.5 |
| UK | 67.6 | 5 | 63.8 | 64.3 | 74.2 | 67.4 | 12.5 | 86.5 | 49.5 | 100.0 | 100.0 | 100.0 | 100.0 | 100.0 | 83.3 | 93.5 | 96.7 | 55.4 | 32.6 | 60.6 | 75.0 | 60.2 | 56.8 |
| Singapore | 65.7 | 6 | 49.7 | 60.6 | 74.7 | 61.6 | 43.5 | 83.0 | 63.2 | 75.0 | 100.0 | 100.0 | 100.0 | 0.0 | 33.3 | 87.5 | 70.8 | 71.1 | 74.0 | 82.6 | 63.8 | 44.7 | 67.2 |
| France | 65.6 | 7 | 38.9 | 66.2 | 50.2 | 51.8 | 57.5 | 84.6 | 71.1 | 100.0 | 100.0 | 100.0 | 100.0 | 100.0 | 100.0 | 99.5 | 99.9 | 16.7 | 27.7 | 51.3 | 67.3 | 34.8 | 39.6 |
| Canada | 63.5 | 8 | 49.8 | 55.0 | 56.7 | 53.8 | 37.0 | 83.4 | 60.2 | 75.0 | 100.0 | 100.0 | 100.0 | 100.0 | 50.0 | 93.5 | 88.4 | 57.3 | 27.3 | 59.5 | 61.5 | 51.7 | 51.5 |
| Japan | 61.4 | 9 | 47.2 | 72.0 | 24.0 | 47.7 | 72.0 | 84.9 | 78.5 | 100.0 | 100.0 | 100.0 | 100.0 | 0.0 | 16.7 | 99.5 | 73.7 | 77.8 | 31.1 | 43.8 | 42.5 | 33.2 | 45.7 |
| Finland | 60.3 | 10 | 40.7 | 89.6 | 65.2 | 65.2 | 82.5 | 87.8 | 85.2 | 100.0 | 100.0 | 0.0 | 0.0 | 0.0 | 66.7 | 57.5 | 46.3 | 95.7 | 13.5 | 57.2 | 18.3 | 38.2 | 44.6 |
| Netherlands | 58.9 | 11 | 36.3 | 73.1 | 37.8 | 49.1 | 68.0 | 85.1 | 76.5 | 100.0 | 0.0 | 0.0 | 100.0 | 0.0 | 100.0 | 87.5 | 55.4 | 58.7 | 34.5 | 69.7 | 68.8 | 42.5 | 54.8 |
| UAE | 56.4 | 12 | 25.2 | 47.6 | 67.1 | 46.6 | 70.5 | 77.4 | 73.9 | 50.0 | 100.0 | 100.0 | 100.0 | 0.0 | 33.3 | 23.5 | 58.1 | 41.4 | 55.8 | 83.8 | 24.3 | 28.5 | 46.8 |
| Sweden | 56.0 | 13 | 41.2 | 63.6 | 92.4 | 65.7 | 76.0 | 85.9 | 81.0 | 75.0 | 100.0 | 0.0 | 0.0 | 0.0 | 66.7 | 57.5 | 42.7 | 33.9 | 24.5 | 55.1 | 26.5 | 32.0 | 34.4 |
| Australia | 55.0 | 14 | 48.1 | 65.7 | 20.5 | 44.8 | 37.5 | 85.1 | 61.3 | 75.0 | 100.0 | 100.0 | 0.0 | 100.0 | 50.0 | 87.5 | 73.2 | 39.7 | 22.9 | 29.9 | 47.3 | 63.5 | 40.6 |
| Denmark | 54.8 | 15 | 37.4 | 68.9 | 53.1 | 53.1 | 72.5 | 88.8 | 80.6 | 100.0 | 0.0 | 0.0 | 0.0 | 0.0 | 66.7 | 57.5 | 32.0 | 87.2 | 53.5 | 74.8 | 24.0 | 28.2 | 53.5 |
| Switzerland | 54.8 | 16 | 55.3 | 98.3 | 35.4 | 63.0 | 58.5 | 84.7 | 71.6 | 75.0 | 0.0 | 100.0 | 0.0 | 0.0 | 0.0 | 79.5 | 36.4 | 46.0 | 35.8 | 75.4 | 52.8 | 31.5 | 48.3 |
| Norway | 54.7 | 17 | 34.4 | 55.7 | 45.0 | 45.0 | 79.0 | 86.4 | 82.7 | 100.0 | 0.0 | 0.0 | 0.0 | 0.0 | 100.0 | 57.5 | 36.8 | 83.8 | 10.3 | 88.3 | 48.3 | 41.0 | 54.3 |
| S.Arabia | 54.5 | 18 | 18.4 | 35.8 | 27.1 | 27.1 | 94.0 | 74.3 | 84.1 | 75.0 | 100.0 | 100.0 | 100.0 | 100.0 | 33.3 | 79.5 | 84.0 | 11.0 | 22.9 | 22.3 | 11.3 | 47.2 | 22.9 |
| Italy | 53.8 | 19 | 27.1 | 79.0 | 24.7 | 43.6 | 66.0 | 79.2 | 72.6 | 100.0 | 100.0 | 0.0 | 0.0 | 0.0 | 100.0 | 87.5 | 55.4 | 26.9 | 57.2 | 46.8 | 33.3 | 54.7 | 43.7 |
| Spain | 51.7 | 20 | 30.6 | 58.6 | 20.7 | 36.6 | 66.0 | 83.7 | 74.9 | 100.0 | 100.0 | 0.0 | 0.0 | 0.0 | 66.7 | 79.5 | 49.5 | 37.6 | 44.6 | 60.4 | 38.8 | 47.3 | 45.7 |
| Malaysia | 51.2 | 21 | 35.3 | 57.2 | 46.3 | 46.3 | 66.0 | 72.2 | 69.1 | 100.0 | 100.0 | 100.0 | 0.0 | 0.0 | 0.0 | 50.0 | 50.0 | 17.4 | 72.0 | 66.8 | 9.5 | 32.2 | 39.6 |
| Indonesia | 51.0 | 22 | 8.6 | 39.3 | 24.0 | 24.0 | 89.5 | 67.5 | 78.5 | 100.0 | 100.0 | 100.0 | 0.0 | 0.0 | 0.0 | 29.5 | 47.1 | 53.2 | 96.1 | 68.8 | 17.8 | 36.8 | 54.5 |
| Türkiye | 46.7 | 23 | 6.9 | 20.8 | 13.8 | 13.8 | 94.0 | 70.9 | 82.5 | 100.0 | 100.0 | 100.0 | 0.0 | 0.0 | 16.7 | 79.5 | 56.6 | 52.0 | 67.6 | 16.5 | 7.5 | 25.3 | 33.8 |
| Ireland | 46.6 | 24 | 34.8 | 57.4 | 46.1 | 46.1 | 47.0 | 83.1 | 65.1 | 100.0 | 0.0 | 0.0 | 0.0 | 0.0 | 66.7 | 73.5 | 34.3 | 49.4 | 32.5 | 67.5 | 16.5 | 39.3 | 41.0 |
| Thailand | 45.8 | 25 | 25.6 | 19.0 | 22.3 | 22.3 | 80.5 | 72.6 | 76.6 | 100.0 | 0.0 | 100.0 | 0.0 | 0.0 | 16.7 | 7.5 | 32.0 | 31.5 | 96.6 | 81.8 | 12.5 | 40.0 | 52.5 |
| Portugal | 45.4 | 26 | 29.9 | 27.0 | 28.4 | 28.4 | 84.5 | 79.6 | 82.1 | 100.0 | 0.0 | 0.0 | 0.0 | 0.0 | 66.7 | 57.5 | 32.0 | 70.1 | 21.8 | 49.0 | 12.5 | 41.8 | 39.0 |
| Israel | 45.0 | 27 | 42.5 | 48.7 | 69.5 | 53.5 | 28.5 | 75.6 | 52.0 | 100.0 | 0.0 | 100.0 | 0.0 | 0.0 | 33.3 | 57.5 | 41.5 | 47.1 | 33.0 | 31.0 | 27.3 | 26.5 | 33.0 |
| Belgium | 45.0 | 28 | 23.8 | 28.6 | 26.2 | 26.2 | 61.0 | 80.2 | 70.6 | 100.0 | 100.0 | 0.0 | 0.0 | 0.0 | 66.7 | 57.5 | 46.3 | 53.1 | 21.9 | 54.5 | 36.0 | 18.8 | 36.9 |
| Chile | 44.9 | 29 | 14.3 | 52.3 | 33.3 | 33.3 | 93.5 | 76.4 | 84.9 | 100.0 | 0.0 | 0.0 | 0.0 | 0.0 | 50.0 | 29.5 | 25.6 | 25.7 | 56.5 | 61.3 | 11.8 | 24.5 | 35.9 |
| Russia | 44.1 | 30 | 18.2 | 32.0 | 25.1 | 25.1 | 41.0 | 65.0 | 53.0 | 100.0 | 100.0 | 100.0 | 0.0 | 100.0 | 66.7 | 50.0 | 73.8 | 47.5 | 24.6 | 2.4 | 31.3 | 17.3 | 24.6 |
| Mexico | 43.8 | 31 | 12.2 | 10.2 | 11.2 | 11.2 | 89.5 | 66.1 | 77.8 | 75.0 | 100.0 | 100.0 | 0.0 | 0.0 | 16.7 | 29.5 | 45.9 | 72.1 | 84.3 | 13.8 | 10.8 | 20.0 | 40.2 |
| Poland | 43.1 | 32 | 28.2 | 42.4 | 35.3 | 35.3 | 73.0 | 78.1 | 75.6 | 75.0 | 0.0 | 0.0 | 0.0 | 0.0 | 66.7 | 57.5 | 28.5 | 60.0 | 37.3 | 24.6 | 25.8 | 18.2 | 33.2 |
| N.Zealand | 41.8 | 33 | 29.2 | 69.2 | 49.2 | 49.2 | 35.0 | 86.7 | 60.8 | 75.0 | 0.0 | 100.0 | 0.0 | 0.0 | 0.0 | 73.5 | 35.5 | 23.9 | 37.1 | 5.7 | 16.8 | 26.0 | 21.9 |
| Hungary | 41.3 | 34 | 29.2 | 15.8 | 22.5 | 22.5 | 88.0 | 75.0 | 81.5 | 100.0 | 0.0 | 0.0 | 0.0 | 0.0 | 66.7 | 57.5 | 32.0 | 35.1 | 53.8 | 36.3 | 11.0 | 9.8 | 29.2 |
| India | 41.0 | 35 | 27.9 | 31.3 | 40.8 | 33.3 | 23.5 | 65.0 | 44.2 | 50.0 | 0.0 | 0.0 | 100.0 | 0.0 | 0.0 | 79.5 | 32.8 | 45.9 | 68.0 | 40.8 | 63.5 | 49.7 | 53.6 |
| Brazil | 40.8 | 36 | 16.9 | 30.9 | 16.2 | 21.3 | 83.5 | 72.5 | 78.0 | 100.0 | 0.0 | 0.0 | 0.0 | 0.0 | 50.0 | 79.5 | 32.8 | 33.1 | 52.7 | 6.1 | 31.0 | 31.7 | 30.9 |
| Peru | 40.8 | 37 | 22.4 | 13.8 | 18.1 | 18.1 | 94.0 | 68.2 | 81.1 | 75.0 | 0.0 | 0.0 | 0.0 | 0.0 | 16.7 | 7.5 | 14.2 | 53.3 | 88.0 | 76.9 | 3.5 | 26.7 | 49.7 |
| Argentina | 37.2 | 38 | 7.3 | 11.1 | 9.2 | 9.2 | 84.5 | 67.9 | 76.2 | 100.0 | 0.0 | 100.0 | 0.0 | 0.0 | 0.0 | 7.5 | 29.6 | 7.1 | 62.4 | 82.8 | 10.0 | 6.7 | 33.8 |
| Colombia | 37.1 | 39 | 13.0 | 17.8 | 15.4 | 15.4 | 90.5 | 68.7 | 79.6 | 100.0 | 0.0 | 0.0 | 0.0 | 0.0 | 50.0 | 7.5 | 22.5 | 43.4 | 69.0 | 0.0 | 20.3 | 21.8 | 30.9 |
| S.Africa | 33.9 | 40 | 5.4 | 28.9 | 17.1 | 17.1 | 58.0 | 63.6 | 60.8 | 75.0 | 0.0 | 0.0 | 0.0 | 0.0 | 0.0 | 0.0 | 10.7 | 40.6 | 75.4 | 73.3 | 5.0 | 39.8 | 46.8 |

## 2.2. Overall Observations

### Key observation 1:

40 evaluated countries can be categorized into three tiers based on their AGILE Index 2025 Scores, with AI Development Level (Pillar 1) and AI Governance Instruments (Pillar 3) showing relatively wide disparities across countries.

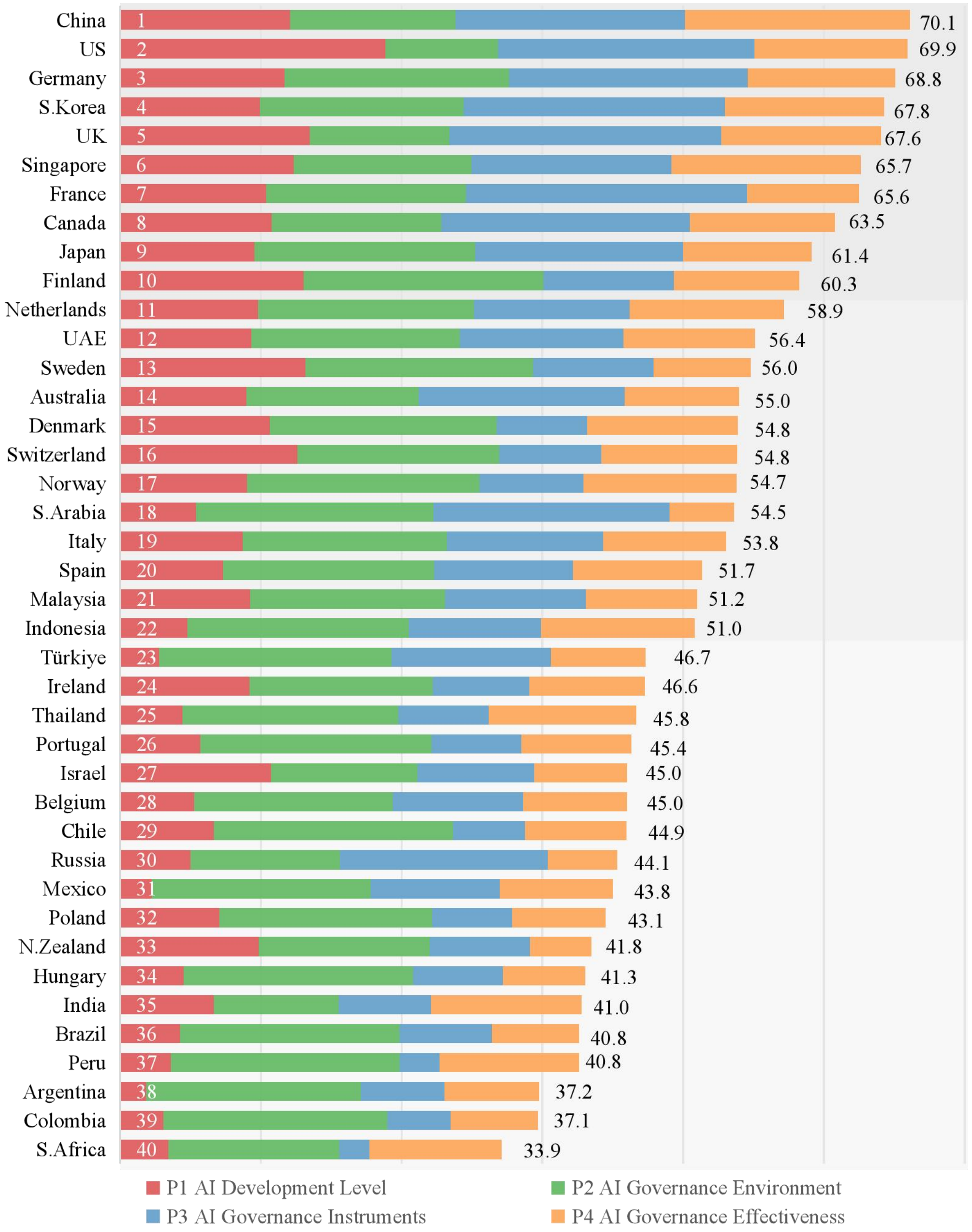

*Figure 2 AGILE Index 2025: Overall Scores and Ranking*

According to the score distribution of the AGILE Index, the 40 evaluated countries can be generally grouped into three tiers. Tier 1 countries are with scores above 60, such as China, the United States, and Germany; Tier 2 countries scores between 50 and 60, such as Netherlands, United Arab Emirates, and Sweden; Tier 3 countries are with scores below 50.

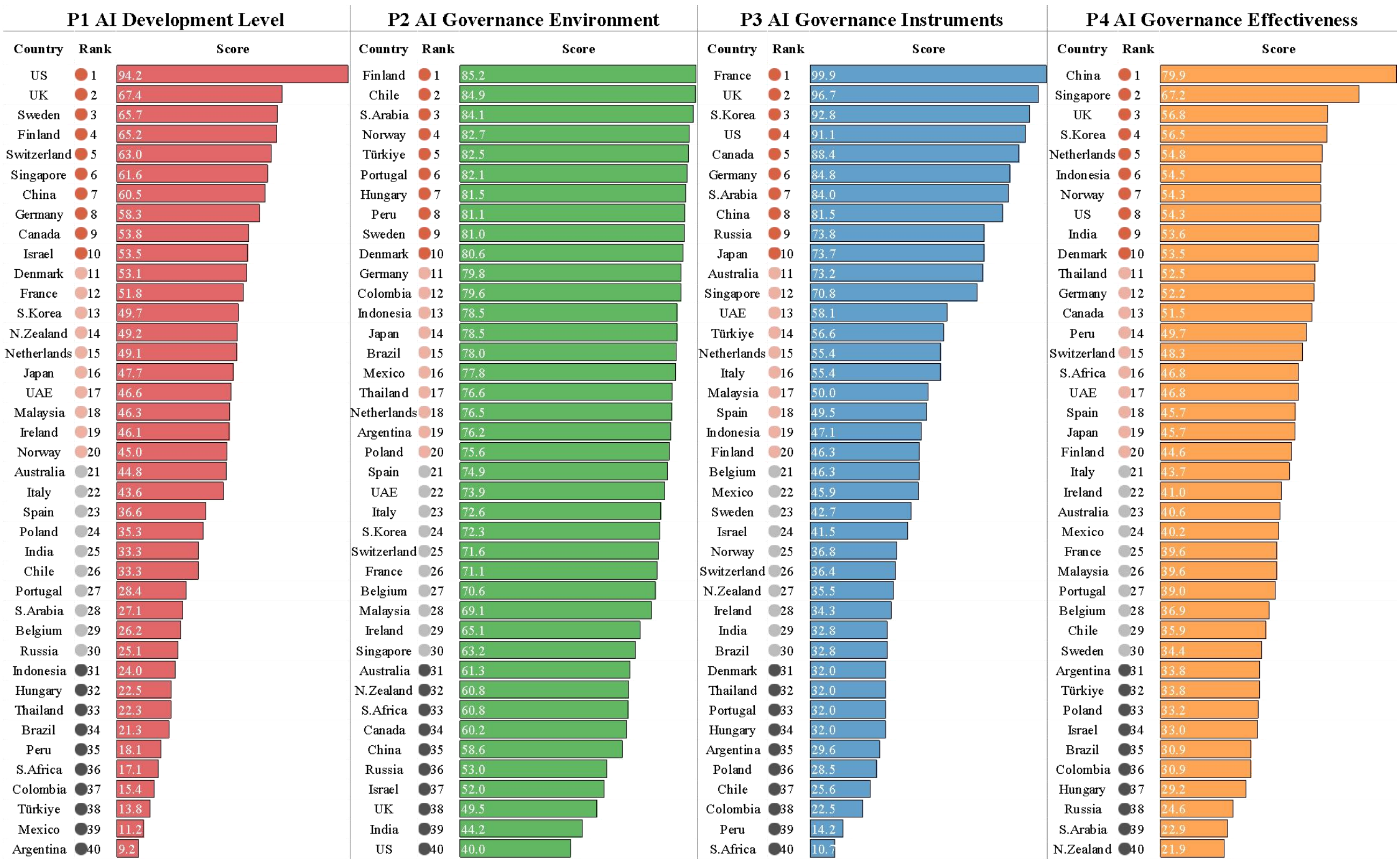

***Figure 3 AGILE Index 2025: Pillar Scores and Rankings***

Among the four pillars, Pillar 1 (AI Development Level) and Pillar 3 (AI Governance Instruments) exhibit significantly higher score dispersion compared to Pillar 2 (Governance Environment) and Pillar 4 (Governance Performance), revealing a more pronounced tiered differentiation. The score gaps between leading and lagging countries are notably wide (P1 max: 94.2, min: 9.2; P3 max: 99.9, min: 10.7). In contrast, the distributions of Pillar 2 and Pillar 4 scores are more concentrated, with smaller differences across tiers than those observed in Pillar 1 and Pillar 3.

## Key observation 2:

Among the 14 countries evaluated last year, the ranking changes reveal a pattern of dynamic shifts among top-tier countries and relative stability among lower-ranked ones, with notable changes such as the position swap between the US and China.

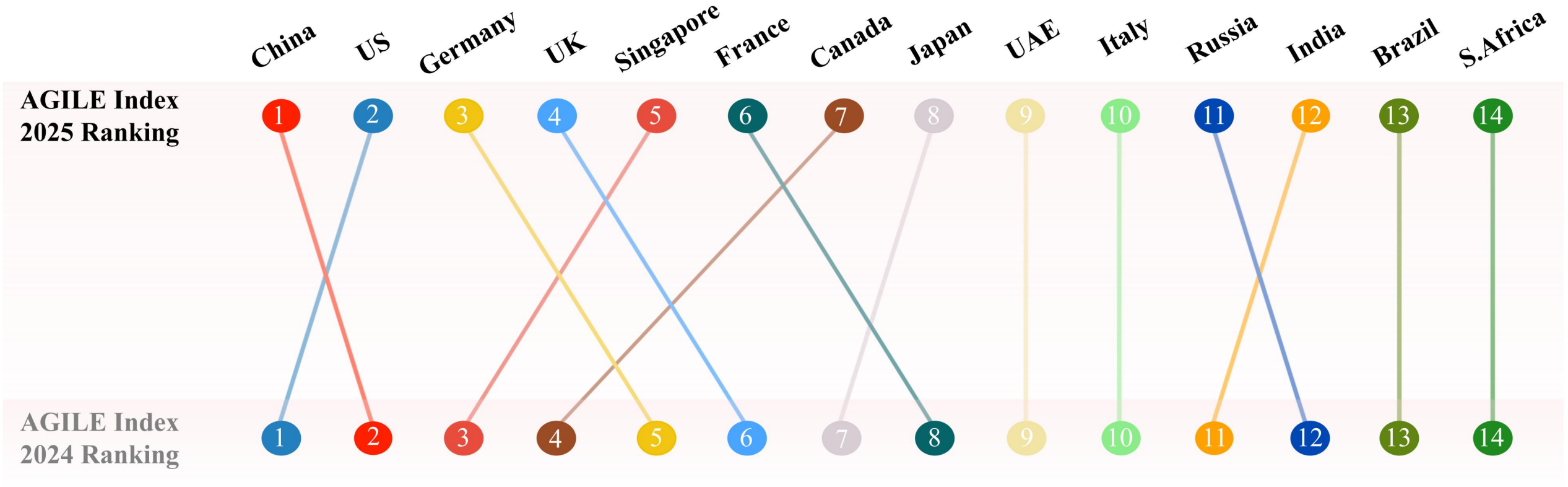

*Figure 4 Relative Ranking Changes within 14 Countries evaluated in AGILE Index 2024*

The 14 countries evaluated last year have experienced changes in their rankings in 2025. Top-ranked countries experienced visible shifts, while the bottom-ranked countries remained relatively stable. The United States dropped to second place primarily due to the impact of its more lenient policy trend on AI legislation. The issuance of US Executive Order 14179 - *"Removing Barriers to American Leadership in Artificial Intelligence"* - stipulates the revocation of Executive Order 14110 - *"Safe, Secure, and Trustworthy Development and Use of Artificial Intelligence"*. However, the new executive order does not possess the same comprehensiveness as Executive Order 14110, resulting to a decrease in its relevant indicator score and further affected its overall ranking. Meanwhile, China has moved into first place with more consistent AI governance policies. Some countries like Singapore and Canada saw ranking fluctuations, while those of Brazil, and South Africa remained stable. Overall, the pattern shows notable changes at the top and relative stability at the bottom.

## Key observation 3:

There remains a generally positive correlation between the AGILE Index total score and GDP per capita across 40 evaluated countries. Countries left behind need to raise the level of preparedness for AI Governance to be more resilient.

This year's AGILE Index continues to demonstrate a positive correlation between countries' per capita GDP and their overall index scores. Generally, countries with higher per capita GDP tend to score higher on the AGILE Index, suggesting that economic development and stronger AI governance may be associated. However, as the scope of evaluation expands, the data also reveals important exceptions: some lower-GDP countries achieve relatively high governance scores, while some wealthier countries do not. This implies that while development may serve as a necessary foundation for governance, it is not the sole determinant of AI governance capacity.

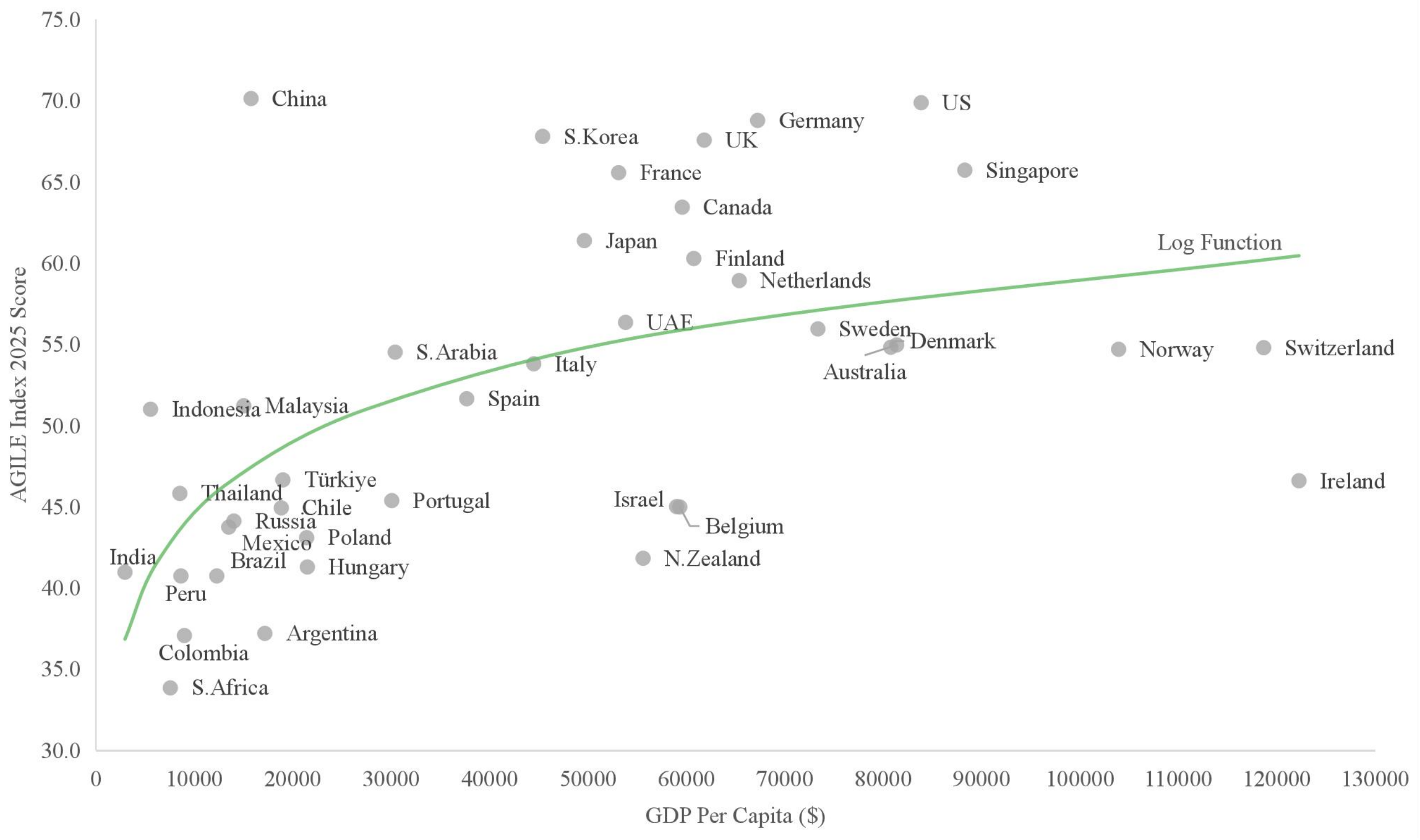

***Figure 5 Positive Relationship between 40 Countries' AGILE Index Score and their GDP Per Capita***

## Key observation 4:

The performance of the 40 countries across the four pillars of AGILE Index reveals four distinct types of AI development and governance status: All-round Leaders, Governance Overachievers, Governance Shortfallers, and Foundation Seekers.

Further analysis of AGILE Index scores of the 40 countries across the four pillars shows a clear stratification, forming four distinct governance types.

- **All-round Leaders:** Countries including United States, Singapore, China, and United Kingdom —score highly and evenly across all four pillars, demonstrating well-rounded strengths.
- **Governance Overachievers:** Compared to the first type, countries like France, South Korea,

and Canada maintain high levels of governance environment and instruments, but relatively lag in AI R&D and overall effectiveness.

- **Governance Shortfallers:** Countries like Ireland, Israel, and New Zealand—feature a mismatch between AI development and governance instruments. These countries experience rapid progress in AI research and application but lag in areas such as governance tool development and policy framework building.
- **Foundation Seekers:** Represented by countries like India and South Africa, this type scores relatively lower across all four pillars, with particular weaknesses in Pillar 1 and Pillar 3. This reflects a dual shortfall in AI development level and AI governance instruments, highlighting the urgent need to strengthen foundational development.

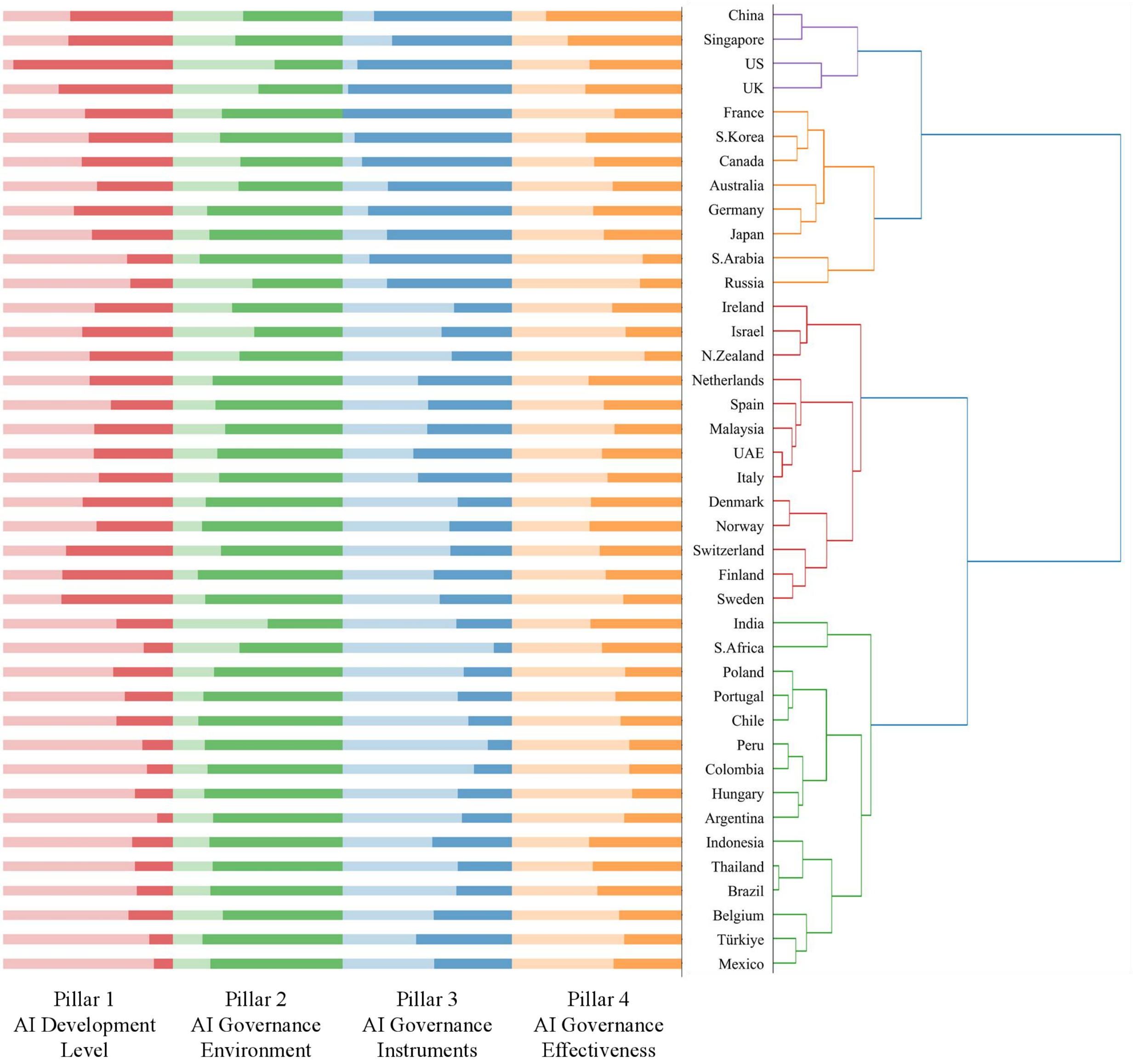

***Figure 6 Four Types of 40 Countries' AGILE Index Pillar Score Distribution***

## Key observation 5:

High-income countries group show a clear advantage over upper-middle and lower-middle income countries group in both AI Development Level (Pillar 1) and AI Governance Instruments (Pillar 3), while the latter tend to perform slightly better with lower AI risk exposure and higher social acceptance of AI.

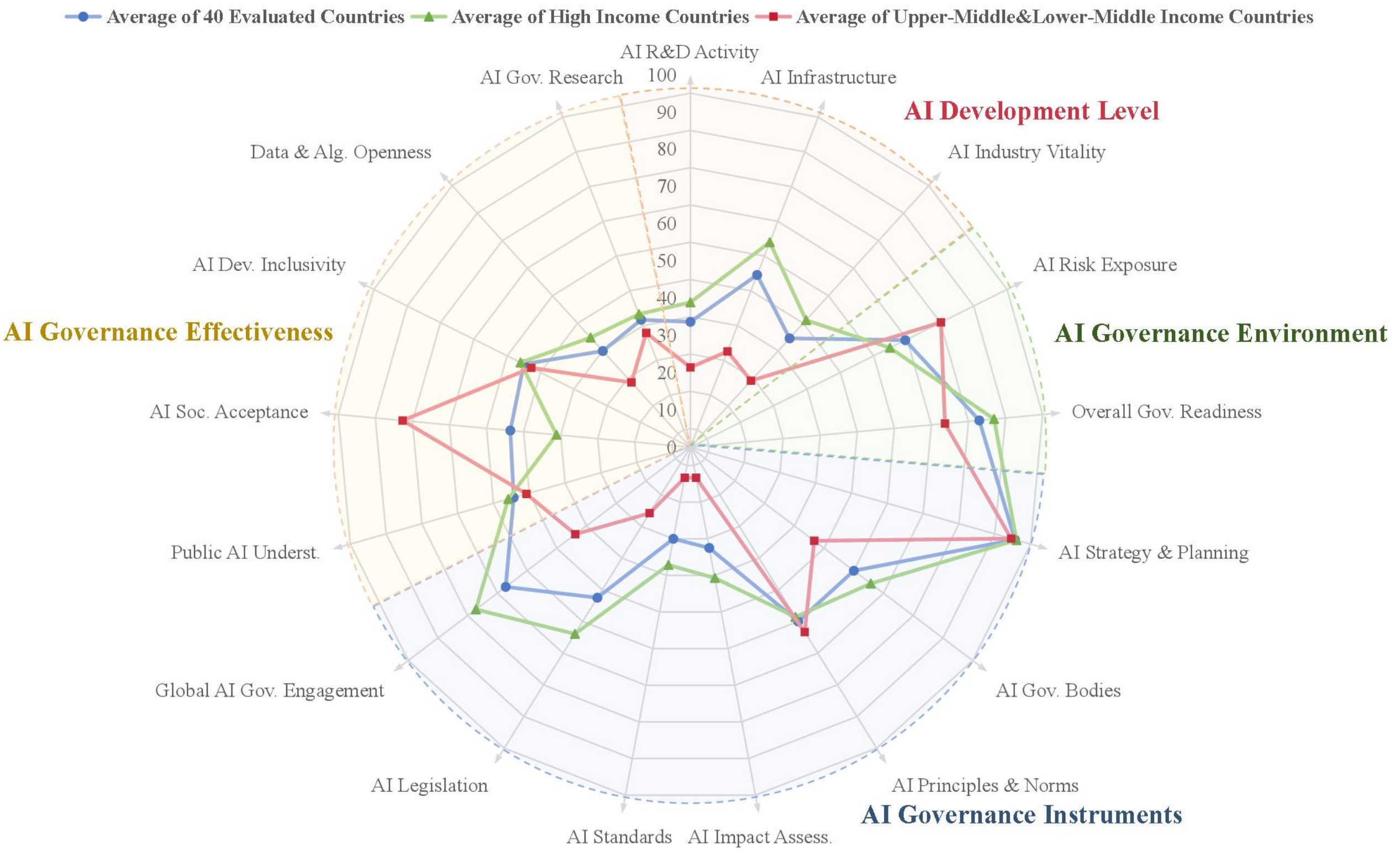

***Figure 7 AGILE Index Score Average by Dimension for High-income Countries, Upper-Middle & Lower-Middle Income Countries, and Both Groups***

The 40 countries are divided into two groups based on per capita income levels[1]: high-income countries versus upper-middle&lower-middle income countries. The average scores for high-income countries group across the four pillars — AI Development Level (Pillar 1), AI Governance Environment (Pillar 2), AI Governance Instruments (Pillar 3), and AI Governance Effectiveness (Pillar 4)—are 48, 71, 58 and 43, respectively. For upper-middle&lower-middle income countries group, the corresponding averages are 24, 72, 38, and 46.

[1] Note: The categorization of countries into high-income and upper-middle income, and lower-middle income country follows the World Bank's income classifications. https://blogs.worldbank.org/zh/opendata/world-bank-country-classifications-by-income-level-for-2024-2025

High-income countries outperform their counterparts significantly in Pillar 1 and Pillar 3. However, upper-middle&lower-middle income countries have a slight advantage in Pillar 2 and Pillar 4. This reflects the former's strengths in technical advancement and governance infrastructure, while the latter benefits from lower AI risk exposure and higher societal acceptance of AI—factors that contribute to their relative strengths in governance environment and effectiveness.

# III. Analysis & Observation

## 3.1. Pillar 1: AI Development Level

### Pillar 1 Overview:

AGILE index assesses AI development levels across three dimensions: AI Research & Development Activity, AI Infrastructure, and AI Industry Vitality.

*Table 2 AI Development Level of 40 Countries in Total*

| 420K+ | 200K+ | 16K+ | 375 | 11.5K+ | 8K+ |
|---|---|---|---|---|---|
| **Researchers** authored AI-related publication(s) | AI-Related **Publications** | Granted AI **Patents** | Large-scale AI **Systems** Developed | Aggregate RMAX Performance [EFlop/s] of Top500 **Supercomputers** | **Data Centers** |
| *From April 2024 to March 2025* | | | *As of March 2025* | | |

In summary, among the 40 countries assessed, a total of over 420,000 researchers authored more than 200,000 AI-related publications, and over 16,000 patents were granted—all within the past year. By March 2025, these countries have developed 375 large-scale AI systems, amassed a combined super-computing power of over 11.5k EFlop/s, and have over 8,000 hosted data centers to facilitate various AI R&D activities.

### Observation 1.1:

While the United States and China each exhibit distinct strengths in AI development, other countries collectively enrich the global AI ecosystem through specialized advantages.

China and the United States each exhibit distinct strengths in AI development. Together, the two countries account for approximately 60% of the total AI-related publications, active AI researchers, and large-scale AI systems developed over the past year. Meanwhile, China maintains a leading position in granted AI patents, while the United States dominates in super-computing and data center infrastructure, along with private investment and the number of newly funded AI companies.

In addition, the UK, Germany, Japan, and Canada—secure top-tier positions in the global AI ecosystem via differentiated strengths: Germany performs well in AI research (5.7% of AI-related publications, with 5.58% of active AI researchers) and super-computing (3.43% of Top 500

Supercomputer performance aggregate); the UK performs well in AI R&D and infrastructure (5.33% of large-scale AI systems, 4.69% of data centers) and AI industry vitality (3.16% of private investment in AI, 6.72% of newly funded AI companies); Japan showcases technical implementation with super-computing (8.08% of Top 500 Supercomputer performance aggregate). Canada maintains a balanced performance in AI development (2.97% of publications, 2.14% of researchers, 3.15% of data centers, 2.02% of private investment, and 3.65% of newly funded AI companies). Meanwhile, India stands out within BRICS: it holds the second-largest share across 5 of the 8 AI indicators, with a notably substantial presence in these areas.

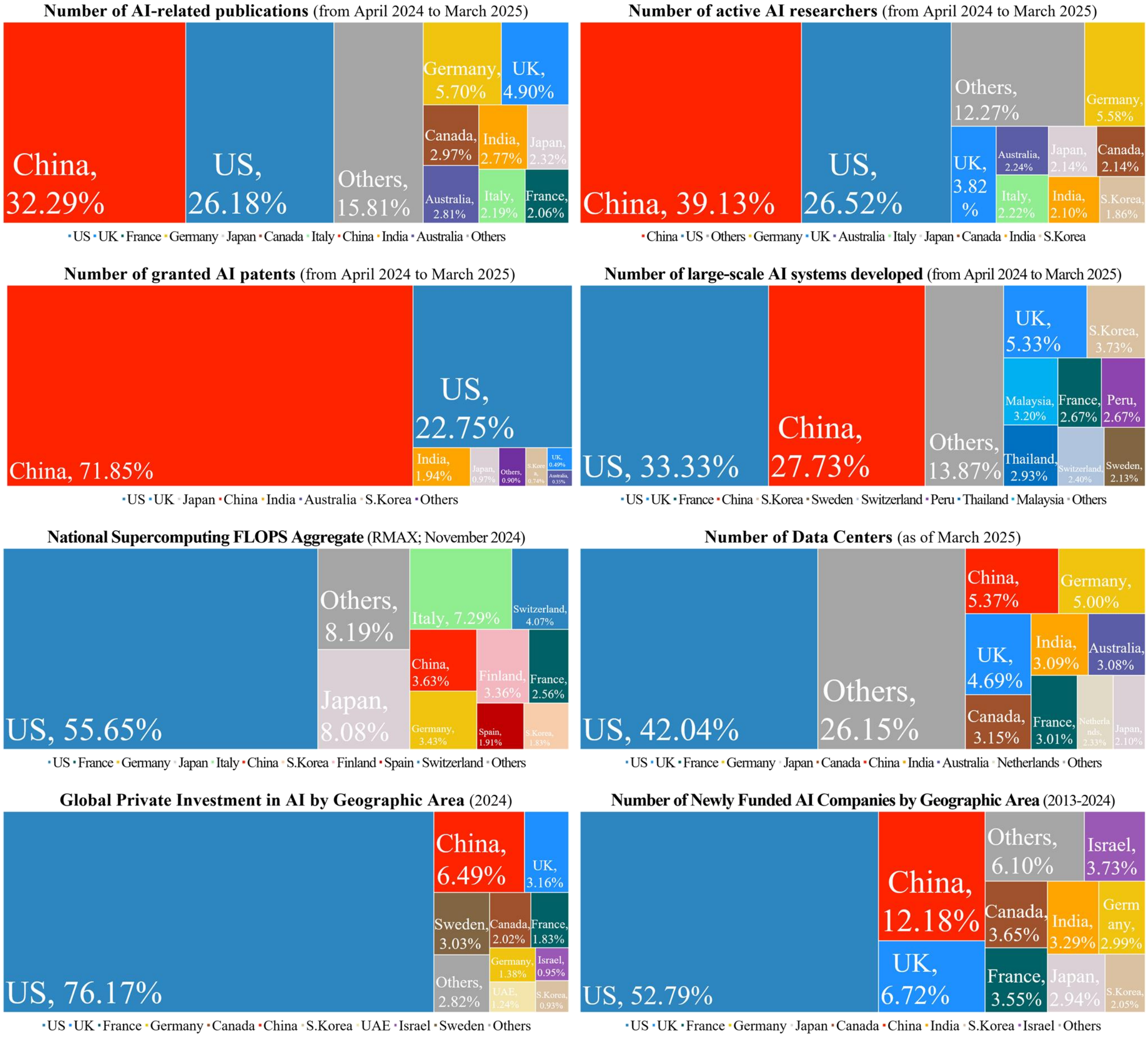

***Figure 8 Share of Key Indicators in AI Development Level Among 40 Countries***

*Data source: DBLP, Our World in Data, Data Center Map, The TOP500 List of Supercomputer, ICT Development Index 2024, Artificial Intelligence Index Report 2025*

## Observation 1.2:

Since 2010, the number of Generative AI (GenAI) patents globally has shown exponential growth, with total patents increased by about 30 times and applications by about 25 times. The growth rate accelerated significantly after 2018.

From 2010 to 2023, both the total number of GenAI total patents and GenAI applications' patents show a rapid growth trend. The total number of GenAI total patents increased from 1,169 in 2010 to 36,389 in 2023, growing by about 31.13 times. The number of GenAI applications' patents rose from 633 in 2010 to 16,046 in 2023, with a growth of approximately 25.35 times. After 2018, the growth rate of both accelerated significantly and began to grow exponentially, indicating that R&D and patent-related activities in the GenAI field entered a rapid expansion phase, with continuous intensification of innovation activities.

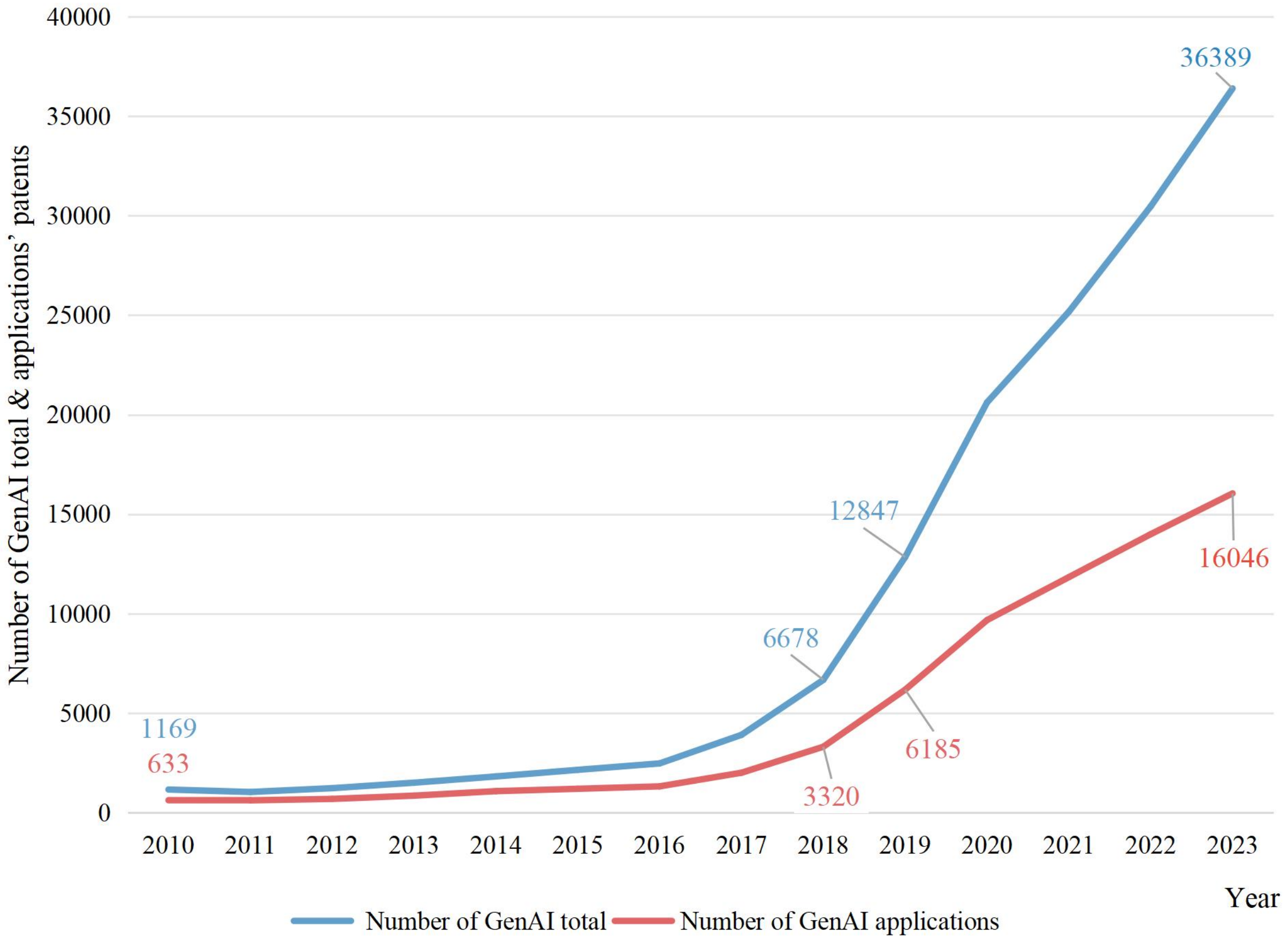

***Figure 9 Annual Trends of GenAI Total Patent & Applications' Patent Counts of 40 Countries***

*Data source: WIPO Patent Landscape Report on Generative AI 2024 (Dataset)*

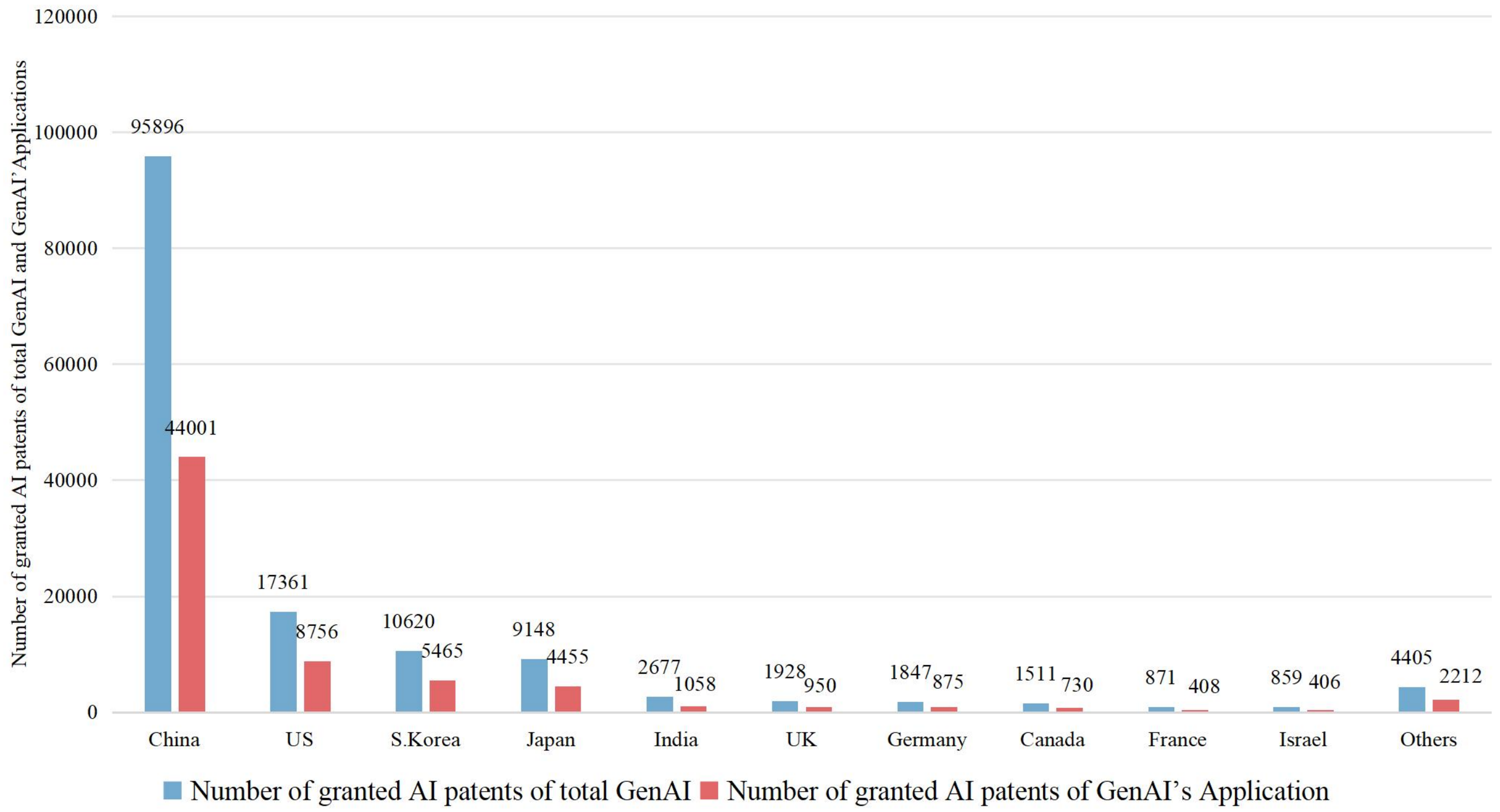

***Figure 10 Granted GenAI Patents by Countries (1997-2023)***

*Data source: WIPO Patent Landscape Report on Generative AI 2024 (Dataset)*

Global granted GenAI patents are highly concentrated, with China accounting for 65% of the total GenAI — exceeding the combined share of the US, South Korea, and Japan. In application-related GenAI patents, China's proportion reaches 63%, demonstrating dominant advantages in both overall quantity and practical implementation. Other countries show significant gaps, with the US ranking second at about 12% and others below 8% in both GenAI patents' number of the total and application.

# 3.2. Pillar 2: AI Governance Environment

## Observation 2.1:

Documented AI incidents roughly doubled globally in 2024 compared to 2023. The US saw an 80% increase, while other countries grew at a faster rate of 110%.

The number of recorded AI risk incidents in 2024 maintained a rapid growth trend compared to 2023. In 2024, the number of recorded AI risk incidents increased by about 4,000, with a growth rate of 104% compared to 2023, indicating an urgent need for a robust AI governance system to keep pace with the rapid development of technology.

Although the number of AI risk incidents in the United States is also rapidly increasing, the growth rate (80% from 2023 to 2024) is slightly lower compared to the total (104% from 2023 to 2024). In 2023, the United States accounted for 31% of total AI risk incidents, but this percentage decreased to 27% in 2024. The share from other countries is on the rise.

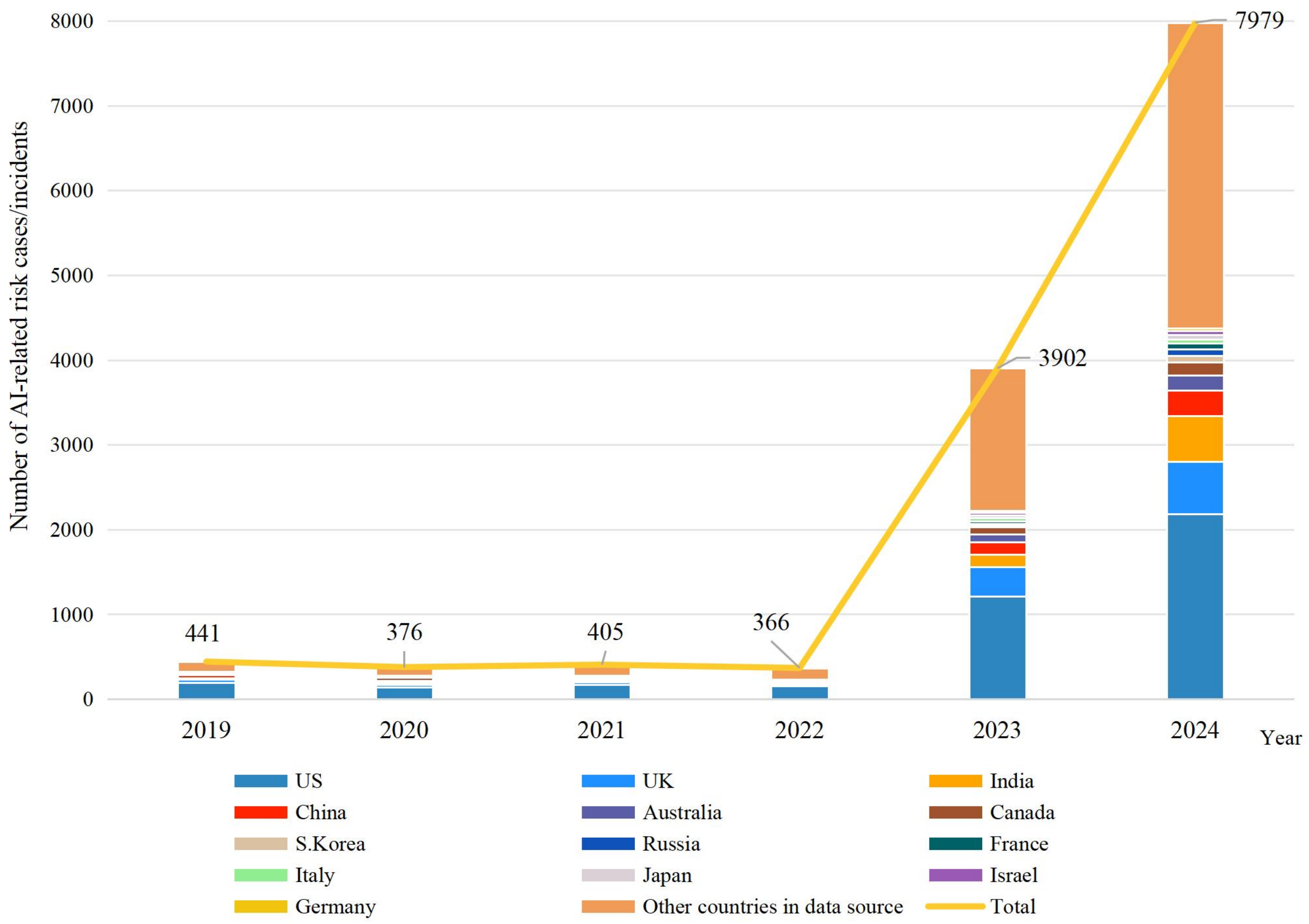

***Figure 11 Number of Documented AI Incidents (2017-2024)***

*Data source: OECD AI Incidents Monitor (OECD AIM)*

## Observation 2.2:

## Risk incidents pertaining to AI Safe & Security, Human Rights, and Data Governance are more numerous, accounting for half of all reported AI incidents.

According to the AI risk incident cases recorded by the OECD AIM up to October 2024, the most frequently reported risks relate to Robustness & Digital Security, Respect for Human Rights, and Privacy & Data Governance, indicating that technical and operational safety issues remain the most prominent concerns in current AI development. However, relatively fewer incidents are associated with Human Well-being, Democracy, and Human Autonomy—areas often linked to longer-term and systemic impacts. This highlights a dual imperative: while addressing immediate safety risks remains essential, it is equally important to anticipate and mitigate less visible but potentially profound risks to societal values. A balanced governance approach must therefore integrate both short-term risk containment and long-term ethical foresight, laying a sustainable foundation for AI development.

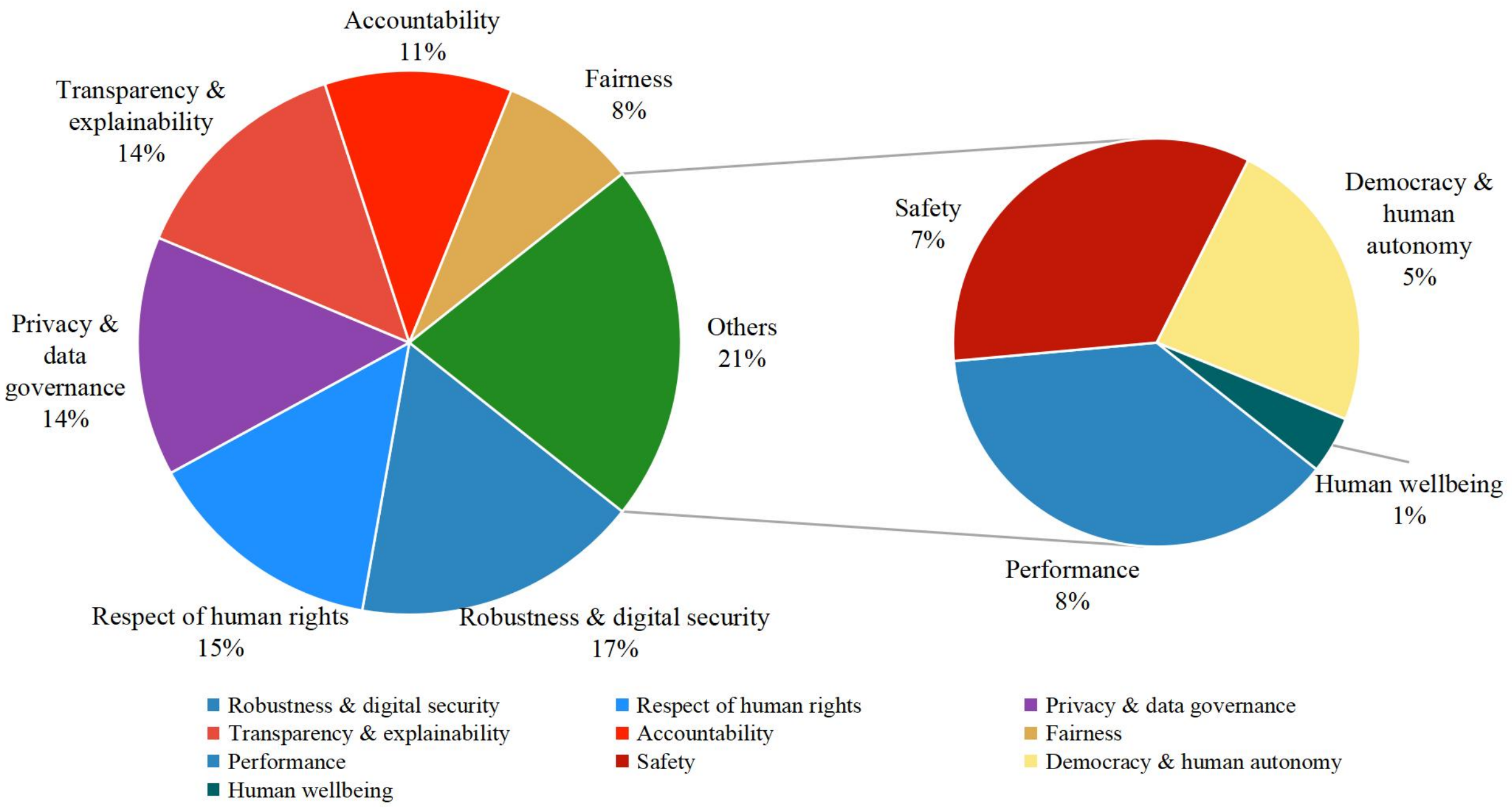

***Figure 12 AI Risk Incident Distribution by topics***

*Data source: OECD AIM*

## Observation 2.3:

Although high-income countries tend to score higher in overall governance — indicating stronger innate readiness and basis for AI governance, practical experience still reveals significant room for proactive efforts to advance AI governance across diverse national contexts.

In the three-dimensional evaluation system of countries' overall governance readiness (including overall assessment of the level of governance in general, the overall level of digital governance, and the overall process of sustainable development in the country), among all the countries assessed, the majority of high-income countries outperformed upper-middle and lower-middle-income countries. For example, high-income countries such as Denmark and Finland, with their mature governance systems, advanced digital infrastructure, and a favorable progress to sustainable development, occupy high-dimensional advantages in the chart, becoming the “first echelon” in governance readiness. However, the advantage in governance readiness of high-income countries is the result of accumulated development and does not constitute a “threshold” for AI governance—the equality of technological iteration and the diversity of governance needs allow every country to have the opportunity to develop according to its own needs in AI governance. For instance, China, as an upper-middle-income country, was at the relatively lower end of the overall governance readiness, but still secured a high performance in AI governance instruments and effectiveness pillars.

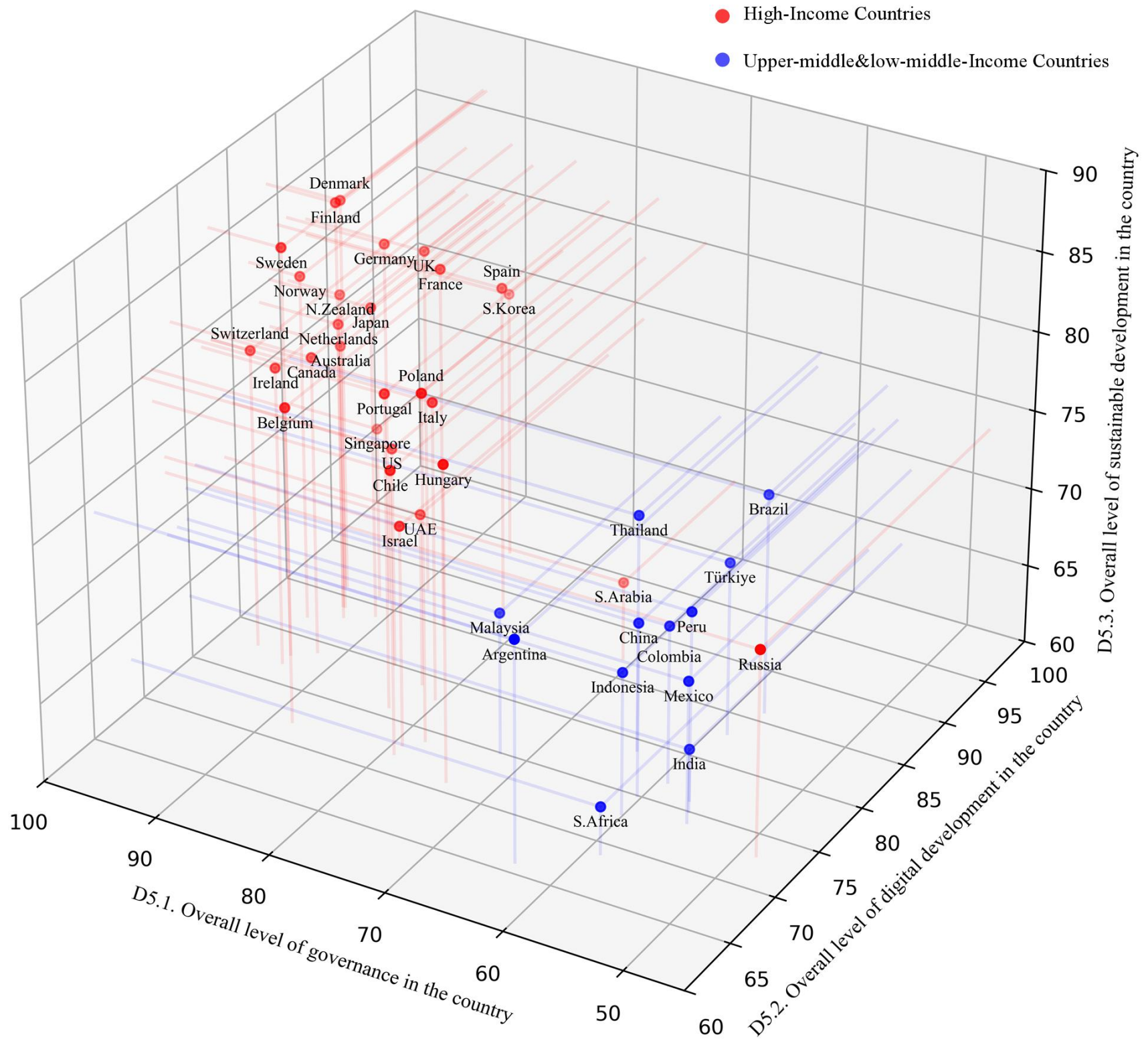

***Figure 13 Composition of 40 Countries’ Overall Governance Readiness Scores (From Relevant Indices)***

*Data source: Worldwide Governance Indicators, Human Development Index, Global Cybersecurity Index, Effective Governance of Data, GovTech Maturity Index, E-Government Development Index, E-Participation Index, 2025 Sustainable Development Index*

# 3.3. Pillar 3: AI Governance Instruments

## Pillar 3 Overview:

AI governance relies on a diverse array of tools and approaches, each serving distinct but complementary roles. The effective coordination and deployment of these instruments are essential for robust AI regulation and sustainable AI innovation. Strategic documents and ethical principles offer high-level direction, while technical assessments and standards provide the scientific basis for regulation. Legislation and institutional frameworks operationalize governance, defining enforceable rules and responsibilities. Building on this foundation, global cooperation mechanisms foster dialogue and inclusiveness, uniting diverse perspectives to ensure the responsible development of AI for the benefit of humanity.

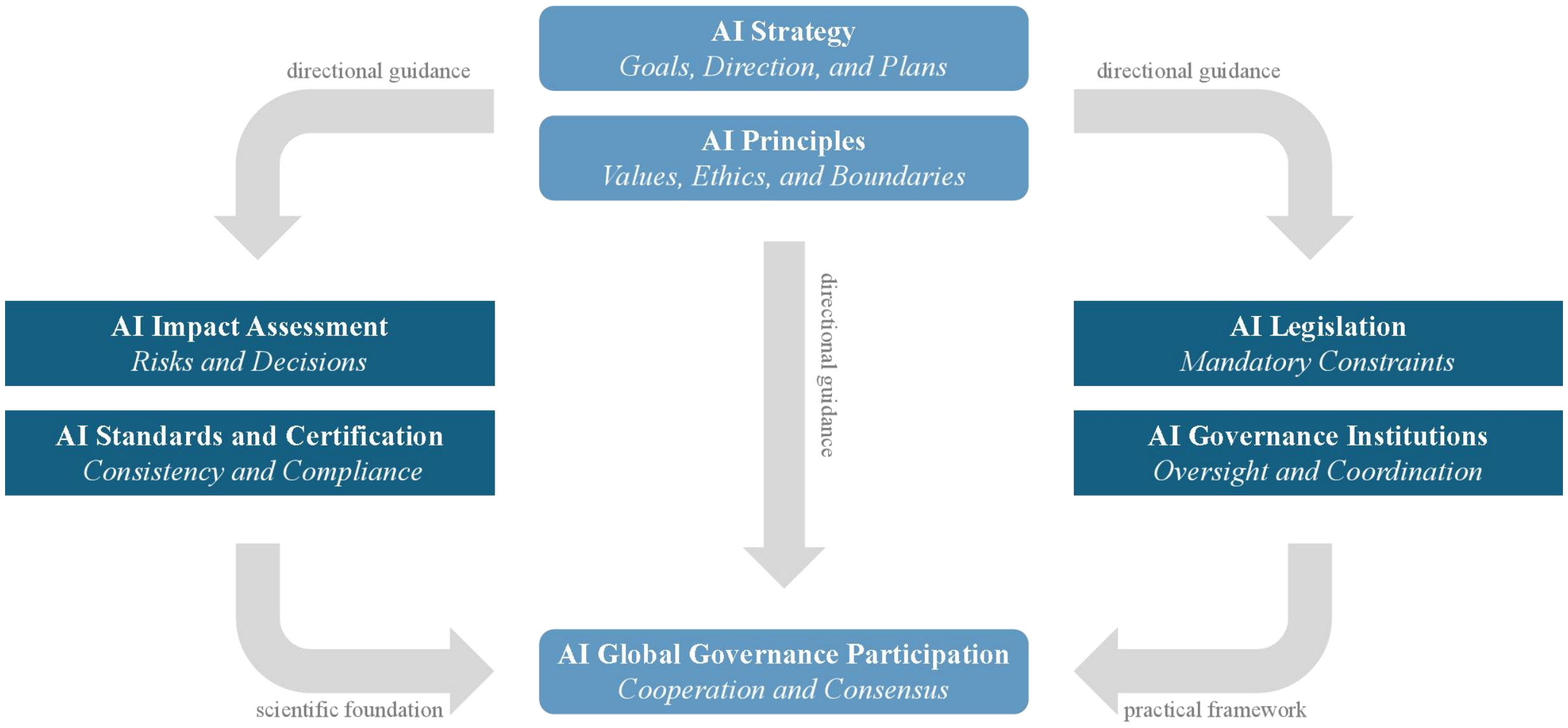

***Figure 14 AI Governance Instruments: A Holistic View***

## Observation 3.1:

### All 40 countries evaluated have published national-level AI strategies, with varying structural approaches to strategy formulation.

The 40 countries assessed have generally performed well in the AI strategy dimension. All the countries currently evaluated have developed national AI strategies, including South Africa, which had not yet developed one in the previous evaluation. In terms of strategic time frames, Switzerland releases an annual updated digital strategy to guide AI development, while countries such as France, Italy, Malaysia, and Turkey have formulated short-term strategies spanning 2-5 years. Indonesia's AI strategy, on the other side, covers a 25-year period. In terms of strategic structure, countries like Argentina, Ireland, and the United States, have adopted modular strategies focusing on different development pillars; countries such as China, Italy, and Peru have adopted a vertical structure presenting a systematic framework of current status-goals-action implementation; and countries like France, India, and South Africa have chosen more comprehensive and hybrid strategies.

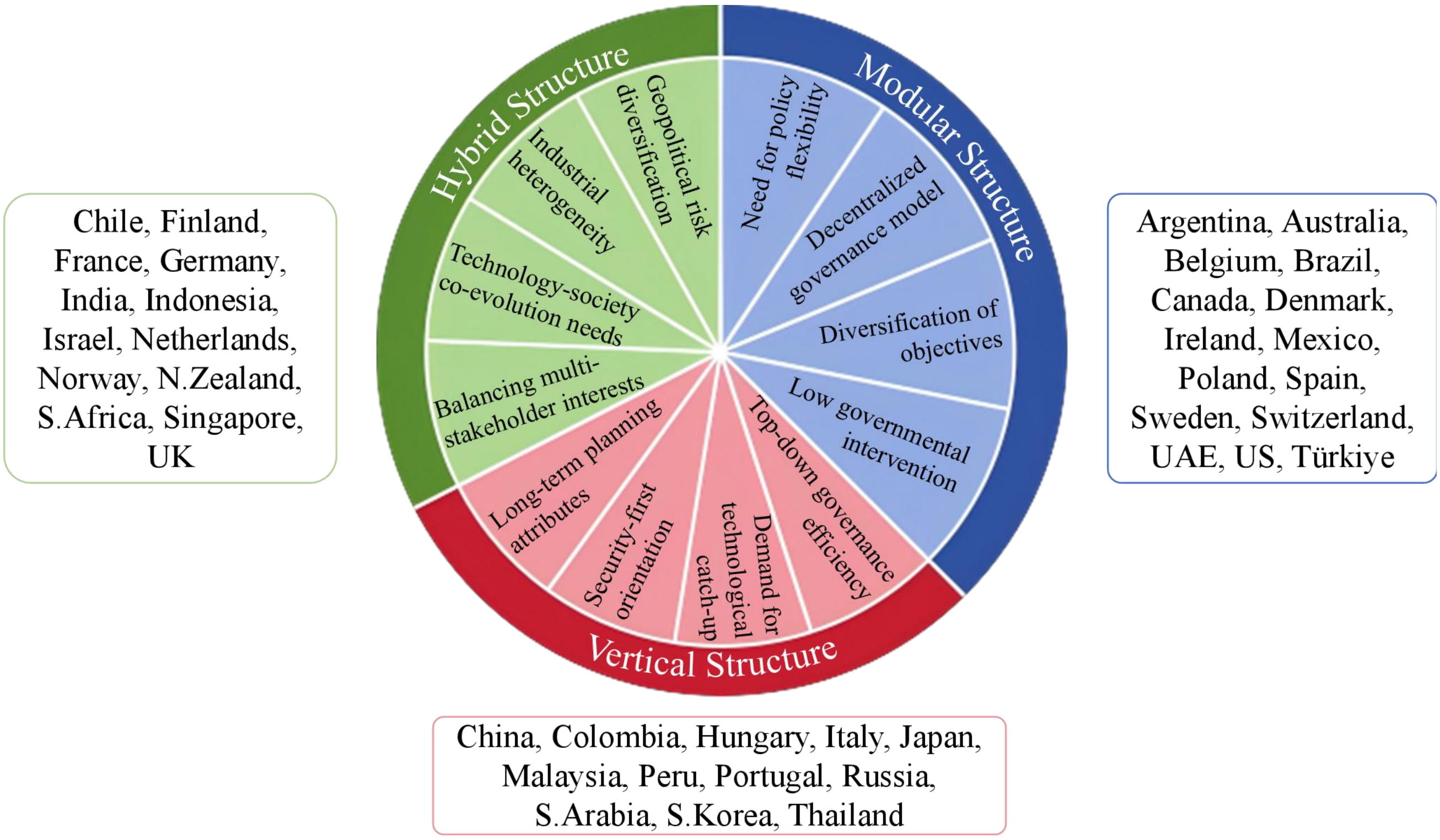

***Figure 15 National AI Strategy Structure Types Among 40 Evaluated Countries***

## Observation 3.2:

Since 2024, the legislation on AI has shown a clear accelerating trend. Some countries have enacted national general regulations on AI, while others have formulated special regulations for vertical fields of AI.

Among the 40 countries included in the assessment, the landscape of AI legislation is characterized by the coexistence of national general regulations and special regulations for vertical fields. A total of 26 countries have either enacted or are in the process of formulating comprehensive national laws and regulations on AI. Among them, 14 countries in the European Union and Norway are all enforcing the *Artificial Intelligence Act*, the world' s first comprehensive AI law, which came into effect on August 1st, 2024. South Korea passed the *Basic Act on Artificial Intelligence* on December 26th, 2024, with plans for implementation in January 2026. Additionally, more than ten countries, including China, France, Japan, Peru, and Turkey are currently drafting their own national AI laws.

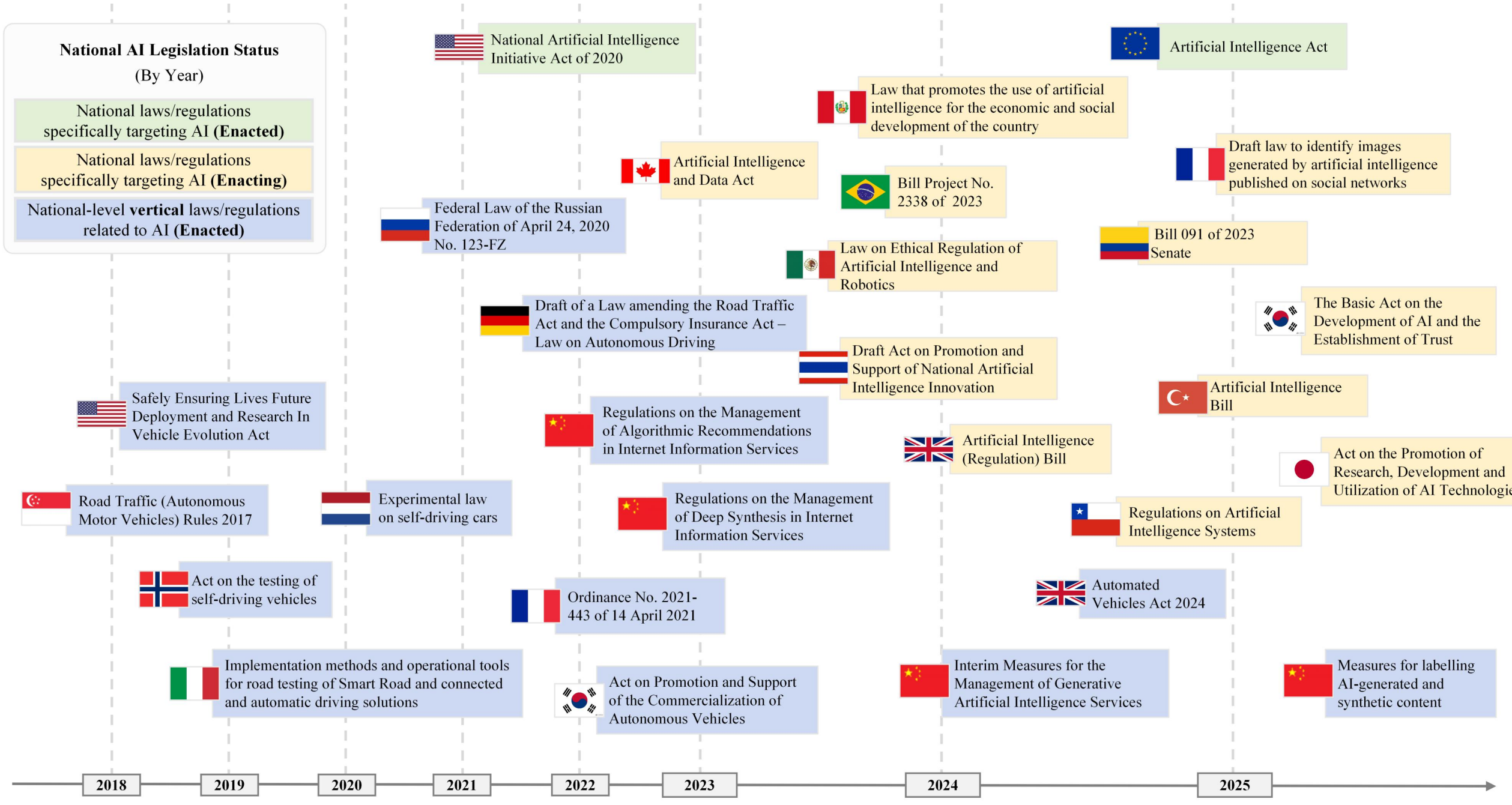

***Figure 16 National AI Legislation Status (By Year)***

While advancing comprehensive legislation, countries are actively integrating AI governance into existing legal frameworks through supplementary clauses or amendments. Currently, 24 out of the 40 countries have established data or information protection laws directly related to AI. 11 of the 40 countries have enacted national AI regulations focusing on specific sectors, primarily in two areas: autonomous driving and generative artificial intelligence.

From a legislative trend perspective, the legislation in the field of autonomous driving was initiated as early as 2017-2018 in the United States, Italy, and Norway. Since 2023, China has started to launch vertical legislation in the field of generative AI, which has become a new governance focus. This reflects the trend that AI governance is moving towards professionalization and segmentation of specific domains.

## Observation 3.3:

All 40 countries have participated in various forms of global AI governance mechanisms, with France, Japan, South Korea, and Singapore showing the highest levels of involvement.

All 40 countries evaluated have participated in various global AI governance mechanisms to varying degrees, highlighting the importance of international cooperation in AI governance. Among them,

France, Japan, South Korea, and Singapore were the most active, participating in all global AI governance events covered by the AGILE Index. Among the 40 countries, all except the United States and Israel have signed UNESCO's *Recommendation on the Ethics of Artificial Intelligence*. The *Global Digital Compact* adopted at the UN Summit of the Future is the most widely supported AI governance document so far, with only Russia not signing it due to sovereignty and security concerns. Outside of UN mechanisms, key global governance frameworks include the AI Safety Summit held in UK, AI Seoul Summit held in South Korea, and AI Action Summit held in France; REAIM Summit led by the Netherlands and South Korea; and the International Network of AI Safety Institutes led by the United States. The global landscape of AI governance has exhibited the distinct characteristics of parallel multilateral collaboration and pluralistic participation. While there remains a gap in achieving a fully inclusive participation and a coordinated response mechanism, core issues such as AI safety and capacity-building have already garnered widespread attention and responses.

***Table 3 Overview of International AI Governance Mechanisms Participation by Countries***

| | International AI Governance Mechanisms/Initiatives | | | | | | | |
|---|---|---|---|---|---|---|---|---|
| **Country** | **UNESCO: Recommendation on the Ethics of Artificial Intelligence (2021)** | **AI Safety Summit: Bletchley Declaration (2023)** | **REAIM 2023 Summit: Call to Action (2023)** | **AI Seoul Summit: Seoul Ministerial Statement (2024)** | **United Nations: Global Digital Compact (2024)** | **REAIM 2024 Summit: Blueprint for Action (2024)** | **International Network of AI Safety Institutes: Initial Members (2024)** | **AI Action Summit: Statement on Inclusive and Sustainable AI for People and the Planet (2025)** |
| France | ✓ | ✓ | ✓ | ✓* | ✓ | ✓ | ✓ | ✓ |
| Japan | ✓ | ✓ | ✓ | ✓* | ✓ | ✓ | ✓ | ✓ |
| Singapore | ✓ | ✓ | ✓ | ✓* | ✓ | ✓ | ✓ | ✓ |
| S.Korea | ✓ | ✓ | ✓ | ✓* | ✓ | ✓ | ✓ | ✓ |
| Canada | ✓ | ✓ | ✓ | ✓* | ✓ | × | ✓ | ✓ |
| Germany | ✓ | ✓ | ✓ | ✓* | ✓ | ✓ | × | ✓ |
| Netherlands | ✓ | ✓ | ✓ | ✓ | ✓ | ✓ | × | ✓ |
| UK | ✓ | ✓ | ✓ | ✓* | ✓ | ✓ | ✓ | × |
| Australia | ✓ | ✓ | × | ✓* | ✓ | × | ✓ | ✓ |
| Italy | ✓ | ✓ | ✓ | ✓* | ✓ | × | × | ✓ |
| US | × | ✓ | ✓ | ✓* | ✓ | ✓ | ✓ | × |
| Chile | ✓ | ✓ | × | ✓ | ✓ | × | × | ✓ |
| China | ✓ | ✓ | ✓ | × | ✓ | × | × | ✓ |
| India | ✓ | ✓ | × | ✓ | ✓ | × | × | ✓ |
| Indonesia | ✓ | ✓ | × | ✓ | ✓ | × | × | ✓ |
| Spain | ✓ | ✓ | × | ✓ | ✓ | × | × | ✓ |
| Switzerland | ✓ | ✓ | × | ✓ | ✓ | × | × | ✓ |
| Brazil | ✓ | ✓ | × | × | ✓ | × | × | ✓ |
| Ireland | ✓ | ✓ | × | × | ✓ | × | × | ✓ |
| Mexico | ✓ | × | × | ✓ | ✓ | × | × | ✓ |
| N.Zealand | ✓ | × | × | ✓ | ✓ | × | × | ✓ |
| S.Arabia | ✓ | ✓ | × | ✓ | ✓ | × | × | × |
| Türkiye | ✓ | ✓ | × | ✓ | ✓ | × | × | × |
| UAE | ✓ | ✓ | × | ✓ | ✓ | × | × | × |
| Belgium | ✓ | × | × | × | ✓ | × | × | ✓ |
| Colombia | ✓ | × | × | × | ✓ | × | × | ✓ |
| Denmark | ✓ | × | × | × | ✓ | × | × | ✓ |
| Finland | ✓ | × | × | × | ✓ | × | × | ✓ |
| Hungary | ✓ | × | × | × | ✓ | × | × | ✓ |
| Israel | × | ✓ | × | ✓ | ✓ | × | × | × |
| Norway | ✓ | × | × | × | ✓ | × | × | ✓ |
| Peru | ✓ | × | × | × | ✓ | × | × | ✓ |
| Poland | ✓ | × | × | × | ✓ | × | × | ✓ |
| Portugal | ✓ | × | × | × | ✓ | × | × | ✓ |
| S.Africa | ✓ | × | × | × | ✓ | × | × | ✓ |
| Sweden | ✓ | × | × | × | ✓ | × | × | ✓ |
| Thailand | ✓ | × | × | × | ✓ | × | × | ✓ |
| Argentina | ✓ | × | × | × | ✓ | × | × | × |
| Malaysia | ✓ | × | × | × | ✓ | × | × | × |
| Russia | ✓ | × | × | × | × | × | × | × |

Note: Countries marked with * are also signatories to the Seoul Declaration for safe, innovative and inclusive AI (2024).

# 3.4. Pillar 4: AI Governance Effectiveness

## Observation 4.1:

### The public' s level of awareness, trust, and optimism regarding AI applications and services shows an overall negative correlation with GDP per capita.

The relationship between GDP per capita and public awareness of AI applications within products and services reveals an negative pattern: Higher income levels correlate with declining awareness of AI's technical features and social implications. Developing economies—notably China, Turkey, Indonesia, Mexico, and Malaysia demonstrate above-average awareness levels relative to peers. Among high-income countries, Asian economies such as South Korea and Singapore exhibit stronger awareness than the general trend observed in developed countries.

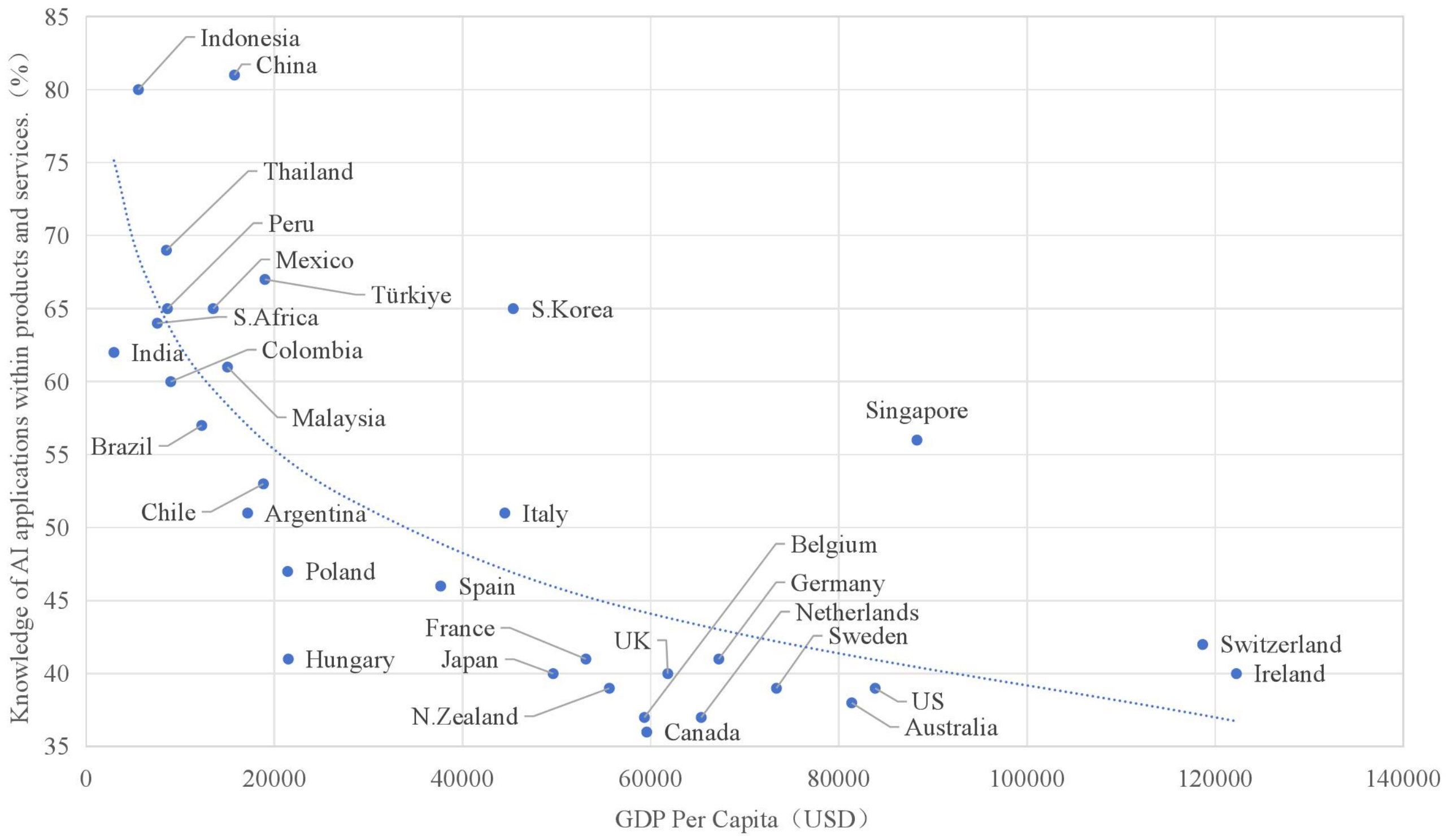

*Figure 17 GDP Per Capita & Awareness of AI Applications Within Products and Services*

*Data Source: THE IPSOSAI MONITOR 2024*

Public trust in AI applications and public positive attitudes toward AI development are notably higher in emerging economies such as China, Indonesia, Thailand, and Mexico. In contrast, the public in countries such as Sweden, the United States, France, Belgium, and Canada tend to adopt more cautious stances toward AI in these areas. In terms of public positive or neutral anticipation for AI usage, most countries view AI favorably for its role in enhancing innovation and improving efficiency — particularly in areas such as personal entertainment choices, time spent on tasks, personal health, and work. However, a cautious outlook persists regarding its ethical use and real-world risks, especially in domains like public trust, the job market, and the governance of online disinformation.

***Table 4 Public Attitudes and Expectations Toward AI (Percentage of Agreement)***

| Dimension | D14.3. Public trust in AI applications | | D14.1.Public positive attitude towards AI development | | D14.2. Public positive or neutral anticipation for AI usage | | | | | | | |
|---|---|---|---|---|---|---|---|---|---|---|---|---|
| | How much do you agree or disagree with the following?（%） | | | | | Public positive or neutral anticipation for AI usage in the following aspects in the next 3-5 years.(%) | | | | | | |
| Country \ Q/A | I trust that companies that use AI will protect my personal data. | I trust AI to not discriminate or show bias towards any group of people. | Products and services using AI make me excited. | Products and services using AI have more benefits than drawbacks. | Products and services using AI will profoundly change my daily life in the next 3-5 years. | My entertainment options | The amount of time it takes me to get things done | My health | My job | Economic condition | The amount of disinformation on the internet | The job market |
| China | 66 | 76 | 80 | 83 | 86 | 92 | 92 | 89 | 89 | 87 | 70 | 63 |
| Indonesia | 66 | 68 | 76 | 80 | 80 | 89 | 93 | 90 | 87 | 84 | 73 | 69 |
| Thailand | 68 | 73 | 76 | 77 | 78 | 90 | 89 | 88 | 85 | 82 | 73 | 72 |
| Mexico | 64 | 76 | 70 | 70 | 76 | 87 | 90 | 84 | 83 | 81 | 69 | 68 |
| Peru | 62 | 74 | 67 | 70 | 76 | 88 | 90 | 86 | 81 | 78 | 73 | 69 |
| Colombia | 50 | 66 | 63 | 68 | 77 | 88 | 88 | 80 | 76 | 72 | 65 | 62 |
| S.Africa | 62 | 72 | 61 | 62 | 76 | 80 | 86 | 79 | 76 | 69 | 63 | 58 |
| Malaysia | 52 | 62 | 68 | 63 | 71 | 82 | 82 | 83 | 79 | 72 | 61 | 68 |
| Singapore | 56 | 60 | 64 | 66 | 79 | 85 | 85 | 78 | 70 | 77 | 54 | 59 |
| Türkiye | 46 | 61 | 70 | 69 | 76 | 78 | 81 | 73 | 77 | 64 | 58 | 63 |
| S.Korea | 37 | 51 | 73 | 66 | 79 | 83 | 87 | 80 | 74 | 70 | 61 | 49 |
| Argentina | 46 | 60 | 52 | 57 | 67 | 85 | 84 | 79 | 78 | 70 | 66 | 65 |
| India | 60 | 63 | 63 | 62 | 65 | 68 | 70 | 69 | 66 | 68 | 64 | 64 |
| Chile | 43 | 59 | 46 | 60 | 69 | 82 | 86 | 79 | 72 | 58 | 65 | 58 |
| Brazil | 45 | 56 | 56 | 56 | 62 | 78 | 81 | 76 | 76 | 67 | 57 | 58 |
| Italy | 58 | 61 | 49 | 53 | 60 | 76 | 78 | 76 | 74 | 62 | 51 | 53 |
| Hungary | 64 | 64 | 47 | 51 | 64 | 71 | 75 | 75 | 67 | 64 | 41 | 45 |
| Spain | 48 | 52 | 45 | 50 | 59 | 79 | 81 | 76 | 73 | 61 | 51 | 49 |
| Germany | 43 | 48 | 46 | 47 | 59 | 77 | 79 | 73 | 73 | 62 | 46 | 53 |
| Netherlands | 44 | 38 | 40 | 36 | 63 | 76 | 81 | 81 | 81 | 65 | 44 | 56 |
| Switzerland | 43 | 43 | 42 | 42 | 55 | 78 | 78 | 78 | 80 | 64 | 46 | 52 |
| N.Zealand | 39 | 43 | 43 | 48 | 64 | 76 | 80 | 77 | 78 | 61 | 37 | 42 |
| Poland | 45 | 53 | 44 | 44 | 56 | 74 | 76 | 71 | 70 | 67 | 43 | 42 |
| Japan | 27 | 38 | 47 | 48 | 63 | 75 | 77 | 76 | 70 | 63 | 47 | 49 |
| Ireland | 42 | 42 | 40 | 45 | 59 | 76 | 74 | 72 | 76 | 63 | 44 | 47 |
| UK | 41 | 44 | 38 | 46 | 58 | 73 | 79 | 75 | 74 | 59 | 41 | 45 |
| Australia | 32 | 42 | 39 | 44 | 61 | 74 | 68 | 73 | 71 | 55 | 39 | 42 |
| Sweden | 35 | 33 | 36 | 43 | 52 | 69 | 77 | 72 | 79 | 59 | 34 | 45 |
| US | 33 | 41 | 34 | 39 | 58 | 67 | 75 | 73 | 73 | 51 | 39 | 44 |
| France | 35 | 41 | 39 | 41 | 57 | 69 | 75 | 67 | 68 | 51 | 38 | 45 |
| Belgium | 40 | 37 | 33 | 38 | 61 | 69 | 74 | 70 | 73 | 53 | 39 | 39 |
| Canada | 28 | 41 | 37 | 40 | 61 | 73 | 75 | 68 | 73 | 50 | 37 | 40 |

*Data Source: THE IPSOS AI MONITOR 2024; the figures in the table reflect the percentage responses of online adults under the age of 75 to the survey questions.*

## Observation 4.2:

A growing share of women are participating in AI research, evidenced by a continuous decline in the male-to-female ratio among AI researchers, which reached 2.2:1 in 2025.

According to overall trends in statistical analysis of the DBLP, the share of females in the AI research field has been steadily growing. The male-to-female author ratio dropped from 4.0:1 in 2017 to 2.2:1 in 2025, reflecting the growing participation of women in AI-related scientific research.

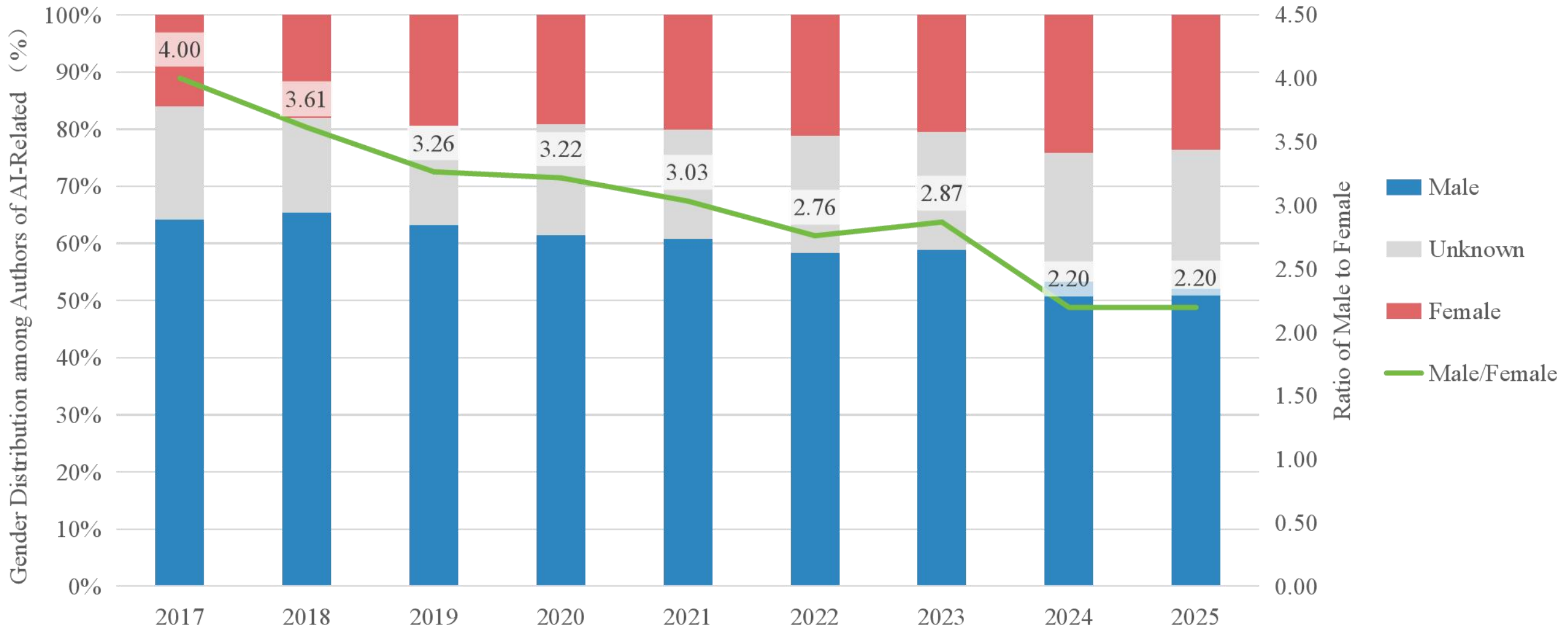

***Figure 18 Gender Equality in AI-Related Publications From 2017 to 2025***

*Data Source: Based on statistical analysis of the DBLP literature database; Gender information is inferred from author names in DBLP based on country-specific naming conventions. Results may be affected by ambiguities in gender-neutral or culturally variable names.*

In the 40 countries evaluated, Thailand has the most balanced male-to-female author ratio, followed by China, Indonesia, and Singapore; countries in North America and Western Europe do not hold a distinct advantage. Some developed countries, such as Germany and Japan, exhibit significant gender imbalance among their AI researchers, according to the analysis.

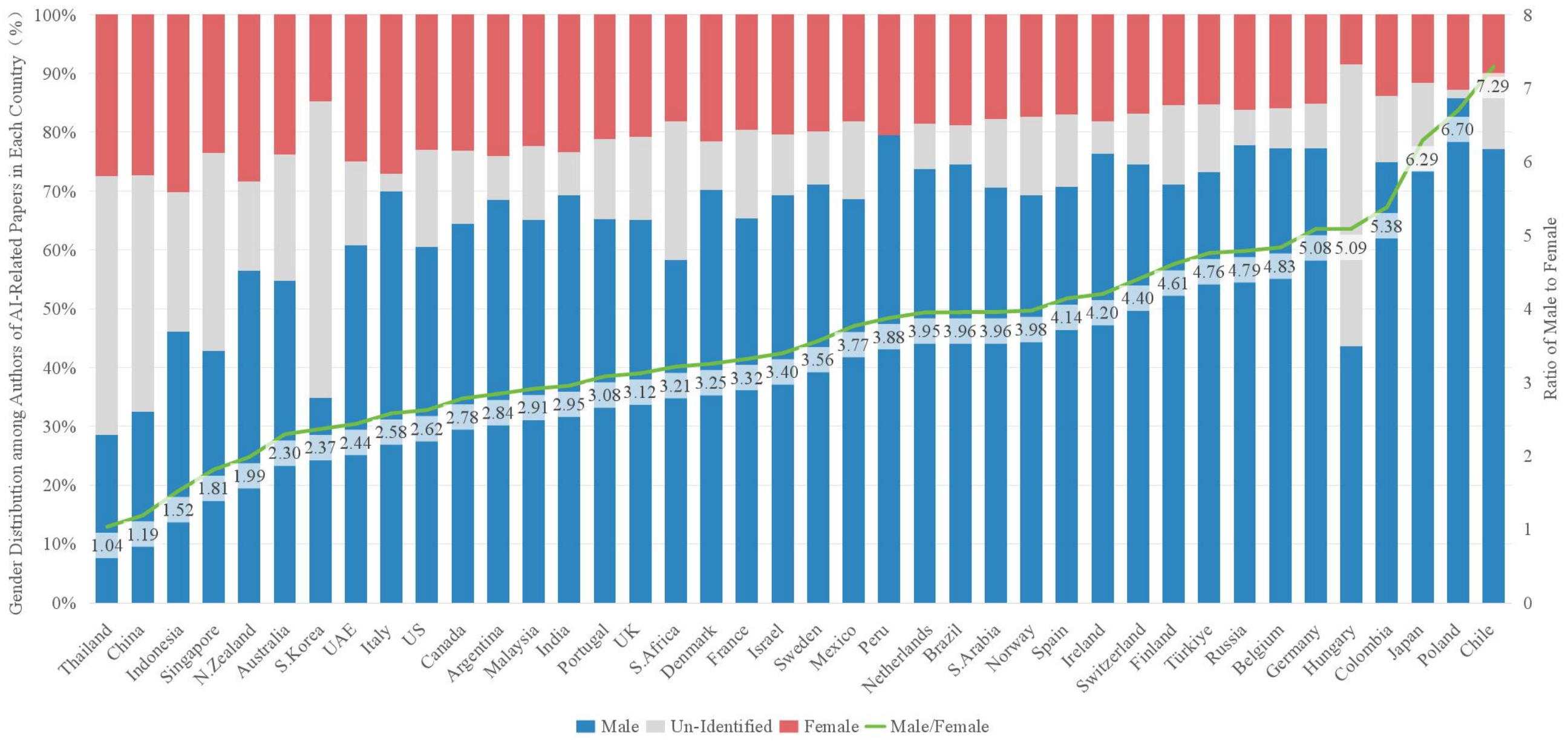

***Figure 19 Gender Equality in AI-Related Publications Across 40 Countries***

*Data Source: Based on statistical analysis of the DBLP literature database (From April 2024 to March 2025)*

## Observation 4.3:

### Economic development, as reflected by GDP per capita, is generally positively correlated with the digital inclusion of social vulnerable groups to a certain extent, but higher economic development does not necessarily result in high digital inclusiveness.

Across countries with varying GDP per capita levels, there are observable trends in the internet access and usage of vulnerable groups (the 55-74-year-old demographic and individuals from households in the lowest income quintile). Generally, countries with higher per capita GDP tend to have a higher share of internet users among these two groups, reflecting a positive correlation between economic development (as indicated by GDP per capita) and the digital inclusion of vulnerable groups. However, some developed countries such as the United States, Australia, and Israel show a notable gap between their performance in this regard and the expectations arising from their economic strength. This further illustrates that economic and technological advantages do not necessarily benefit vulnerable groups, and truly inclusive technological development requires sustained support from social policies and governance efforts.

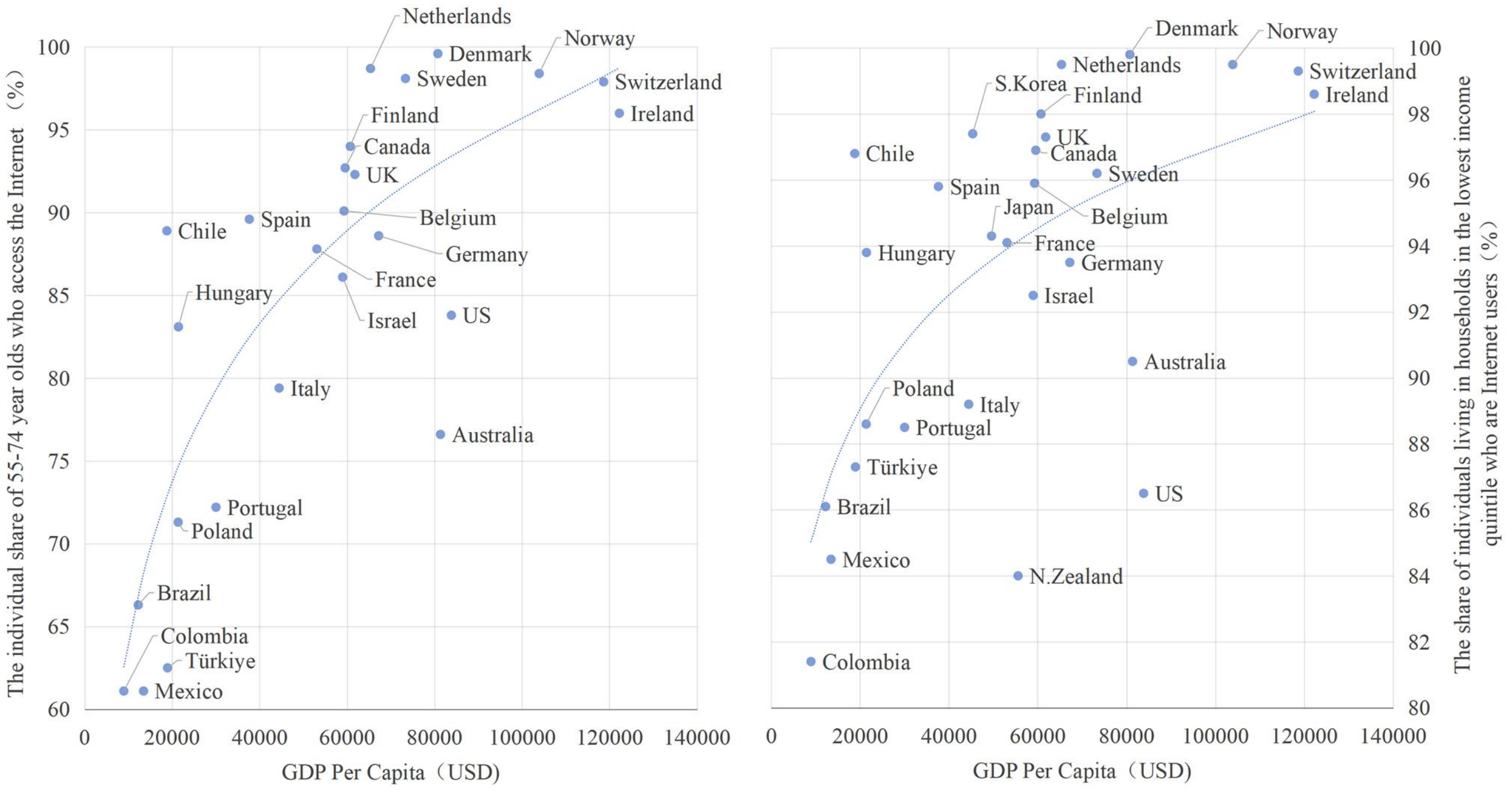

***Figure 20 GDP per Capita & Internet Usage and Access among Vulnerable Population Groups***

*Data Source: OECD Going Digital Toolkit*

## Observation 4.4:

China and the United States represent the highest shares in the openness of impactful AI models and datasets, together account for more than 70% of the total observed among all 40 countries.

Based on statistical analysis of the Hugging Face open-source community, China has 2,014 impactful AI models (50.3%), ranking first globally, while the United States ranks second with 995 models (24.8%). In terms of AI datasets, the United States leads with 922 impactful models, followed closely by China with 912 models. This data indicates that China and the United States dominate the AI open ecosystem, not only in terms of quantity but also in playing a significant role in the development of global AI research and applications. Although North America and Europe still hold advantageous positions, with Canada, the United Kingdom, France, Germany, and Switzerland all ranking in the top ten for AI models and datasets. India has shown excellent performance in AI data openness, becoming the only developing country other than China to enter the top ten. Brazil, Chile, and Indonesia have also gained more attention for the Global South in the field of artificial intelligence through their practices in AI-related scientific research and applications.

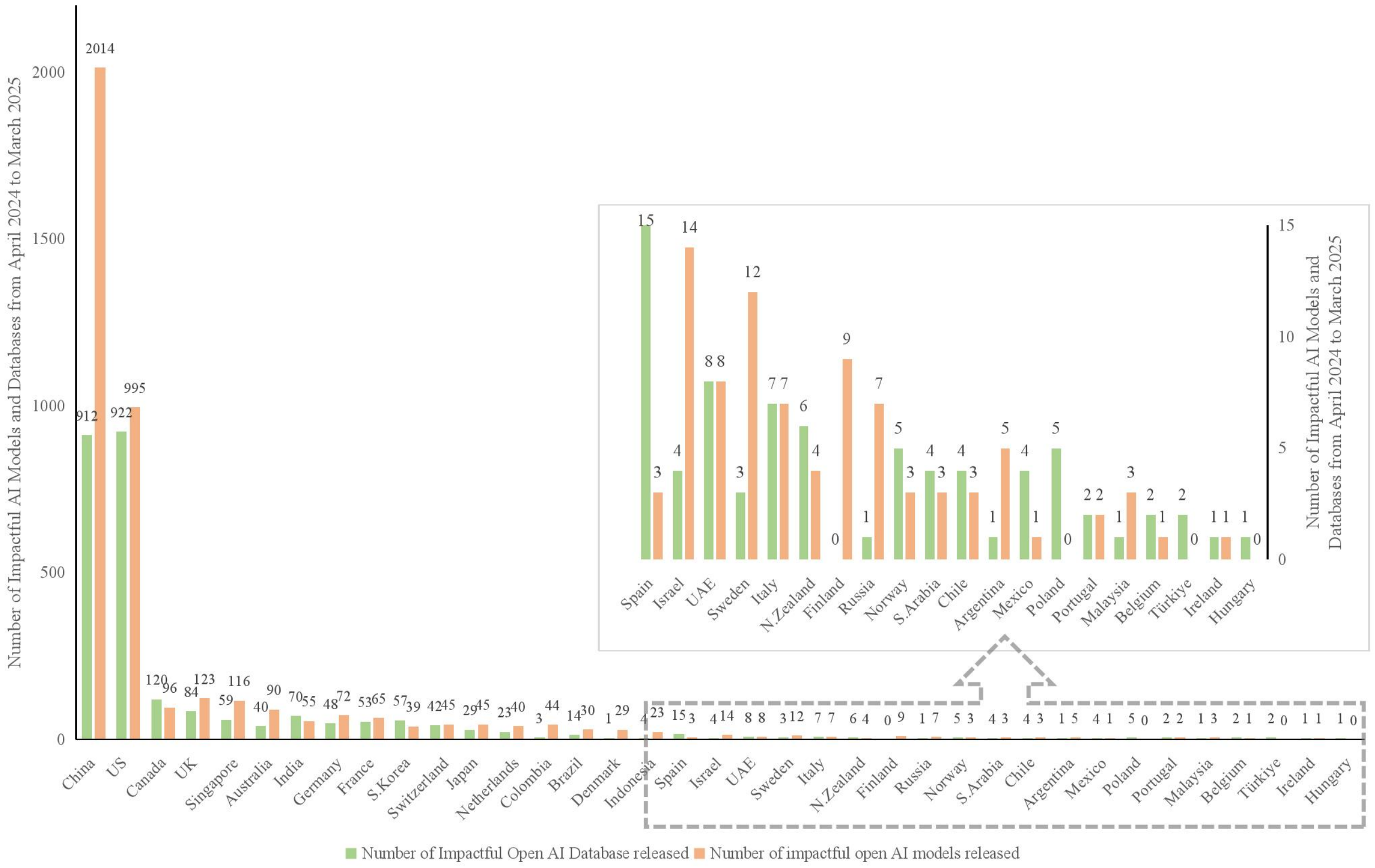

***Figure 21 Number of Impactful Open AI Database＆Models Released***

*Data Source: Based on statistical analysis of the Hugging Face Open-Source Community (As of March 2025)*

## Observation 4.5

North America and Western Europe lead the open-source AI ecosystem, with the United States far ahead at 30.43% of the 40 countries' total. Meanwhile, China and India also show strong engagement.

From the perspective of the activity of contributors to popular AI projects on GitHub, The United States, with a significant lead of 202,118 (30.43%) contributors, demonstrates its comprehensive advantages in terms of technological investment, developer ecosystem, and educational system. Several Western European developed countries are also among the top in terms of activity, such as Germany (82,912; 12.48%), the Netherlands (73,558; 11.07%), and Norway (58,525; 8.81%). Meanwhile, China (62,789; 9.45%) and India (14,815; 2.23%) rank 4th and 8th respectively—indicating that they have also laid a solid foundation in programming talent scale, education popularization, and technical community activity.

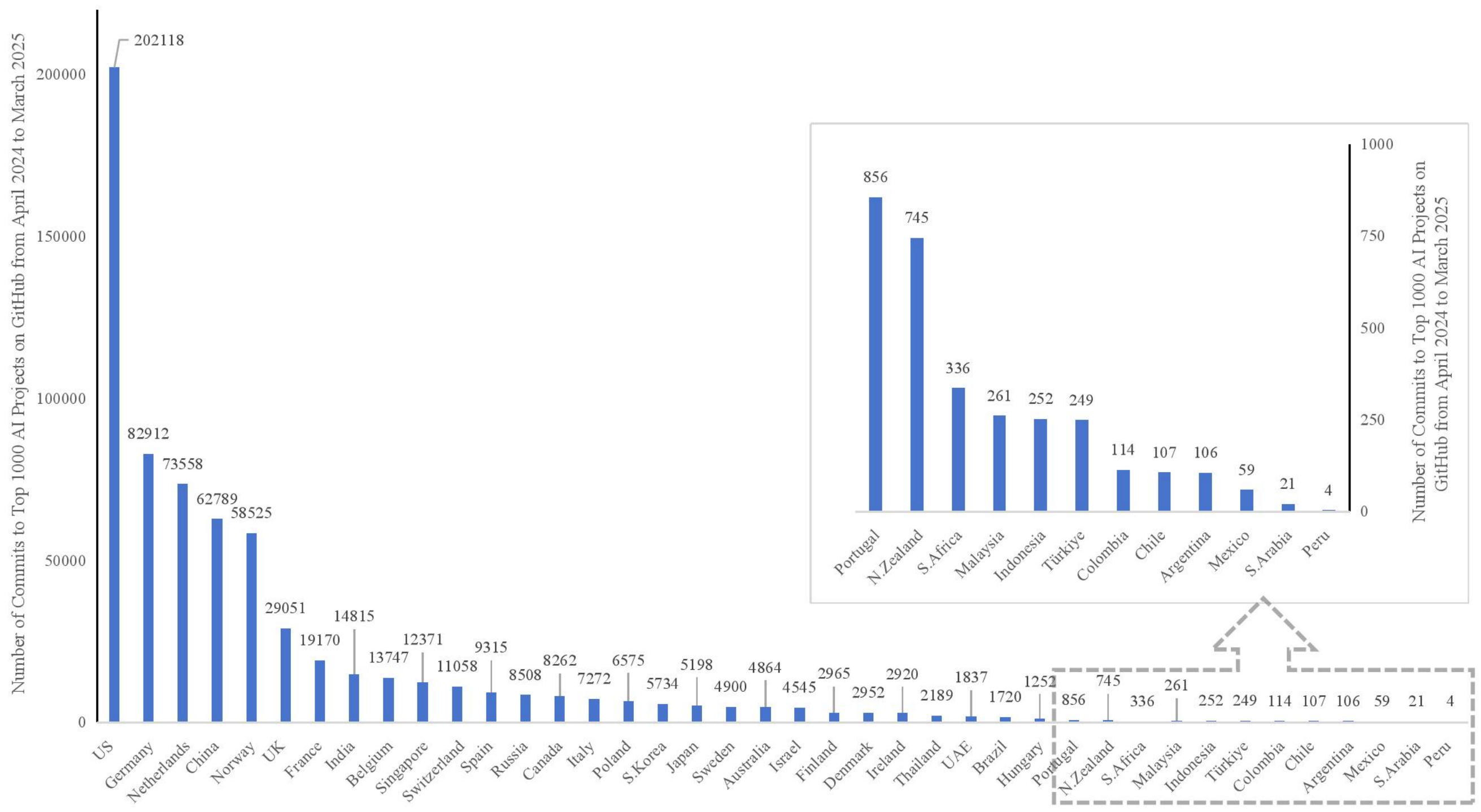

***Figure 22 Total Contributions to GitHub's Top 1000 AI Projects (Commits, by Country)***

*Data Source: Based on statistical analysis of GitHub Commits (As of March 2025)*

## Observation 4.6:

The AI research community is focusing more on governance-related issues, as shown by a rising share of AI governance publications in total AI publications—up to about 14% in 2024. Among the 40 countries, China and the US together account for about 54% of contributions, over half.

Despite some fluctuations in the overall number of AI publications in recent years, the share of research focused on governance has continued to rise—from 10.4% in 2020 to 14.4% in 2025. Governance is gradually becoming an integral part of the broader AI research landscape.

The number of Al governance publications in these 40 countries is mainly contributed by the United States and China, accounting for about 28% and 26% respectively. Germany, the United Kingdom, and Australia have also made important contributions. Compared with last year, Germany has moved from the fourth-ranking country in terms of proportion to the third, surpassing the United Kingdom.

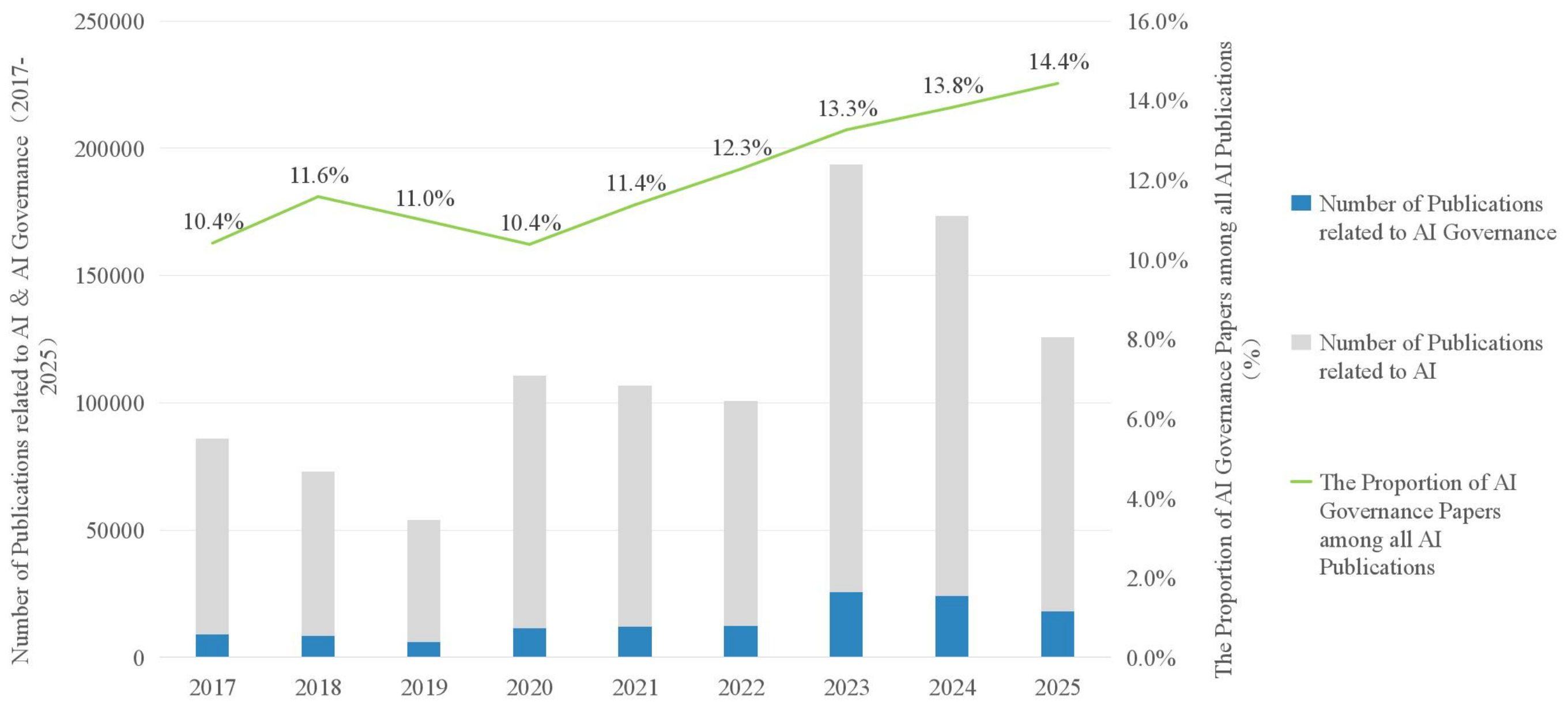

***Figure 23 The Proportion of AI Governance Papers among all AI Publications (2017-2025)***

*Data Source: Based on statistical analysis of the DBLP literature database (As of March 2025)*

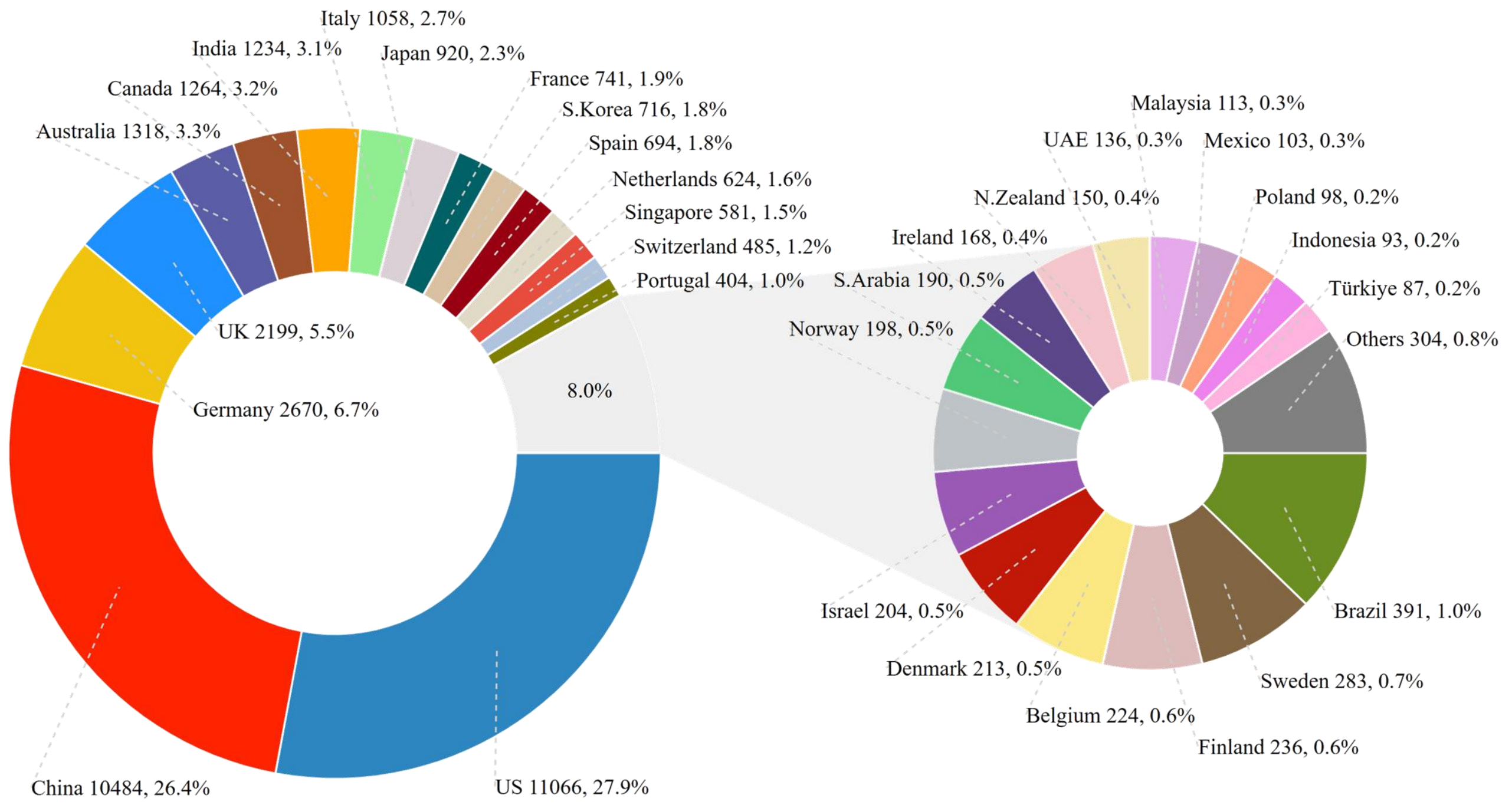

***Figure 24 Proportion of AI Governance-Related Publications Across 40 Countries***

*Data source: Based on statistical analysis of the DBLP literature database (From April 2024 to March 2025)*

## Observation 4.7:

International co-authorship in AI governance research is most prevalent among China, the US, Canada, Germany, the UK, and Australia, accounting for about 70% of all co-authored publications within the 40 countries.

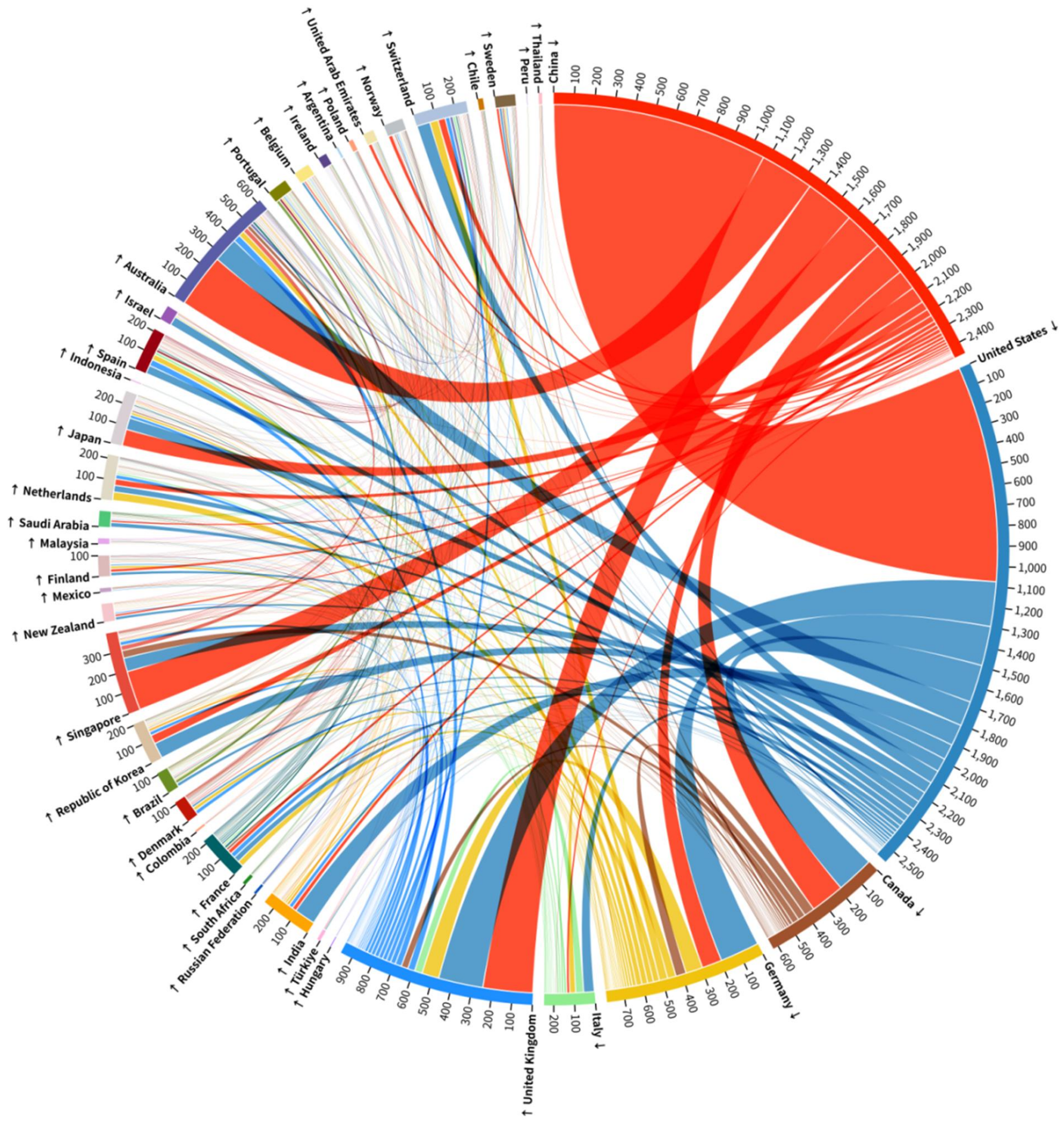

*Figure 25 The Number of International Co-Authorships in AI Governance Publications within the 40 Countries*

*Data source: Based on statistical analysis of the DBLP literature database (From April 2024 to March 2025)*

Analysis of collaborations in AI governance-related publications reveals a clear pattern of significant international cooperation. China, the US, Canada, Germany, the UK, and Australia together lead in such collaboration, accounting for about 70% of all co-authored publications within the 40 countries. Meanwhile, among the total AI governance papers from these 40 countries, only 14.7% were identified as involving international collaborations with the 40 countries, underscoring the need for deepening collaboration on AI governance research.

## Observation 4.8:

The United States and China lead in AI research for the Sustainable Development Goals (SDGs), together accounting for approximately 60% of all SDG-related AI publications among the 40 countries, with other countries also made significant contributions to various goals.

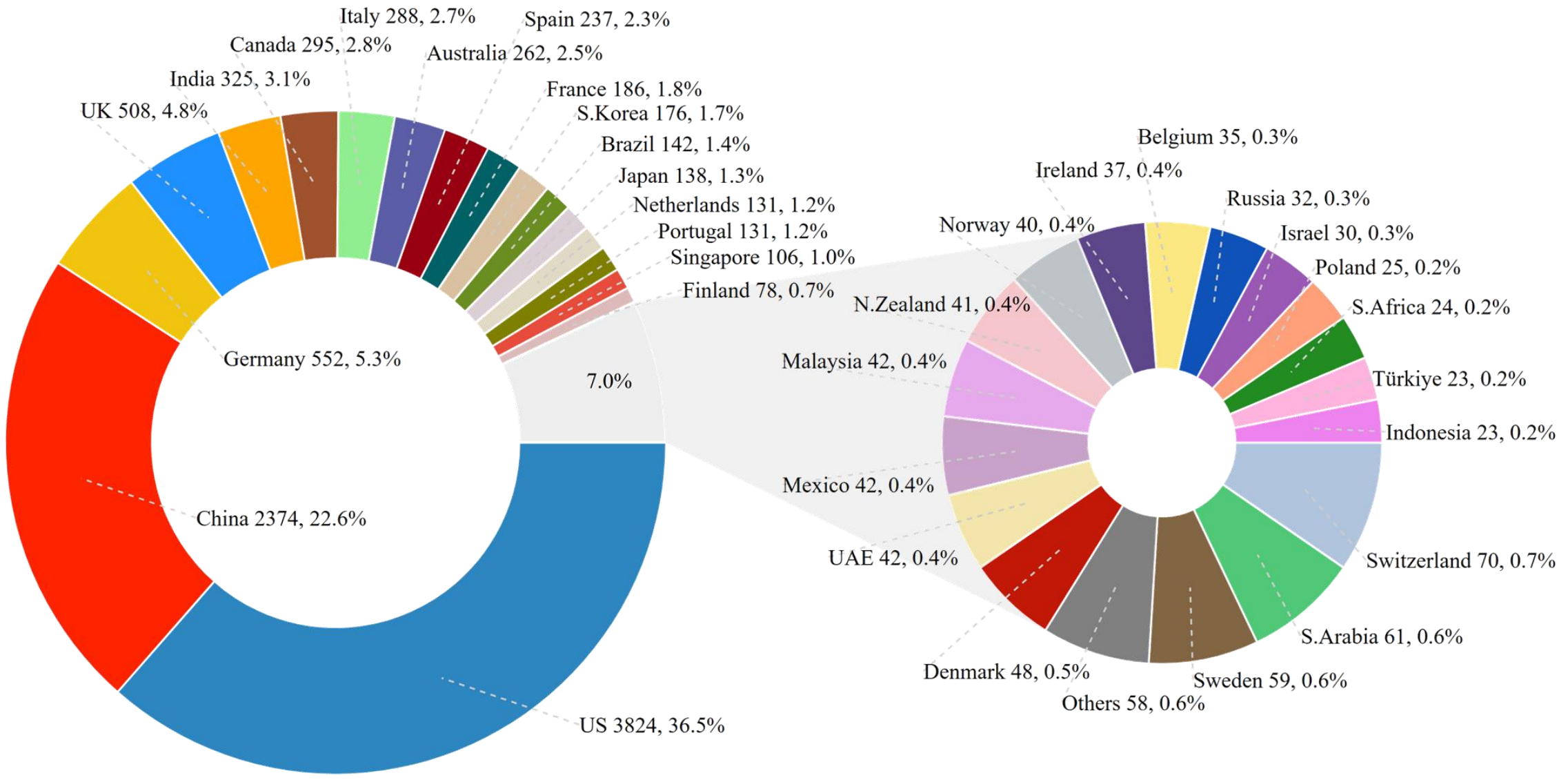

***Figure 26 Country Distribution of AI for SDGs Publications***

*Data source: Based on statistical analysis of the DBLP literature database (From April 2024 to March 2025)*

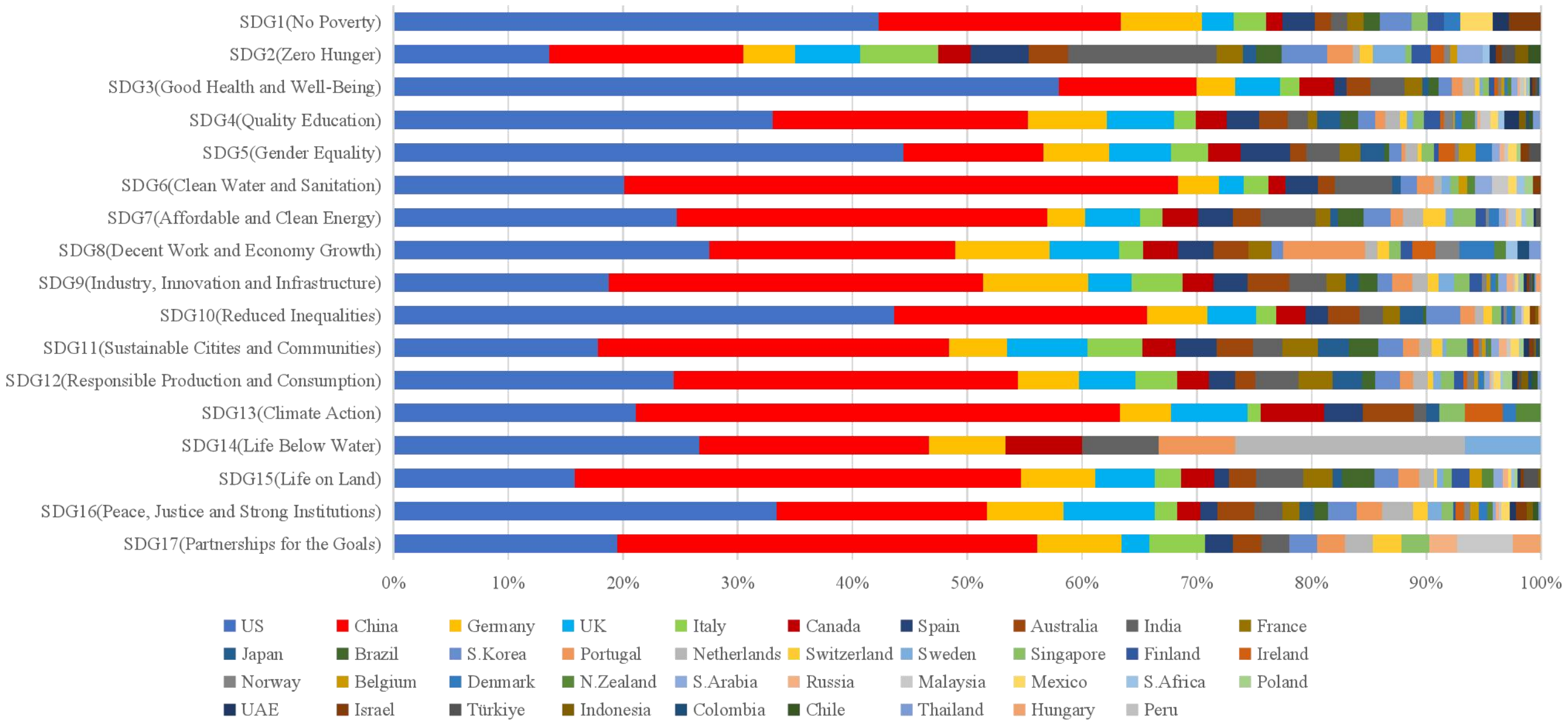

***Figure 27 Country Distribution of AI Research in 17 SDGs***

*Data source: Based on statistical analysis of the DBLP literature database (From April 2024 to March 2025)*

The 40 countries have made varying contributions to advancing AI for the achievement of sustainable development goals, showcasing their distinct efforts and priorities in AI for Good. Overall, the United States and China together account for 59.1% of AI for sustainable development goals publications among these 40 countries. Germany and the United Kingdom also play a significant role in nearly all fields where AI intersects with the SDGs. Meanwhile, other countries have made important contributions in different goals. For instance, India has made considerable contributions to research in fields such as AI and SDG 2 (Zero Hunger), SDG 6 (Clean Water and Sanitation), SDG 7 (Affordable and Clean Energy), and SDG 14 (Life Below Water); the Netherlands has also demonstrated notable performance in the field of AI and SDG 14.

## Observation 4.9:

SDG3 (Good Health and Well-Being), SDG11 (Sustainable Cities and Communities), and SDG9 (Industry, Innovation and Infrastructure) have received widespread research attention. Meanwhile, High-Income Countries have allocated higher AI research attention to social-level SDGs, while Upper-Middle&Lower-Middle Income Countries have relatively higher AI research attention to economically and environmentally related SDGs.

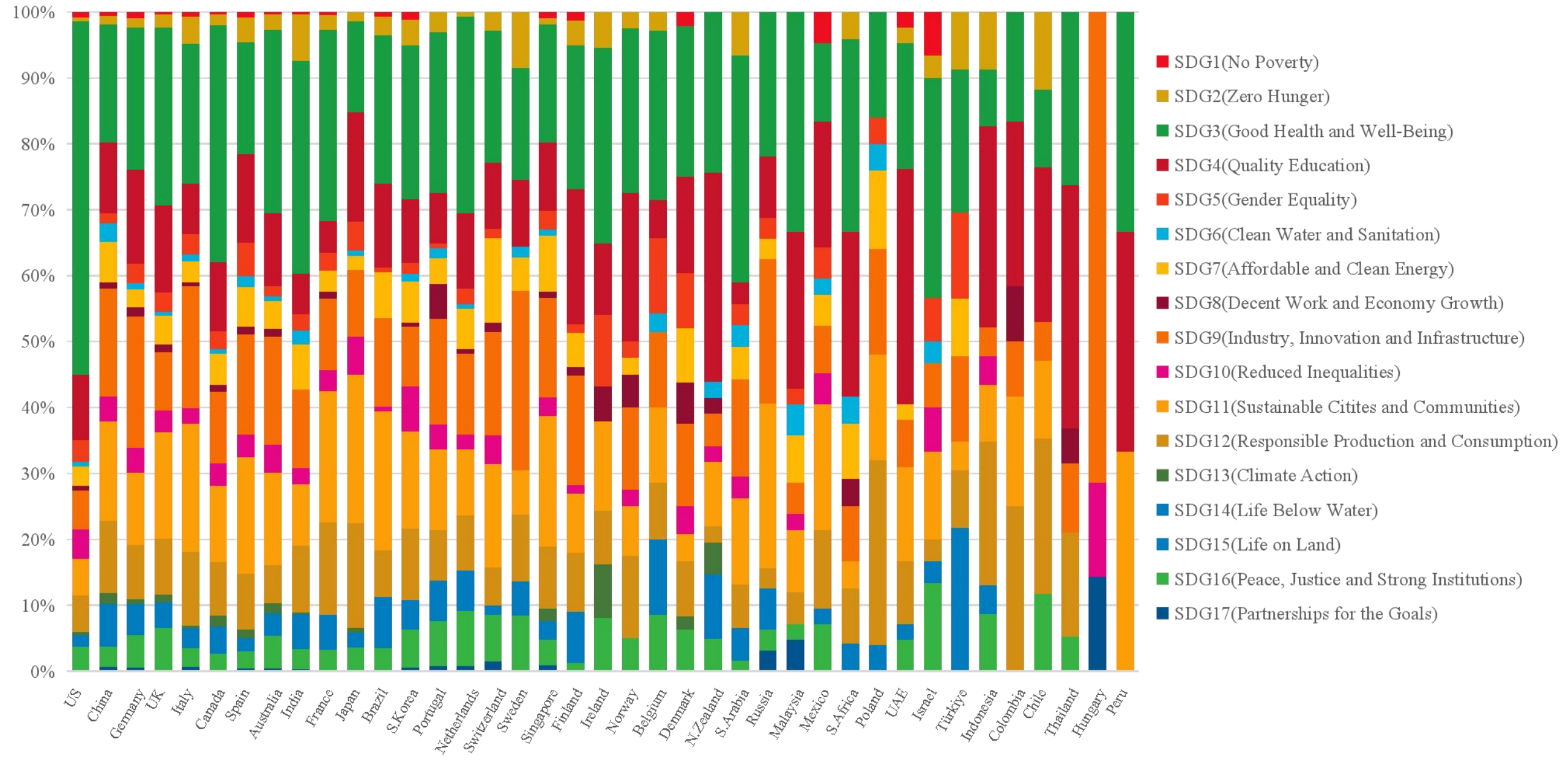

***Figure 28 AI for SDGs Research Distribution in Each Country***

*Data source: Based on statistical analysis of the DBLP literature database (From April 2024 to March 2025)*

All 40 countries show a relevant consistent focus in the literature on AI’ s role in advancing sustainable development goals. Among these, SDG 3 (Good Health and Well-Being) is the most popular, followed by SDG 11 (Sustainable Cities and Communities), SDG 9 (Industry, Innovation, and Infrastructure), SDG 4 (Quality Education), and SDG 12 (Responsible Consumption and Production).

Meanwhile, a comparison of the distribution of research attention across the 17 Sustainable Development Goals between the two country groups shows that the high-income country group has relatively higher attention to goals such as SDG 3 (Good Health and Well-being), SDG 5 (Gender Equality), and SDG 16 (Peace, Justice and Strong Institutions) than the upper-middle&lower-middle income country group. In contrast, the latter have relatively higher research attention to goals including SDG 6 (Clean Water and Sanitation), SDG 7 (Affordable and Clean Energy), SDG 9 (Industry, Innovation and Infrastructure), and SDG 11 (Sustainable Cities and Communities).

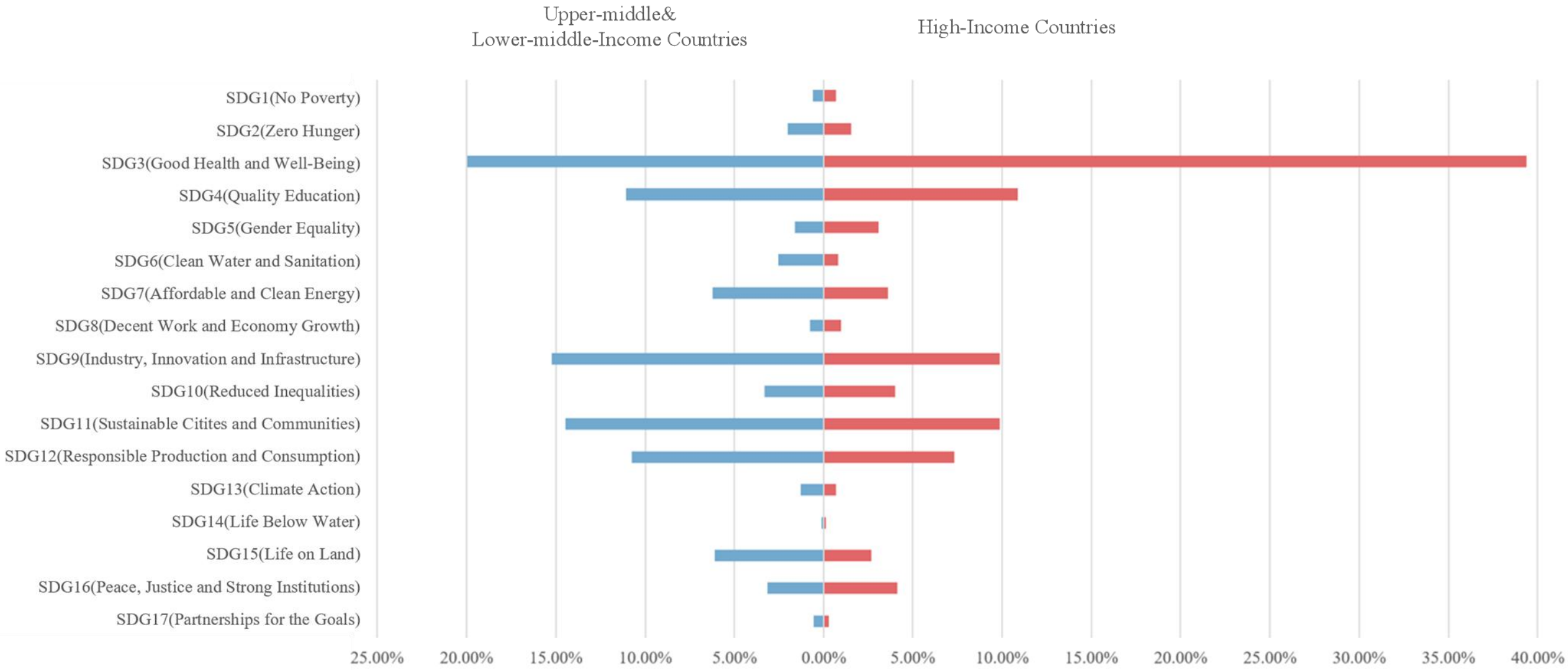

***Figure 29 Comparison of AI for SDGs Publication Topics between High-Income Countries and Upper-Middle&Lower-Middle-Income Countries***

*Data source: Based on statistical analysis of the DBLP literature database (From April 2024 to March 2025; the horizontal axis represents the proportion of AI publications on each SDG by the two country groups relative to all SDG-related AI publications during the period)*

# IV. Country Profile

# Argentina

| AGILE Index Ranking | Population(2024) | GDP Per Capita(2024) | Country Group |
|---|---|---|---|
| 38/40 | 46 Million | 17,144 $ | Upper-Middle Income |

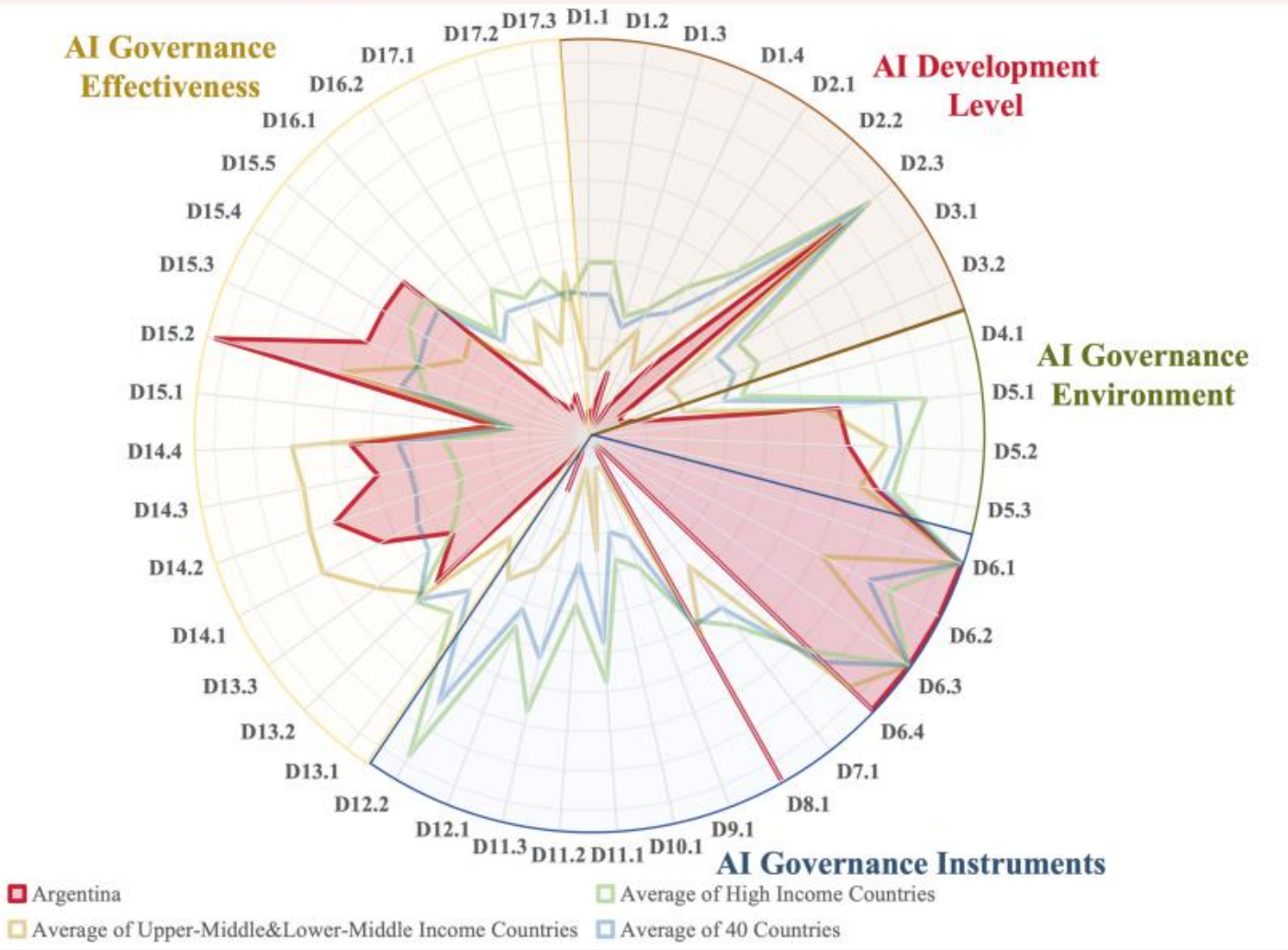

Pillar Score and Ranking

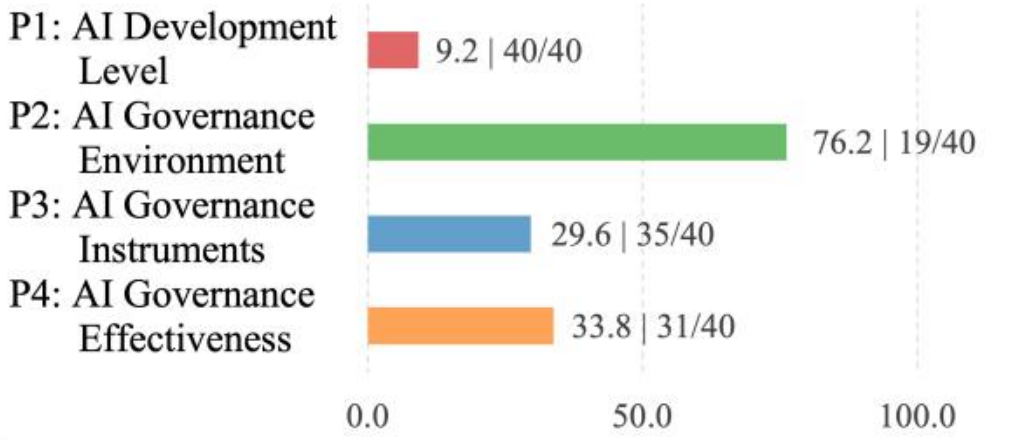

**Highest Rankings in:**

D15 AI Development Inclusivity (Ranking: 3/40)
D4 AI Risk Exposure (Ranking: 9/40)

**Lowest Rankings in:**

D13 Public Understanding of AI (Ranking: 40/40)
D17 AI Governance Research Activity (Ranking: 40/40)

# Australia

| AGILE Index Ranking | Population(2024) | GDP Per Capita(2024) | Country Group |
|---|---|---|---|
| 14/40 | 27 Million | 81,355 $ | High Income |

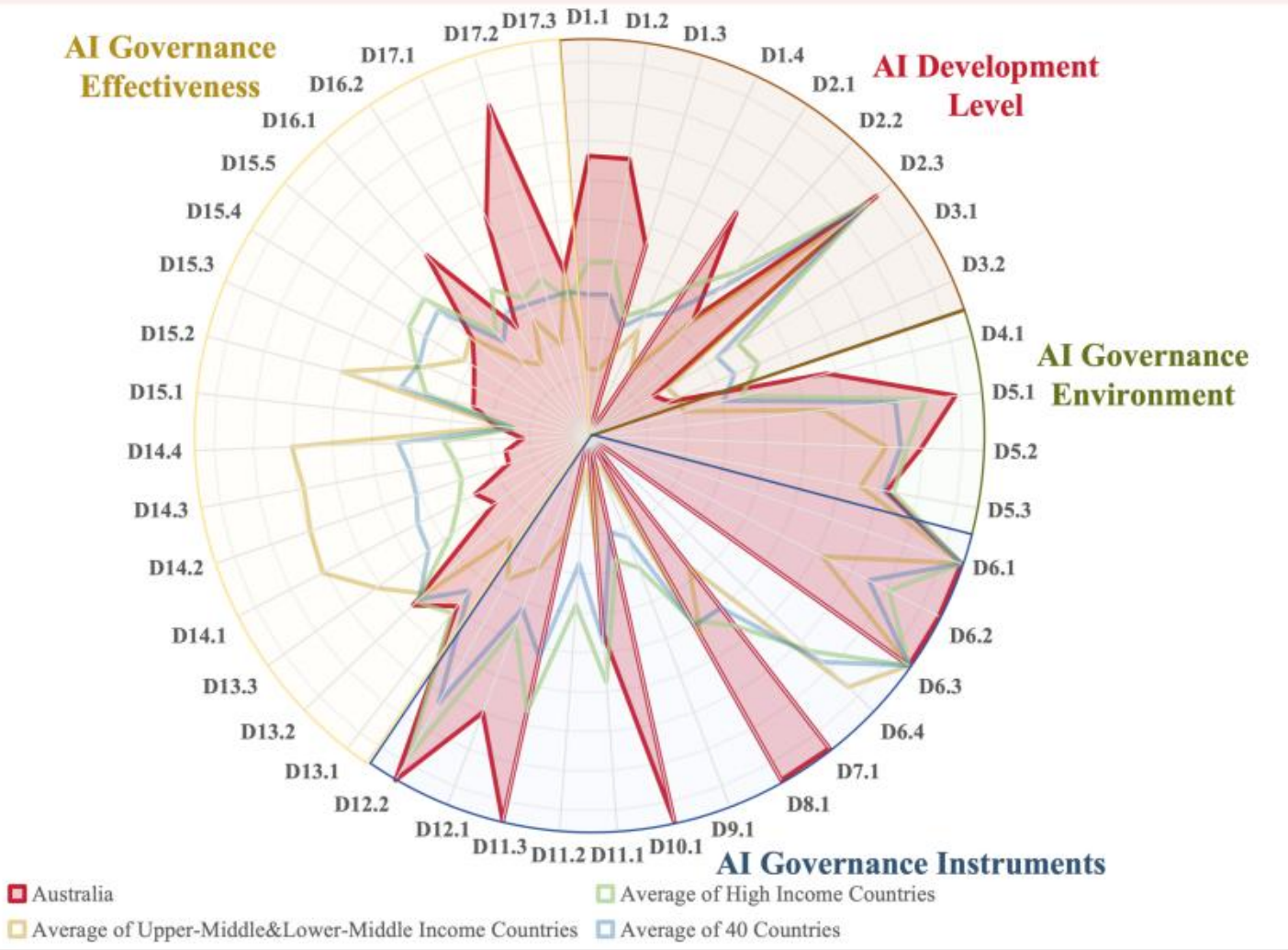

Pillar Score and Ranking

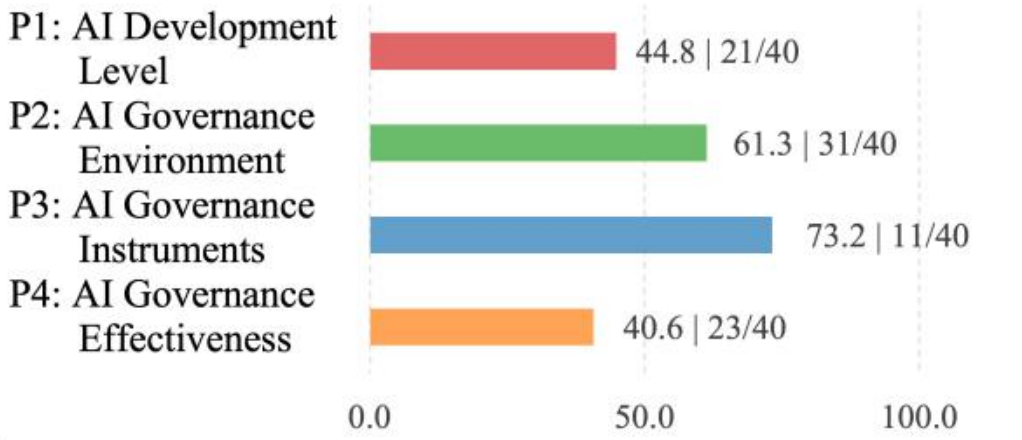

**Highest Rankings in:**

D17 AI Governance Research Activity (Ranking: 3/40)
D1 AI Research and Development Activity (Ranking: 9/40)

**Lowest Rankings in:**

D14 AI Social Acceptance (Ranking: 35/40)
D4 AI Risk Exposure (Ranking: 34/40)

# Belgium

| AGILE Index Ranking | Population(2024) | GDP Per Capita(2024) | Country Group |
|---|---|---|---|
| 28/40 | 12 Million | 59,317 $ | High Income |

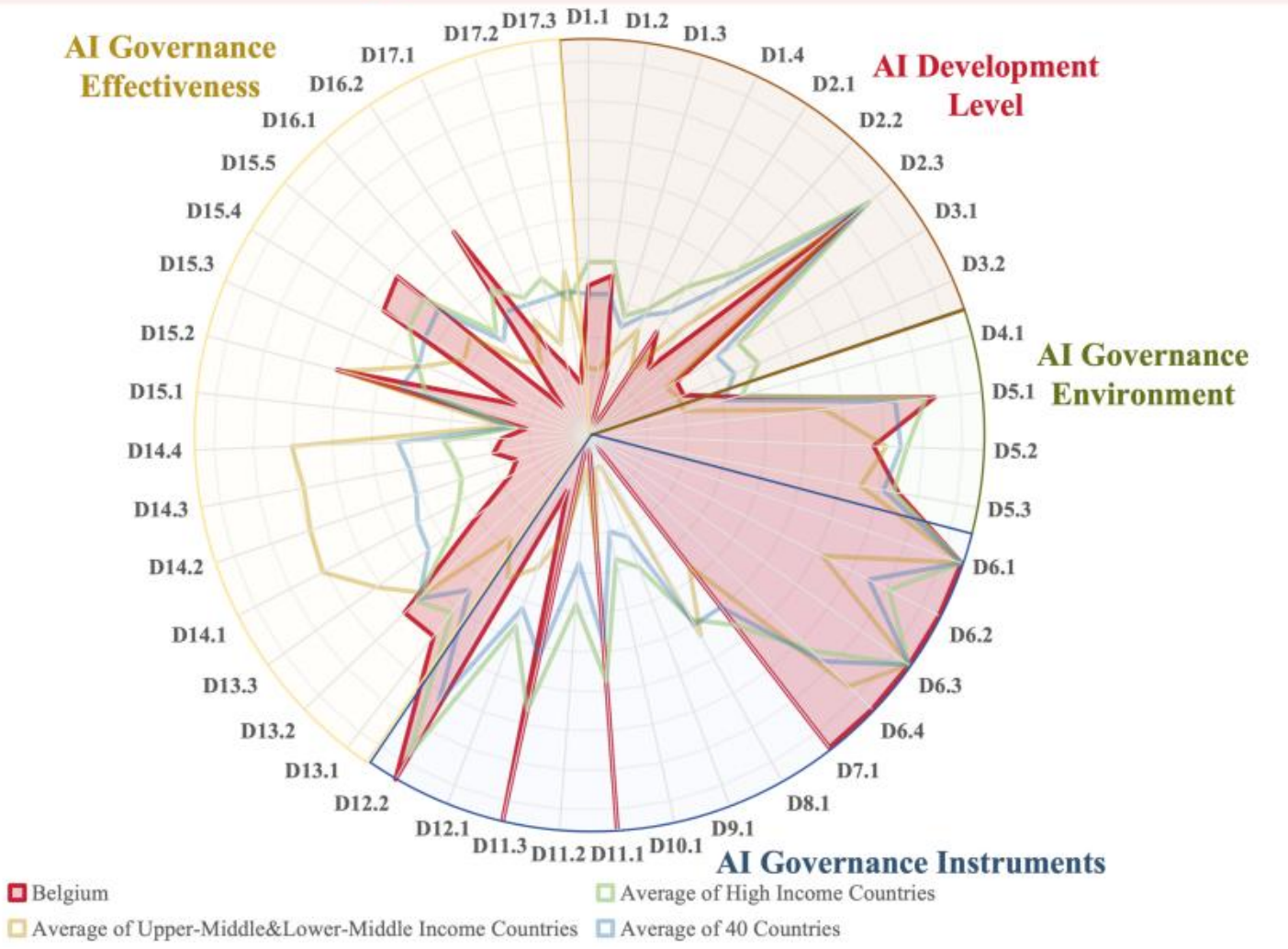

Pillar Score and Ranking

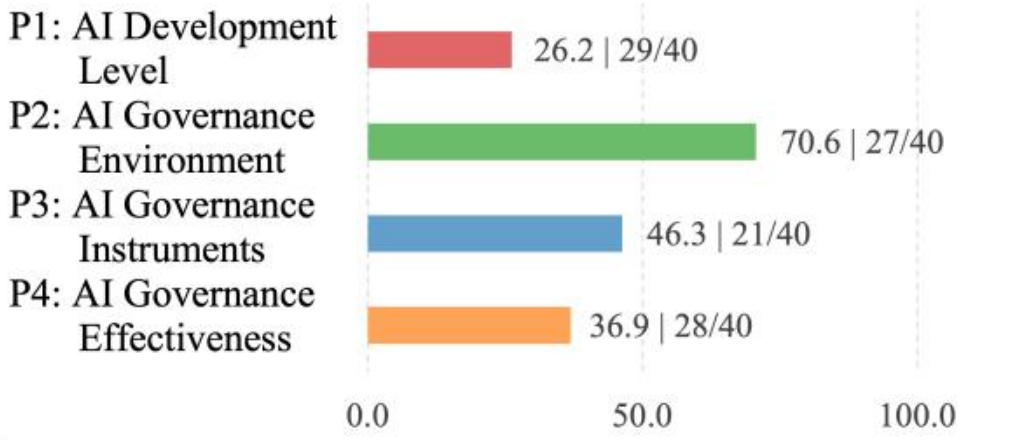

**Highest Rankings in:**

D13 Public Understanding of AI (Ranking: 16/40)
D16 Data & Algorithm Openness (Ranking: 16/40)

**Lowest Rankings in:**

D14 AI Social Acceptance (Ranking: 36/40)
D17 AI Governance Research Activity (Ranking: 36/40)

# Brazil

| AGILE Index Ranking | Population(2024) | GDP Per Capita(2024) | Country Group |
|---|---|---|---|
| 36/40 | 212 Million | 12,263 $ | Upper-Middle Income |

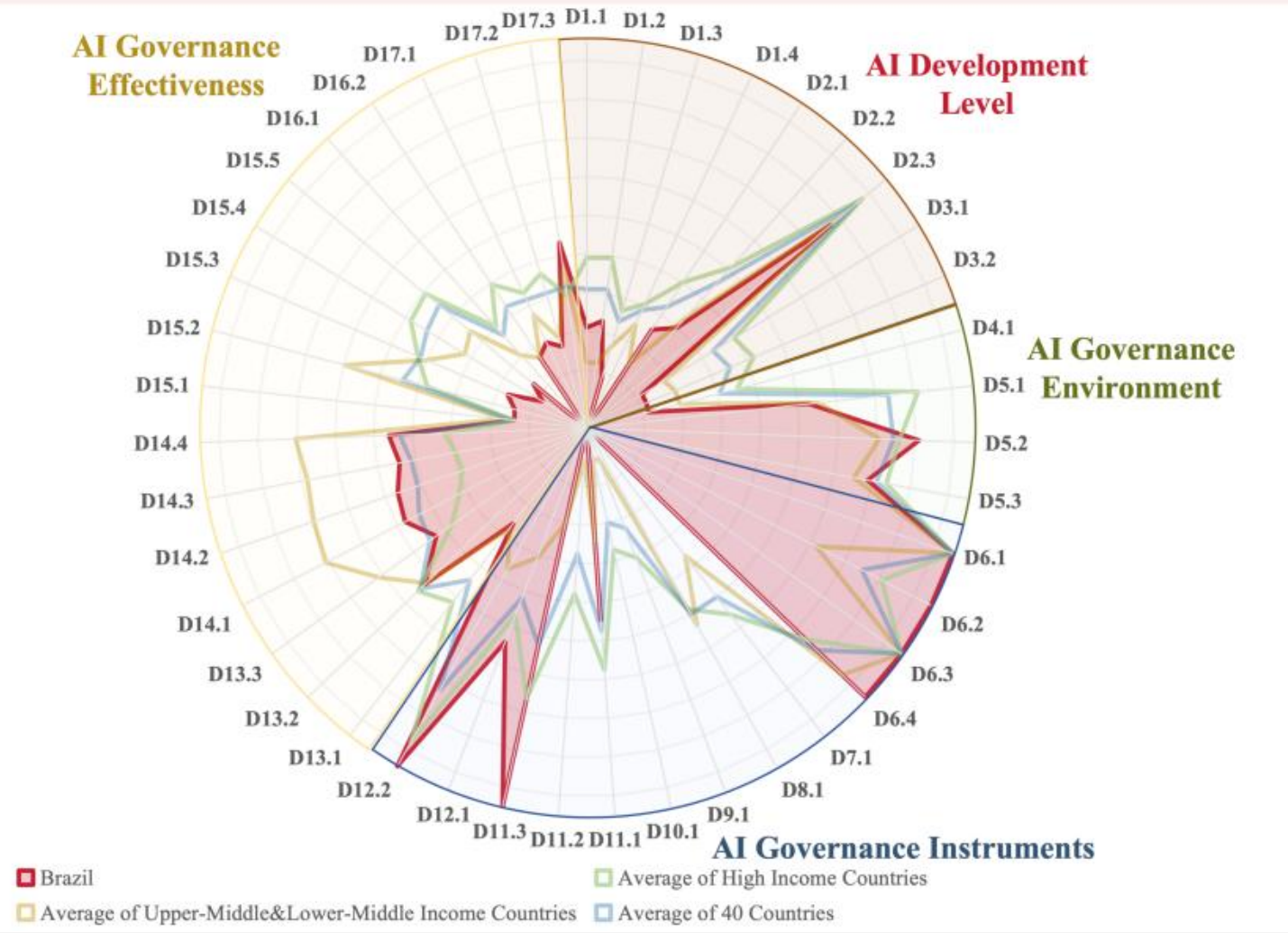

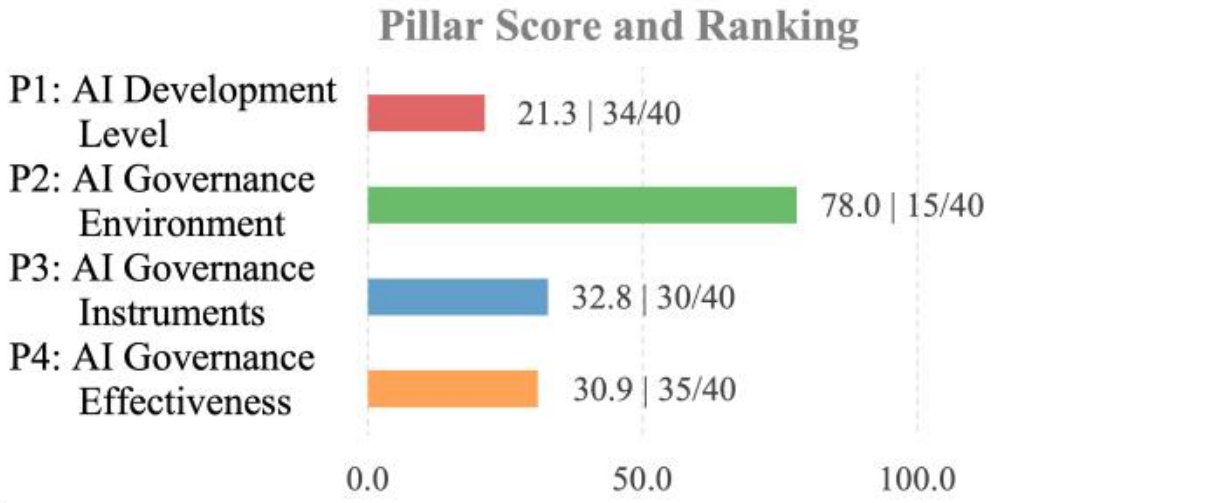

**Highest Rankings in:**

D4 AI Risk Exposure (Ranking: 11/40)
D14 AI Social Acceptance (Ranking: 18/40)

**Lowest Rankings in:**

D15 AI Development Inclusivity (Ranking: 37/40)
D3 AI Industry Vitality (Ranking: 36/40)

# Canada

| AGILE Index Ranking | Population(2024) | GDP Per Capita(2024) | Country Group |
|---|---|---|---|
| 8/40 | 40 Million | 59,564 $ | High Income |

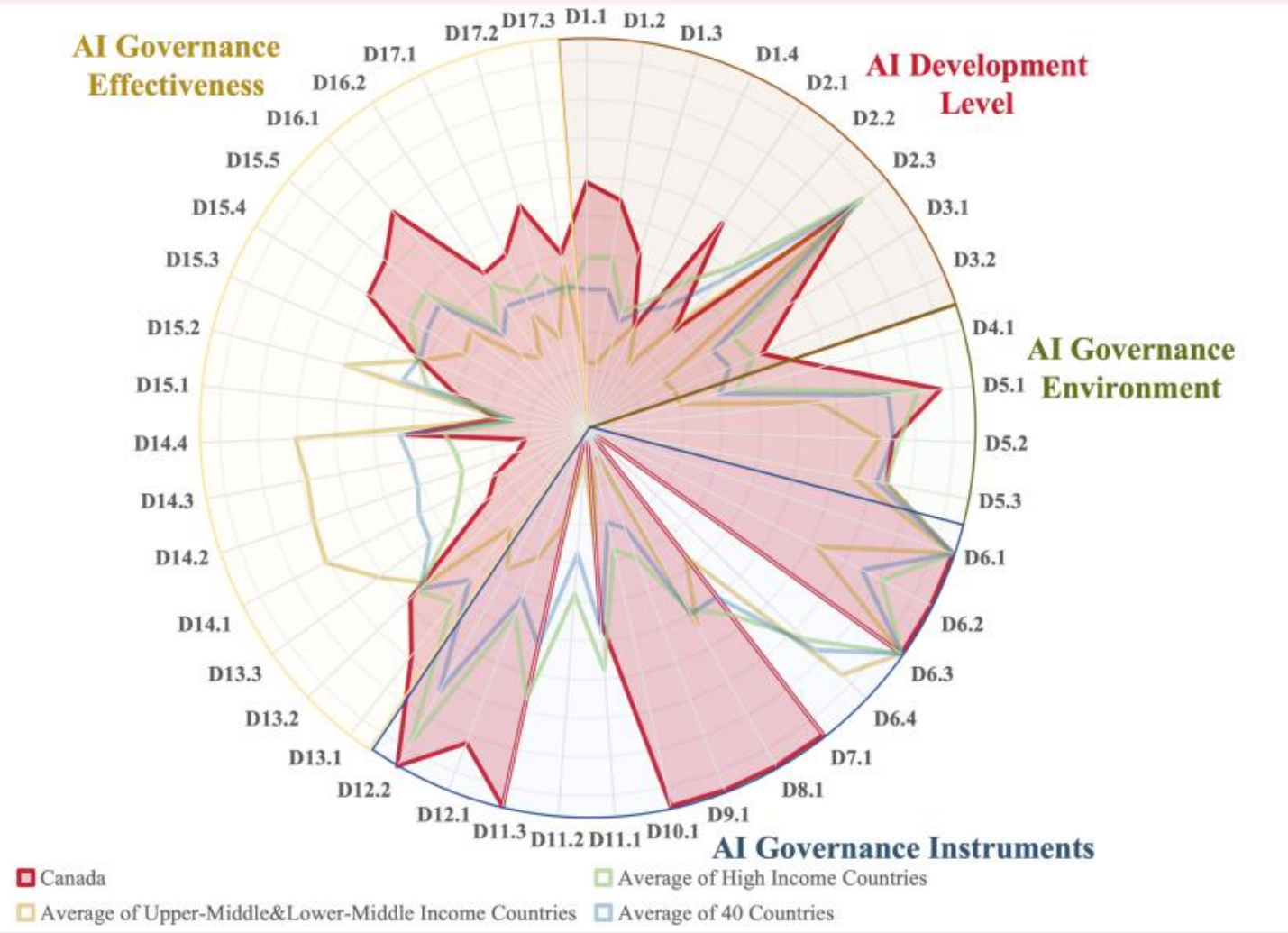

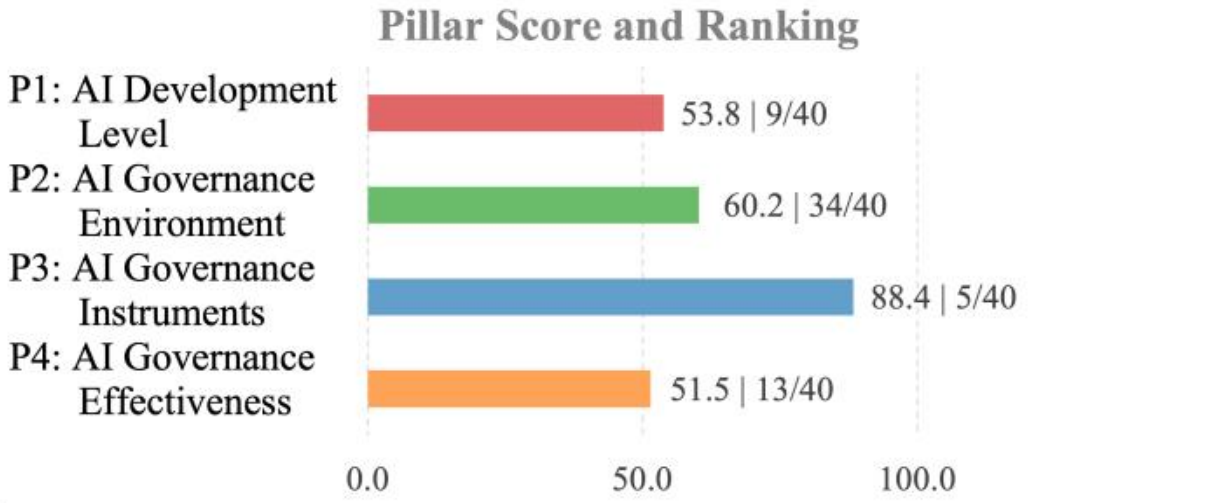

**Highest Rankings in:**

P3 Governance Instruments (Ranking: 5/40)
D1 AI Research and Development Activity (Ranking: 7/40)

**Lowest Rankings in:**

D4 AI Risk Exposure (Ranking: 35/40)
D14 AI Social Acceptance (Ranking: 31/40)

# Chile

| AGILE Index Ranking | Population(2024) | GDP Per Capita(2024) | Country Group |
|---|---|---|---|
| 29/40 | 20 Million | 18,820 $ | High Income |

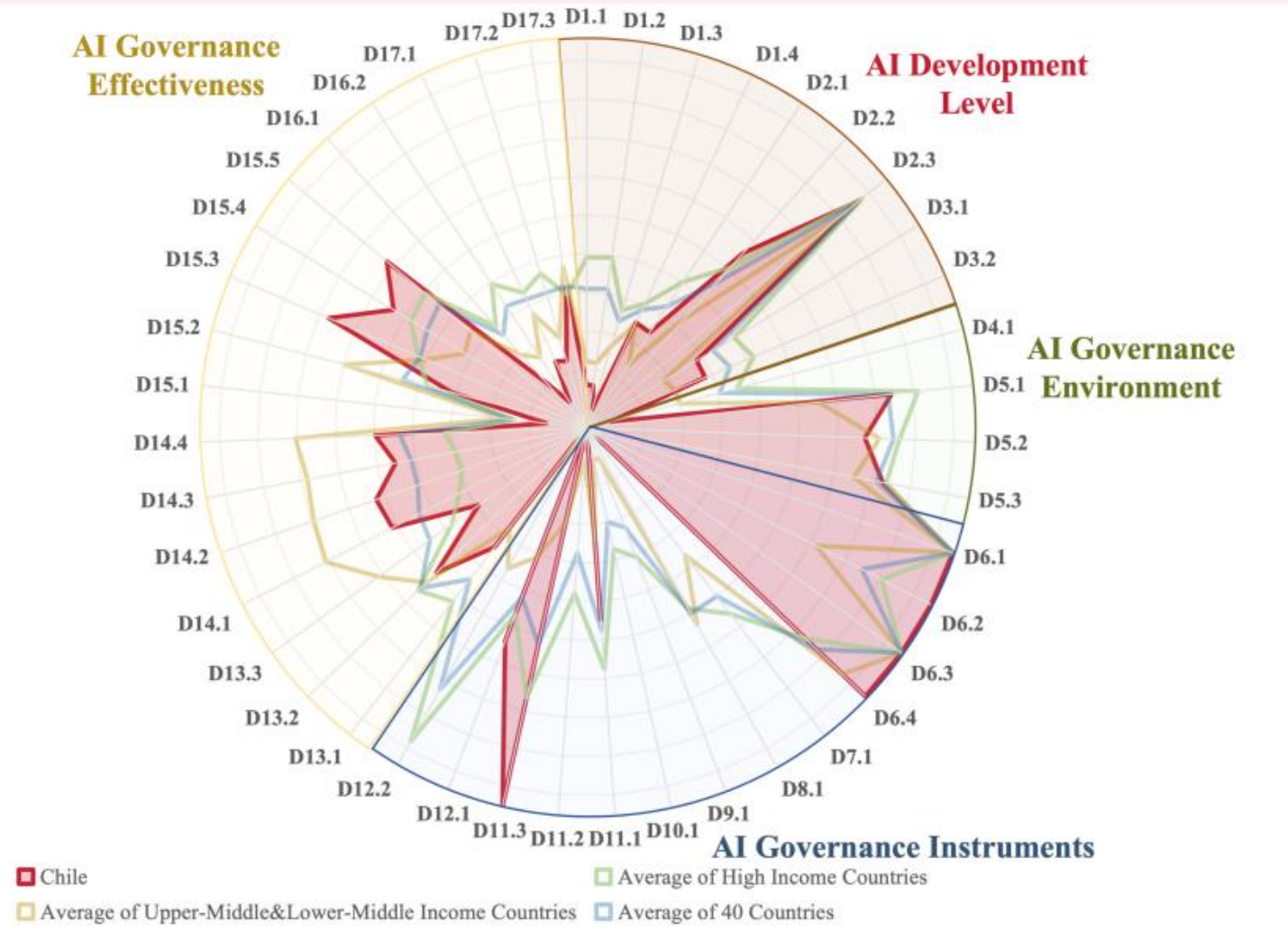

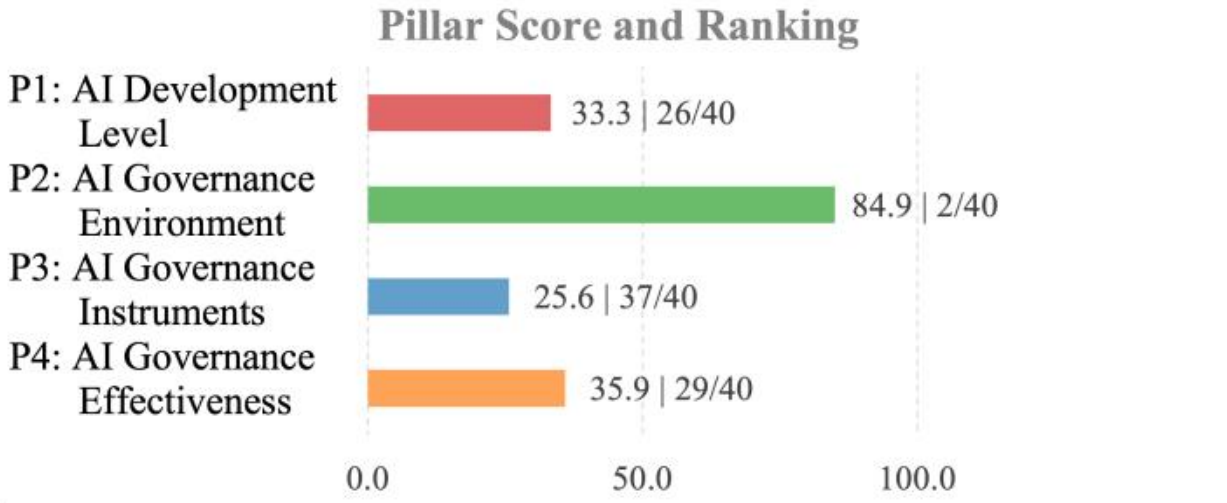

**Highest Rankings in:**

D4 AI Risk Exposure (Ranking: 4/40)
D14 AI Social Acceptance (Ranking: 14/40)

**Lowest Rankings in:**

P3 Governance Instruments (Ranking: 37/40)
D13 Public Understanding of AI (Ranking: 35/40)

# China

| AGILE Index Ranking | Population(2024) | GDP Per Capita(2024) | Country Group |
|---|---|---|---|
| 1/40 | 1,451 Million | 15,745 $ | Upper-Middle Income |

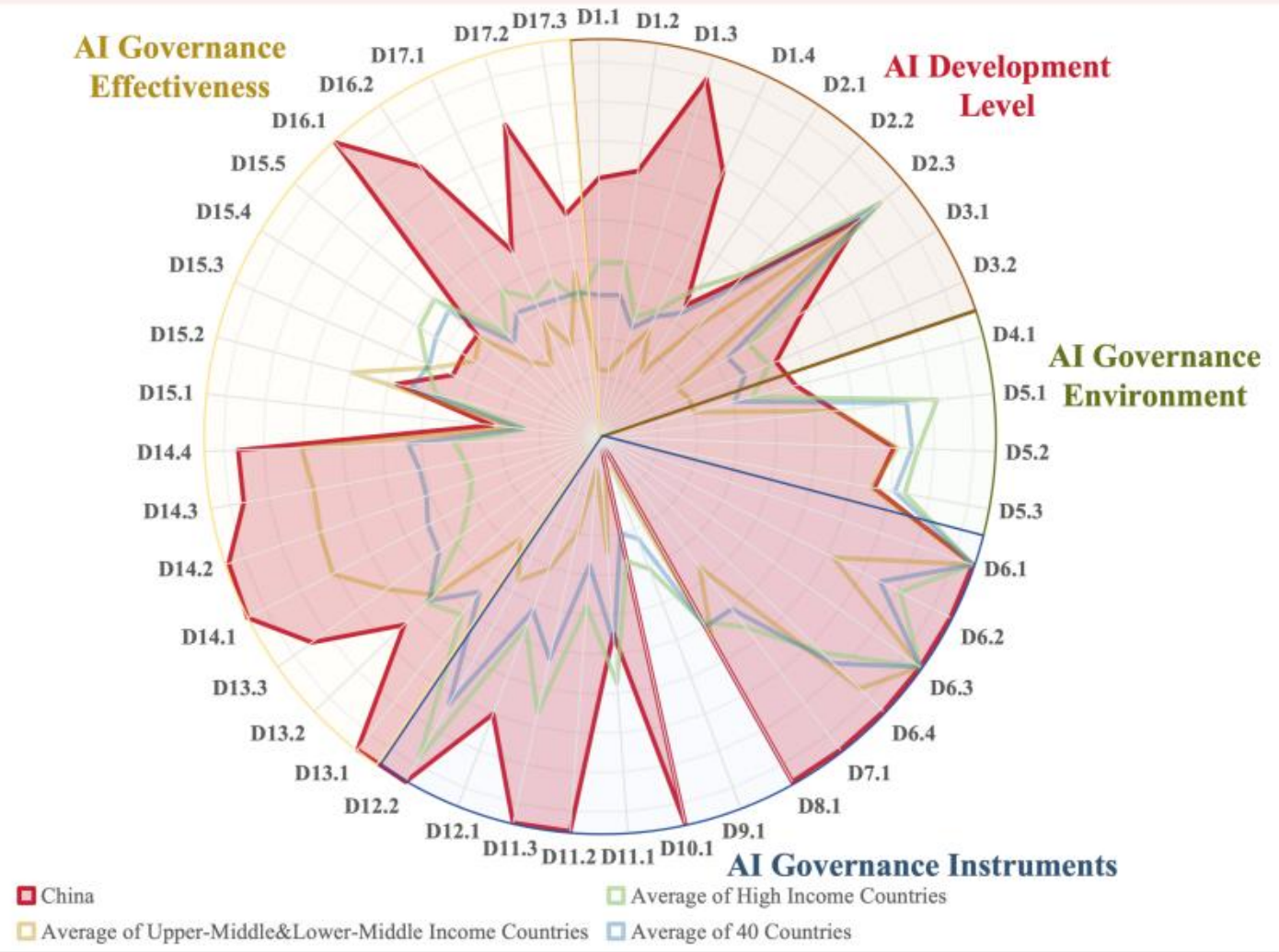

Pillar Score and Ranking

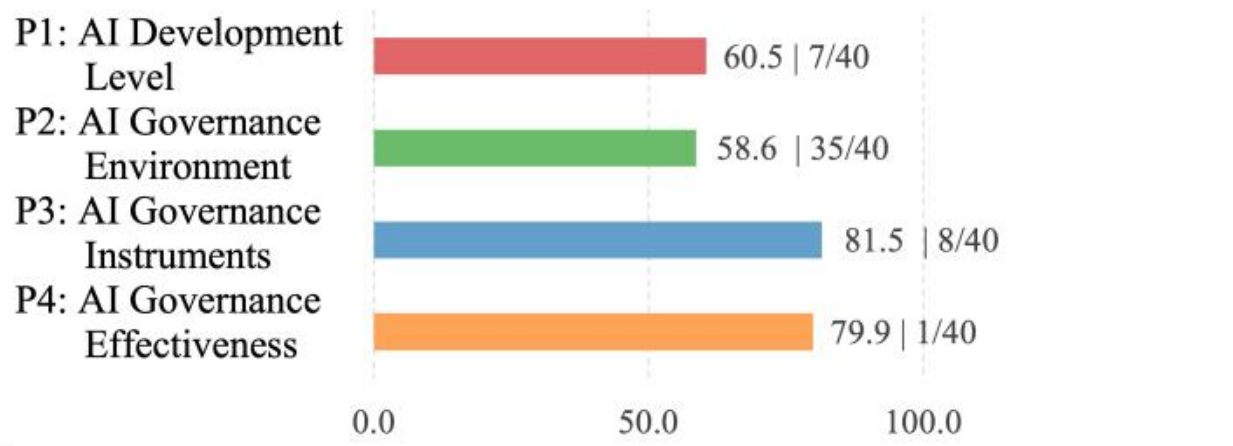

**Highest Rankings in:**
D13 Public Understanding of AI (Ranking: 1/40)
D14 AI Social Acceptance (Ranking: 1/40)

**Lowest Rankings in:**
D5 Overall Governance Readiness (Ranking: 32/40)
D4 AI Risk Exposure (Ranking: 30/40)

# Colombia

| AGILE Index Ranking | Population(2024) | GDP Per Capita(2024) | Country Group |
|---|---|---|---|
| 39/40 | 53 Million | 8,973 $ | Upper-Middle Income |

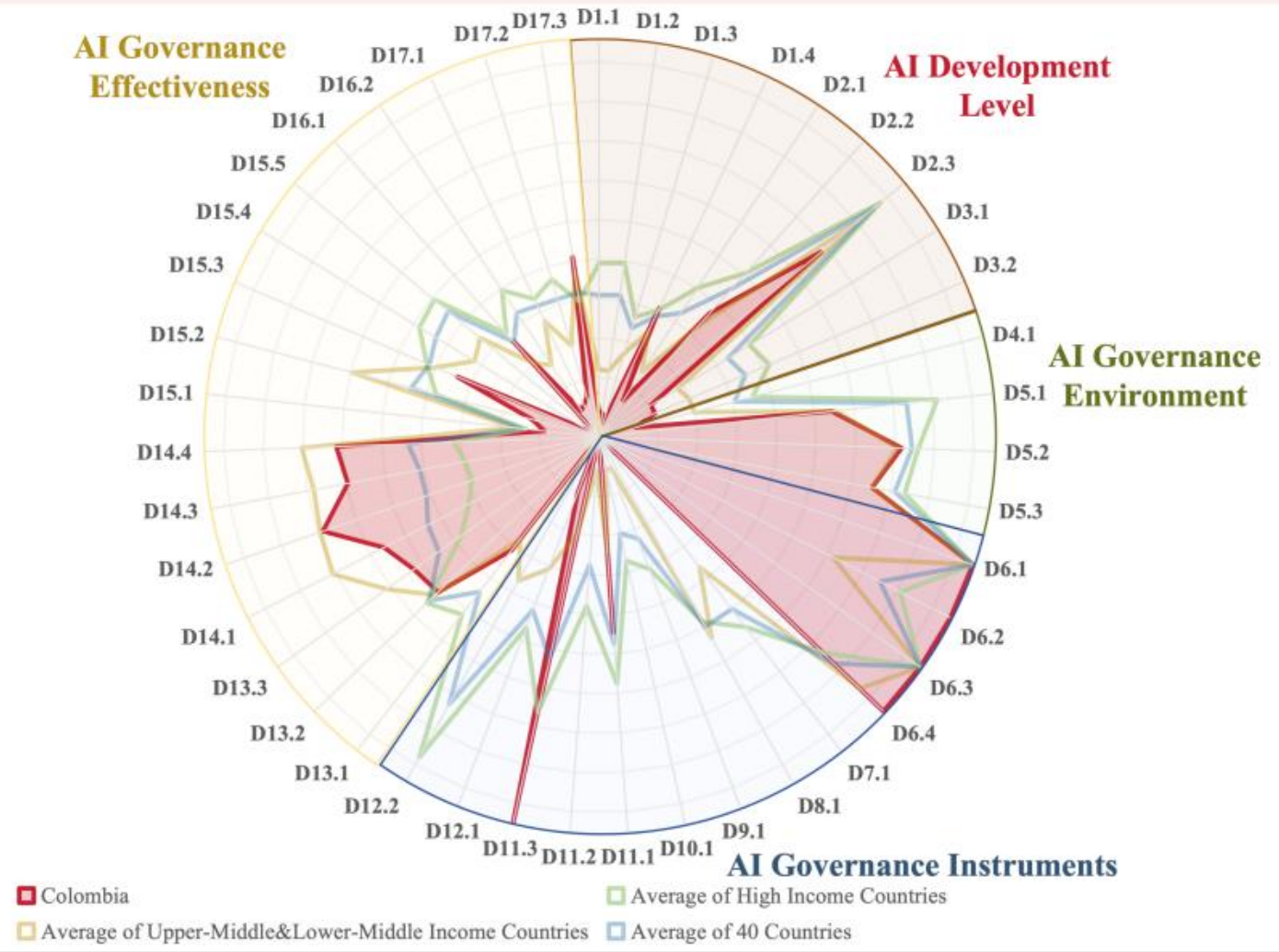

Pillar Score and Ranking

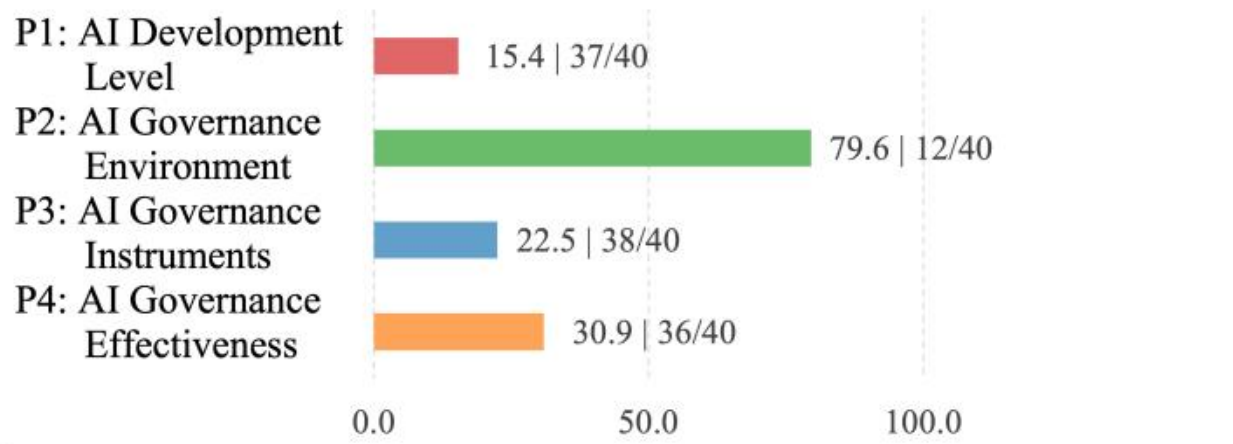

**Highest Rankings in:**
D4 AI Risk Exposure (Ranking: 5/40)
D14 AI Social Acceptance (Ranking: 9/40)

**Lowest Rankings in:**
D15 AI Development Inclusivity (Ranking: 40/40)
P3 Governance Instruments (Ranking: 38/40)

# Denmark

| AGILE Index Ranking | Population(2024) | GDP Per Capita(2024) | Country Group |
|---|---|---|---|
| 15/40 | 5,977 Thousand | 80,759 $ | High Income |

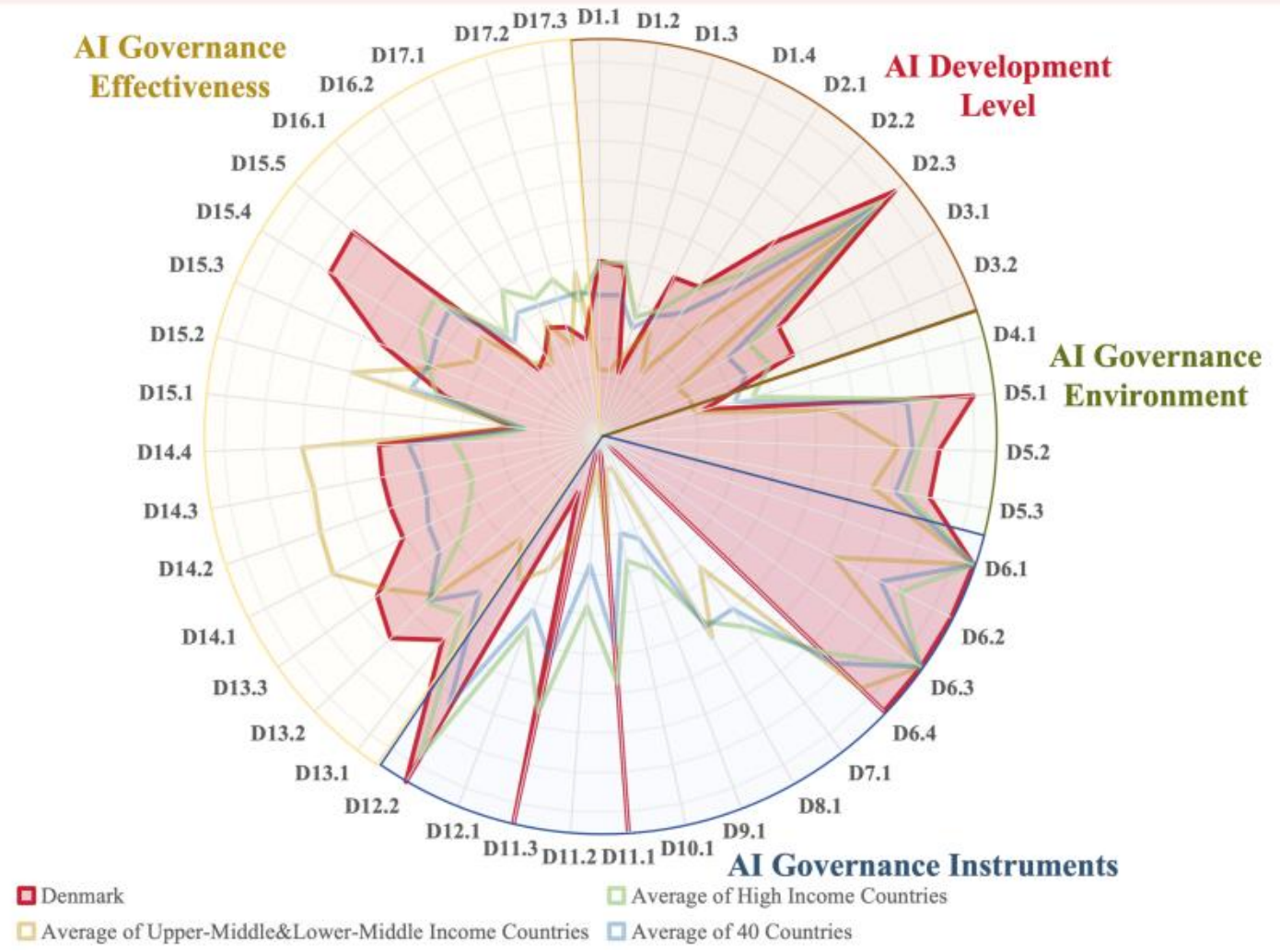

Pillar Score and Ranking

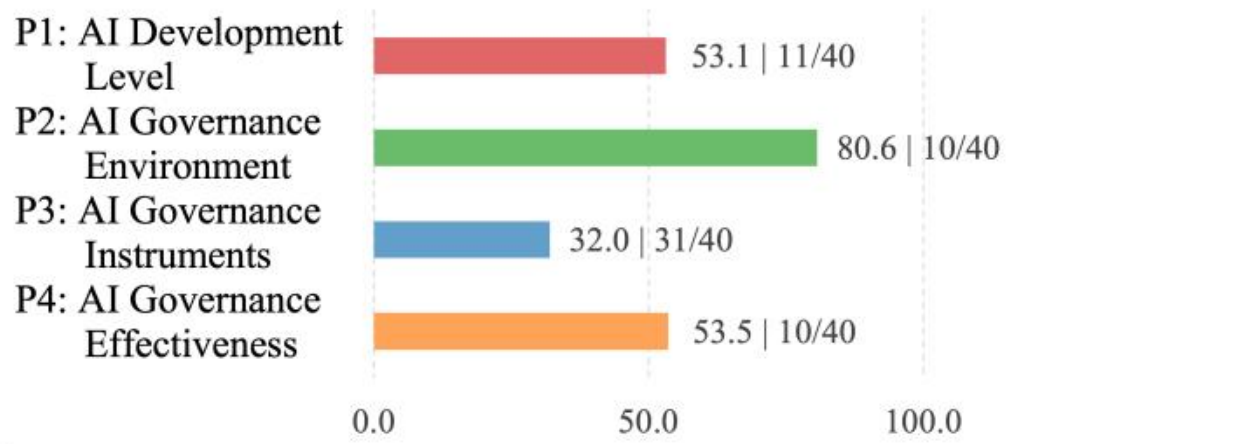

**Highest Rankings in:**
D5 Overall Governance Readiness (Ranking: 1/40)
D13 Public Understanding of AI (Ranking: 4/40)

**Lowest Rankings in:**
P3 Governance Instruments (Ranking: 31/40)
D17 AI Governance Research Activity (Ranking: 28/40)

# Finland

| AGILE Index Ranking | Population(2024) | GDP Per Capita(2024) | Country Group |
|---|---|---|---|
| 10/40 | 5,617 Thousand | 60,739 $ | High Income |

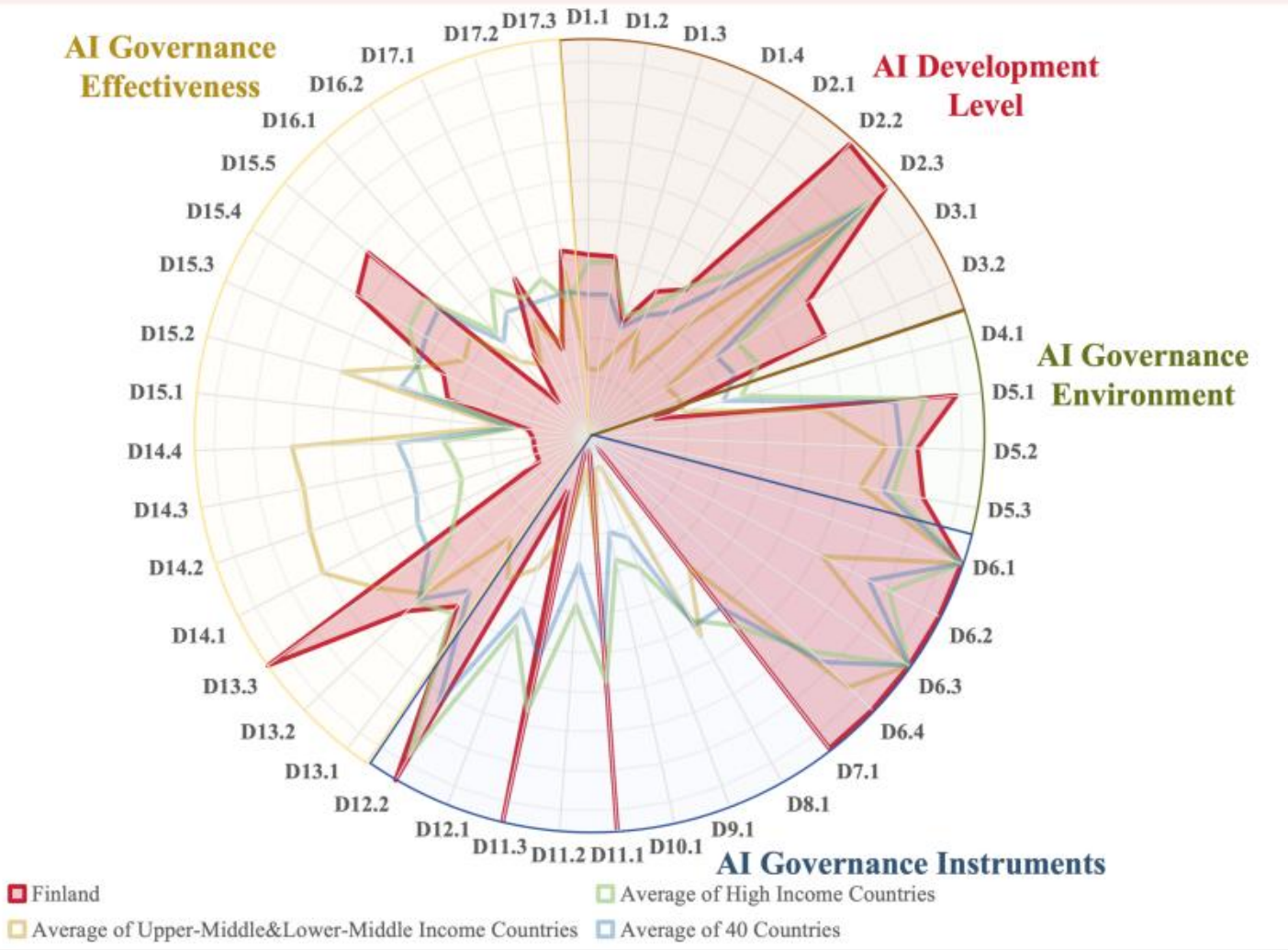

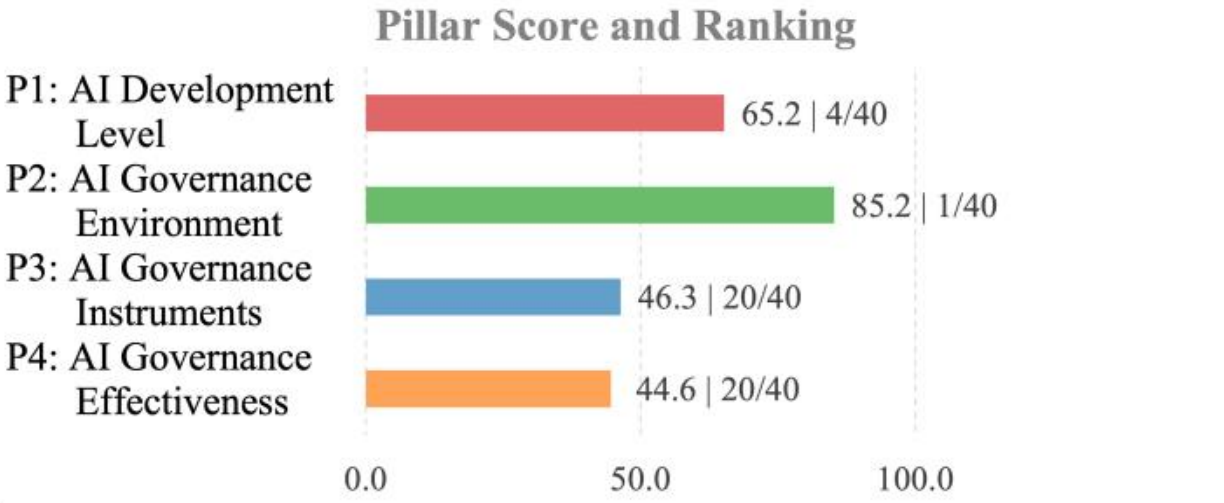

**Highest Rankings in:**

D5 Overall Governance Readiness (Ranking: 2/40)
D2 AI Infrastructure (Ranking: 3/40)

**Lowest Rankings in:**

D14 AI Social Acceptance (Ranking: 39/40)
D16 Data & Algorithm Openness (Ranking: 24/40)

# France

| AGILE Index Ranking | Population(2024) | GDP Per Capita(2024) | Country Group |
|---|---|---|---|
| 7/40 | 67 Million | 53,101 $ | High Income |

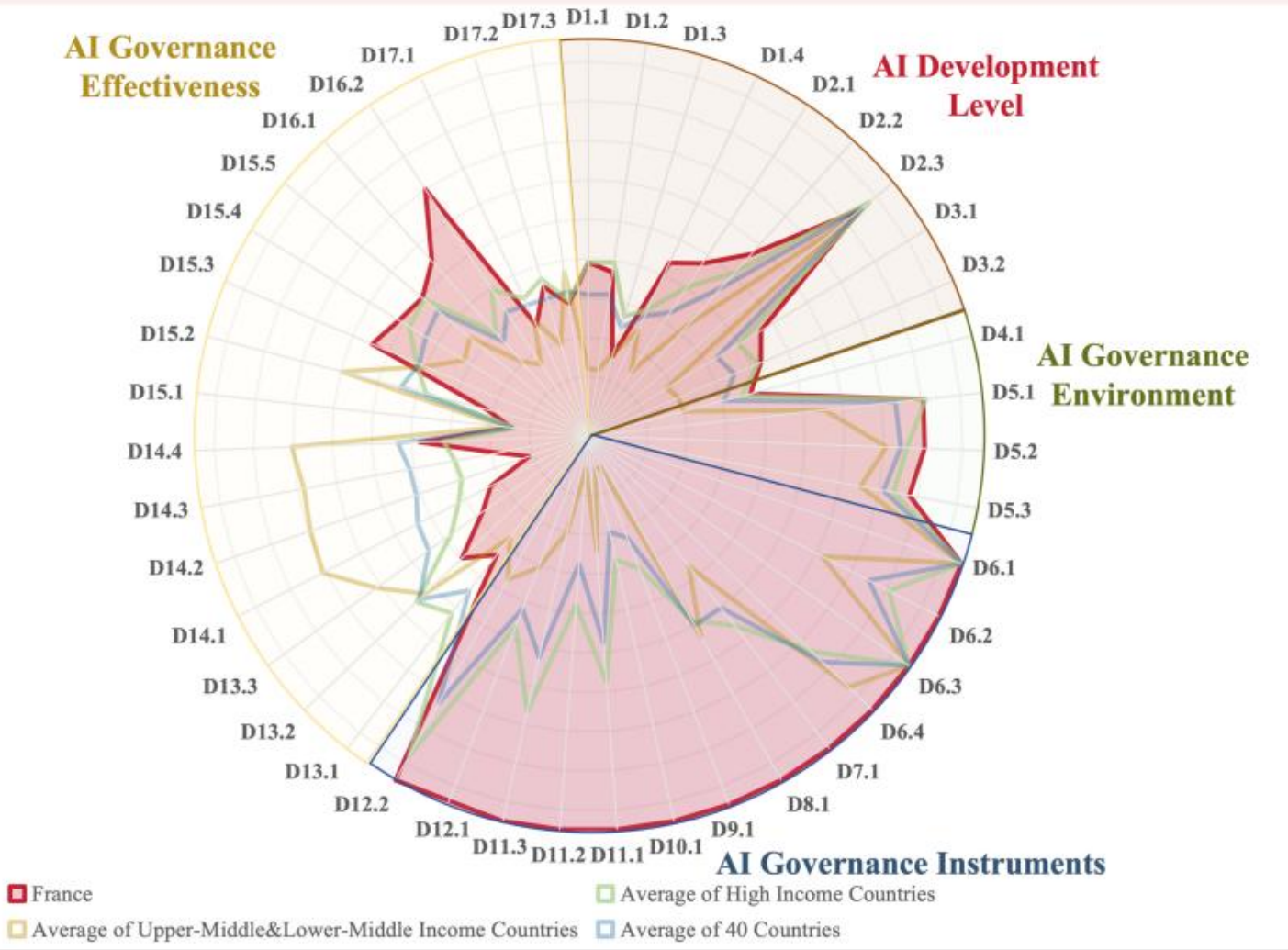

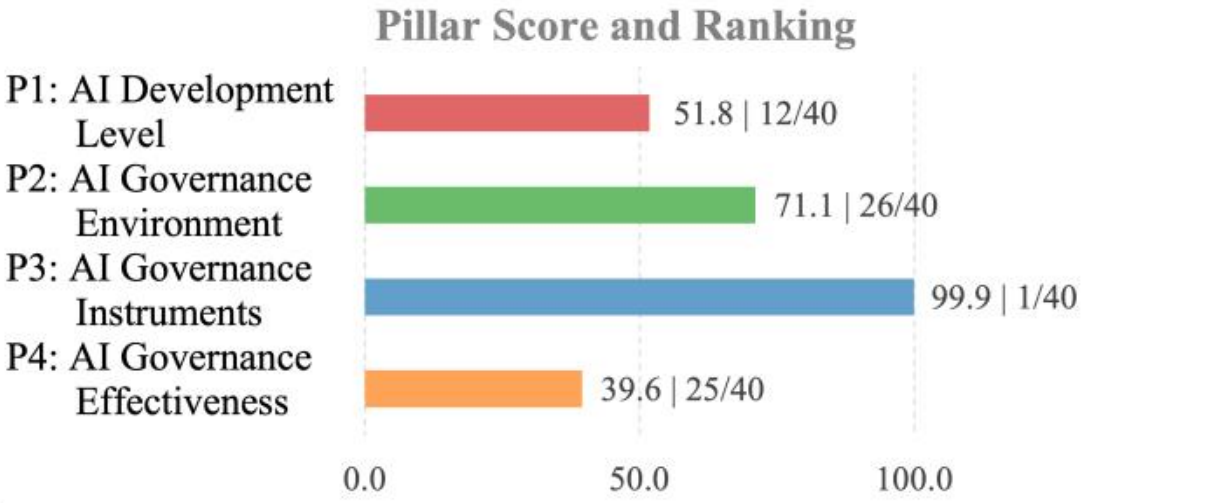

**Highest Rankings in:**

P3 Governance Instruments (Ranking: 1/40)
D16 Data & Algorithm Openness (Ranking: 6/40)

**Lowest Rankings in:**

D13 Public Understanding of AI (Ranking: 38/40)
D14 AI Social Acceptance (Ranking: 30/40)

# Germany

| AGILE Index Ranking | Population(2024) | GDP Per Capita(2024) | Country Group |
|---|---|---|---|
| 3/40 | 72 Million | 67,226 $ | High Income |

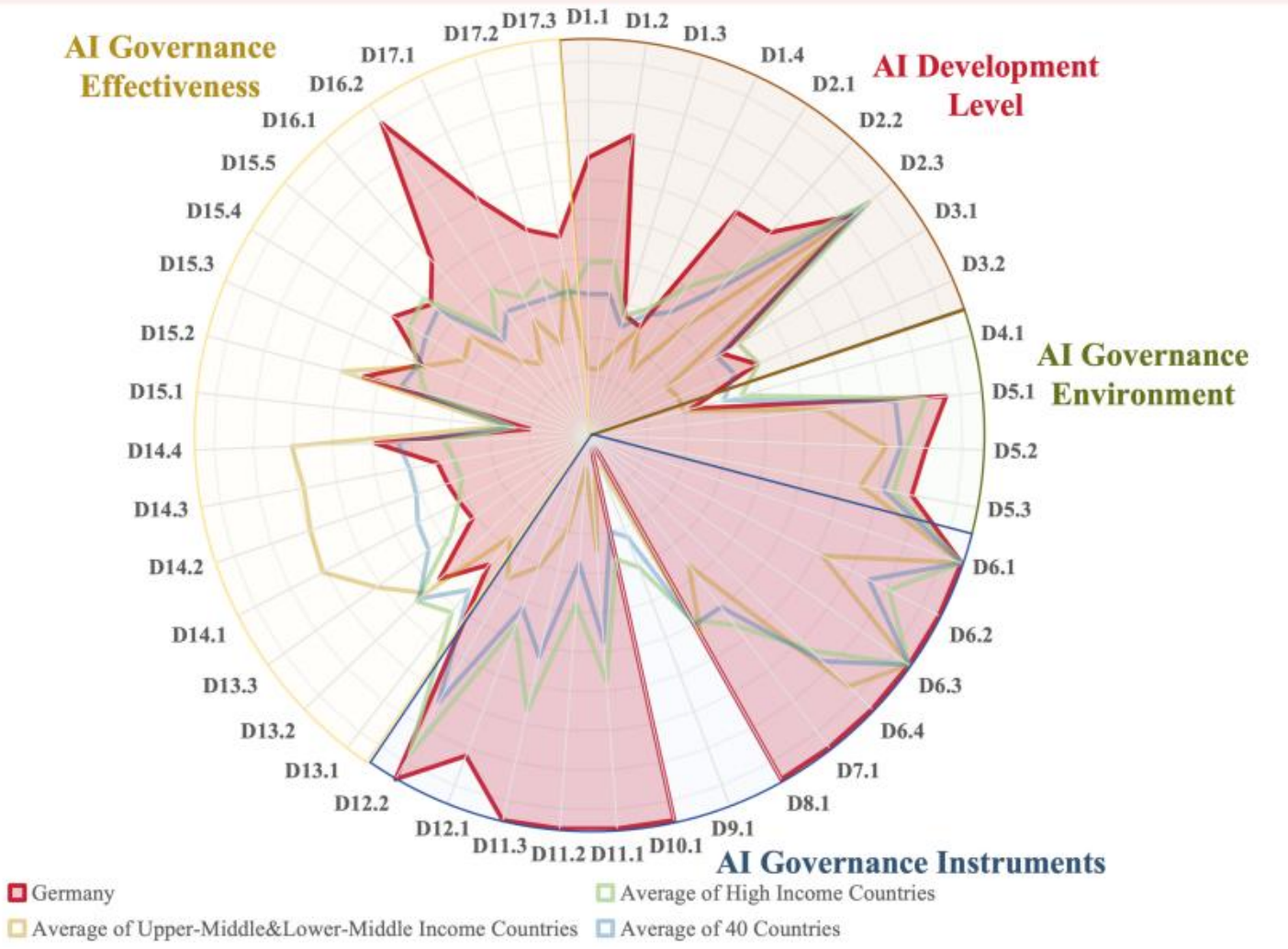

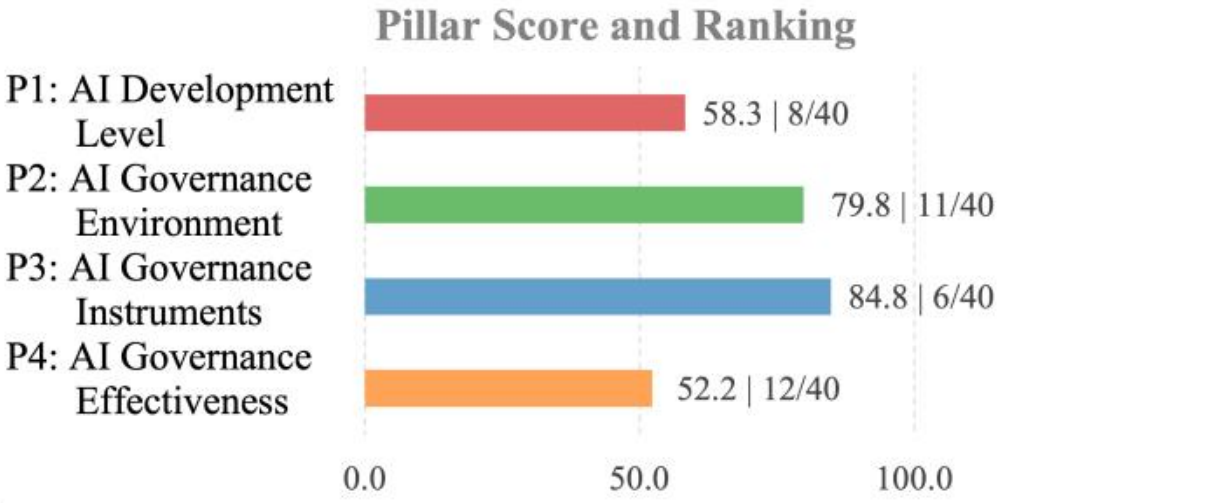

**Highest Rankings in:**

D5 Overall Governance Readiness (Ranking: 3/40)
D16 Data & Algorithm Openness (Ranking: 3/40)

**Lowest Rankings in:**

D13 Public Understanding of AI (Ranking: 33/40)
D14 AI Social Acceptance (Ranking: 21/40)

# Hungary

| AGILE Index Ranking | Population(2024) | GDP Per Capita(2024) | Country Group |
|---|---|---|---|
| 34/40 | 10 Million | 21,468 $ | High Income |

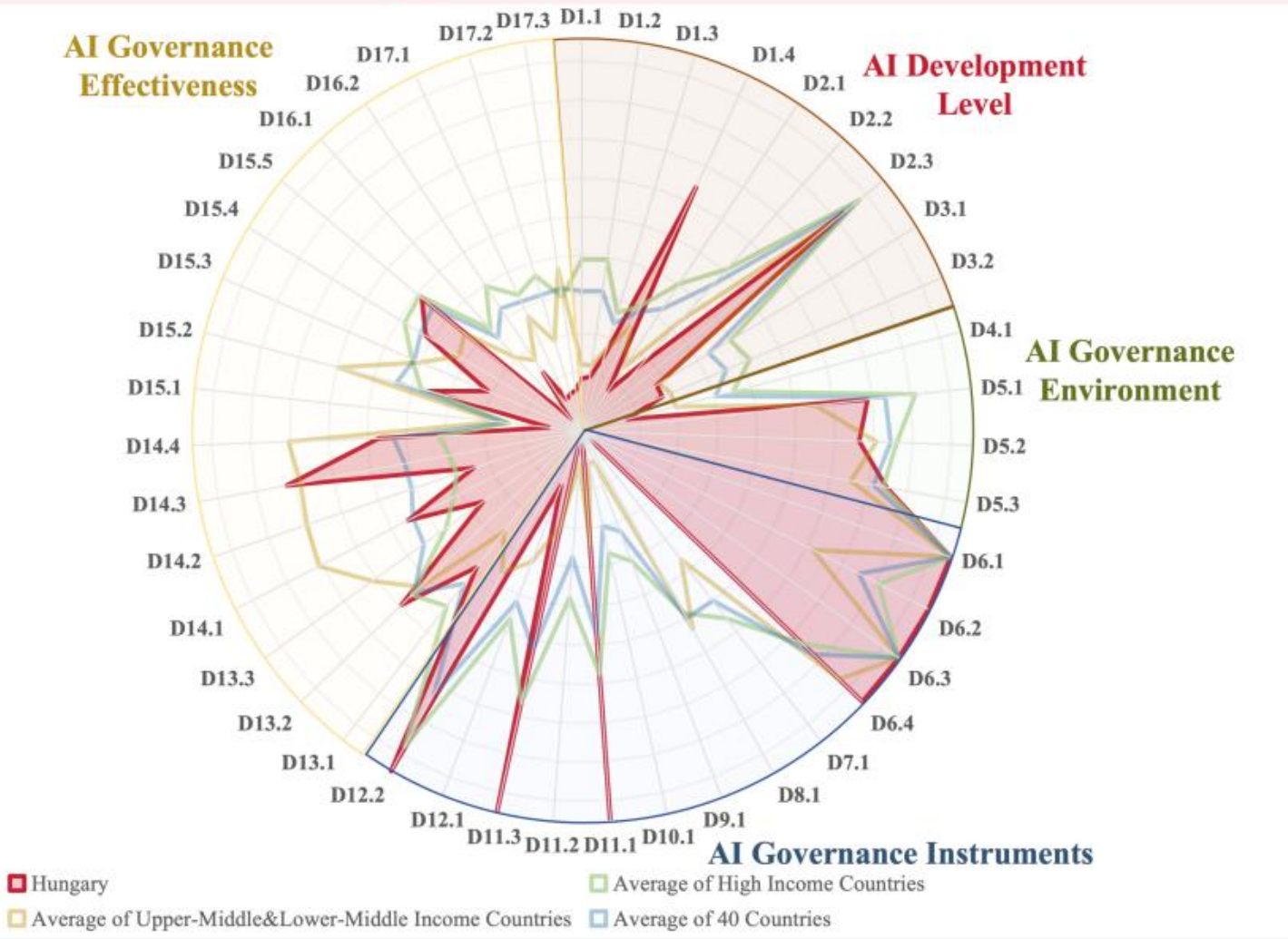

Pillar Score and Ranking

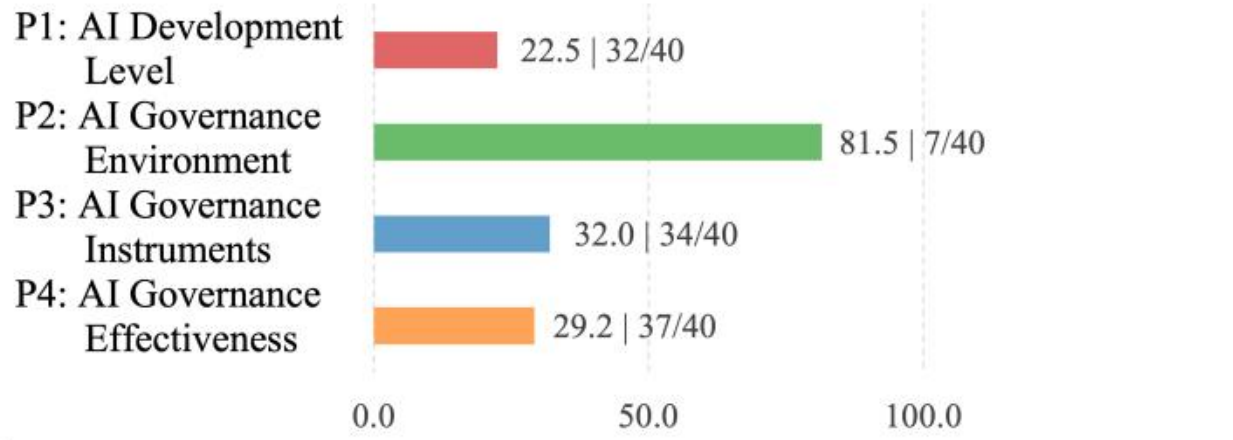

**Highest Rankings in:**
D4 AI Risk Exposure (Ranking: 8/40)
D14 AI Social Acceptance (Ranking: 17/40)

**Lowest Rankings in:**
D17 AI Governance Research Activity (Ranking: 39/40)
D2 AI Infrastructure (Ranking: 37/40)

# India

| AGILE Index Ranking | Population(2024) | GDP Per Capita(2024) | Country Group |
|---|---|---|---|
| 35/40 | 1451 Million | 2,932 $ | Lower-Middle Income |

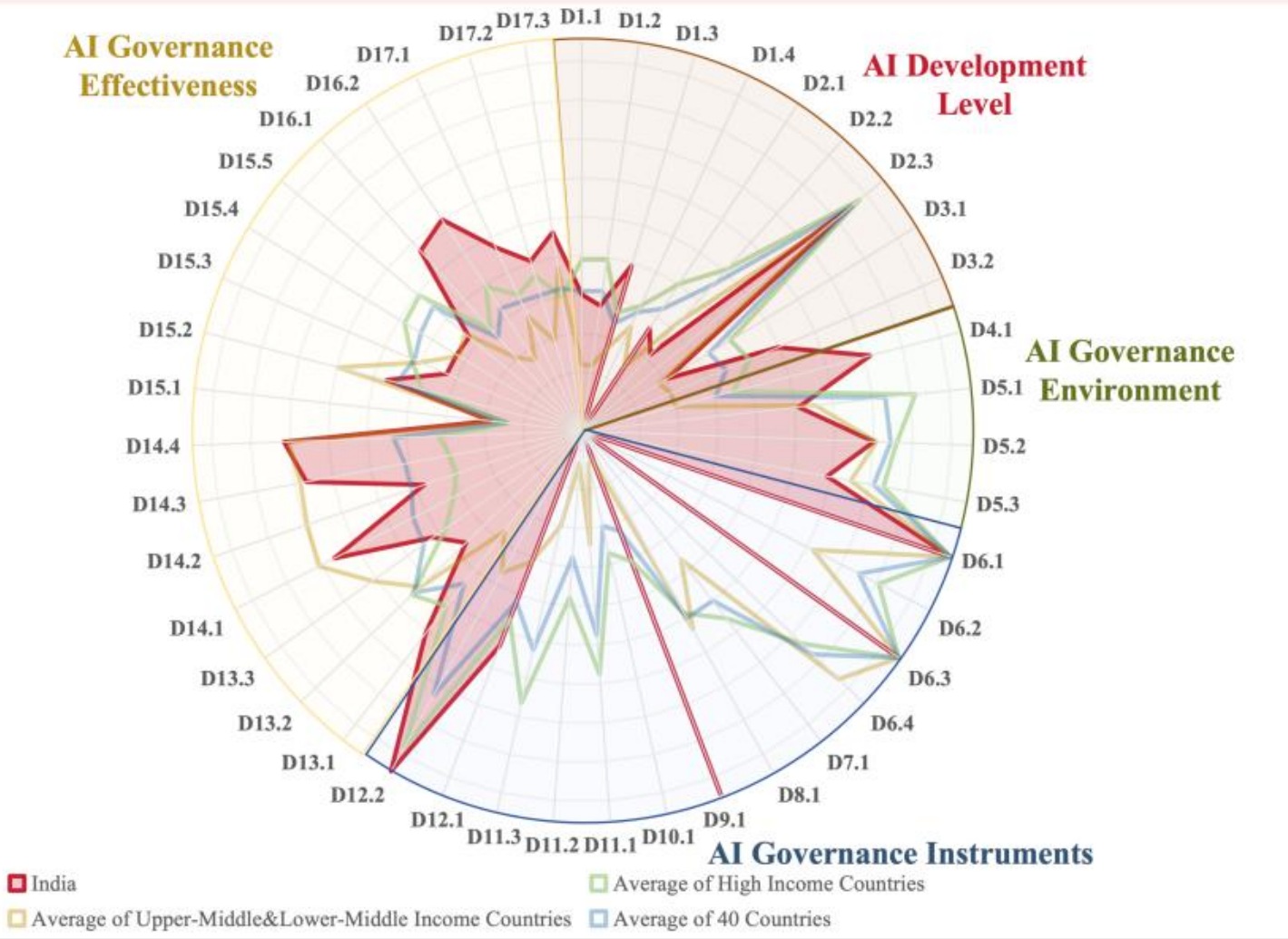

Pillar Score and Ranking

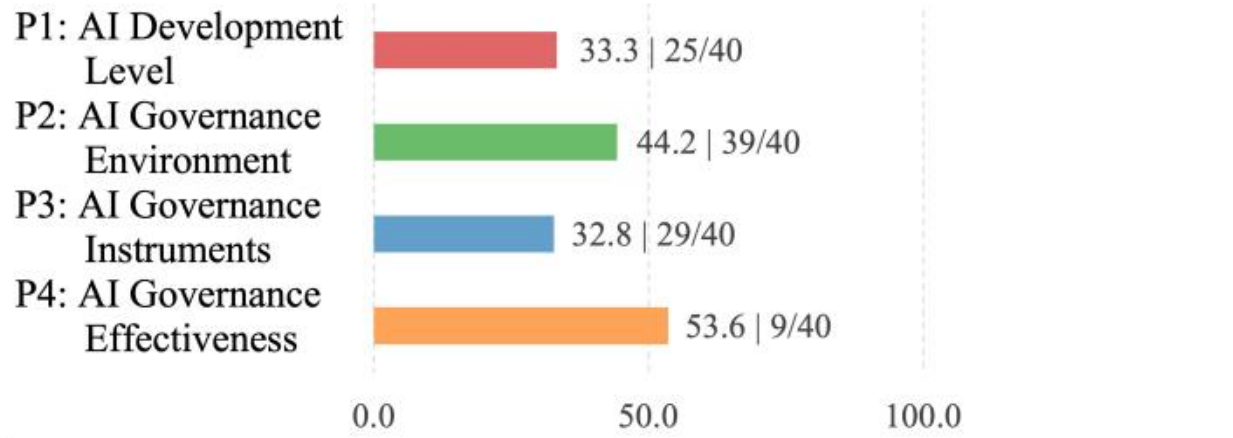

**Highest Rankings in:**
D16 Data & Algorithm Openness (Ranking: 7/40)
D17 AI Governance Research Activity (Ranking: 8/40)

**Lowest Rankings in:**
D5 Overall Governance Readiness (Ranking: 39/40)
D4 AI Risk Exposure (Ranking: 38/40)

# Indonesia

| AGILE Index Ranking | Population(2024) | GDP Per Capita(2024) | Country Group |
|---|---|---|---|
| 22/40 | 283 Million | 5,531 $ | Upper-Middle Income |

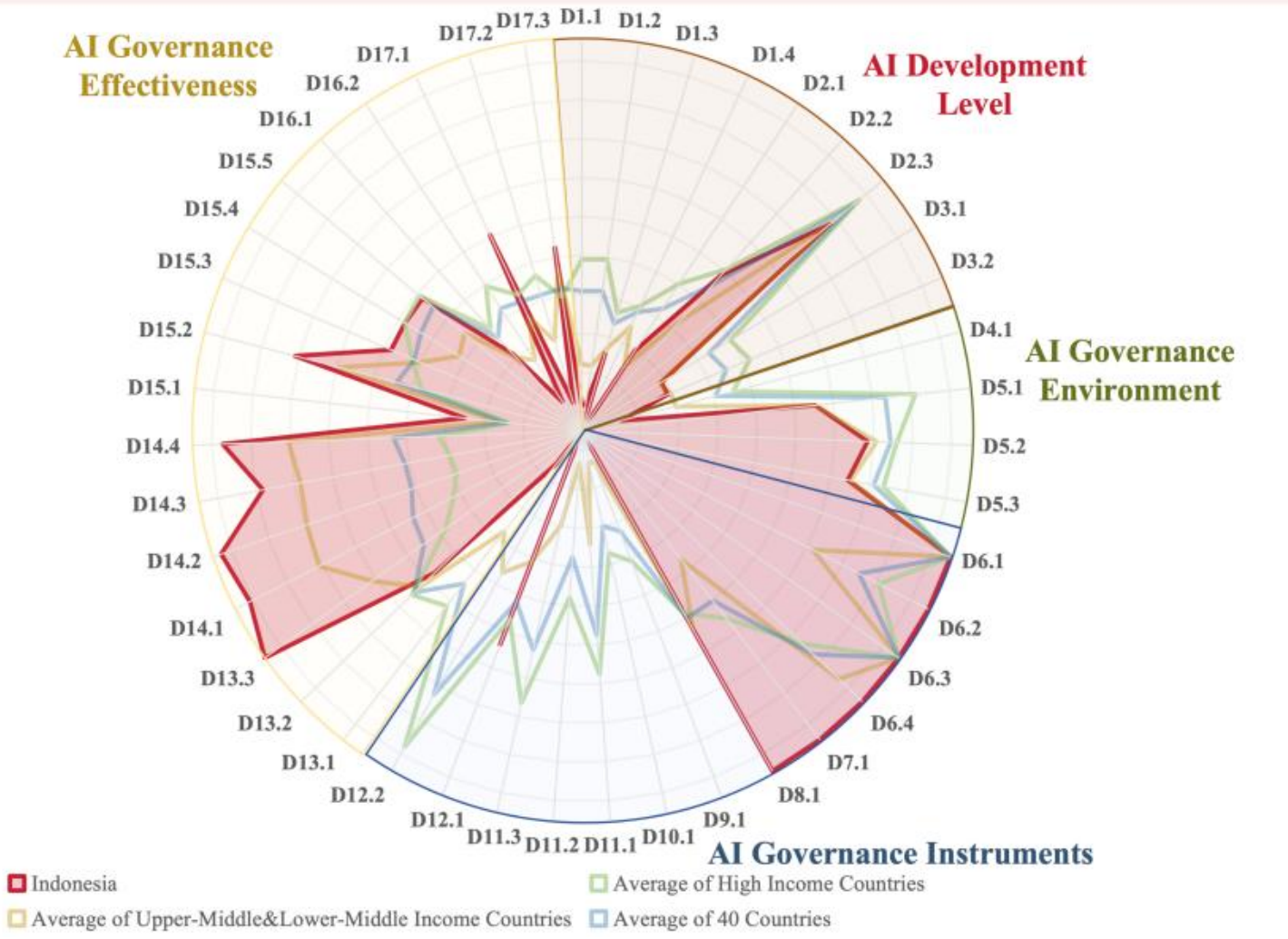

Pillar Score and Ranking

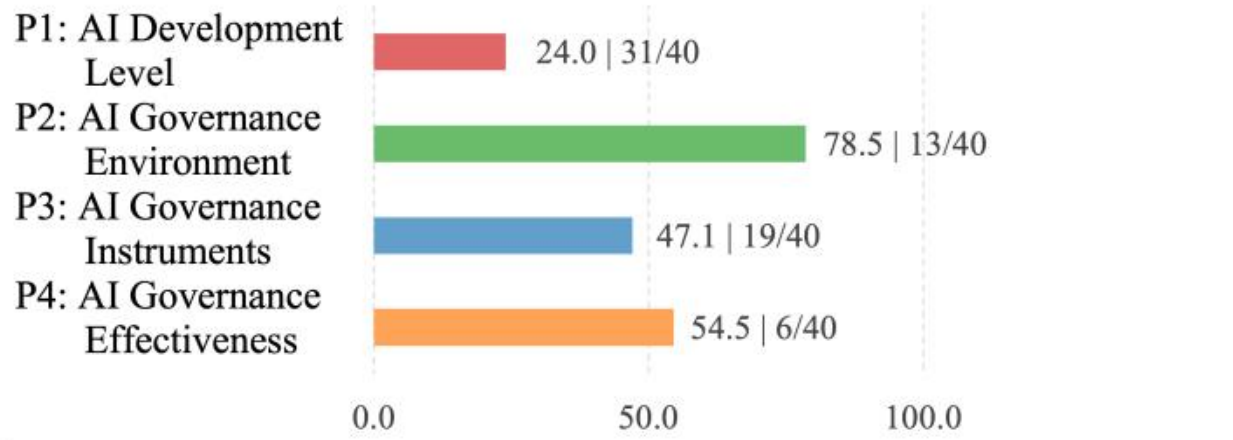

**Highest Rankings in:**
D14 AI Social Acceptance (Ranking: 3/40)
D4 AI Risk Exposure (Ranking: 6/40)

**Lowest Rankings in:**
D1 AI Research and Development Activity (Ranking: 37/40)
D5 Overall Governance Readiness (Ranking: 36/40)

# Ireland

| AGILE Index Ranking | Population(2024) | GDP Per Capita(2024) | Country Group |
|---|---|---|---|
| 24/40 | 5,255 Thousand | 122,243 $ | High Income |

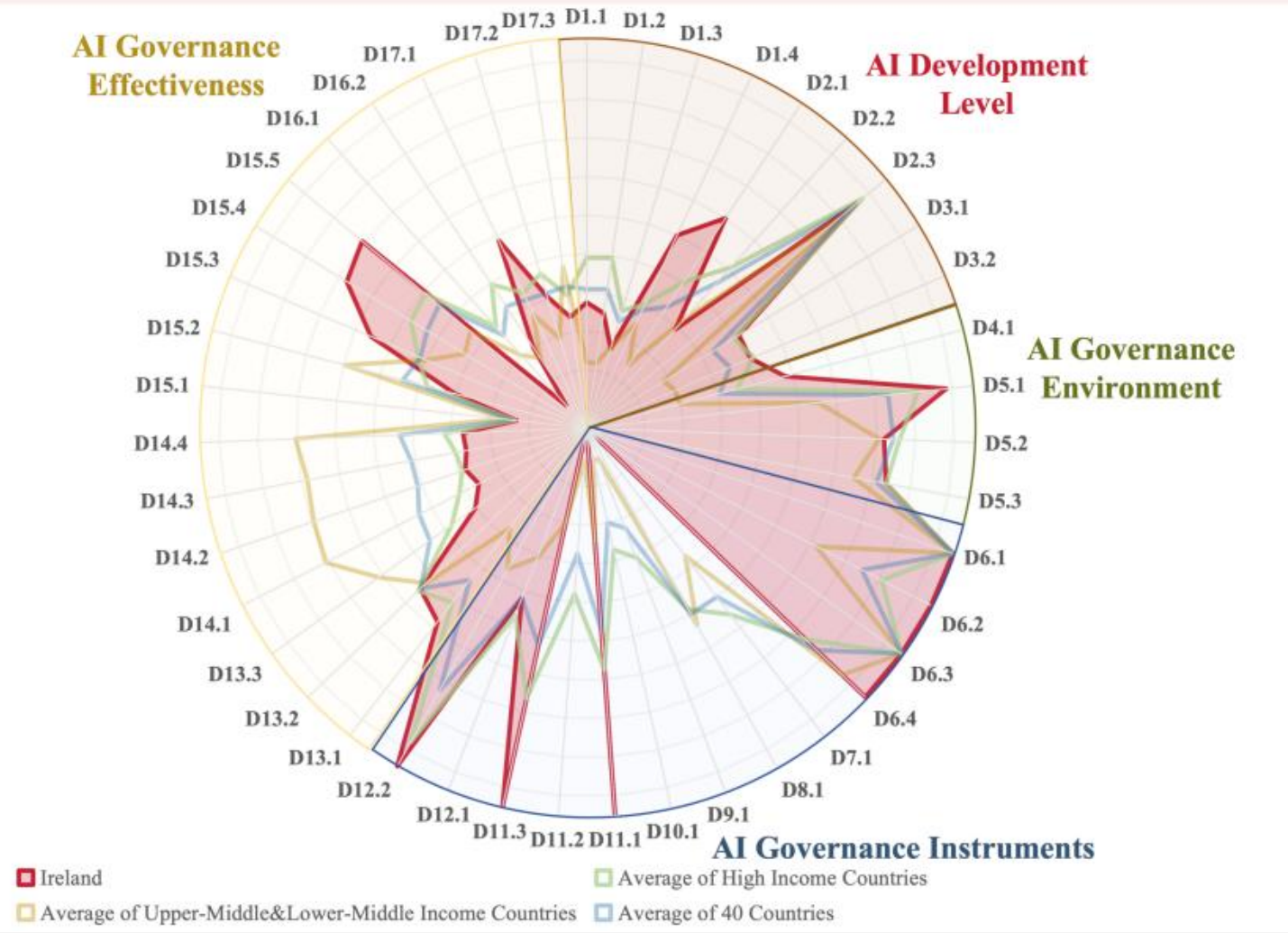

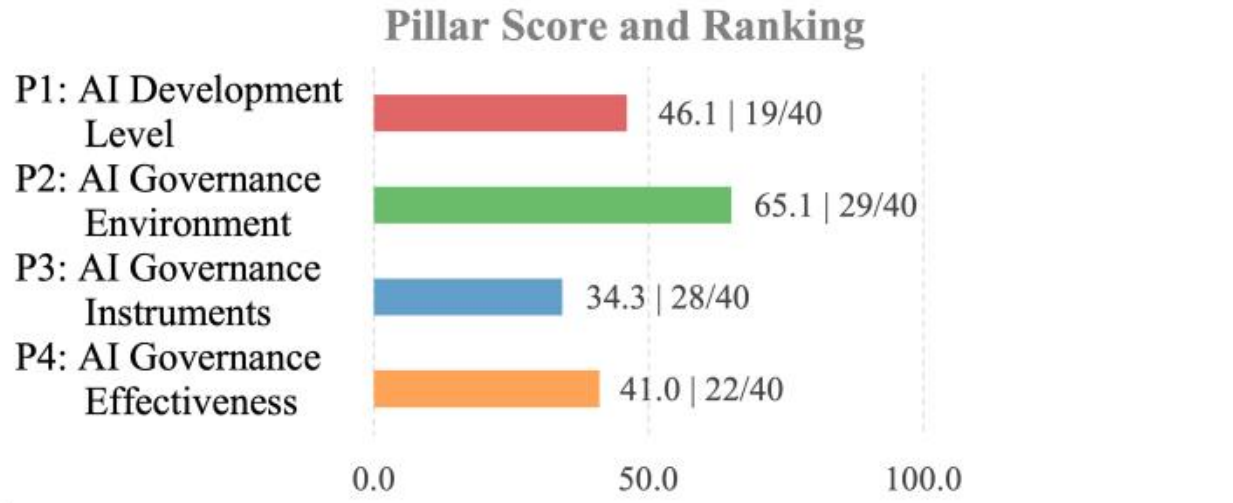

**Highest Rankings in:**

D15 AI Development Inclusivity (Ranking: 12/40)
D3 AI Industry Vitality (Ranking: 14/40)

**Lowest Rankings in:**

D4 AI Risk Exposure (Ranking: 31/40)
D16 Data & Algorithm Openness (Ranking: 28/40)

# Israel

| AGILE Index Ranking | Population(2024) | GDP Per Capita(2024) | Country Group |
|---|---|---|---|
| 27/40 | 5,255 Thousand | 68,992 $ | High Income |

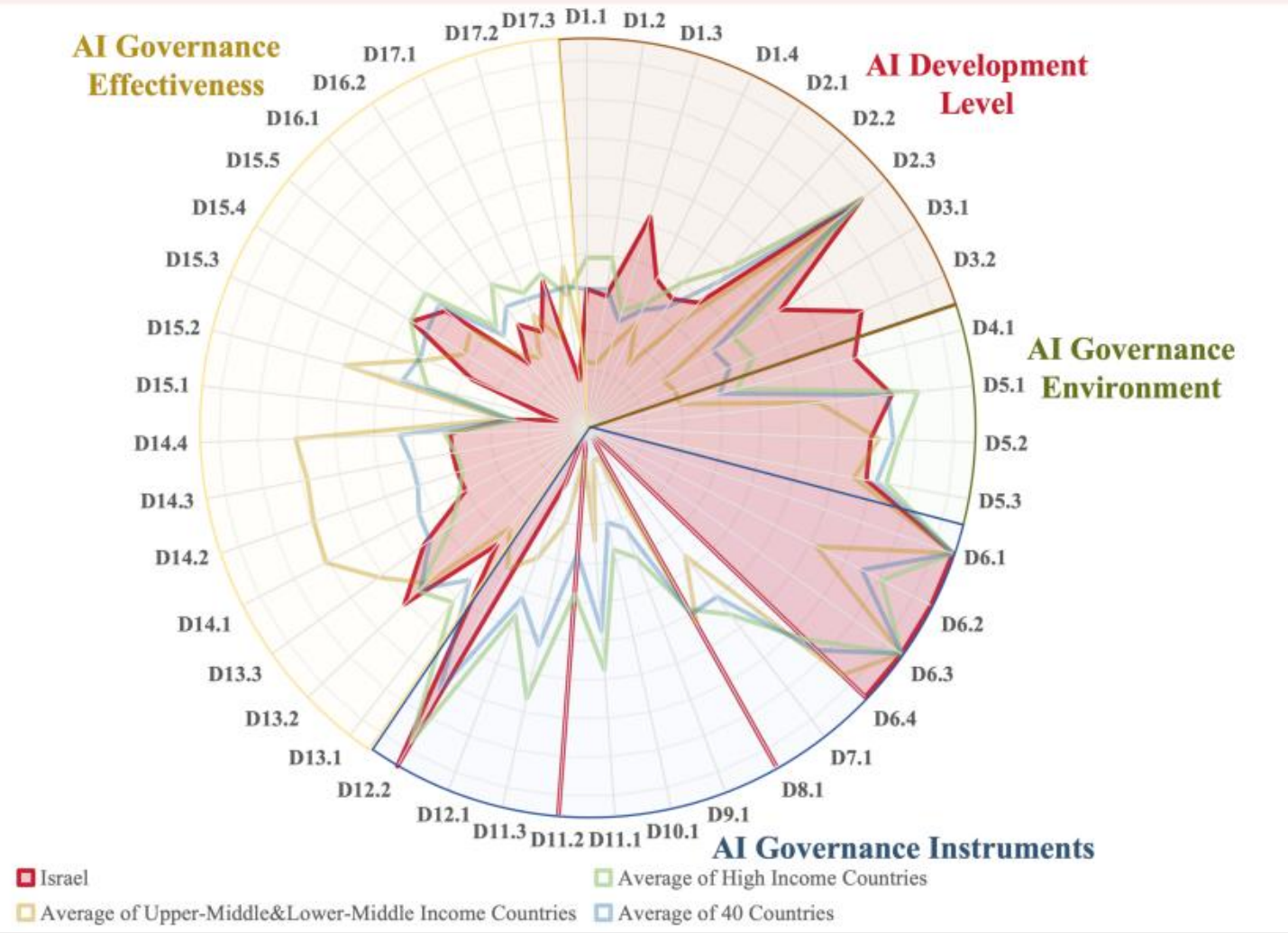

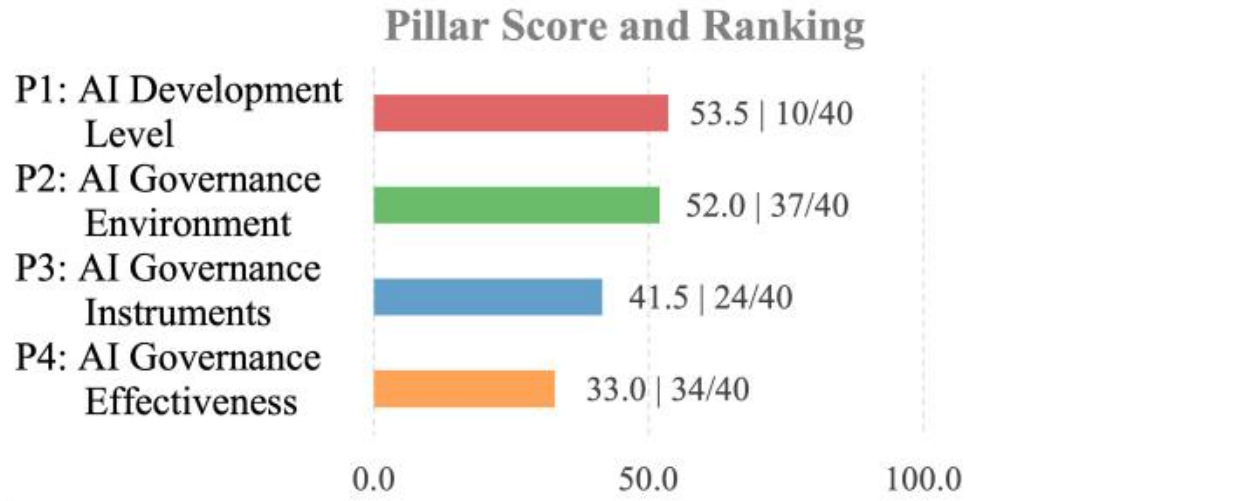

**Highest Rankings in:**

D3 AI Industry Vitality (Ranking: 5/40)
D1 AI Research and Development Activity (Ranking: 11/40)

**Lowest Rankings in:**

D4 AI Risk Exposure (Ranking: 37/40)
D17 AI Governance Research Activity (Ranking: 30/40)

# Italy

| AGILE Index Ranking | Population(2024) | GDP Per Capita(2024) | Country Group |
|---|---|---|---|
| 19/40 | 59 Million | 44,479 $ | High Income |

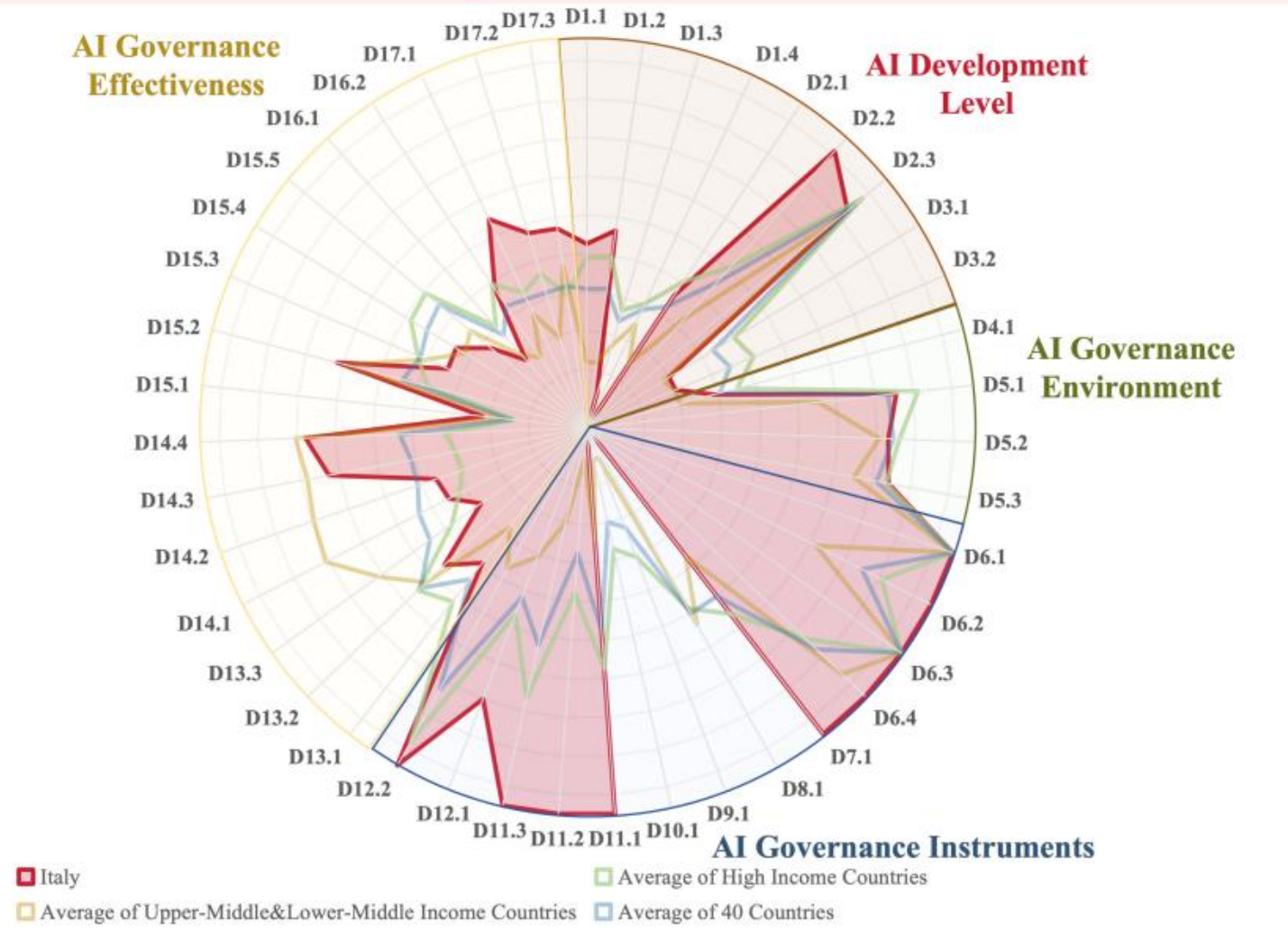

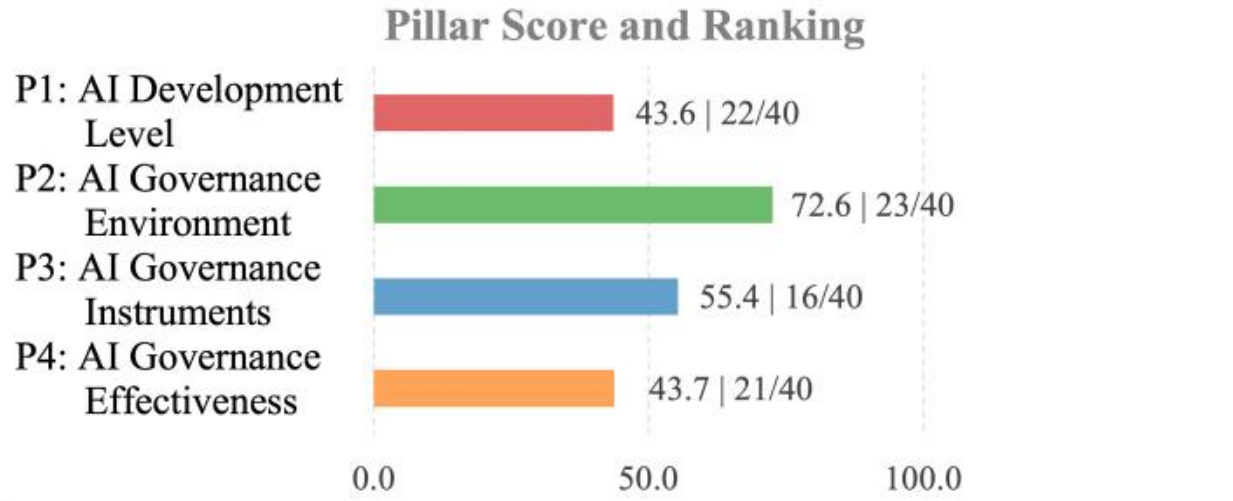

**Highest Rankings in:**

D2 AI Infrastructure (Ranking: 4/40)
D17 AI Governance Research Activity (Ranking: 6/40)

**Lowest Rankings in:**

D13 Public Understanding of AI (Ranking: 34/40)
D3 AI Industry Vitality (Ranking: 27/40)

# Japan

| AGILE Index Ranking | Population(2024) | GDP Per Capita(2024) | Country Group |
|---|---|---|---|
| 9/40 | 124 Million | 49,620 $ | High Income |

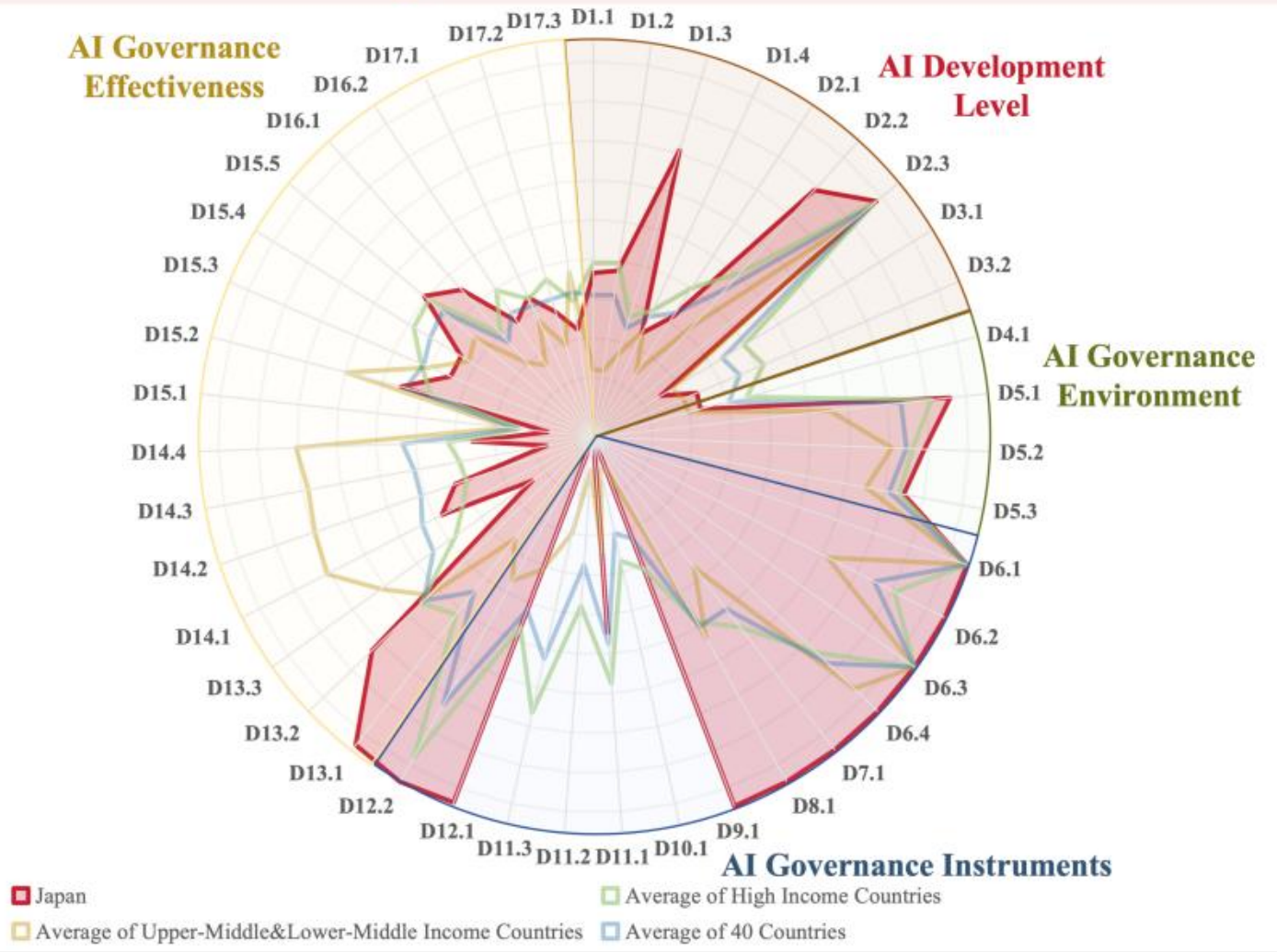

Pillar Score and Ranking

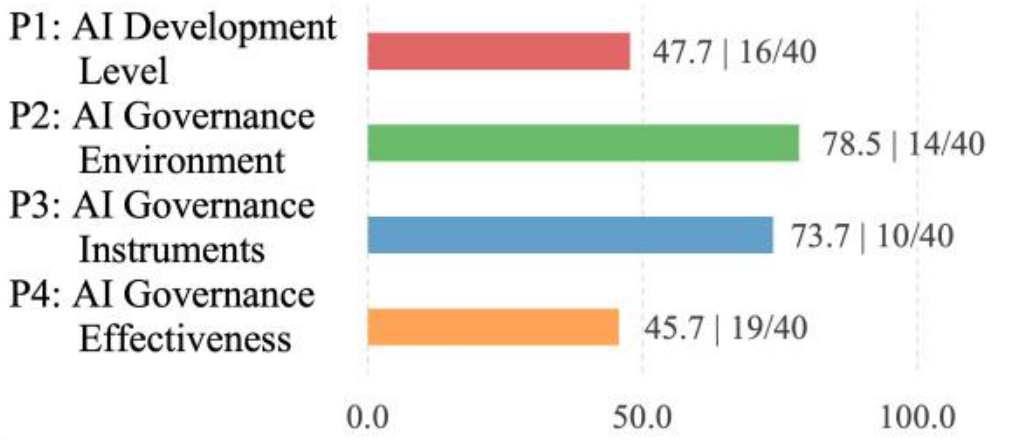

**Highest Rankings in:**

D13 Public Understanding of AI (Ranking: 6/40)
D2 AI Infrastructure (Ranking: 7/40)

**Lowest Rankings in:**

D14 AI Social Acceptance (Ranking: 29/40)
D3 AI Industry Vitality (Ranking: 28/40)

# Malaysia

| AGILE Index Ranking | Population(2024) | GDP Per Capita(2024) | Country Group |
|---|---|---|---|
| 21/40 | 36 Million | 15,003 $ | Upper-Middle Income |

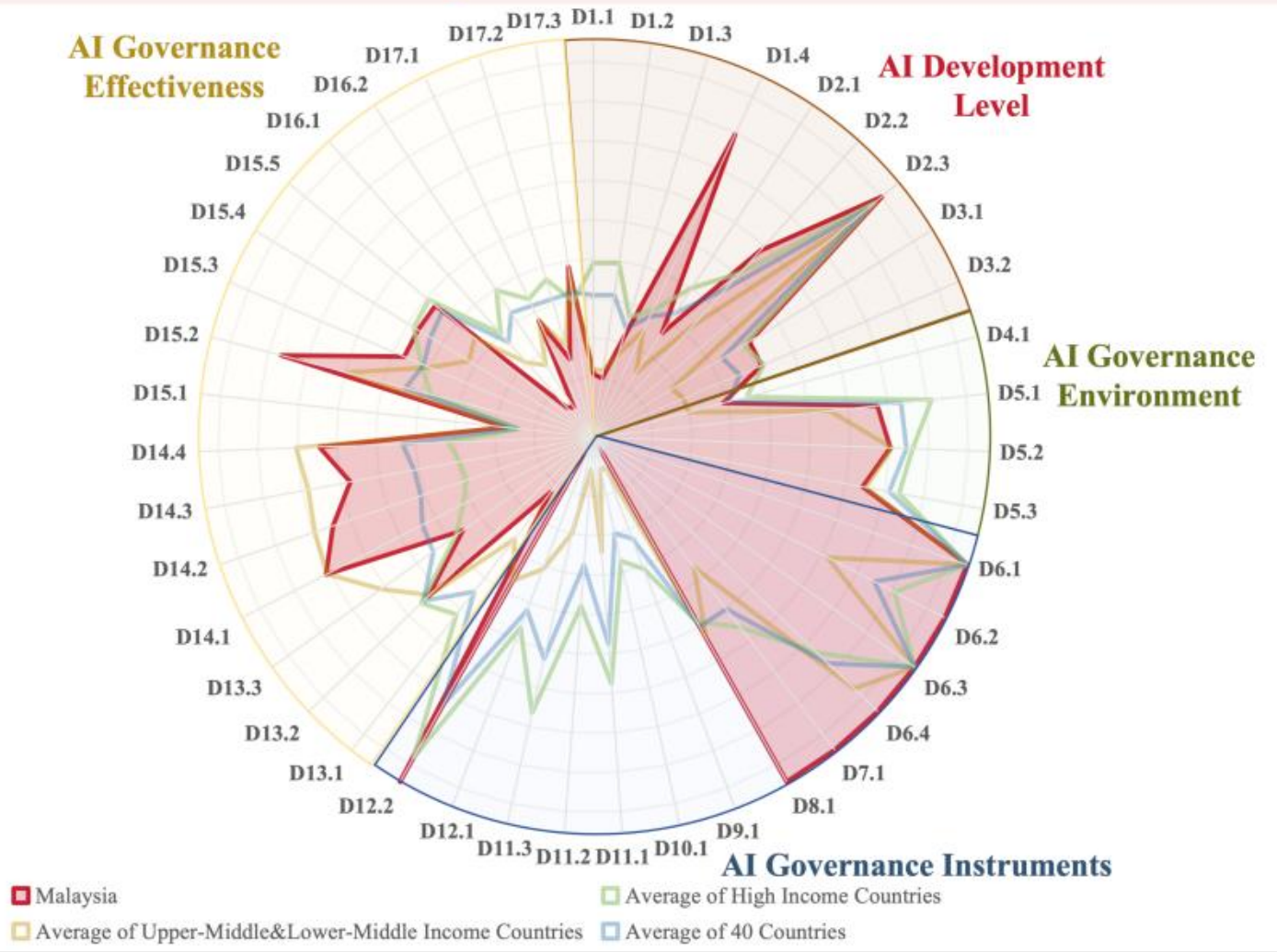

Pillar Score and Ranking

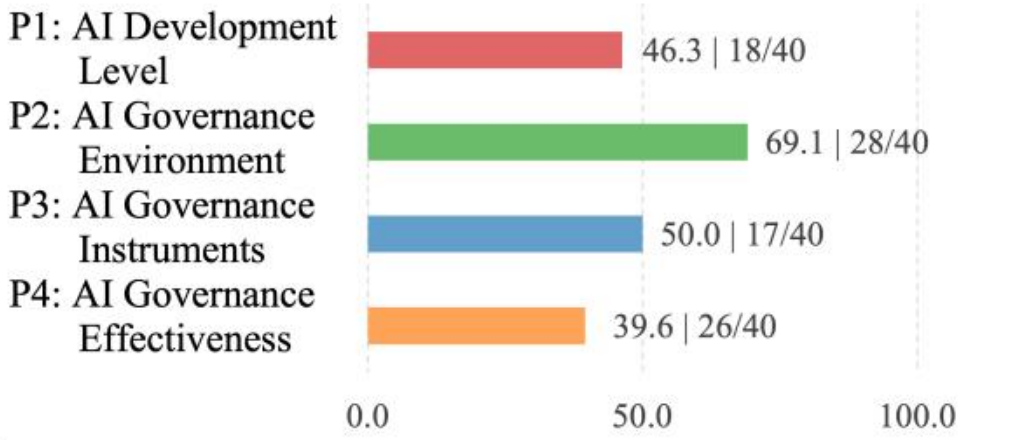

**Highest Rankings in:**

D14 AI Social Acceptance (Ranking: 8/40)
D3 AI Industry Vitality (Ranking: 13/40)

**Lowest Rankings in:**

D13 Public Understanding of AI (Ranking: 37/40)
D16 Data & Algorithm Openness (Ranking: 37/40)

# Mexico

| AGILE Index Ranking | Population(2024) | GDP Per Capita(2024) | Country Group |
|---|---|---|---|
| 31/40 | 131 Million | 13,482 $ | Upper-Middle Income |

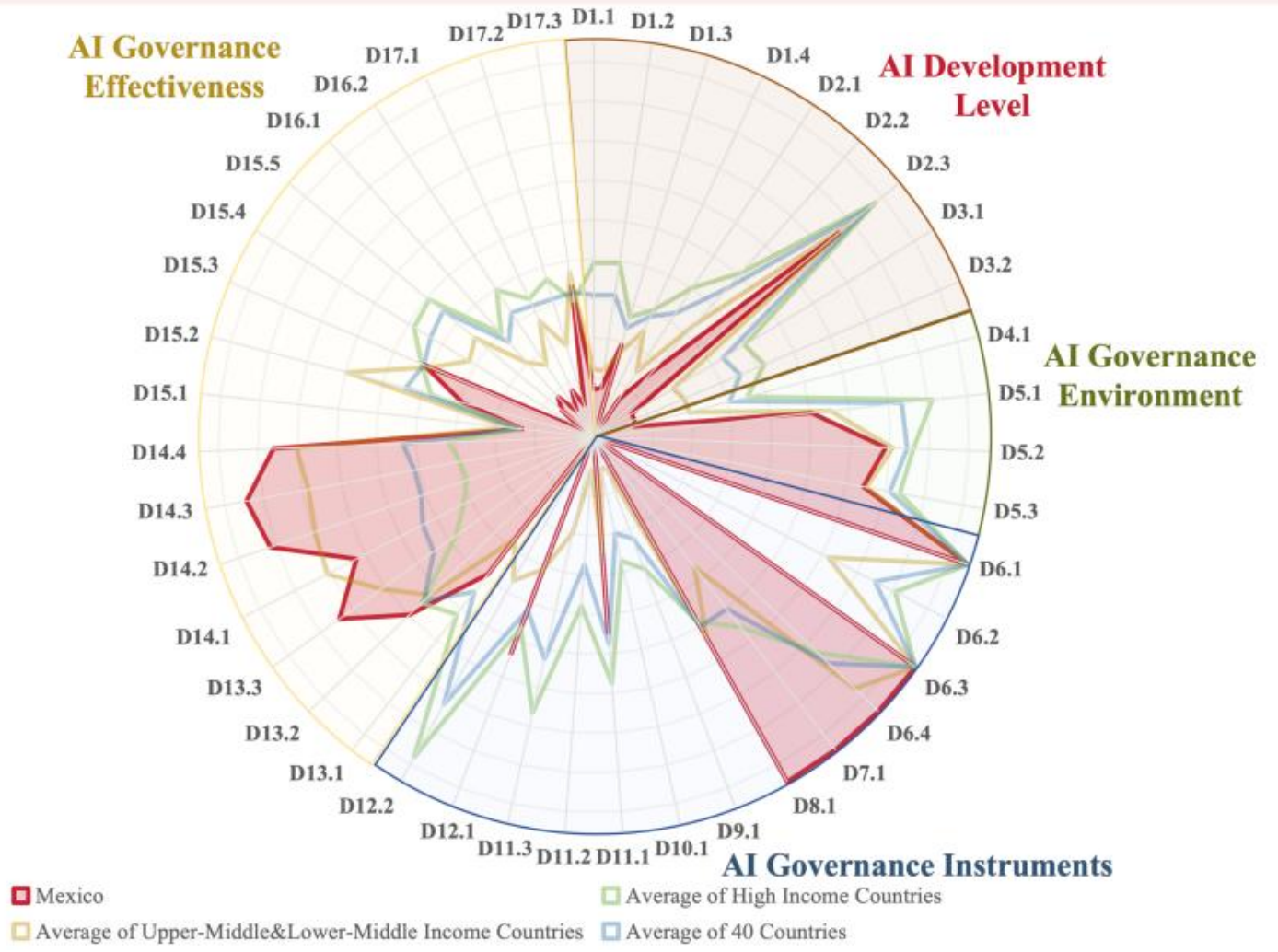

Pillar Score and Ranking

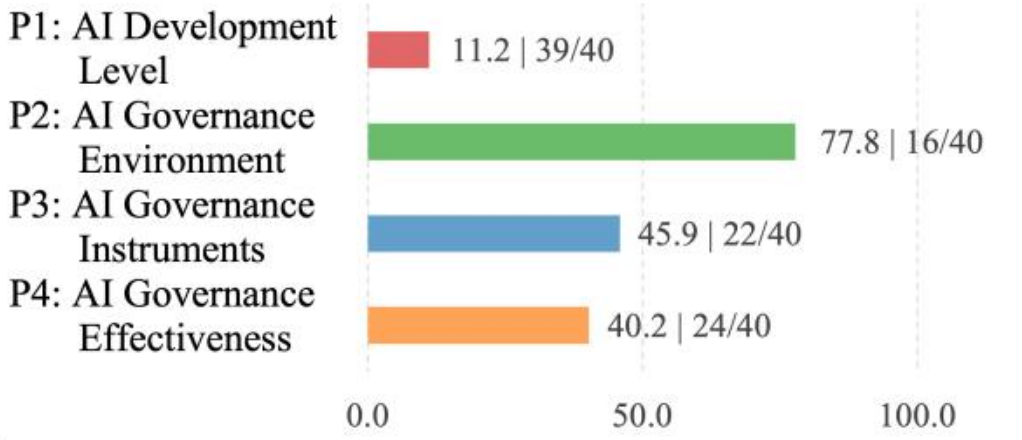

**Highest Rankings in:**

D14 AI Social Acceptance (Ranking: 5/40)
D4 AI Risk Exposure (Ranking: 6/40)

**Lowest Rankings in:**

D2 AI Infrastructure (Ranking: 40/40)
D3 AI Industry Vitality (Ranking: 39/40)

# Netherlands

| AGILE Index Ranking | Population(2024) | GDP Per Capita(2024) | Country Group |
|---|---|---|---|
| 11/40 | 18 Million | 65,357 $ | High Income |

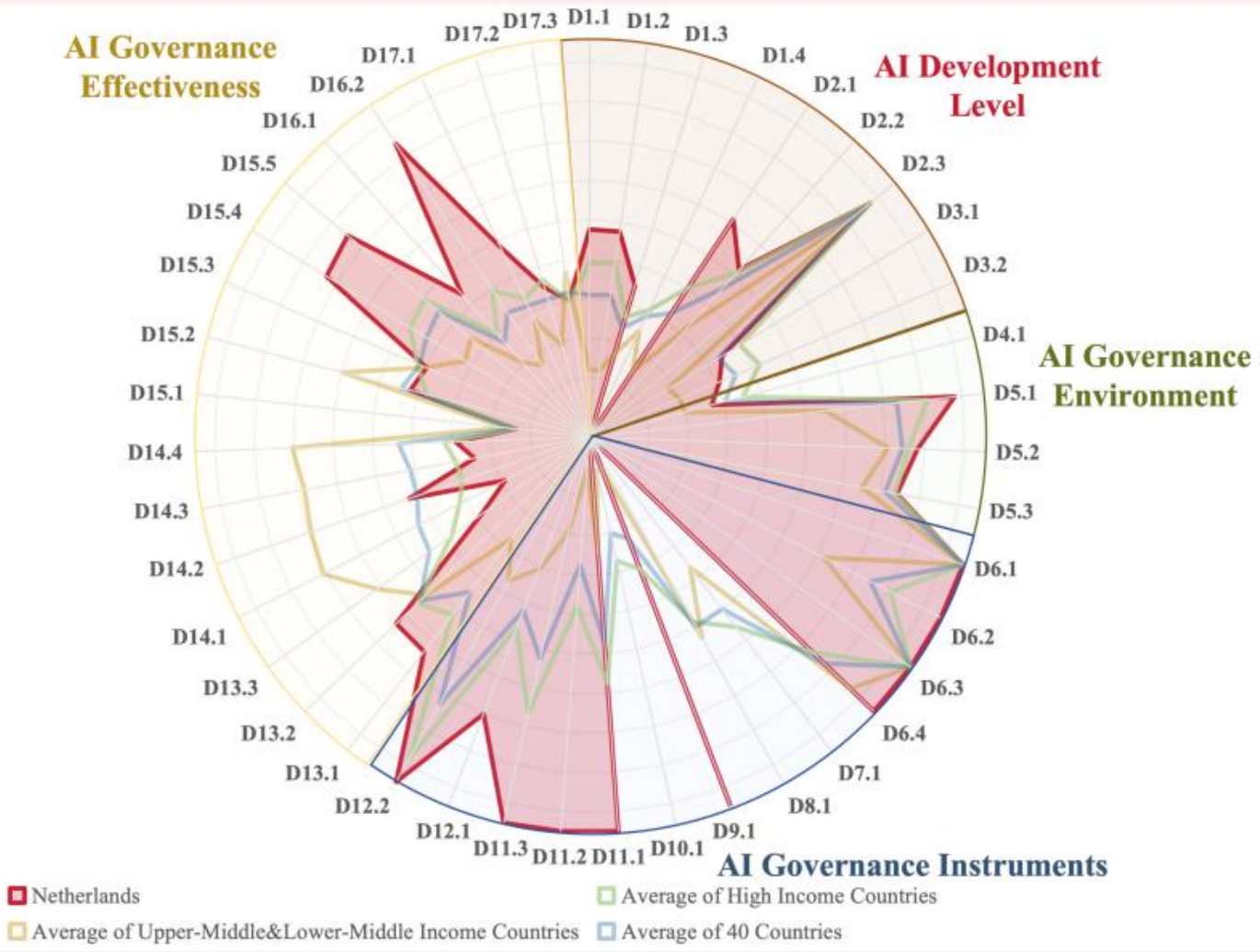

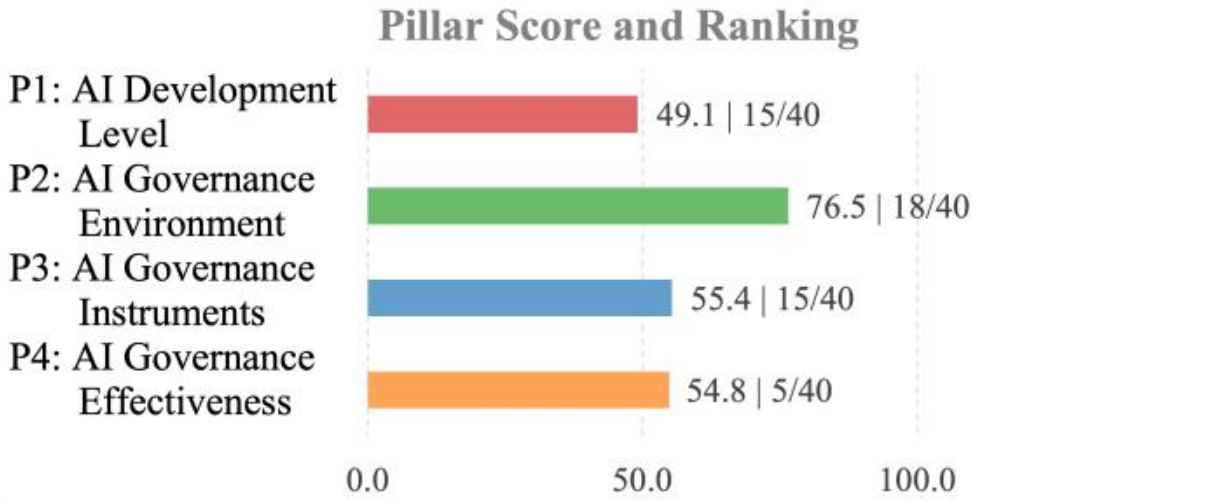

**Highest Rankings in:**

D16 Data & Algorithm Openness (Ranking: 5/40)
D2 AI Infrastructure (Ranking: 6/40)

**Lowest Rankings in:**

D14 AI Social Acceptance (Ranking: 26/40)
D4 AI Risk Exposure (Ranking: 21/40)

# New Zealand

| AGILE Index Ranking | Population(2024) | GDP Per Capita(2024) | Country Group |
|---|---|---|---|
| 33/40 | 5,214 Thousand | 55,591 $ | High Income |

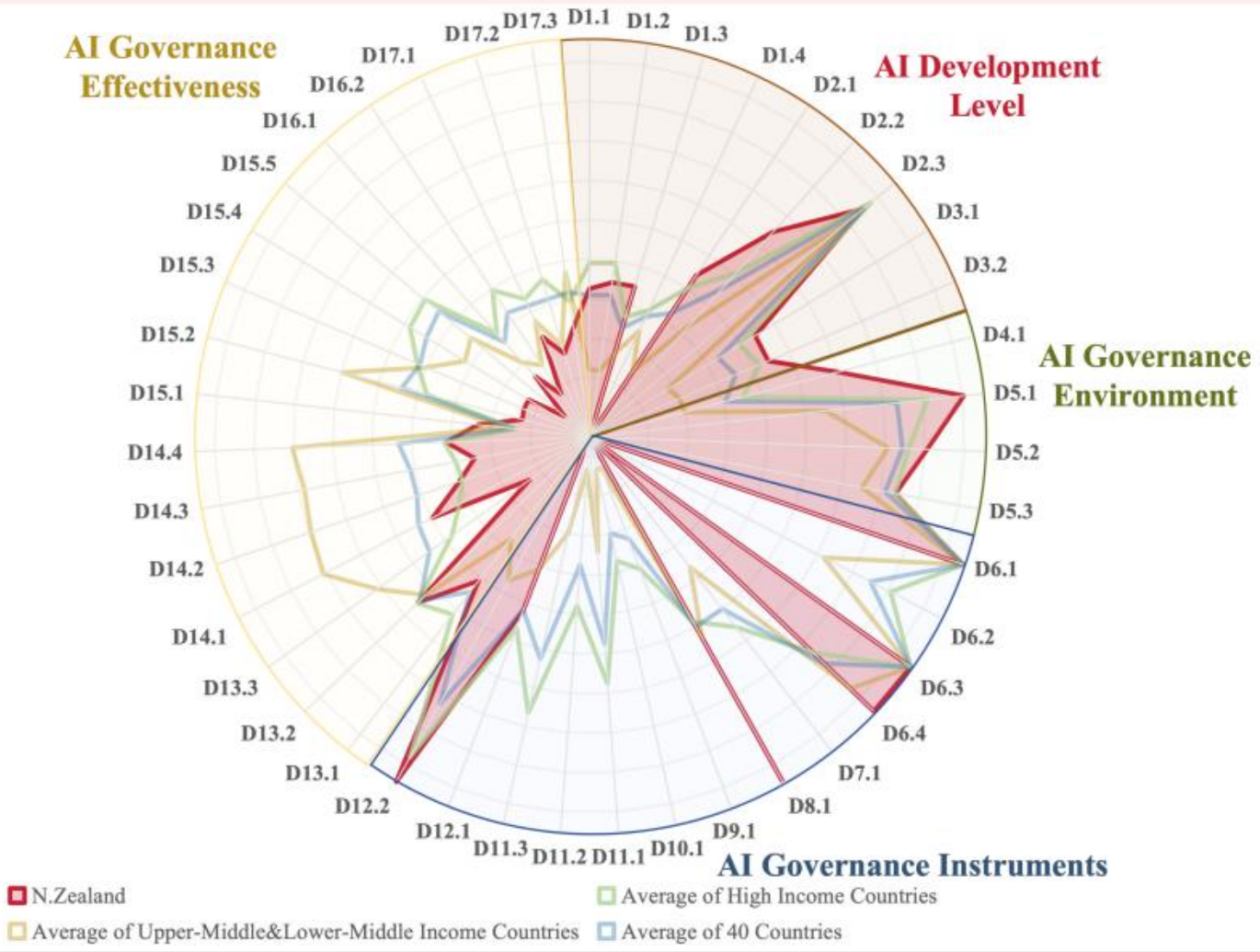

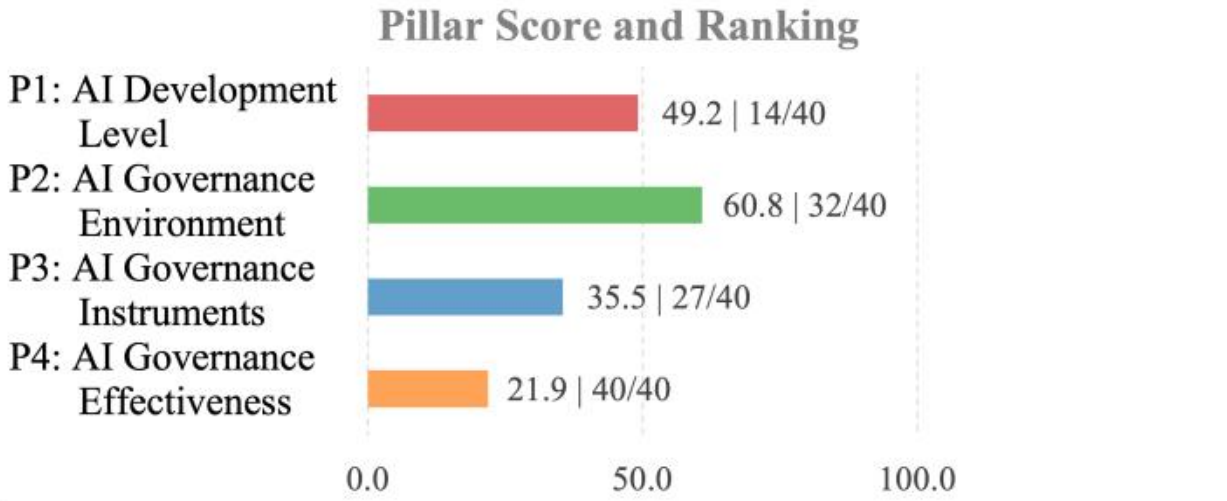

**Highest Rankings in:**

D5 Overall Governance Readiness (Ranking: 4/40)
D2 AI Infrastructure (Ranking: 8/40)

**Lowest Rankings in:**

D15 AI Development Inclusivity (Ranking: 38/40)
D4 AI Risk Exposure (Ranking: 36/40)

# Norway

| AGILE Index Ranking | Population(2024) | GDP Per Capita(2024) | Country Group |
|---|---|---|---|
| 17/40 | 5,577 Thousand | 103,915 $ | High Income |

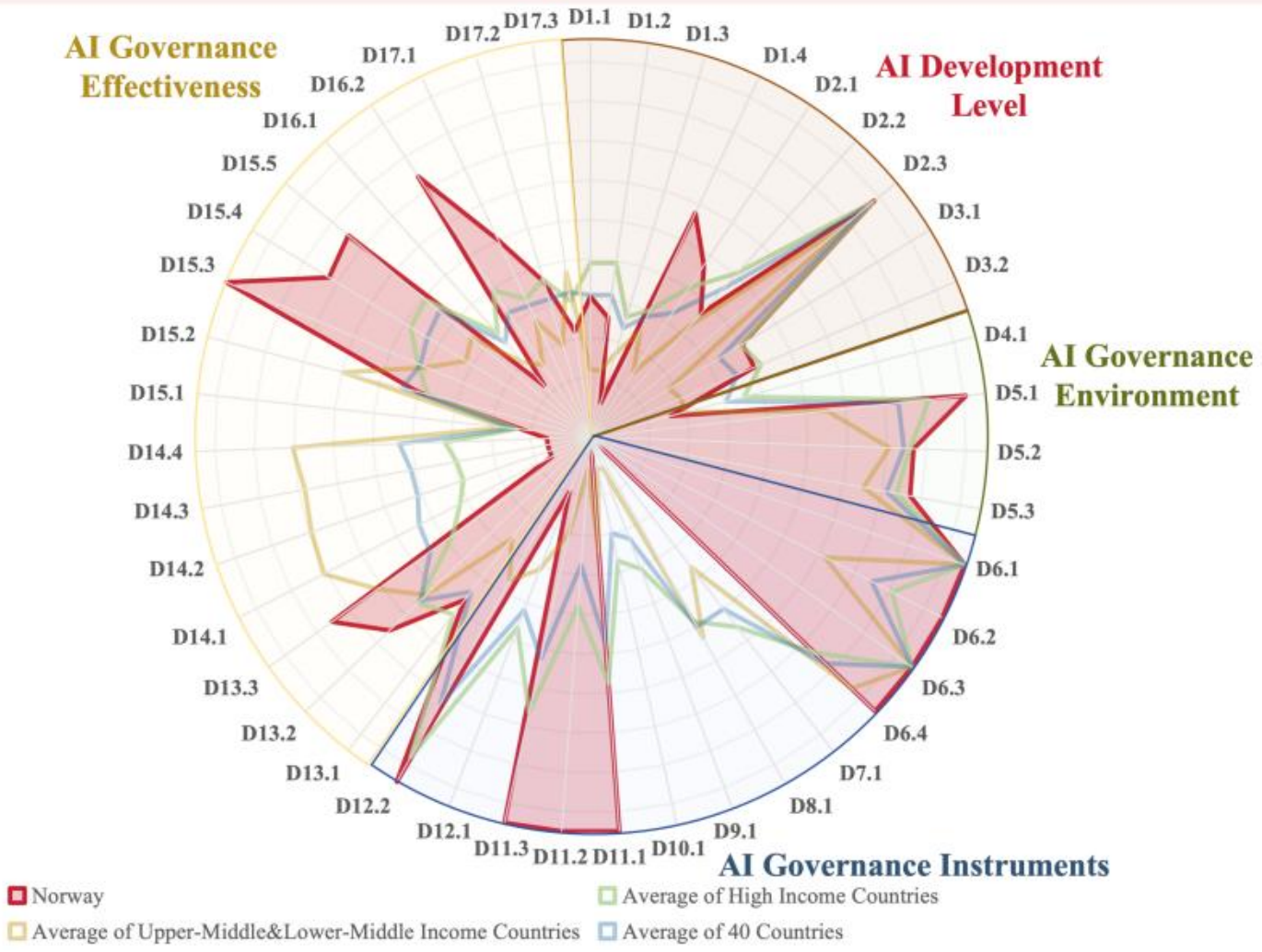

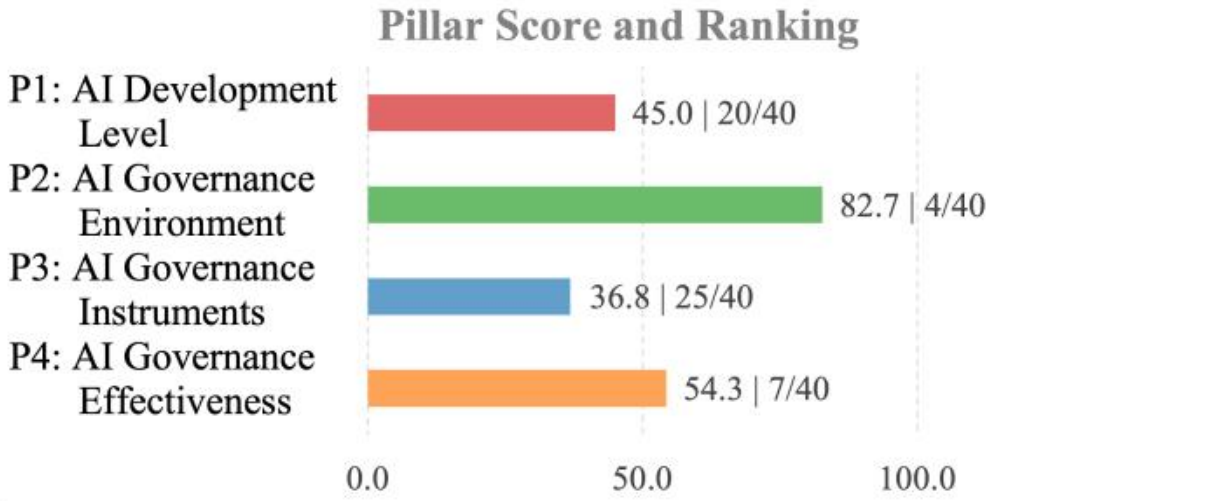

**Highest Rankings in:**

D15 AI Development Inclusivity (Ranking: 1/40)
D13 Public Understanding of AI (Ranking: 5/40)

**Lowest Rankings in:**

D14 AI Social Acceptance (Ranking: 40/40)
P3 Governance Instruments (Ranking: 25/40)

# Peru

| AGILE Index Ranking | Population(2024) | GDP Per Capita(2024) | Country Group |
|---|---|---|---|
| 37/40 | 34 Million | 8,620 $ | Upper-Middle Income |

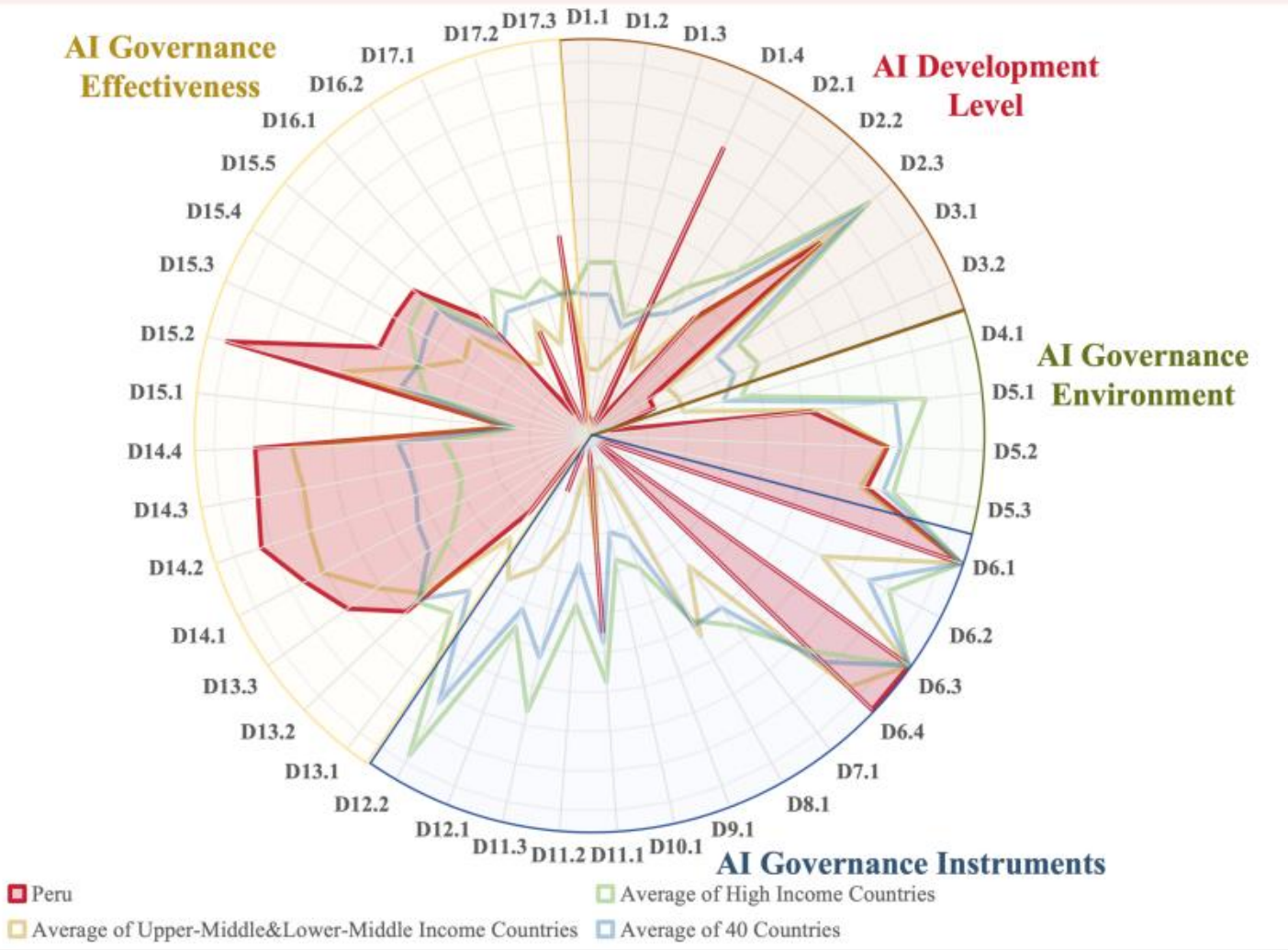

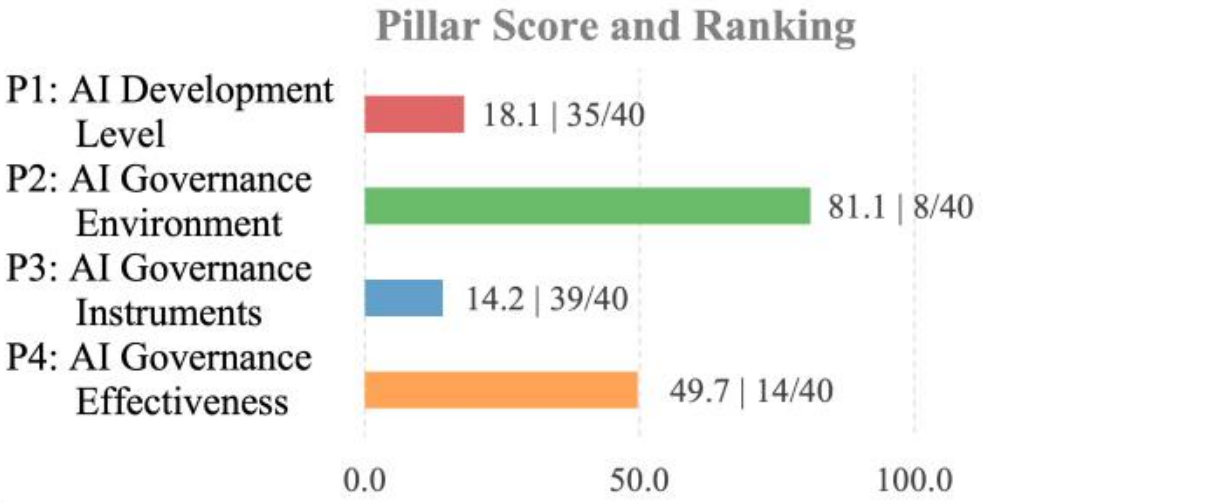

**Highest Rankings in:**

D4 AI Risk Exposure (Ranking: 1/40)
D14 AI Social Acceptance (Ranking: 4/40)

**Lowest Rankings in:**

D16 Data & Algorithm Openness (Ranking: 40/40)
P3 Governance Instruments (Ranking: 39/40)

# Poland

| AGILE Index Ranking | Population(2024) | GDP Per Capita(2024) | Country Group |
|---|---|---|---|
| 32/40 | 39 Million | 21,397 $ | High Income |

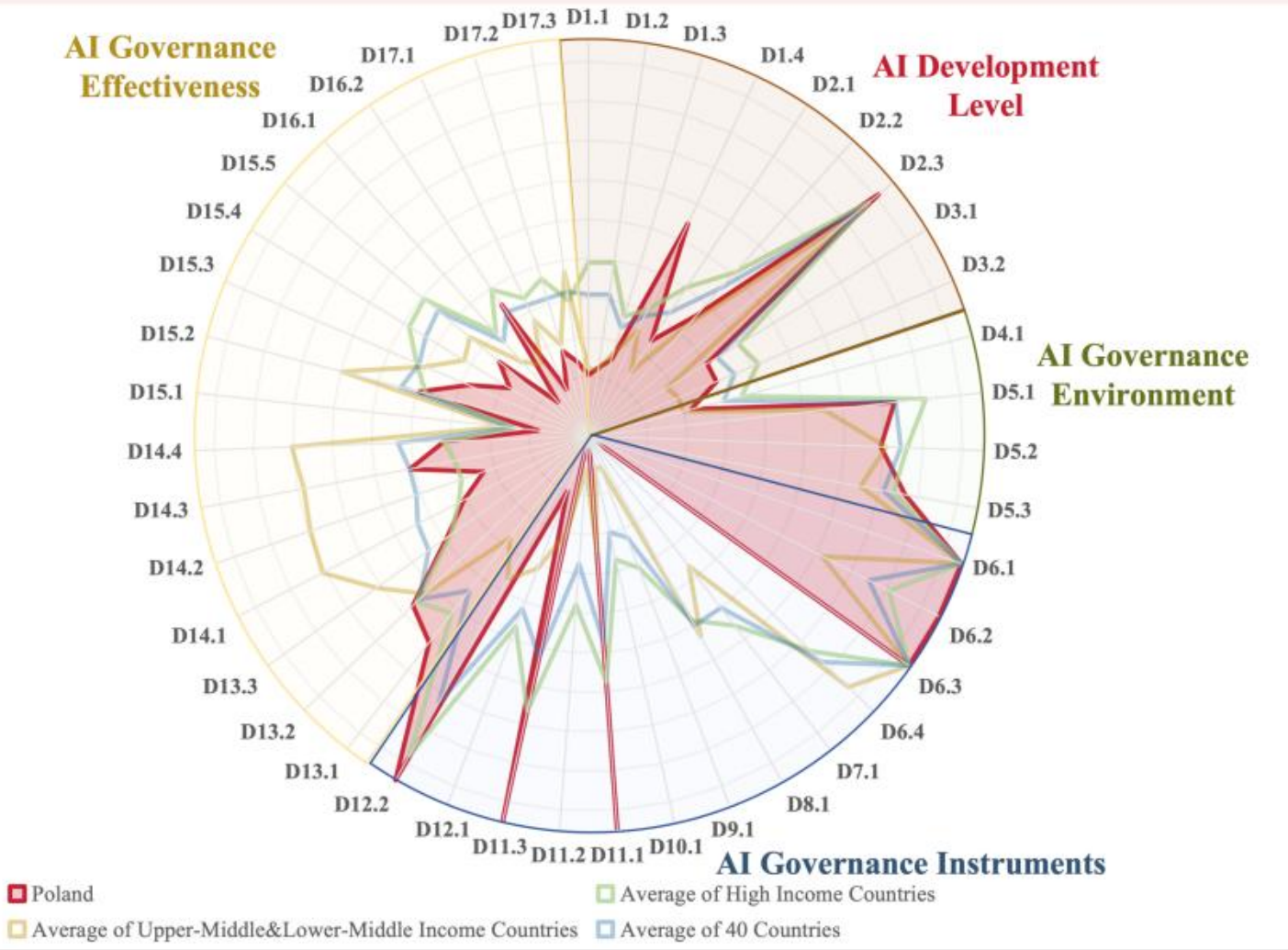

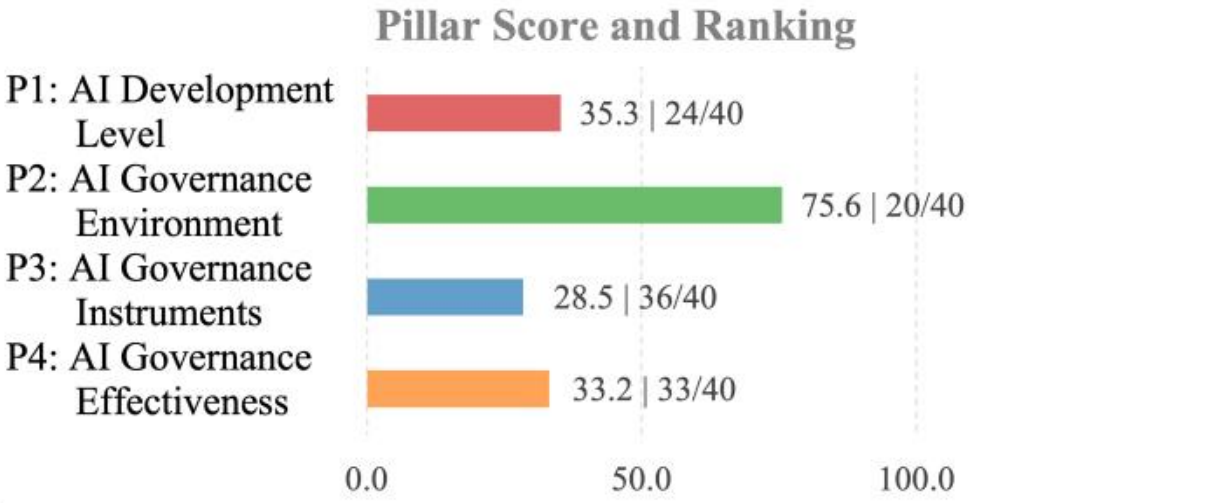

**Highest Rankings in:**

D13 Public Understanding of AI (Ranking: 10/40)
D4 AI Risk Exposure (Ranking: 16/40)

**Lowest Rankings in:**

D17 AI Governance Research Activity (Ranking: 37/40)
P3 Governance Instruments (Ranking: 36/40)

# Portugal

| AGILE Index Ranking | Population(2024) | GDP Per Capita(2024) | Country Group |
|---|---|---|---|
| 26/40 | 10 Million | 30,048 $ | High Income |

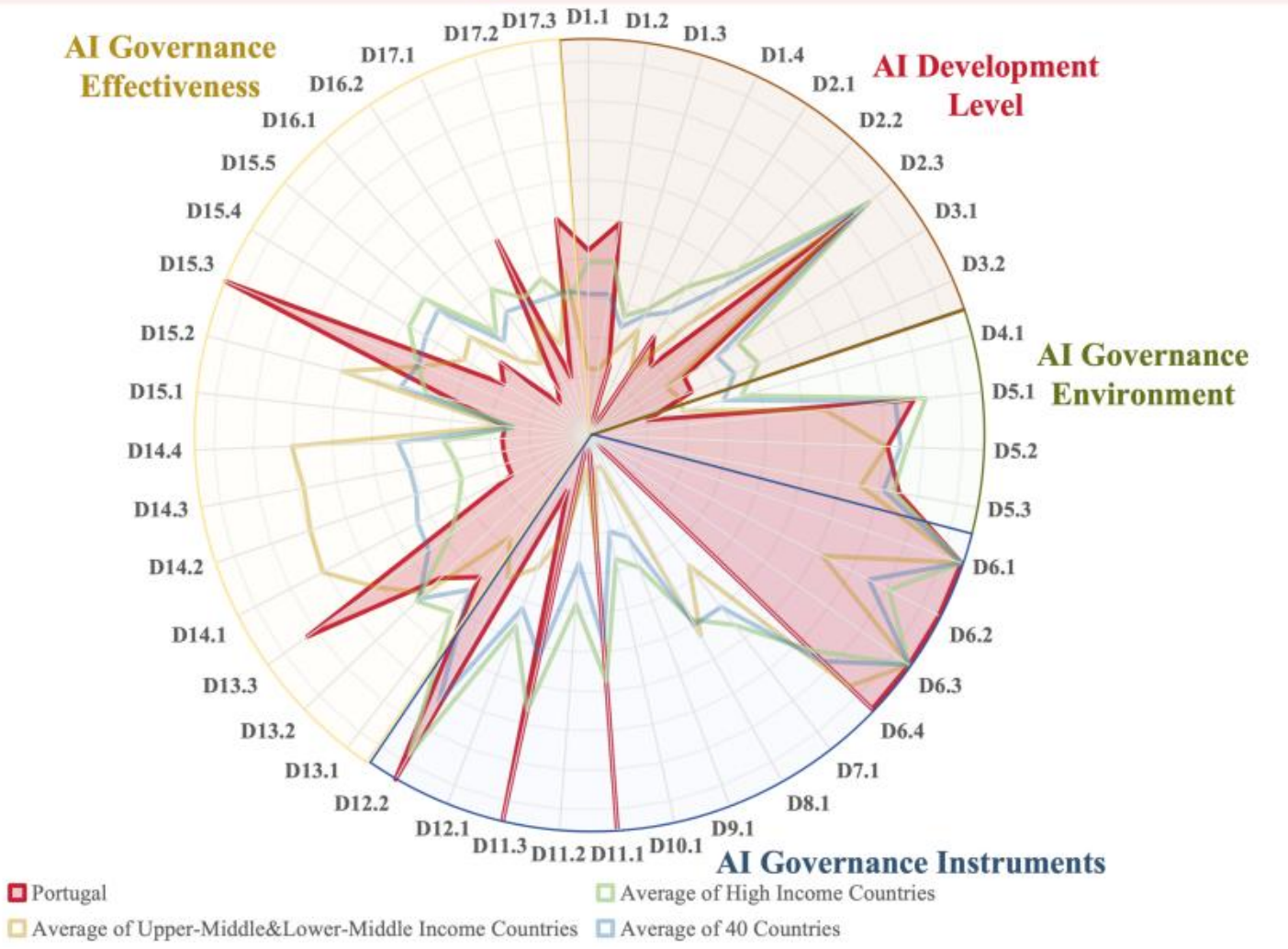

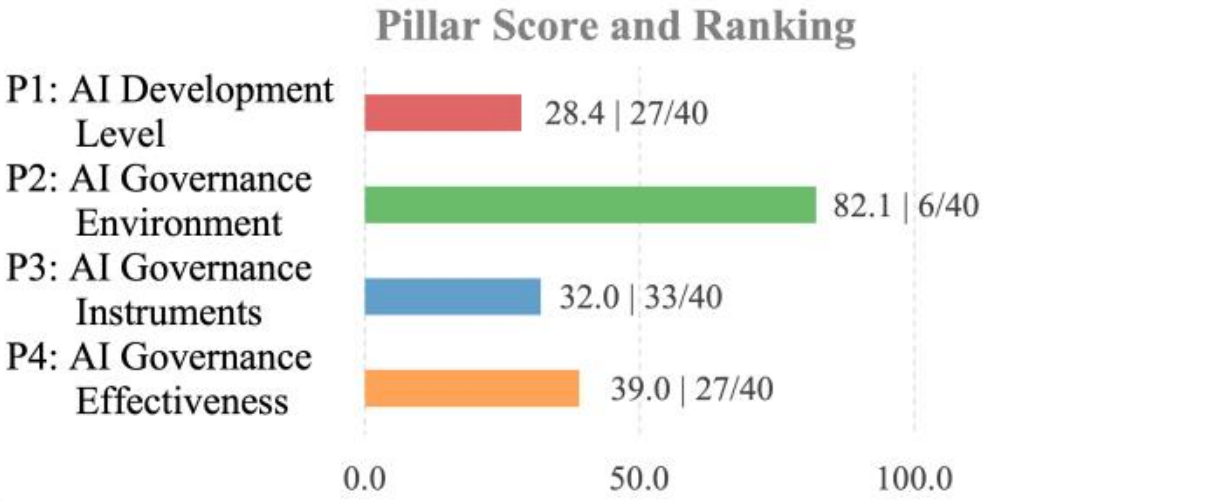

**Highest Rankings in:**

D4 AI Risk Exposure (Ranking: 9/40)
D13 Public Understanding of AI (Ranking: 9/40)

**Lowest Rankings in:**

D14 AI Social Acceptance (Ranking: 37/40)
D2 AI Infrastructure (Ranking: 33/40)

# Republic of Korea

| AGILE Index Ranking | Population(2024) | GDP Per Capita(2024) | Country Group |
|---|---|---|---|
| 4/40 | 52 Million | 45,377 $ | High Income |

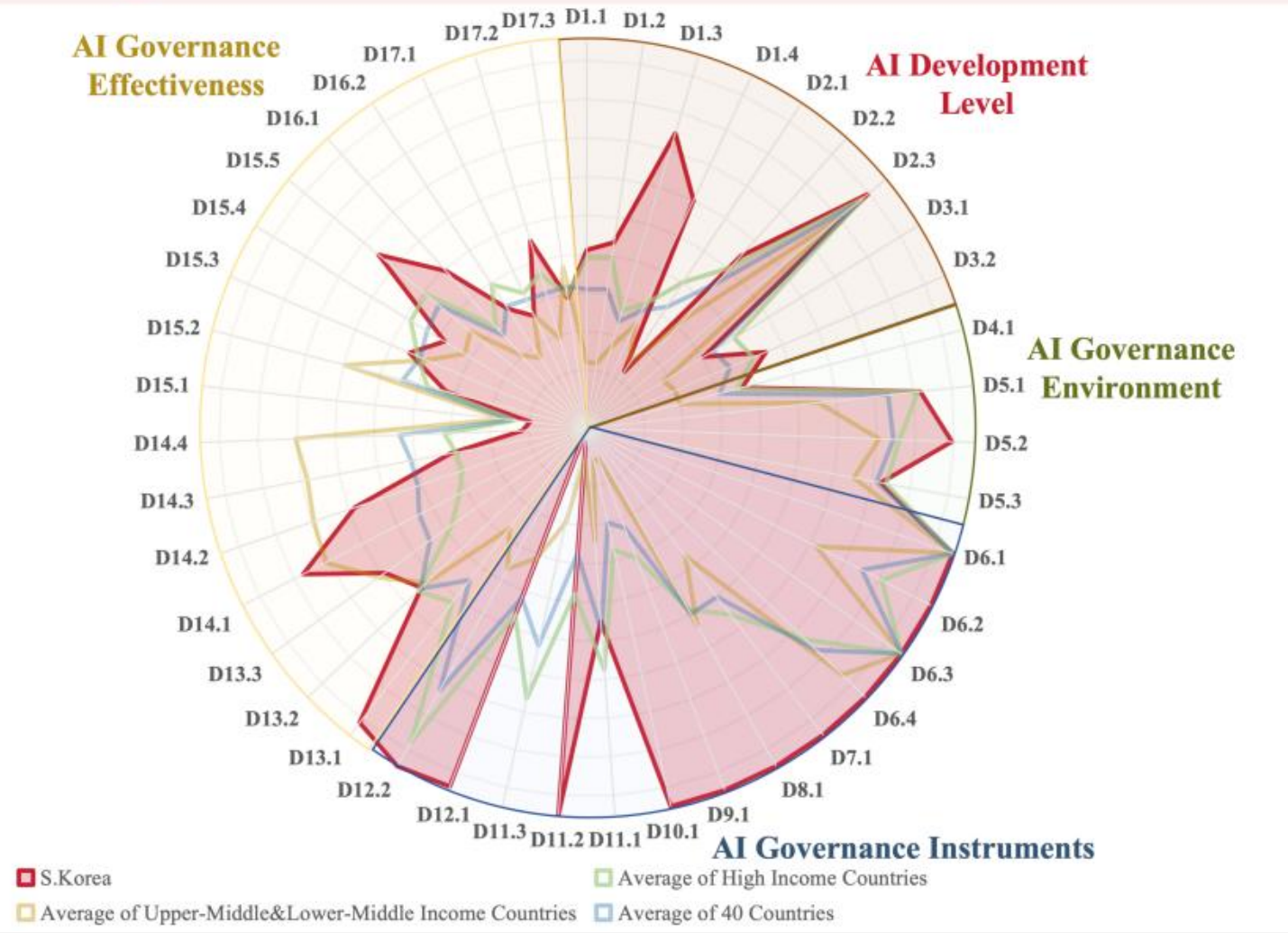

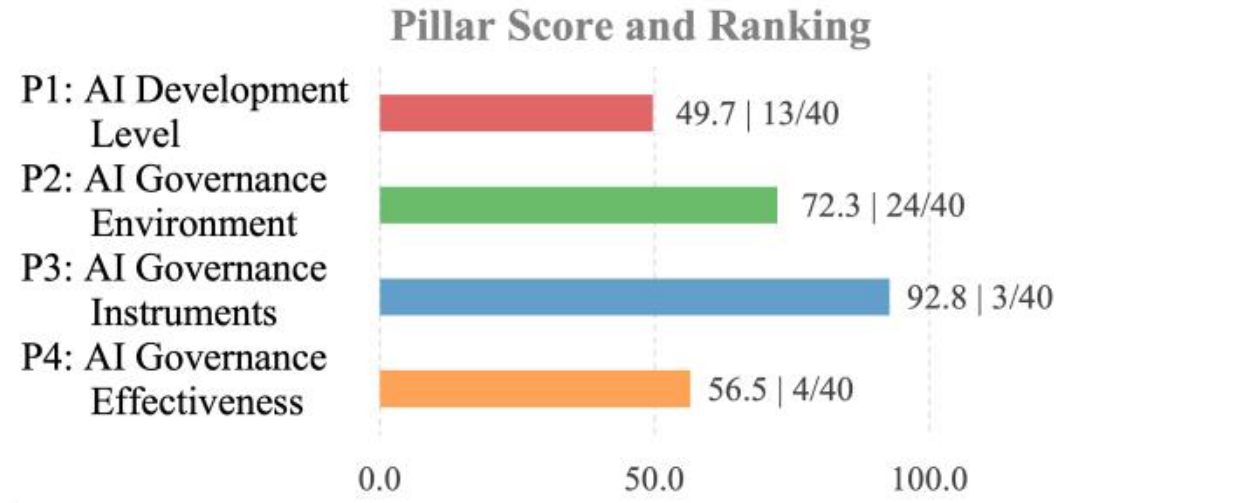

**Highest Rankings in:**

D13 Public Understanding of AI (Ranking: 2/40)
P3 AI Governance Instruments (Ranking: 3/40)

**Lowest Rankings in:**

D4 AI Risk Exposure (Ranking: 26/40)
D2 AI Infrastructure (Ranking: 24/40)

# Russian Federation

| AGILE Index Ranking | Population(2024) | GDP Per Capita(2024) | Country Group |
|---|---|---|---|
| 30/40 | 145 Million | 14,001 $ | High Income |

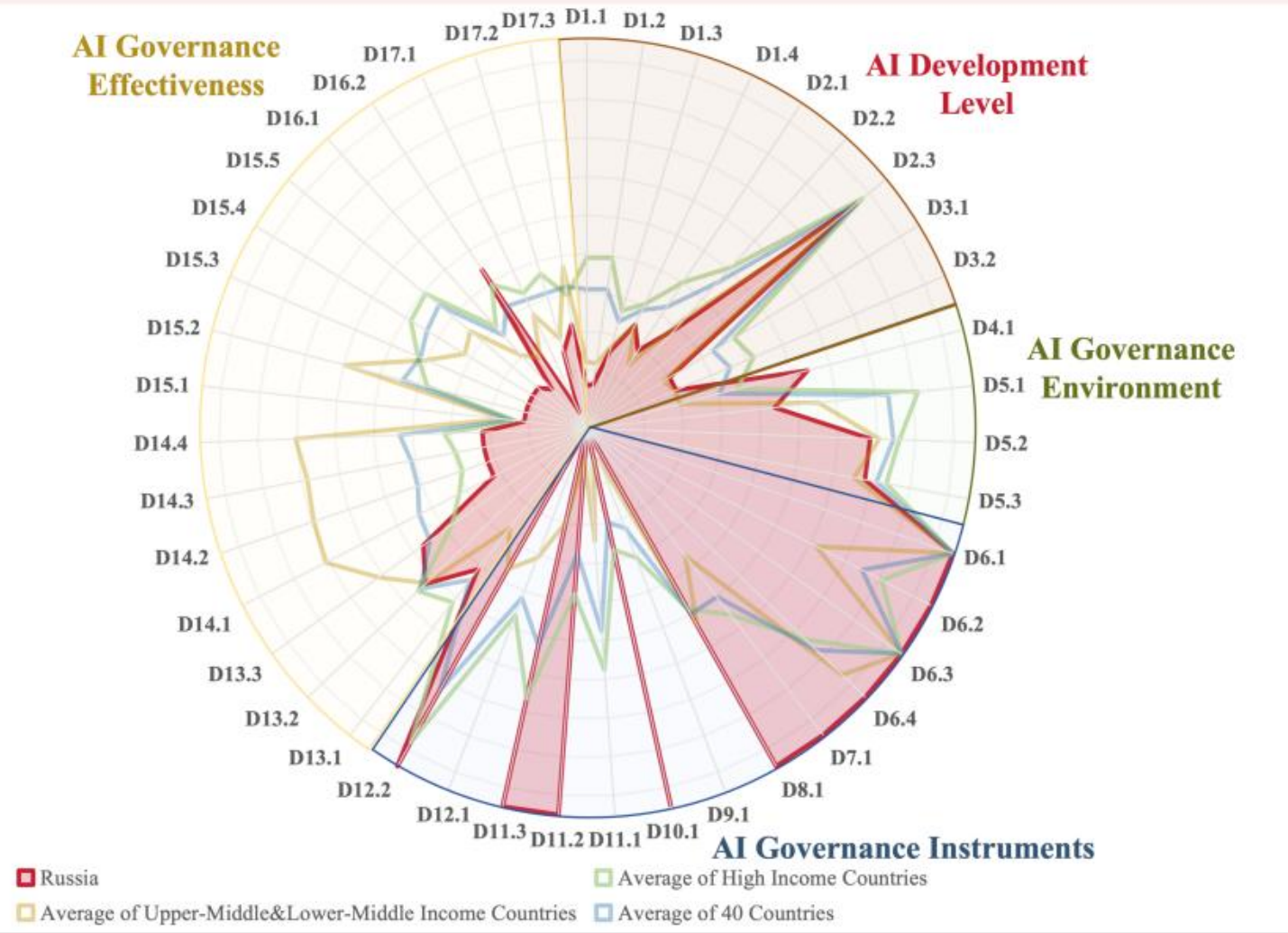

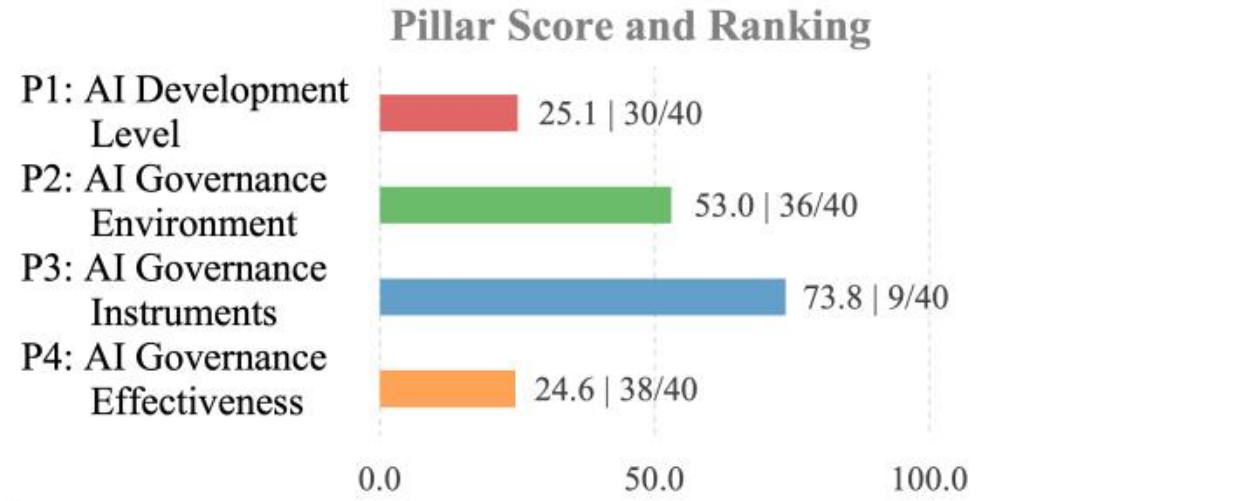

**Highest Rankings in:**

P3 Governance Instruments (Ranking: 9/40)
D16 Data & Algorithm Openness (Ranking: 18/40)

**Lowest Rankings in:**

D15 AI Development Inclusivity (Ranking: 39/40)
D17 AI Governance Research Activity (Ranking: 38/40)

# Saudi Arabia

| AGILE Index Ranking | Population(2024) | GDP Per Capita(2024) | Country Group |
|---|---|---|---|
| 18/40 | 34 Million | 30,393 $ | High Income |

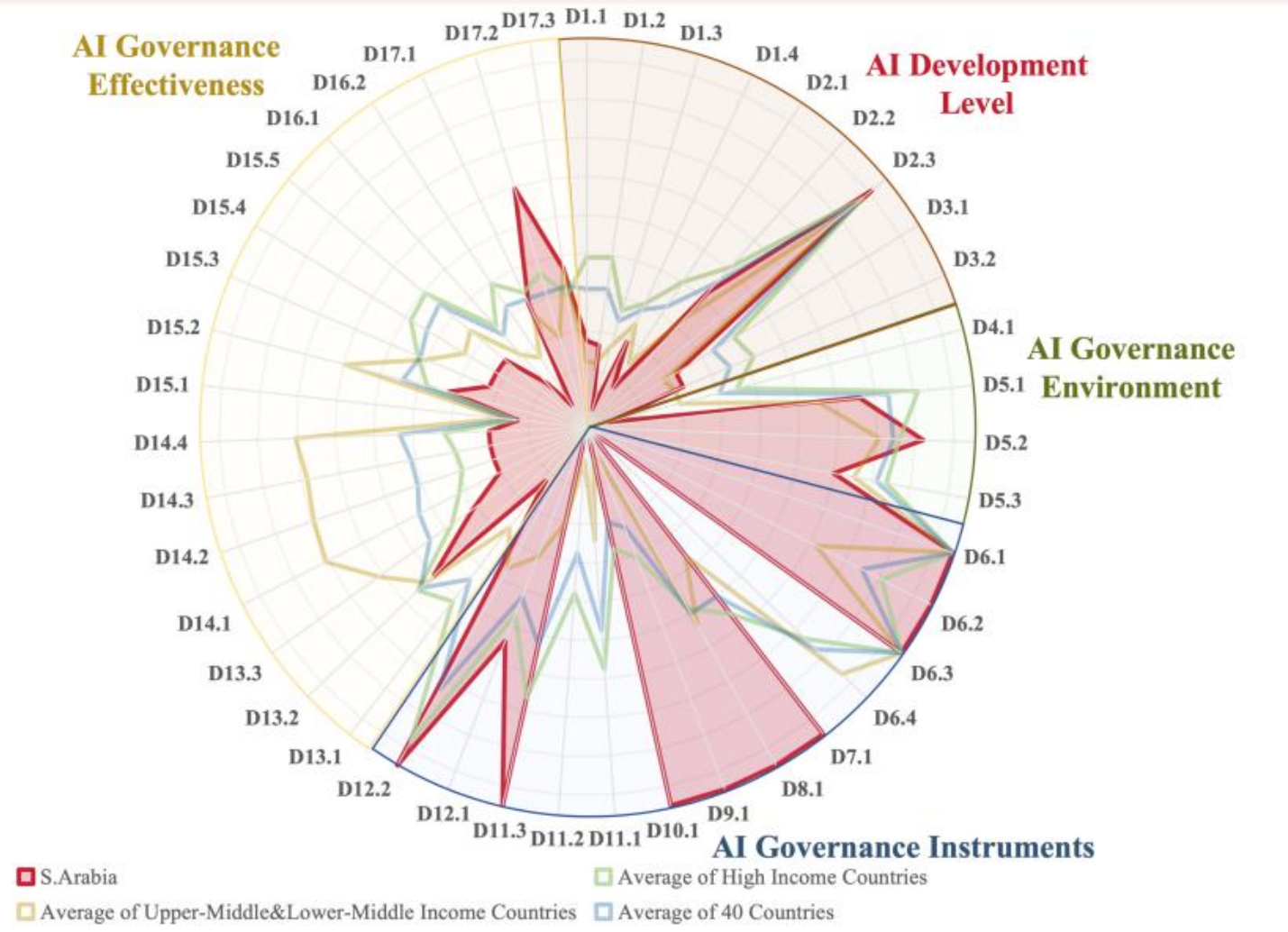

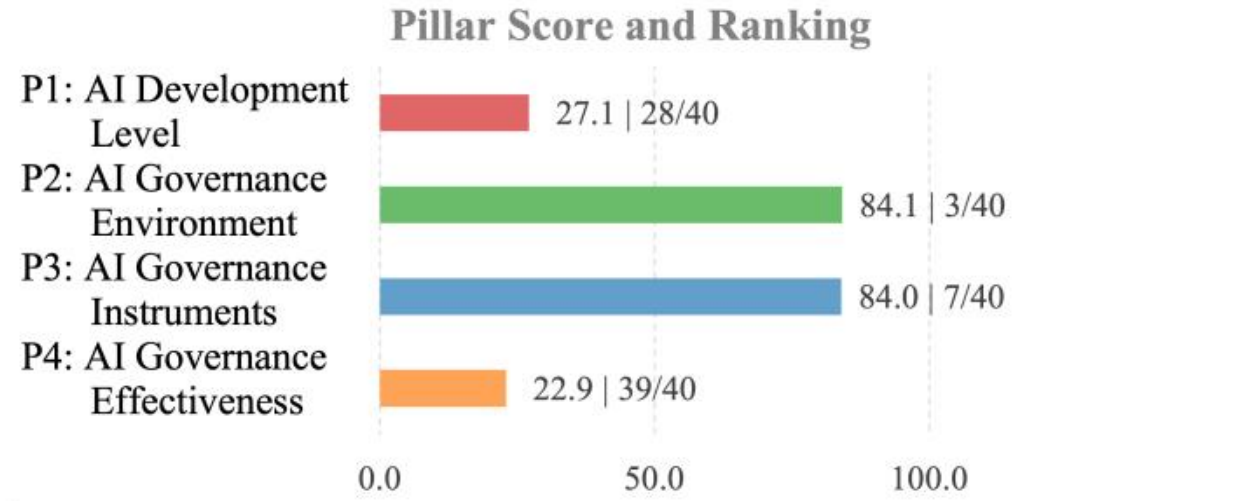

**Highest Rankings in:**

D4 AI Risk Exposure (Ranking: 1/40)
P3 Governance Instruments (Ranking: 7/40)

**Lowest Rankings in:**

D13 Public Understanding of AI (Ranking: 39/40)
D14 AI Social Acceptance (Ranking: 33/40)

# Singapore

| AGILE Index Ranking | Population(2024) | GDP Per Capita(2024) | Country Group |
|---|---|---|---|
| 6/40 | 5,832 Thousand | 88,285 $ | High Income |

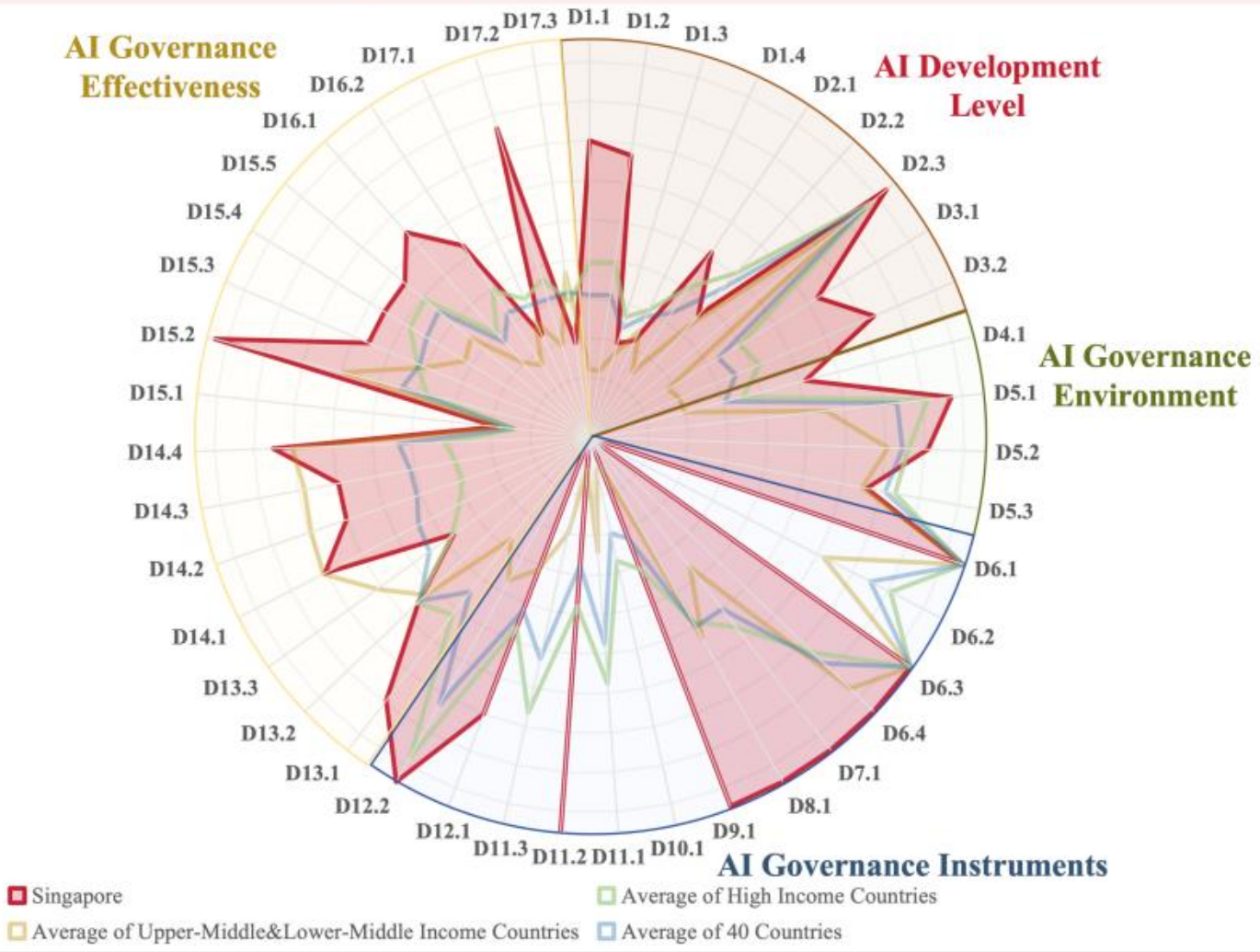

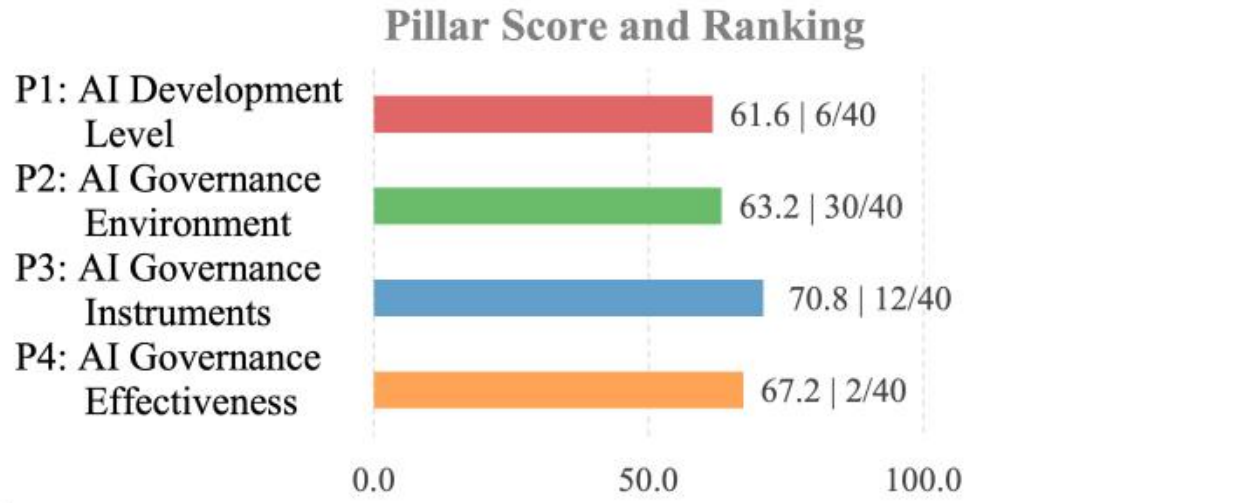

**Highest Rankings in:**

D3 AI Industry Vitality (Ranking: 3/40)

D15 AI Development Inclusivity (Ranking: 4/40)

**Lowest Rankings in:**

D4 AI Risk Exposure (Ranking: 32/40)

D5 Overall Governance Readiness (Ranking: 17/40)

# South Africa

| AGILE Index Ranking | Population(2024) | GDP Per Capita(2024) | Country Group |
|---|---|---|---|
| 40/40 | 64 Million | 7,540 $ | Upper-Middle Income |

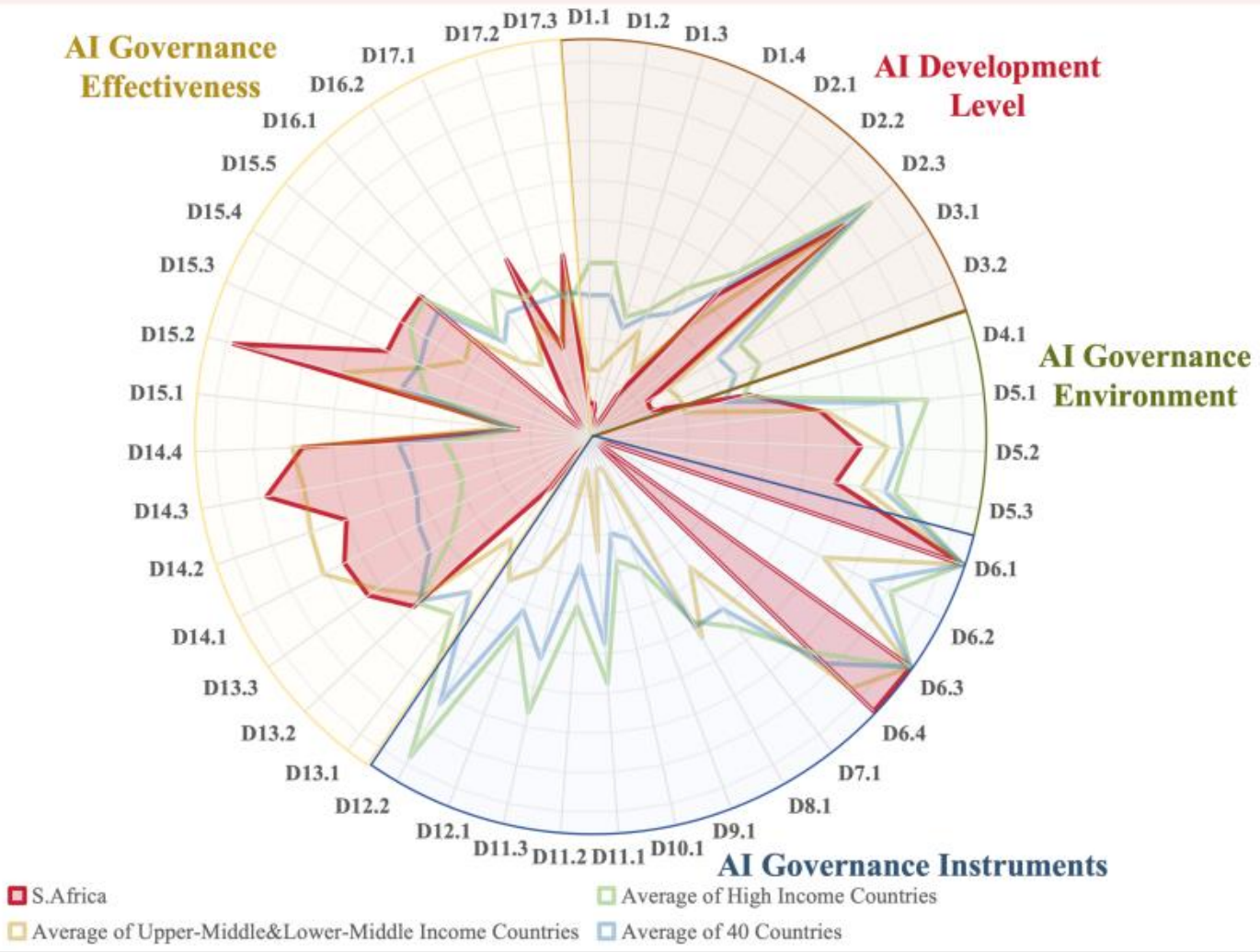

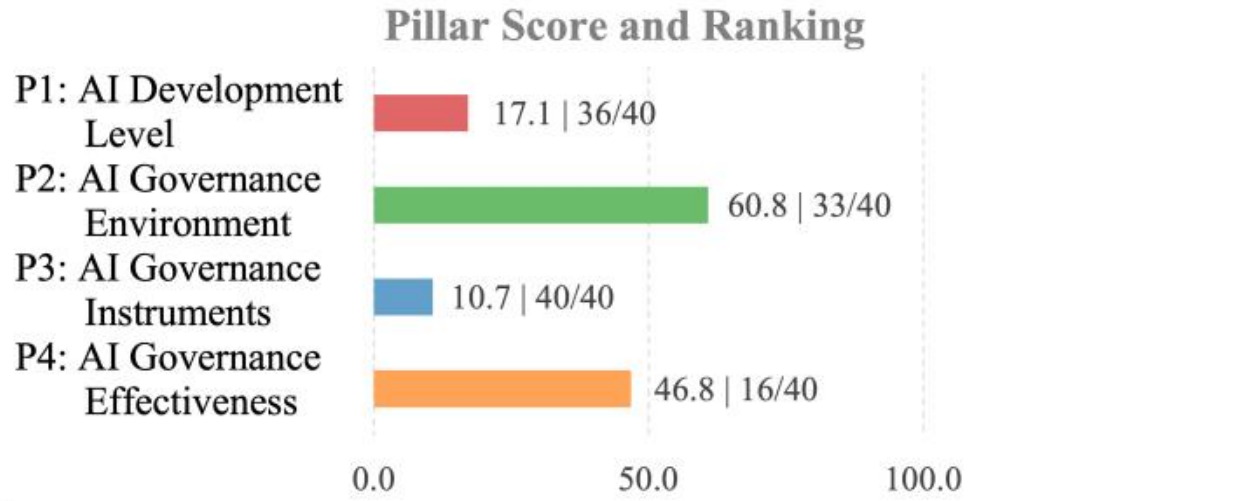

**Highest Rankings in:**

D14 AI Social Acceptance (Ranking: 6/40)

D15 AI Development Inclusivity (Ranking: 9/40)

**Lowest Rankings in:**

D1 AI Research and Development Activity (Ranking: 40/40)

D5 Overall Governance Readiness (Ranking: 40/40)

P3 Governance Instruments (Ranking: 40/40)

# Spain

| AGILE Index Ranking | Population(2024) | GDP Per Capita(2024) | Country Group |
|---|---|---|---|
| 20/40 | 48 Million | 37,663 $ | High Income |

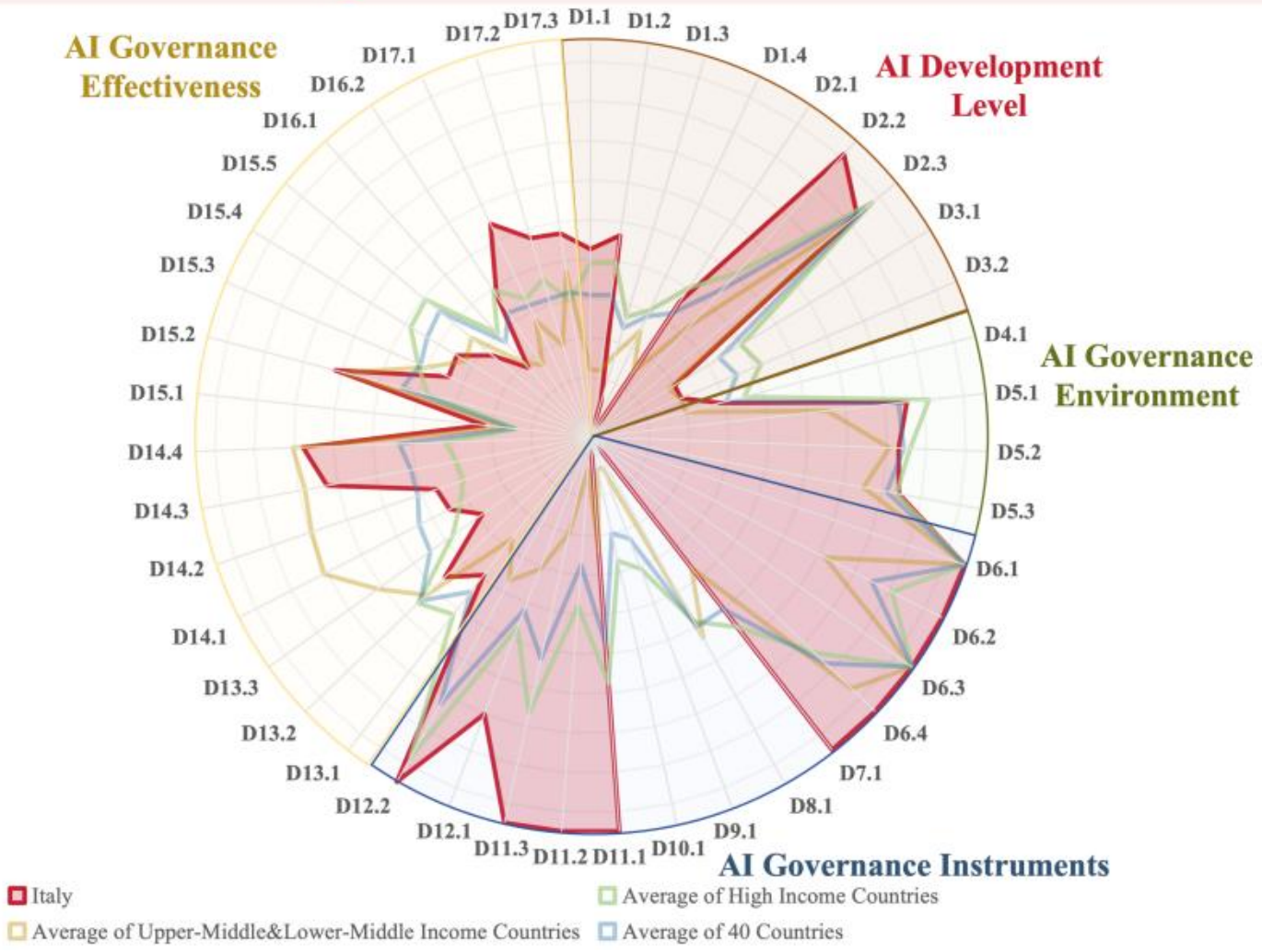

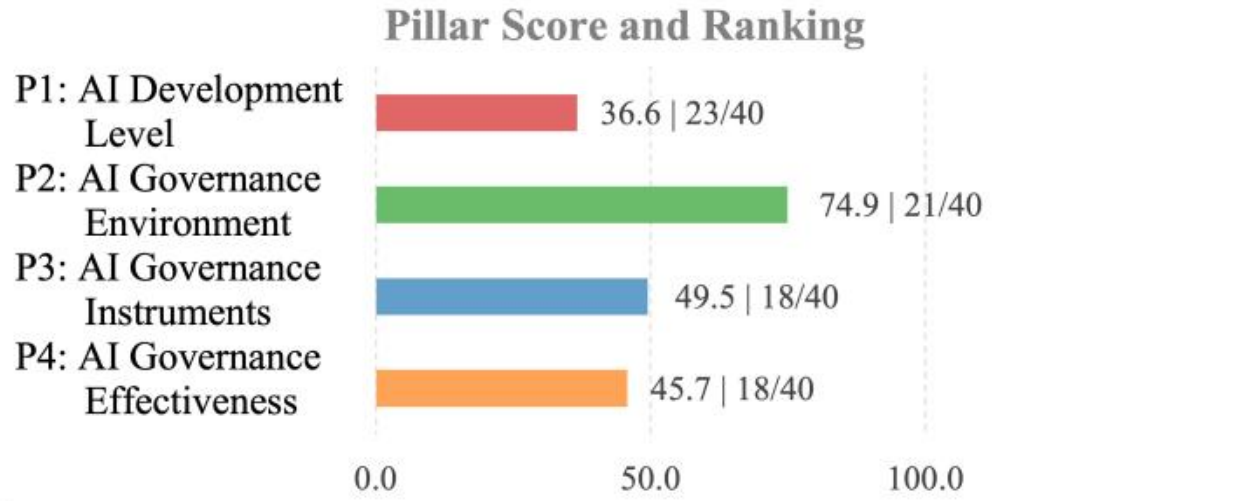

**Highest Rankings in:**

D17 AI Governance Research Activity (Ranking: 9/40)

D5 Overall Governance Readiness (Ranking: 14/40)

**Lowest Rankings in:**

D3 AI Industry Vitality (Ranking: 32/40)

D13 Public Understanding of AI (Ranking: 28/40)

# Sweden

| AGILE Index Ranking | Population(2024) | GDP Per Capita(2024) | Country Group |
|---|---|---|---|
| 13/40 | 18 Million | 73,353 $ | High Income |

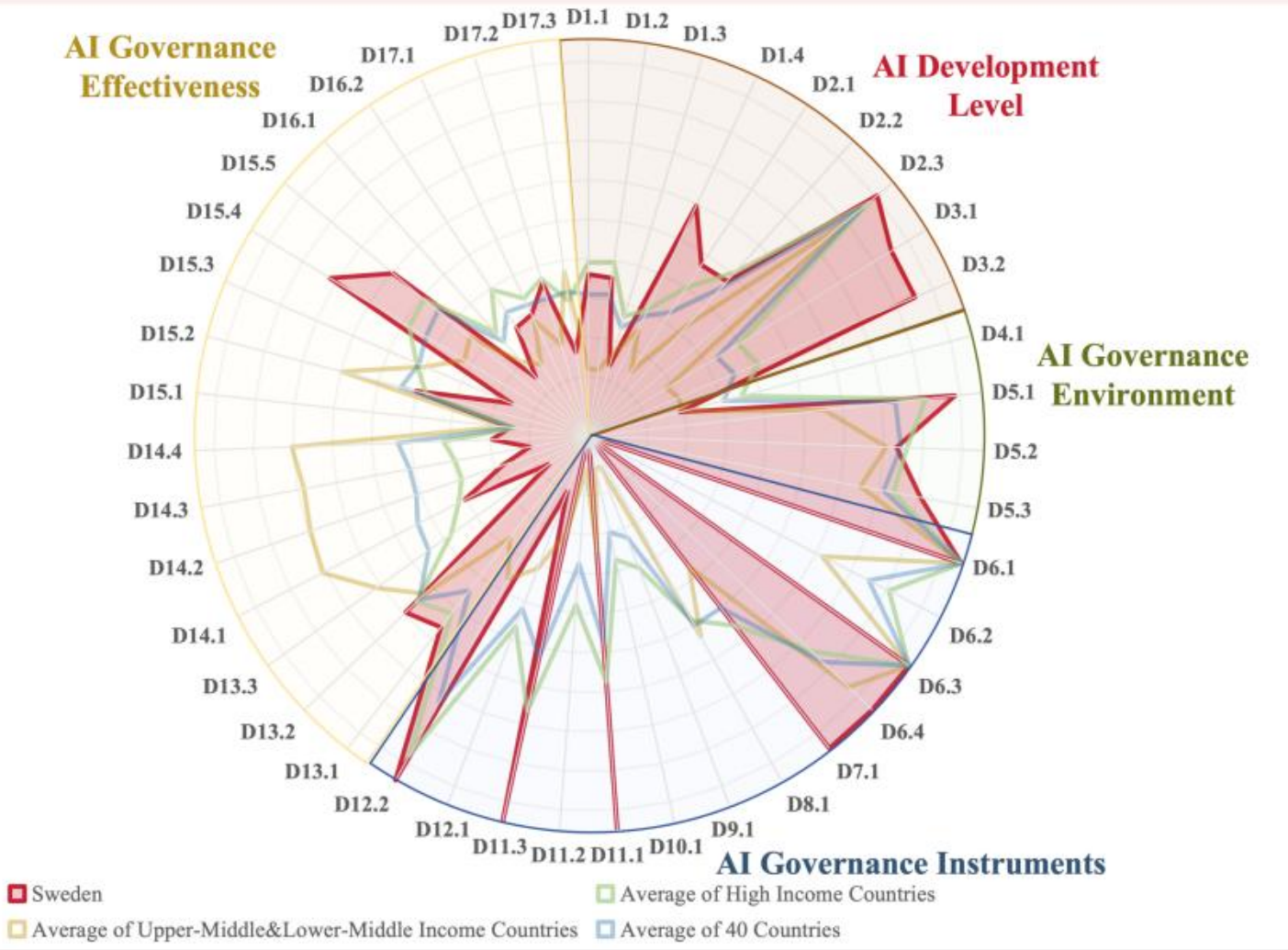

Pillar Score and Ranking

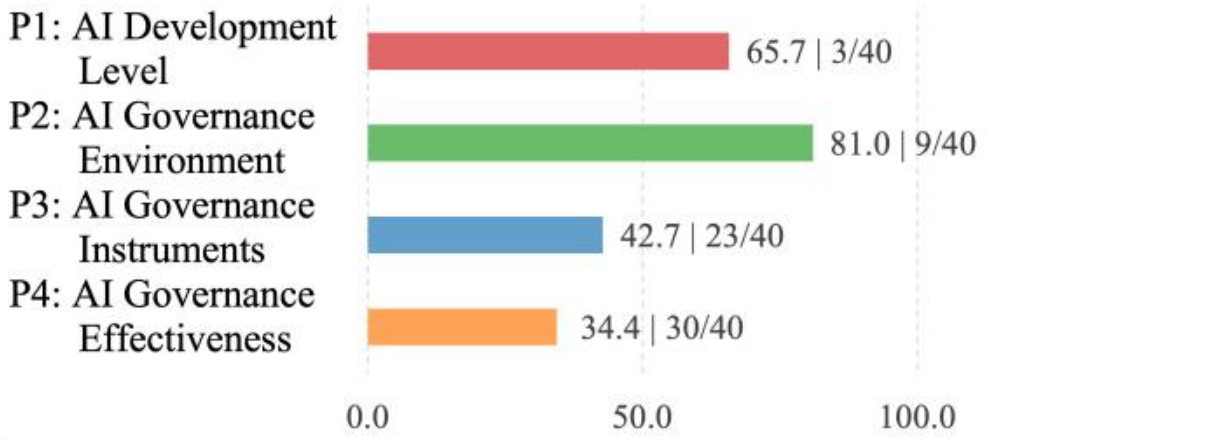

**Highest Rankings in:**
D3 AI Industry Vitality (Ranking: 2/40)
D5 Overall Governance Readiness (Ranking: 8/40)

**Lowest Rankings in:**
D14 AI Social Acceptance (Ranking: 34/40)
D13 Public Understanding of AI (Ranking: 30/40)

# Switzerland

| AGILE Index Ranking | Population(2024) | GDP Per Capita(2024) | Country Group |
|---|---|---|---|
| 16/40 | 8,922 Thousand | 118,656 $ | High Income |

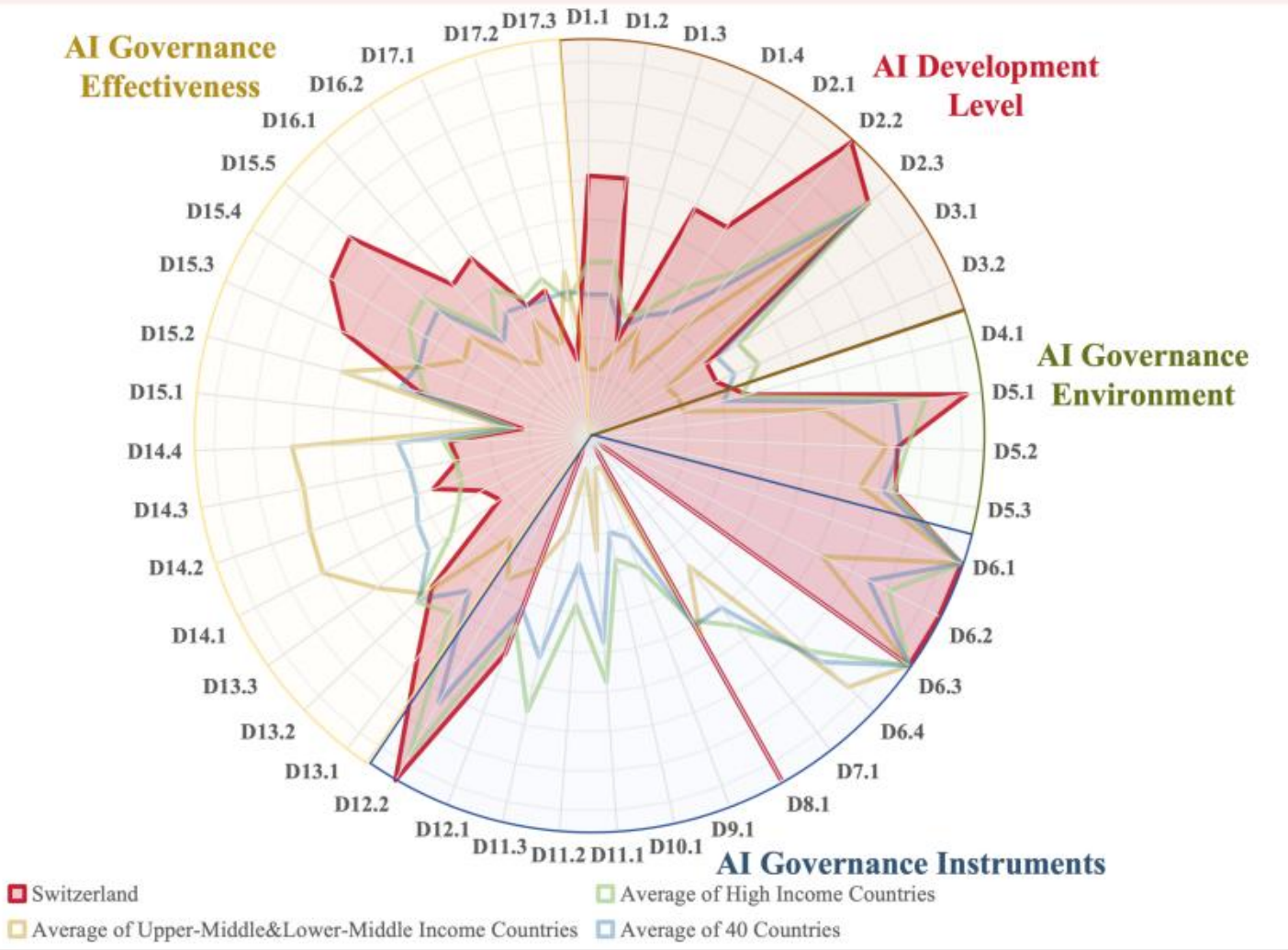

Pillar Score and Ranking

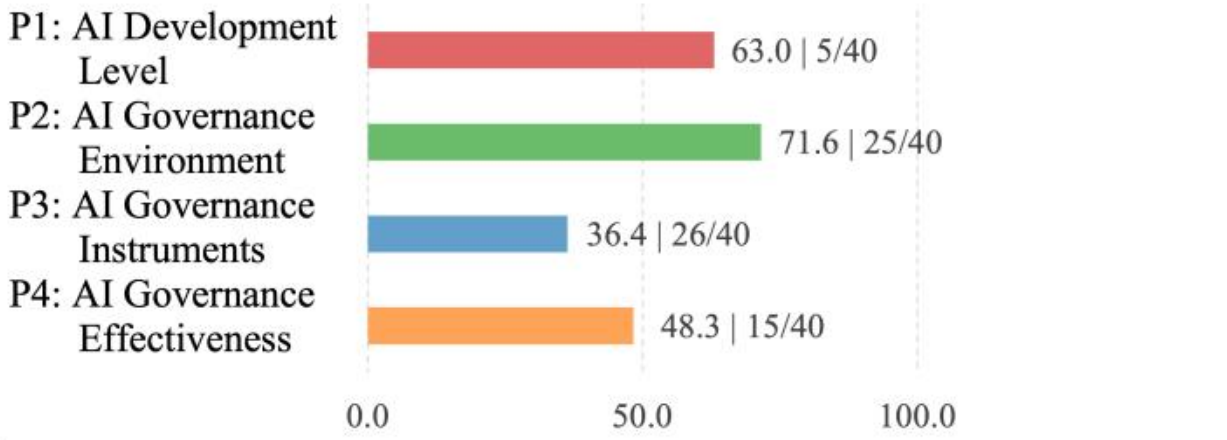

**Highest Rankings in:**
D2 AI Infrastructure (Ranking: 2/40)
D1 AI Research and Development Activity (Ranking: 5/40)

**Lowest Rankings in:**
D4 AI Risk Exposure (Ranking: 26/40)
D17 AI Governance Research Activity (Ranking: 26/40)

# Thailand

| AGILE Index Ranking | Population(2024) | GDP Per Capita(2024) | Country Group |
|---|---|---|---|
| 25/40 | 72 Million | 8,508 $ | Upper-Middle Income |

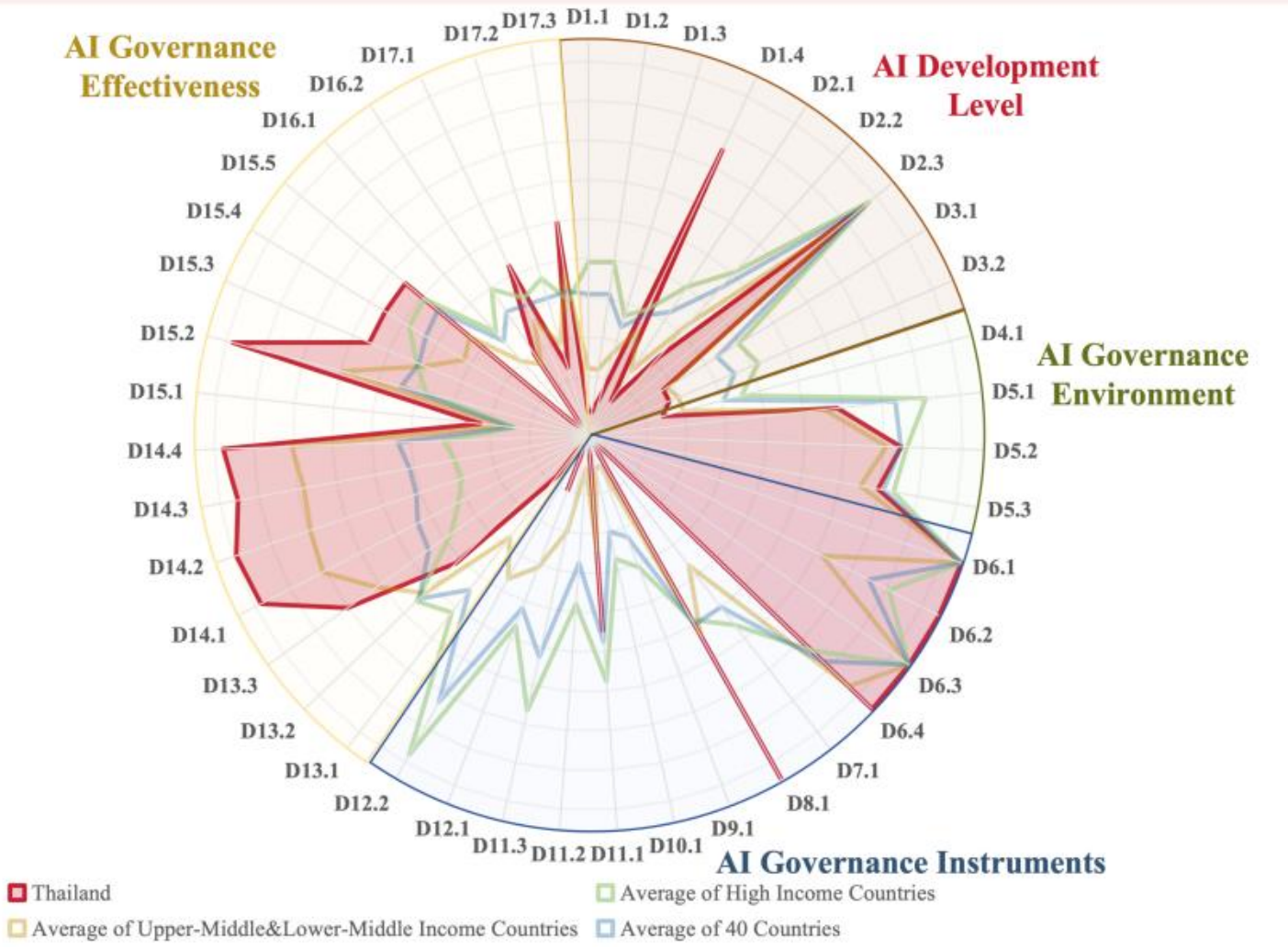

Pillar Score and Ranking

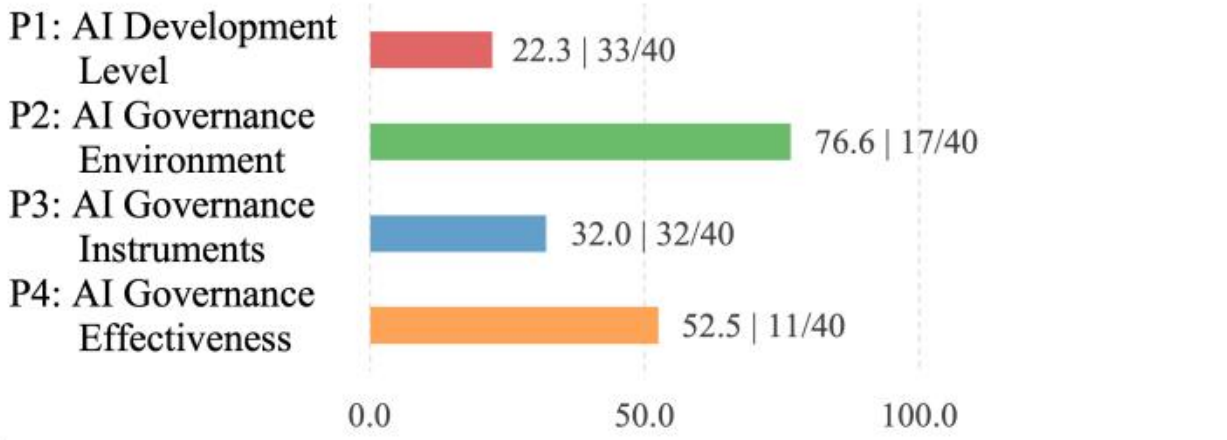

**Highest Rankings in:**
D14 AI Social Acceptance (Ranking: 2/40)
D15 AI Development Inclusivity (Ranking: 5/40)

**Lowest Rankings in:**
D2 AI Infrastructure (Ranking: 35/40)
D13 Public Understanding of AI (Ranking: 32/40)

# Türkiye

| AGILE Index Ranking | Population(2024) | GDP Per Capita(2024) | Country Group |
|---|---|---|---|
| 23/40 | 87 Million | 18,981 $ | Upper-Middle Income |

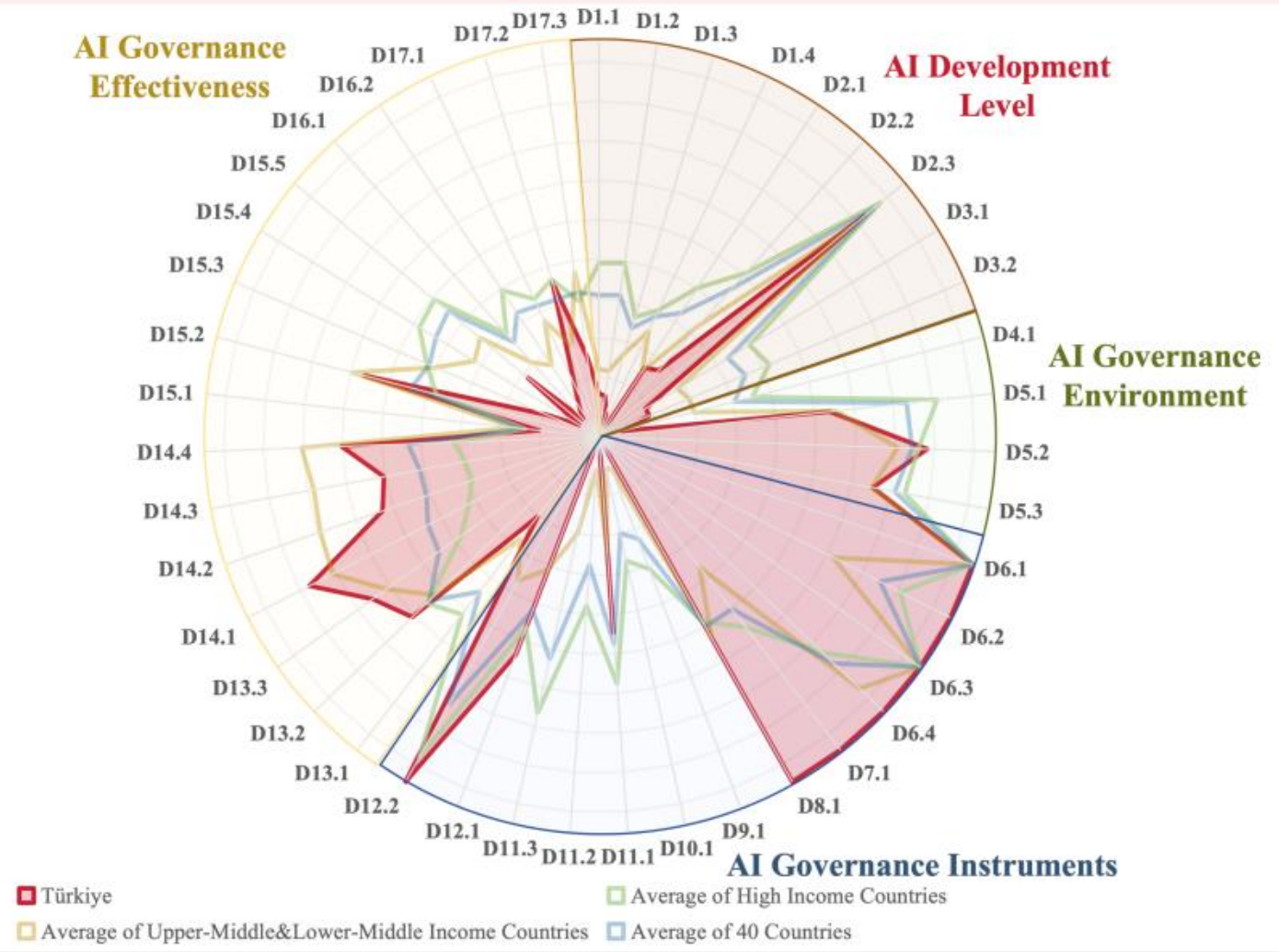

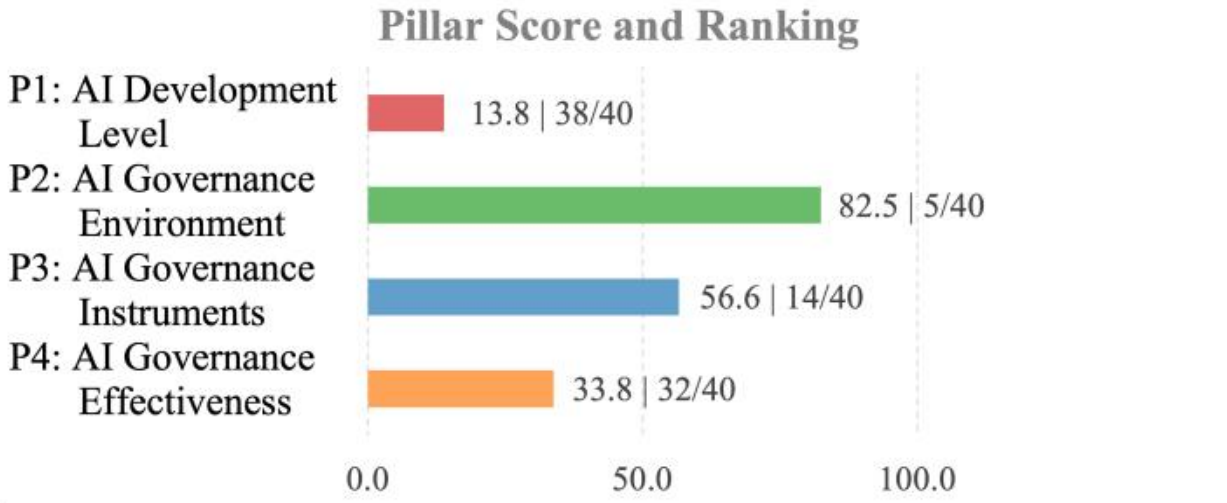

**Highest Rankings in:**

D4 AI Risk Exposure (Ranking: 1/40)

D14 AI Social Acceptance (Ranking: 11/40)

**Lowest Rankings in:**

D1 AI Research and Development Activity (Ranking: 39/40)

D16 Data & Algorithm Openness (Ranking: 38/40)

# United Arab Emirates

| AGILE Index Ranking | Population(2024) | GDP Per Capita(2024) | Country Group |
|---|---|---|---|
| 12/40 | 11 Million | 53,808 $ | High Income |

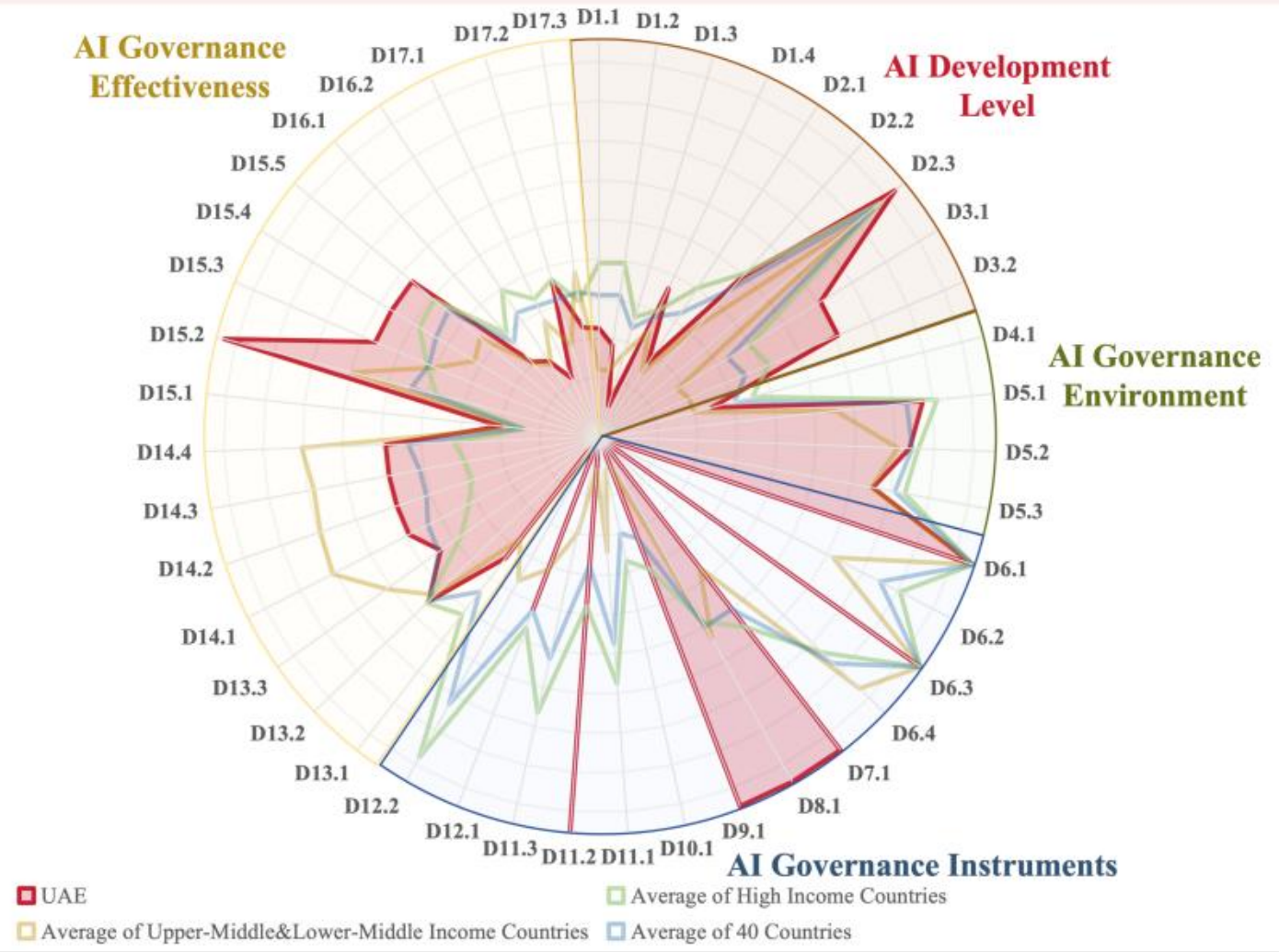

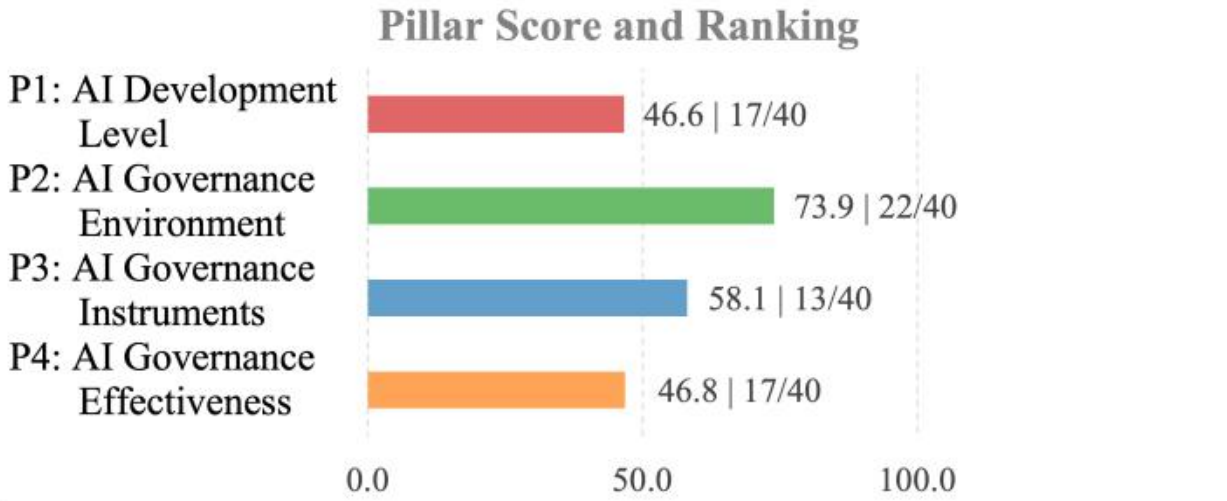

**Highest Rankings in:**

D15 AI Development Inclusivity (Ranking: 2/40)

D3 AI Industry Vitality (Ranking: 6/40)

**Lowest Rankings in:**

D1 AI Research and Development Activity (Ranking: 28/40)

D17 AI Governance Research Activity (Ranking: 27/40)

# United Kingdom

| AGILE Index Ranking | Population(2024) | GDP Per Capita(2024) | Country Group |
|---|---|---|---|
| 5/40 | 69 Million | 61,804 $ | High Income |

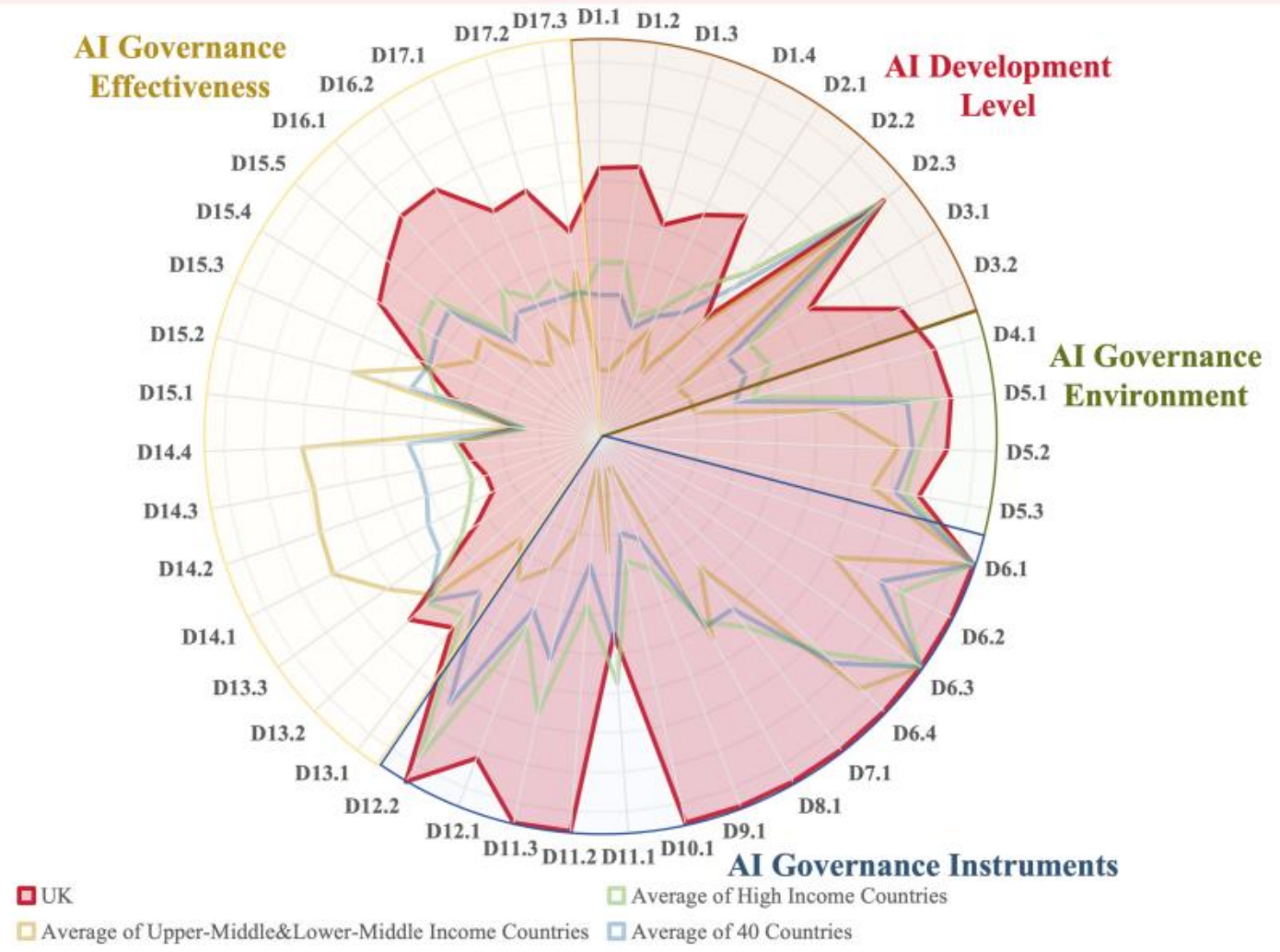

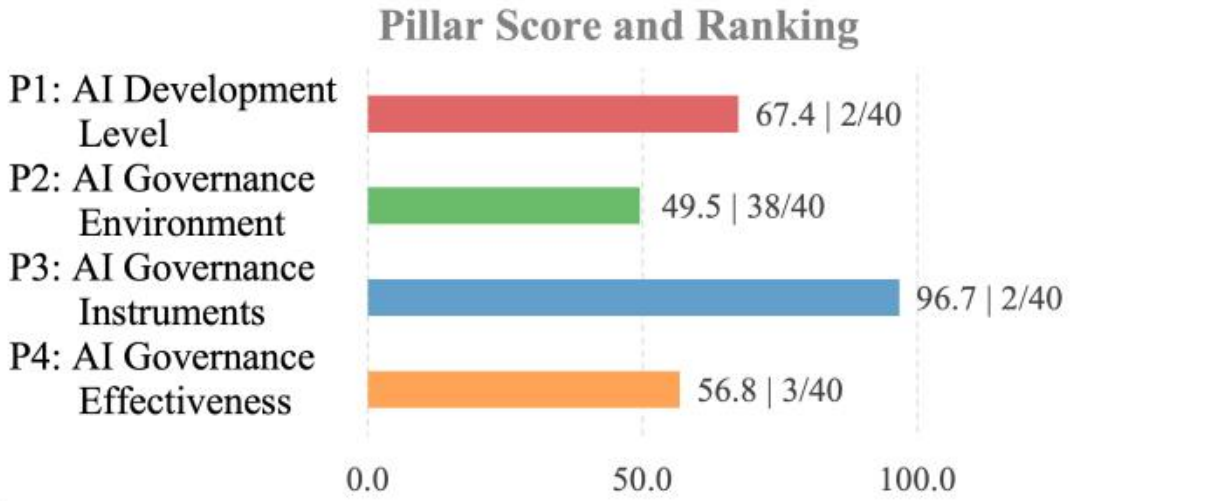

**Highest Rankings in:**

P3 Governance Instruments (Ranking: 2/40)

D1 AI Research and Development Activity (Ranking: 3/40)

**Lowest Rankings in:**

D4 AI Risk Exposure (Ranking: 39/40)

D14 AI Social Acceptance (Ranking: 27/40)

# United States

| AGILE Index Ranking | Population(2024) | GDP Per Capita(2024) | Country Group |
|---|---|---|---|
| 2/40 | 345 Million | 83,845 $ | High Income |

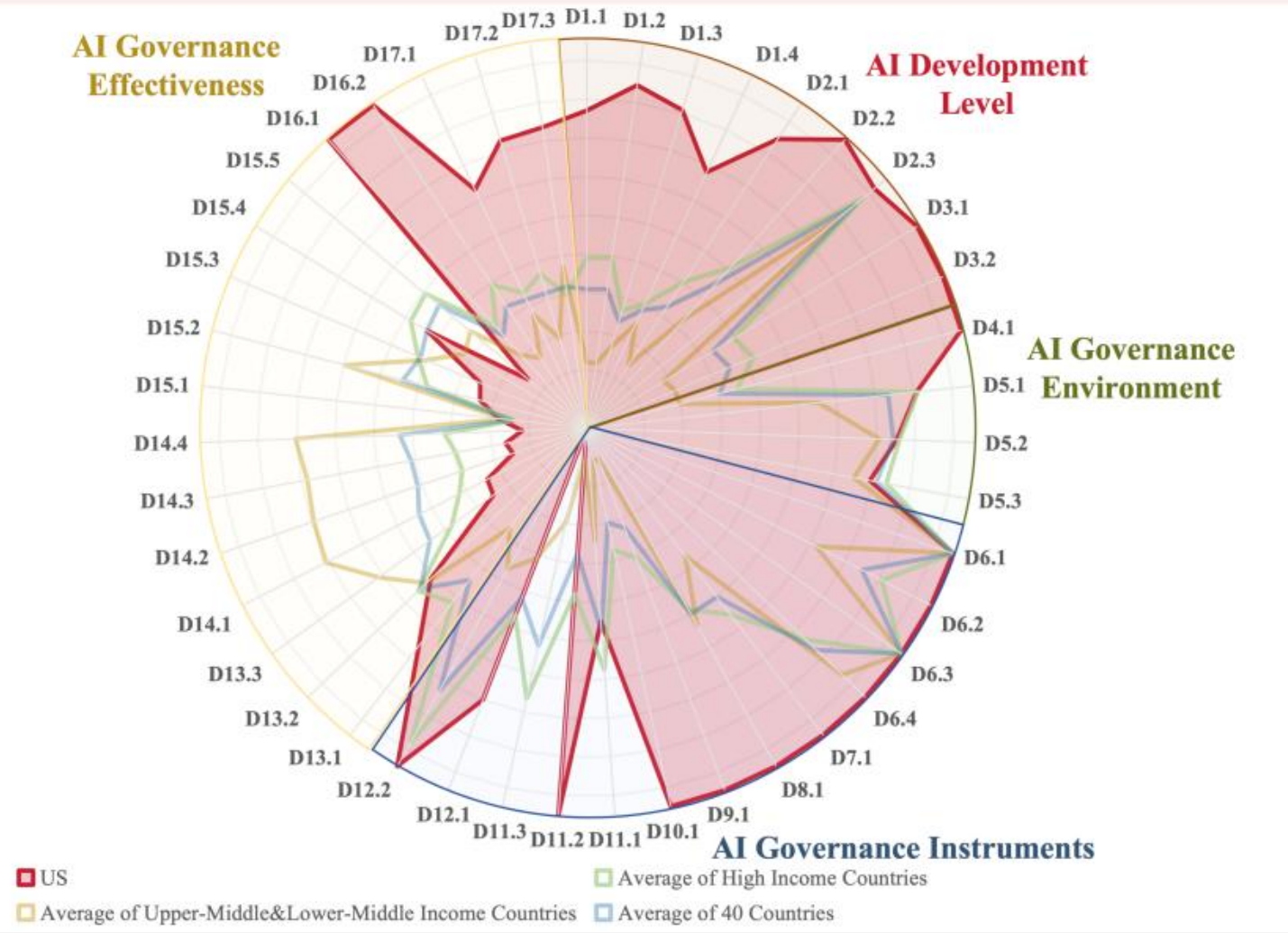

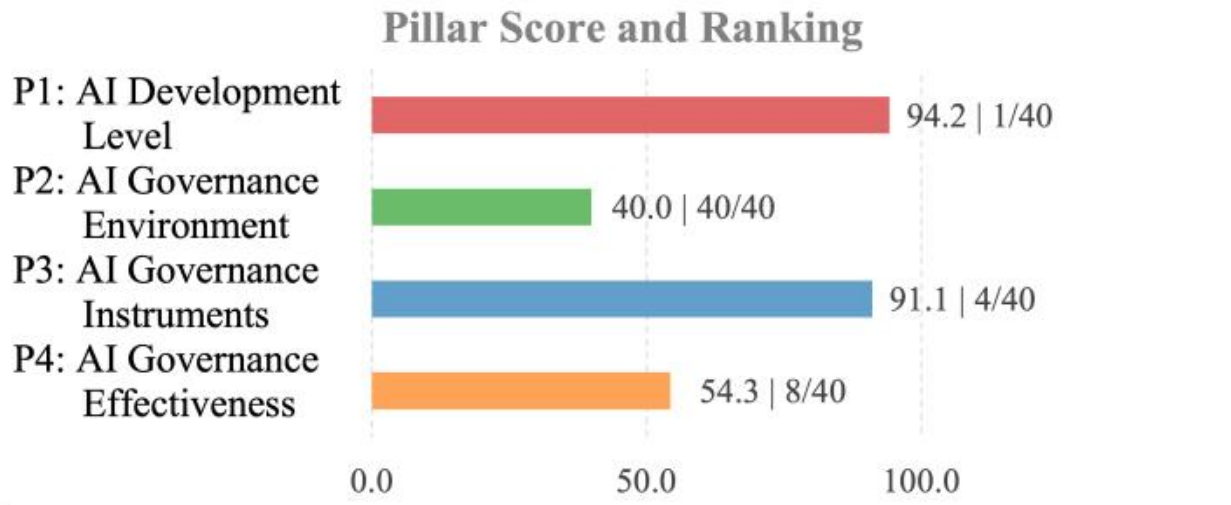

**Highest Rankings in:**

D1 AI Research and Development Activity (Ranking: 1/40)
D2 AI Infrastructure (Ranking: 1/40)
D3 AI Industry Vitality (Ranking: 1/40)
D16 Data & Algorithm Openness (Ranking: 1/40)
D17 AI Governance Research Activity(Ranking: 1/40)

**Lowest Rankings in:**

D4 AI Risk Exposure (Ranking: 40/40)
D14 AI Social Acceptance (Ranking: 38/40)

# Appendix

# 1. Indicator System and Data Sources

*Table 5 AGILE Index 2025: Dimensions and Indicators (in Detail)*

| Pillars | Dimensions | Content of Evaluation | Referencing Articles from UNESCO's AI Rec. and RAM[2] | Indicators |
|---|---|---|---|---|
| P1. AI Development Level | D1. AI Research and Development Activity | Assessment of countries' level of activity in AI-related R&D | AI Rec. A83, RAM4.2.1 | D1.1. Number of AI-related publications & national per capita |
| | | | | D1.2. Number of active AI researchers & its ratio to total population |
| | | | | D1.3. Number of granted AI patents & national per capita |
| | | | | D1.4. Number of large-scale AI systems developed & its ratio to GDP |
| | D2. AI Infrastructure | Assessment of the level of deployment and access to AI technologies and digital ecosystem infrastructure in each country | AI Rec. A59, A80, RAM6.2.1, 6.2.3 | D2.1. Number of data centers & its ratio to total population |
| | | | | D2.2. National supercomputing FLOPS aggregate & its ratio to total population |
| | | | | D2.3. Overall level of ICT development in the country |
| | D3. AI Industry Vitality | Assessment of the level of activity in AI-related industries in each country | AI Rec. A117, RAM5.2.3 | D3.1. Private investment in AI & its ratio to GDP |
| | | | | D3.2. Number of newly funded AI companies & its ratio to GDP |
| P2. AI Governance Environment | D4. AI Risk Exposure | Assessment of the level of exposure to AI-related ethical and safety risks in each country | AI Rec. A50 | D4.1. Number of AI-related risk cases/incidents & its ratio to GDP |
| | D5. Overall Governance Readiness | Assessment of countries' preparedness and implementation capabilities in AI governance | AI Rec. A54, RAM2.2.2, RAM2.2.8 | D5.1. Overall level of governance in the country |
| | | | | D5.2. Overall level of digital governance in the country |
| | | | | D5.3. Overall level of sustainable development in the country |
| P3. AI Governance Instruments | D6. AI Strategy & Planning | Assessment of the development of AI | AI Rec. A56, A71, RAM 2.2.1 | D6.1. Whether an AI strategy has been released in the country |
| | | | | D6.2. Whether the AI strategy has implementation plans |

[2] The references in this column indicate the supporting articles from the 'IV. Areas of policy action' section of UNESCO's *Recommendation on the Ethics of Artificial Intelligence* (hereafter referred to as the AI Rec.)as well as the UNESCO's *Readiness Assessment Methodology* (hereafter referred to as the RAM) that closely correspond to the evaluation content of the specific AGILE Index Dimension.

| | | | | |
|---|---|---|---|---|
| | | strategy/planning/ roadmap in each country | | D6.3. Whether the AI strategy mentions training or skills upgrading |
| | | | | D6.4. Whether the AI strategy has an ethical component |
| | D7. AI Governance Bodies | Assessment of the establishment of AI governance institutions or bodies in each country | AI Rec. A58, RAM1.4 | D7.1. Whether AI governance bodies have been established or designated in the country |
| | D8. AI Principles & Norms | Assessment of the development of AI governance principles and norms in each country | AI Rec. A48 | D8.1. Whether governments have issued national-level AI principles or norms |
| | D9. AI Impact Assessment | Assessment of the development of AI impact assessment tools/frameworks in each country | AI Rec. A50 | D9.1. Whether governments have introduced AI impact assessment mechanisms |
| | D10. AI Standards & Certification | Assessment of the establishment of AI standards/certification mechanisms in each country | AI Rec. A64 | D10.1. Whether governments have developed national-level standards and certification mechanisms for AI |
| | D11. AI Legislation Status | Assessment of the enactment status of AI laws and related regulations in each country | AI Rec. A133, RAM2.2.2 | D11.1. Whether countries have enacted or are in the process of enacting comprehensive national laws or regulations specifically targeting AI |
| | | | | D11.2. Whether countries have enacted national-level vertical laws or regulations specifically targeting AI |
| | | | | D11.3. Whether countries have implemented national-level data/information protection laws pertaining to AI |
| | D12. Global AI Governance Engagement | Assessment of the degree of countries' participation in international AI governance | AI Rec. A80, RAM6.2.2 | D12.1. Participation level in international AI governance actions/mechanisms |
| | | | | D12.2. Participation level in ISO AI standardization |
| P4. AI Governance Effectiveness | D13. Public Understanding of AI | Assessment of the public's AI competence and AI risk awareness in each country | AI Rec. A101, RAM4.2.1, 4.2.2 | D13.1. Public foundations of AI literacy |
| | | | | D13.2. Public discussion of AI |
| | | | | D13.3. Public awareness of AI's impact |
| | D14. AI Social Acceptance | Assessment of the degree of public acceptance in AI technologies and applications in each country | AI Rec. A39, RAM3.2.2 & 3.3.4 | D14.1.Public positive attitude towards AI development |
| | | | | D14.2. Public positive or neutral anticipation for AI usage |
| | | | | D14.3. Public trust in AI applications |
| | | | | D14.4. Enterprises' willingness to adopt AI |
| | D15. AI Development Inclusivity | Assessment of the inclusiveness relating to AI | AI Rec. A91, A105, RAM3.2.1 | D15.1. Gender ratio of active AI researchers |
| | | | | D15.2. Gender ratio of internet usage |
| | | | | D15.3. Share of young females who can program |
| | | R&D and applications in each country | | D15.4. Share of the aged using the internet |
| | | | | D15.5. Share of the low-income internet users |
| | D16. Data & Algorithm Openness | Assessment of the level of open source and openness of AI data and algorithms in each country | AI Rec. A75, A76 | D16.1. Number of impactful open AI models and datasets released |
| | | | | D16.2. The level of contributions in the AI developer community |
| | D17. AI Governance Research Activity | Assessment of countries' activity of research in AI governance | AI Rec. A131, RAM3.2.3, RAM4.2.1 | D17.1. Number of publications on AI Governance topics & their proportion in AI-related publications |
| | | | | D17.2. Number of publications on AI Safety&Security topics & their proportion in AI-related publications |
| | | | | D17.3. Number of publications on AI for SDGs topics & their proportion in AI-related publications |

***Table 6 The Newly Added Indicators in AGILE Index 2025 Compared with 2024 Edition***

| Pillar | Dimension | Newly Added Indicators |
|---|---|---|
| P1. AI Development Level | D2. AI Infrastructure | D2.3 Overall level of ICT development in the country |
| P2. AI Governance Environment | D5. Overall Governance Readiness | D5.2. Overall level of digital governance in the country |
| P3. AI Governance Instruments | D6. AI Strategy & Planning | D6.4 Whether the AI strategy has an ethical component |
| P4. AI Governance Effectiveness | D13. Public Understanding of AI | D13.3. Public awareness of AI's impact |
| | D14. AI Social Acceptance | D14.2. Public positive or neutral anticipation for AI usage |
| | | D14.3. Public trust in AI applications |
| | D15. AI Development Inclusivity | D15.2. Gender ratio of internet usage |
| | | D15.3. Share of young females who can program |
| | D17. AI Governance Research Activity | D17.2. Total number & the proportion of literature on AI safety topics |

# P1. AI Development Level

### D1. AI Research and Development Activity

AI Research and Development Activity refers to the level of activity in artificial intelligence-related research and development in various countries. According to UNESCO's *Recommendation on the Ethics of Artificial Intelligence*, Article 83, "Member States should encourage international cooperation and collaboration in the field of AI to bridge geo-technological lines." This recommendation aligns with Dimension 1, which involves assessing the level of AI development to facilitate comparative analysis of technological gaps among different countries and regions.

Dimension 1 currently covers four indicators:

- **D1.1. Number of AI-related publications & its ratio to total population**
  - Data Source: Based on statistical analysis of the DBLP Computer Science Bibliography literature database (Data from April 2024 to March 2025)
- **D1.2. Number of active AI researchers & its ratio to total population**
  - Data Source: Based on statistical analysis of the DBLP Computer Science Bibliography literature database; targeting on researchers who have authored AI-related publication(s) in DBLP during the past year (Data from April 2024 to March 2025)
- **D1.3. Number of granted AI patents & its ratio to total population**
  - Data Source: Based on statistical analysis of the World Intellectual Property Organization (WIPO)[3] PATENTSCOPE database (Data from April 2024 to March 2025); WIPO Patent Landscape Report on Generative AI 2024 (Dataset)
- **D1.4. Number of large-scale AI systems developed & its ratio to GDP**
  - Data Source: Epoch (2025)—with major processing by Our World in Data[4] (Cumulative number of large-scale AI systems by country; Country refers to "the location of the primary organization with which the authors of a large-scale AI systems are affiliated"; Web data accessed as of March 2025)

**D2. AI Infrastructure**

AI Infrastructure refers to the foundational technology and digital ecosystem for artificial intelligence. According to UNESCO's *Recommendation on the Ethics of Artificial Intelligence*, Article 59, "Member States should foster the development of, and access to, a digital ecosystem for ethical and inclusive development of AI systems at the national level...Such an ecosystem includes, in particular, digital technologies and infrastructure..." and Article 80, "Member States should work through international organizations to provide platforms for international cooperation on AI for development, including... infrastructure, and facilitating multi-stakeholder collaboration...". These recommendations align with Dimension 2.

Dimension 2 currently covers three indicators:

- **D2.1. Number of data centers & its ratio to total population**
  - Data Source: Data Center Map database[5] (Web data accessed as of March 2025)
- **D2.2. National supercomputing FLOPS aggregate & its ratio to total population**
  - Data Source: Based on statistical analysis of the TOP500 List of Supercomputer[6] (November 2024, The 64th TOP500 List; Sum of Rpeak & Rmax by country)
- **D2.3. Overall level of ICT development in the country**

3 https://www.wipo.int/portal/en/index.html

4 https://ourworldindata.org/grapher/cumulative-number-of-large-scale-ai-systems-by-country.

5 https://www.datacentermap.com/datacenters/

6 https://www.top500.org/lists/top500/list/2024/11/

- Data Source: International Telecommunication Union(ITU)'s ICT Development Index (IDI) 2024

**D3. AI Industry Vitality**

The AI Industry Vitality refers to the activity of a country in artificial intelligence-related industries. According to UNESCO's *Recommendation on the Ethics of Artificial Intelligence*, Article 117, "Member States should support collaboration agreements among governments, academic institutions, vocational education and training institutions, industry, workers' organizations and civil society to bridge the gap of skillset requirements to align training programmes and strategies with the implications of the future of work and the needs of industry, including small and medium enterprises," this recommendation is consistent with Dimension 3.

Dimension 3 currently covers two indicators:

- **D3.1. Private investment in AI & its ratio to GDP**
  - Data Source: *Artificial Intelligence Index Report 2025*—Stanford University &QUID[7] (Global private investment in AI by geographic area, 2024)
- **D3.2. Number of newly funded AI companies & its ratio to GDP**
  - Data Source: *Artificial Intelligence Index Report 2025*—Stanford University &QUID (Number of newly funded AI companies by geographic area, 2024)

## P2. AI Governance Environment

**D4. AI Risk Exposure**

AI Risk Exposure refers to the degree of exposure to ethical and safety risks and issues related to AI in various countries. The more issues there are, the higher the urgency for AI governance in that country. Therefore, this dimension has a negative impact on the background. According to UNESCO's *Recommendation on the Ethics of Artificial Intelligence*, Article 50, "Member States should introduce frameworks for impact assessments, such as ethical impact assessment, to identify and assess benefits, concerns and risks of AI systems, as well as appropriate risk prevention, mitigation and monitoring measures, among other assurance mechanisms," this recommendation is consistent with Dimension 4.

Dimension 4 currently has one indicator:

- **D4.1. Number of AI-related risk cases/incidents & its ratio to GDP**
  - Data Source: The AI incidents data are from multiple sources including the OECD AI Incidents Monitor (AIM)[8], the AI Incident Database (AIID)[9], the AI, Algorithmic, and Automation Incidents and Controversies Repository (AIAAIC)[10], and the AI Governance Observatory from AI Governance Online (AIGO)[11] (Web data assessed as of December 2024)

7 https://40006059.fs1.hubspotusercontent-na1.net/hubfs/40006059/Stanford_HAI_2024_AI-Index-Report.pdf

8 https://oecd.ai/en/incidents

9 https://incidentdatabase.ai/

10 https://www.aiaaic.org/aiaaic-repository

11 https://www.ai-governance.online/ai-governance-observatory

**D5. Overall Governance Readiness**

AI Governance Readiness refers to the favourable conditions in a country for governing AI and utilizing AI to achieve the United Nations Sustainable Development Goals. According to UNESCO's *Recommendation on the Ethics of Artificial Intelligence*, Article 54, "Member States should ensure that AI governance mechanisms are inclusive, transparent, multidisciplinary, multilateral (this includes the possibility of mitigation and redress of harm across borders) and multi-stakeholder. In particular, governance should include aspects of anticipation, and effective protection, monitoring of impact, enforcement and redress", this recommendation aligns with Dimension 5.

Dimension 5 currently covers three indicators:

- **D5.1. Overall level of governance in the country**
    - Data Source: Worldwide Governance Indicators[12] (WGI)—World Bank; Data accessed on 10/30/2024); The Human Development Index (HDI) 2022[13]—the United Nations Development Programme
- **D5.2. Overall level of digital governance in the country**
    - Data Source: 2024 Global Cybersecurity Index (v5)[14]—International Telecommunication Union (Country score); Global Data Barometer first edition (How are countries doing on Governance[15]); GovTech Maturity Index (GTMI) Data Dashboard[16]—World Bank (Data assessed as of March 2025); E-Government Development Index 2024; E-Participation Index 2024[17]—UN E-Government Knowledgebase
- **D5.3. Overall level of sustainable development in the country**
    - Data Source: *Sustainable Development Report 2024*[18]—the SDG Transformation Center

## P3. AI Governance Instruments

**D6. AI Strategy & Planning**

AI Strategy & Planning refers to the overall plans formulated by governments of various countries for the development and application of artificial intelligence. In Article 56 of UNESCO's *Recommendation on the Ethics of Artificial Intelligence*, it is stated: "Member States are encouraged to develop national and regional AI strategies..."; and in Article 71: "Member States should work to develop data governance strategies...". This recommendation is aligned with the direction assessed in Dimension 7, which evaluates whether AI-related strategies have been established.

12 https://www.worldbank.org/en/publication/worldwide-governance-indicators/interactive-data-access
13 https://hdr.undp.org/data-center/human-development-index#/indicies/HDI
14 https://www.itu.int/en/ITU-D/Cybersecurity/Pages/Global-Cybersecurity-Index.aspx
15 https://globaldatabarometer.org/module/governance/
16 https://www.worldbank.org/en/programs/govtech/gtmi
17 https://publicadministration.un.org/egovkb/Data-Center
18 https://unstats.un.org/sdgs/report/2024/

Dimension 6 currently covers four indicators:

- **D6.1. Whether an AI strategy has been released in the country**
  - Data Source: Desk Research Based on Publicly Available Data (Data as of March 2025)
- **D6.2. Whether the AI strategy has implementation plans**
  - Data Source: Desk Research Based on Publicly Available Data (Data as of March 2025)
- **D6.3. Whether the AI strategy mentions training or skills upgrading**
  - Data Source: Desk Research Based on Publicly Available Data (Data as of March 2025)
- **D6.4. Whether the AI strategy has an ethical component**
  - Data Source: Desk Research Based on Publicly Available Data (Data as of March 2025)

**D7. AI Governance Bodies**

AI Governance Bodies refer to specialized agencies established by governments of various countries to oversee AI governance affairs. According to UNESCO's *Recommendation on the Ethics of Artificial Intelligence*, Article 58, countries should "......consider adding the role of an independent AI Ethics Officer or some other mechanism to oversee ethical impact assessment, auditing and continuous monitoring efforts and ensure ethical guidance of AI systems". This recommendation aligns with the direction assessed in Dimension 7, which evaluates whether specialized agencies responsible for AI governance have been established.

Dimension 7 currently covers one indicator:

- **D7.1. Whether AI governance bodies have been established or designated in the country**
  - Data Source: Desk Research Based on Publicly Available Data (Data as of March 2025)

**D8. AI Principles & Norms**

AI Principles & Norms refer to the principles and norms established by governments of various countries to guide the development, application, and governance of artificial intelligence. According to UNESCO's *Recommendation on the Ethics of Artificial Intelligence*, it is underlined that "......ensure that national AI strategies are guided by ethical principles", and in Article 48, "The main action is for Member States to put in place effective measures, including, for example, policy frameworks or mechanisms." This recommendation aligns with Dimension 8, which evaluates whether principles and norms for guiding AI have been established.

Dimension 8 currently covers one indicator:

- **D8.1. Whether governments have issued national-level AI principles or norms**
  - Data Source: Desk Research Based on Publicly Available Data (Data as of March 2025)

**D9. AI Impact Assessment**

AI Impact Assessment refers to the evaluation of the potential impacts of artificial intelligence systems, including their effects on individuals, society, and the environment. According to UNESCO's *Recommendation on the Ethics of Artificial Intelligence*, Article 50, countries should "introduce frameworks for impact assessments, such as ethical impact assessment, to identify and assess benefits, concerns and risks of AI systems." This recommendation aligns with Dimension 9, which assesses whether tools/frameworks for assessing the impact of artificial intelligence have been developed.

Dimension 9 currently covers one indicator:

- **D9.1. Whether governments have introduced AI impact assessment mechanisms**
    - Data Source: Desk Research Based on Publicly Available Data (Data as of March 2025)

**D10. AI Standards & Certification**

AI Standards & Certification refers to mechanisms for assessing artificial intelligence systems to ensure compliance with relevant ethical and safety standards and to issue certification marks for compliance. According to UNESCO's *Recommendation on the Ethics of Artificial Intelligence*, Article 64, "Member States, international organizations and other relevant bodies should develop international standards that describe measurable, testable levels of safety and transparency, so that systems can be objectively assessed, and levels of compliance determined." This recommendation aligns with Dimension 10, which assesses whether mechanisms for assessing AI systems against standards have been developed.

Dimension 10 currently covers one indicator:

- **D10.1. Whether governments have developed national-level standards and certification mechanisms for AI**
    - Data Source: Desk Research Based on Publicly Available Data (Data as of March 2025)

**D11. AI Legislation Status**

AI legislation refers to national-level documents with binding legal force that countries establish for the development and governance of artificial intelligence. In the legislative dimension, AGILE focuses on three key areas: general national-level AI-related laws and regulations, national-level sector-specific AI-related laws and regulations, and data protection laws and regulations that include AI-related clauses or amendments. UNESCO's *Recommendation on the Ethics of Artificial Intelligence* "Recommends that Member States apply on a voluntary basis the provisions of this Recommendation by taking appropriate steps, including whatever legislative or other measures...," and its Article 133 states, "Data collection and processing should be conducted in accordance with international law, national legislation on data protection and data privacy, and the values and principles outlined in this Recommendation." This recommendation aligns with Dimension 11, which assesses the legal framework related to AI.

Dimension 11 currently covers three indicators:

- **D11.1. Whether countries have enacted or are in the process of enacting comprehensive national laws or regulations specifically targeting AI**
    - Data Source: Desk Research Based on Publicly Available Data (Data as of March 2025)
- **D11.2. Whether countries have enacted national-level vertical laws or regulations specifically targeting AI**
    - Data Source: Desk Research Based on Publicly Available Data (Data as of March 2025)

- **D11.3. Whether countries have implemented national-level data/information protection laws pertaining to AI**
  - Data Source: Desk Research Based on Publicly Available Data (Data as of March 2025)

**D12. Global AI Governance Engagement**

Global AI Governance Engagement refers to the participation of countries in international AI governance affairs through international mechanisms. According to UNESCO's *Recommendation on the Ethics of Artificial Intelligence*, Article 80, countries should "work through international organizations to provide platforms for international cooperation on AI for development." This recommendation aligns with Dimension 12, which assesses the degree of international participation of countries in the field of AI governance.

Dimension 12 currently covers two indicators:

- **D12.1. Participation level in international AI governance actions/mechanisms**
  - Data Source: Desk Research Based on Publicly Available Data: Recommendation on the Ethics of Artificial Intelligence 2021, Bletchley Declaration 2023, Seoul Ministerial Statement 2024, Global Digital Compact 2024, Statement on Inclusive and Sustainable Al for People and the Planet 2025, OECD AI Principles/G20 AI Principles 2019, REAIM Call to Action 2023, REAIM Blueprint for Action 2024, Inaugural Convening of International Network of AI Safety Institutes 2024. (Data as of March 2025)
- **D12.2. Participation level in ISO AI standardization**
  - Data Source: Based on the list of Participating Members of ISO/IEC JTC 1/SC 42 (Artificial intelligence); Web data assessed as of December 2024

## P4. AI Governance Effectiveness

**D13. Public Understanding of AI**

According to UNESCO's *Recommendation on the Ethics of Artificial Intelligence*, Article 101, member states should "work with international organizations, educational institutions and private and non-governmental entities to provide adequate AI literacy education to the public on all levels in all countries in order to empower people and reduce the digital divides and digital access inequalities resulting from the wide adoption of AI systems." This recommendation aligns with Dimension 13, which assesses whether efforts contribute to promoting public awareness of AI.

Dimension 13 currently covers three indicators:

- **D13.1. Public foundations of AI literacy**
  - Data Source: OECD PISA math scores 2018 & 2022; *Coursera Skill Report 2024* (number of Coursera learners)
- **D13.2. Public discussion of AI**
  - Data Source: Based on Google Trends' data of interest over time with AI as a field of study. (Data as of March 2025)

- **D13.3. Public awareness of AI's impact**
  - Data Source: IPSOS AI MONITOR 2024[19] (How much do you agree or disagree with the following? I have a good understanding of what artificial intelligence is; I know which types of products and services use artificial intelligence); OECD Going Digital Toolkit, GDT[20] (Ability of adults to identify online disinformation created by generative AI; Web data assessed as of March 2025)

**D14. AI Social Acceptance**

According to UNESCO's *Recommendation on the Ethics of Artificial Intelligence*, Article 39, "......allows for public scrutiny that can decrease corruption and discrimination, and can also help detect and prevent negative impacts on human rights. Transparency aims at providing appropriate information to the respective addressees to enable their understanding and foster trust." This value aligns with Dimension 13, which assesses whether there are surveys conducted regarding public attitudes towards AI.

Dimension 14 currently covers four indicators:

- **D14.1.Public positive attitude towards AI development**
  - Data Source: IPSOS AI MONITOR 2024 (How much do you agree or disagree with the following? Products and services using artificial intelligence have more benefits than drawbacks; Products and services using artificial intelligence make me excited); OECD GDT (Share of adults who feel AI will have a positive impact on their life; Web data assessed as of March 2025).
- **D14.2. Public positive or neutral anticipation for AI usage**
  - Data Source: IPSOS AI MONITOR 2024 (How much do you agree or disagree with the following? Products and services using artificial intelligence will profoundly change my daily life in the next 3-5 years; Do you think the increased use of artificial intelligence will make the following better, worse or stay the same in the next 3-5 years? The amount of disinformation on the internet, my entertainment options, the amount of time it takes me to get things done, my health, my job, the job market, the economy in ...)
- **D14.3. Public trust in AI applications**
  - Data Source: IPSOS AI MONITOR 2024 (How much do you agree or disagree with the following? I trust that companies that use artificial intelligence will protect my personal data; I trust artificial intelligence to not discriminate or show bias toward any group of people)
- **D14.4. Enterprises' willingness to adopt AI**
  - Data Source: *IBM Global AI Adoption Index 2022*[21] (AI adoption rates around the world: Deployed AI&Exploring AI)

**D15. AI Development Inclusivity**

19 https://www.ipsos.com/en-us/ipsos-ai-monitor-2024

20 https://goingdigital.oecd.org/indicator/81

21 https://www.ibm.com/downloads/documents/us-en/107a02e94a48f5c1

According to UNESCO's *Recommendation on the Ethics of Artificial Intelligence*, Article 91 states, "Member States should encourage female entrepreneurship, participation and engagement in all stages of an AI system life cycle," and Article 105 states, "Member States should promote the participation and leadership of girls and women, diverse ethnicities and cultures, persons with disabilities, marginalized and vulnerable people or people in vulnerable situations, minorities and all persons not enjoying the full benefits of digital inclusion." These recommendations align with Dimension 15, which evaluates whether the development of AI is inclusive of different groups and genders.

Dimension 15 currently covers five indicators:

- **D15.1. Gender ratio of active AI researchers**
  - Data Source: Based on statistical analysis of the DBLP Computer Science Bibliography literature database; targeting on researchers who have authored AI-related publication(s) in DBLP duiring the past year (Data from April 2024 to March 2025)
- **D15.2. Gender ratio of internet usage**
  - Data Source: Digital Gender Gaps[22]—The University of Oxford (Facebook gender gap: Fix & Mobile indicators; Web data assessed as of October 2024); OECD GDT (Disparity in Internet use between men and women: male-female; Web data assessed as of March 2025)
- **D15.3. Share of young females who can program**
  - Data Source: OECD GDT (Women as a share of all 16-24 year-olds who can program; Web data assessed as of March 2025)
- **D15.4. Share of the aged using the internet**
  - Data Source: OECD GDT (Internet user aged 55-74 years; Web data assessed as of March 2025)
- **D15.5. Share of the low-income internet users**
  - Data Source: OECD GDT (Low-income internet user; Web data assessed as of March 2025)

**D16. Data & Algorithm Openness**

According to UNESCO's *Recommendation on the Ethics of Artificial Intelligence*, Article 75 states, "Member States should promote open data," and Article 76 states, "Member States should promote and facilitate the use of quality and robust datasets for training, development and use of AI systems, and exercise vigilance in overseeing their collection and use." This recommendation aligns with Dimension 16, which evaluates whether data, algorithms, and models are open to the public.

Dimension 16 currently covers two indicators:

- **D16.1. Number of impactful open AI models and datasets released**
  - Data Source: Based on statistical analysis of Top 1000 Trending Hugging Face datasets/models, with inspecting authors' GitHub user profiles to identify their countries (Data as of March 2025)
- **D16.2. The level of contributions in the AI developer community**
  - Data Source: Based on Statistical analysis of commit histories in GitHub's top 1,000 AI repositories

22 https://www.digitalgendergaps.org/

by stars (Data as of March 2025)

**D17. AI Governance Research Activity**

According to UNESCO's *Recommendation on the Ethics of Artificial Intelligence*, Article 131 states, "Member States should, according to their specific conditions, governing structures and constitutional provisions, credibly and transparently monitor and evaluate policies, programmes and mechanisms related to ethics of AI, using a combination of quantitative and qualitative approaches……(d) strengthening the research- and evidence-based analysis of and reporting on policies regarding AI ethics; (e) collecting and disseminating progress, innovations, research reports, scientific publications, data and statistics regarding policies for AI ethics……". This recommendation aligns with Dimension 17, which evaluates the quantitative analysis of relevant research on AI governance topics.

Dimension 17 currently covers three indicators:

- **D17.1. Number of publications on AI Governance topics & their proportion in AI-related publications**
  - Data Source: Based on statistical analysis of the DBLP Computer Science Bibliography literature database (Data from April 2024 to March 2025)
- **D17.2. Number of publications on AI Safety&Security topics & their proportion in AI-related publications**
  - Data Source: Based on statistical analysis of the DBLP Computer Science Bibliography literature database (Data from April 2024 to March 2025)
- **D17.3. Number of publications on AI for SDGs topics & their proportion in AI-related publications**
  - Data Source: Based on statistical analysis of the DBLP Computer Science Bibliography literature database (Data from April 2024 to March 2025)

# 2. Data Collection and Index Evaluation Methodology

## 2.1 Data Collection Method for Literature Analysis

To analyze nationality and gender information in the DBLP Computer Science Bibliography, we adopted a multi-step methodology. First, the author's nationality was determined using the institutional address provided in the publication. If no address was available, nationality was inferred through the author's collaboration network. To identify the author's gender, we used the global_gender_predictor package, which is based on the *World Gender Name Dictionary* (Second Edition). When needed, we also collected supplementary metadata such as article titles, abstracts, author names, publication dates, institutional affiliations, publication categories, and links. To identify whether a scientific publication was AI-related, we applied a dual filtering approach based on publisher and keyword data. First, any literature published in AI journals or conferences was classified as AI-related. The names and abbreviations of such venues were sourced from the AMiner AI journal rankings.

Second, we constructed a comprehensive keyword list for various AI subfields (e.g., machine learning, neural networks, reinforcement learning, Bayesian models, Markov processes, etc.). Publications with these keywords in their titles were also identified as AI-related. To classify publications related to AI governance, we developed a separate keyword list that includes terms such as *"for human" "transparency"*, and *"privacy"*. A publication was considered governance-related if these keywords appeared in the title.

## 2.2 Scoring Methodology for AI Strategy & Planning (D6)

When a country has an officially released national-level AI strategy or plan, it scores 100 points for Indicator 6.1 "Whether an AI strategy has been released in the country." If the country has not yet published such a strategy, it scores 0 points. In this dimension, our judgment of the "national-level overall strategy" primarily includes the strategy, approach, roadmap, and plan. Notably, some countries have multiple versions of their AI strategy, but this does not result in a higher score for Indicator 6.1. When scoring Indicators 6.2–6.4, we use the latest or most representative AI strategy of each country as the evaluation target, and, when necessary, consider other AI strategies published by the country.

Indicator 6.2 "Whether the AI strategy has implementation plans" is scored either 0 points or 100 points. This indicator requires the AI strategy to include quantifiable, verifiable target indicators or propose specific, actionable plans with practical value (in terms of content, not structure). When a country's AI strategy includes targets or measures that meet these requirements, it scores 100 points; otherwise, it scores 0 points. If a country does not have an AI strategy, it will receive 0 points for this indicator. We believe that, as a national-level overall plan, an AI strategy should be clear and feasible. If an AI strategy only provides vague and broad opinions, it lacks executability and makes it difficult to assess whether its objectives have been achieved.

Indicator 6.3 "Whether the AI strategy mentions training or skills upgrading" and Indicator 6.4 "Whether the AI strategy has an ethical component" use a similar evaluation approach: if mentioned, the score is 100 points; if not mentioned, the score is 0 points. If a country does not have an AI strategy, both indicators will score 0 points. It is important to note that Indicator 6.3 is not entirely equivalent to talent or education; it requires an emphasis on workforce skill training, public awareness improvement, and the enhancement of enrollment plans for relevant disciplines. Indicator 6.4 requires AI's ethical risks and value pursuits to be specifically addressed.

## 2.3 Scoring Methodology for AI Legislation Status (D11)

The legal systems of different countries vary. For example, countries like the UK and the US follow the common law system, which is primarily based on case law, while countries like Germany and France follow the civil law system, which relies on codified law. In some countries, executive orders or rulings by the highest courts may carry the same legal weight as laws passed by legislative bodies. Furthermore, different countries have variations in the specific processes and terminology used at various stages such as introduction or proposal, review, approval, assent, promulgation, effective date, and implementation date. To account for these differences, the AI legislation dimension acknowledges all types of legally binding legal documents in scoring, including but not limited to laws, acts or bills, statutes, codes, regulations, amendments, decrees, executive orders, and precedents. In this paper, these are generally referred to as laws or regulations.

If a country has implemented national-level comprehensive AI laws and regulations, it will score 100 points for Indicator 11.1 "Whether countries have enacted or are in the process of enacting comprehensive national laws or regulations specifically targeting AI". If a country only has AI laws and regulations in the process of being developed, it will score 50 points. If neither is present, the country will score 0 points.

The laws and regulations that meet the requirements of Indicator 11.1 must satisfy the following conditions: First, the law or regulation should be specifically established to govern or develop AI. Therefore, unless in special circumstances, general laws such as cyber laws or information technology laws, while potentially related to or having regulatory authority over AI, are not included in this category. Second, the law or regulation must be national in scope, rather than regional or local. Therefore, laws enacted by provinces or states are not included. Finally, the law or regulation must address AI in general, rather than focusing on a specific subfield of AI. Legislation concerning such subfields will be evaluated under Indicator 11.2.

It is important to note that "in the process of enacting" requires at least a draft or proposal that is available for review. Some countries (e.g., Indonesia) may have legislative plans but do not yet have a draft or proposal available for public review, in which case the country will score 0 points. If a country's AI comprehensive law has been approved but not yet officially implemented (e.g., South Korea), the country will score 50 points. If a country (e.g., the United States) has both implemented AI comprehensive laws and has AI laws in development, the country will score 100 points. If a country complies with and implements corresponding regional laws, these are considered national-level AI comprehensive laws and the country will score 100 points. However, if the regional law is still under development, the corresponding country will not score 50 points but rather 0 points.

Indicator 11.2 "Whether countries have established national-level vertical laws or regulations specifically addressing AI" mainly examines the legislation on AI vertical fields in various countries. Different areas of AI technology and applications are considered vertical fields, with typical examples being autonomous driving, generative AI, and others. In Indicator 11.2, in addition to newly enacted laws, we also accept the addition of specific provisions, targeted amendments, or the development of targeted implementation rules to existing laws. These laws and regulations tend to focus more on details rather than broad and comprehensive laws, and their formulation and implementation are relatively vague. Therefore, Indicator 11.2 adopts a more general distinction, no longer considering the "in the process of enacting" stage but instead using "published" as the scoring standard. A country will score 100 points if it has established national-level laws and regulations for AI vertical fields. If no such laws exist, the country will score 0 points.

Indicator 11.3 "Whether countries have implemented national-level data/information protection laws pertaining to AI" follows similar scoring rules to Indicator 11.2. We recognize both specially enacted AI data or information laws and the addition of specific provisions or amendments to existing data protection or personal information laws. Countries with recognized AI data or information protection laws will score 100 points, and those that comply with and implement corresponding regional laws are also included. Countries without such laws will score 0 points. General data protection laws without specific provisions targeting AI are not within the scope of this indicator, but if there is clear evidence that such laws have jurisdiction over AI-related infringement cases, they may be considered AI-specific data protection laws.

## 2.4 Score Calculation at Each Level

In processing raw scores, we incorporated data entries and statistics from multiple sources for triangulation, enhancing reliability. This is especially useful when small fluctuations in scarce data can significantly impact scores; multiple data sources can reduce bias. Strong correlations between different data elements allow for mutual supplementation in cases of missing data. Where appropriate, ratio scores were considered to ensure fair comparisons between countries with different baseline statistics (such as population and GDP). Finally, percentile-fit normalization (see below) was used to standardize and average various data. In identifying genders, we combined average level inference, allocating 22.9% of unidentified genders as female, and the identified proportion was then percentile-fit normalized and averaged.

Where appropriate, ratio scores were aggregated to ensure fair comparisons between countries at different baseline factors (such as population and GDP). To compute the indicator score, we will use the average normalization score of the total and the ratio. For example, if a country obtains a normalized score of 5 in total number and 3 in per capita number, then the country's score in this indicator will be 4.

After obtaining indicator scores, we averaged the scores within each dimension and then standardized them to obtain dimension scores. Simple standardization was used to readjust the mean to 50; due to the dispersion of scores based on survey indicators and tools, averages were used without further standardization. We then averaged the dimension scores to obtain pillar scores and averaged the pillar scores to obtain the index score. Here, D4. AI Risk Exposure is a negative factor in the P2 aspect, so [100 - dimension score] was used for averaging.

## 2.5 Score Normalization and Data Imputation

For simple normalization, we use:

$$25 \times \frac{x_n - \mu}{\sigma} + 50$$

Where $\mu$ is the statistical mean of all countries, $\sigma$ is the statistical variance, and $x_n$ is the raw data of the country. After standardization, scores exceeding 0 and 100 were truncated to ensure they remained within the 0-100 range. For percentile-fit normalization, after each simple standardization, we extracted and removed one percentile of scores, then repeated the standardization and extraction on the remaining data until four score quartiles were obtained. This was necessary due to significant clustering in the original data and large magnitude differences, requiring adjustment of the standard deviation for better comparison of data at lower scales.

In the case of missing data within an indicator, the imputation shall be carried out in the following order. First, for indicators that were available in the previous year but are missing in the current year, the current value is estimated by calculating the average growth rate of that indicator across all countries and then multiplying the previous year's data by this growth rate. Second, if there is no reference data from previous years, hierarchical imputation is performed. For missing data under the same indicator, we first calculate the indicator score based on the available data. The resulting score is then used to fill in the missing item's score, after which the indicator score is recalculated. Similarly, if the indicator score is missing, we use the same approach to calculate the dimension score from available indicator scores, use the dimension score to impute the missing indicator score, and then recalculate. If the dimension score is also missing, we apply the same method using the pillar score to impute the dimension score.

# 3. Related Indexes

In recent years, alongside the AGILE Index, several international indices and reports on AI governance have been continuously updated, serving as key references for national policy-making, capacity assessment, and engagement in global dialogue. To offer a holistic view of tools supporting AI governance evaluation, this appendix provides a brief overview of selected updates from other indices and reports for readers' reference.

The table below provides an overview of Government AI Readiness Index (GRI) by Oxford Insights; Artificial Intelligence and Democratic Values Index (AIDVI), by Center for AI and Digital Policy (CAIDP); Global AI Vibrancy Tool (GAIVT) and AI Index Report (HAI AI Index) by Stanford University Human-Centered Artificial Intelligence (Stanford HAI); and Global Artificial Intelligence Index (GAII) by Tortoise Media.

***Table 7 Development Status of AI Governance Indices***

| Index | General Description | Indicator | Data Source Updates | Methodological Adjustments |
|---|---|---|---|---|
| GRI | Evaluates how ready governments are to use AI effectively. | New indicators added; several outdated ones replaced | including Responsible AI | Refined handling of skewed data; outliers preserved |
| AIDVI | Evaluates national AI policies against democratic and ethical standards | Included metric on Council of Europe AI Treaty | Country scores revised based on development | Shifts emphasis to legal-institutional governance |
| GAIVT | Measures national AI ecosystem vibrancy using 8-pillar structure | Continues to apply the previous indicator set | Update expands on previous versions | Stable framework with refined data and operations |
| HAI AI Index | A comprehensive annual review of global AI development trends | Adds new analyses on AI hardware, inference costs, and publication trends | Data on corporate adoption of responsible AI practices | Diversity Section Removed as Independent Chapter |
| GAII | Tracks AI development across innovation, implementation, and investment | Added indicators on Semiconductor Manufacturing, AI Ethics, and AI Startups Acquisition | Enhancing Data Timeliness, Comprehensiveness, and Source Diversity | Adjusted weighting; refined sub-pillar definitions |

*Note: Report abbreviations are used throughout the table.*

These updates involve changes in indicator design, data sources, methodological frameworks, and report structure, reflecting ongoing efforts by various institutions to refine their evaluation focus and improve the utility of their tools.

The overall trend in recent AI governance index developments is that major structural frameworks have remained largely consistent, while more modest adjustments have been made in areas such as governance priorities, data foundations, and evaluation dimensions.

- Gradual expansion of indicator sets: While maintaining the original structural frameworks, several reports have introduced new dimensions of assessment, covering areas such as government execution capacity, legal mechanisms, environmental impact, and practical deployment of AI technologies.
- Increasing dynamism in data sources: Many updates involve the replacement of outdated datasets with more recent inputs from international databases, industry surveys, and platform-generated data, reflecting efforts to enhance the timeliness and contextual relevance of evaluations.
- Stability in methodological design: Despite adjustments in specific indicators, the core methodologies—such as pillar structures, scoring models, and normalization procedures—remain largely consistent, ensuring year-to-year continuity and comparability.
- Emerging role as international governance references: These indices are gradually evolving from static ranking tools into policy-oriented measurement instruments that inform national governance improvements and contribute to the construction of international cooperation frameworks.

# 4. Links to Illustrations

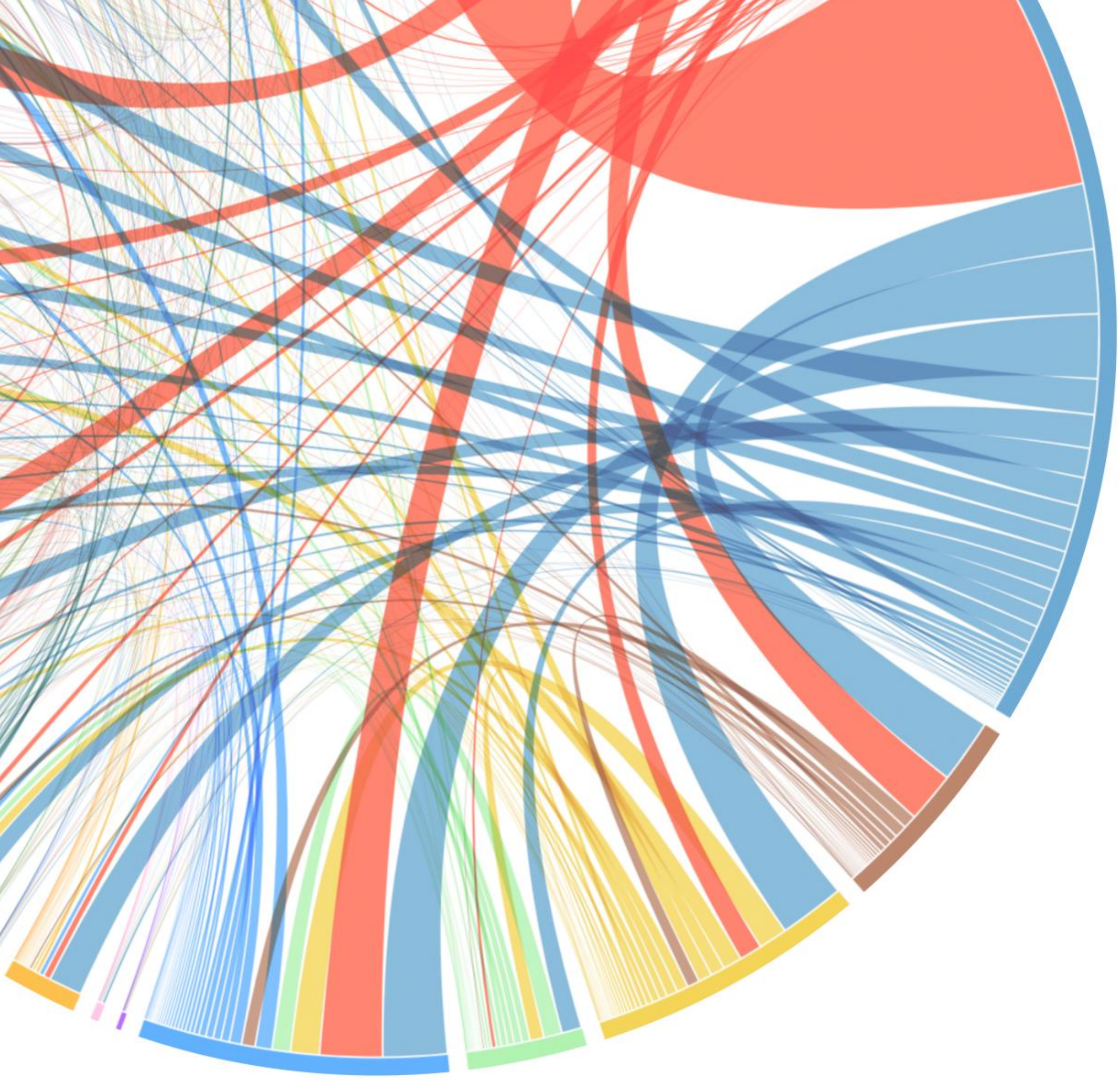

**AGILE Index 2025**

Website: https://agile-index.ai/

Email: contact@long-term-ai.cn